\documentclass[a4paper,11pt,dvipsnames]{article}
\usepackage{jheppub}

\usepackage{outlines} 
\newcommand{\bo}{\begin{outline}} 
\newcommand{\eo}{\end{outline}}

\usepackage{amscd} 
\usepackage{bm} 
\usepackage{bbm} 
\usepackage{mathrsfs} 
\usepackage[normalem]{ulem}
\usepackage{MnSymbol} 
\usepackage{placeins}
\usepackage{amssymb, pifont}

\def\tbf#1{{\textbf{#1}}}

\usepackage{tikz, pgf, float} 
\usetikzlibrary{tikzmark, arrows, calc, shapes, shapes.arrows, decorations.markings, decorations.pathmorphing, cd, positioning}
\tikzset{Dynkin/.style={circle, draw=black, inner sep=0pt, fill=white, minimum size=2mm}}
\tikzset{explain/.style={red, thick, <-, shorten <=3pt}}

\usepackage{multirow} 
\usepackage{colortbl} 
\usepackage{arydshln} 
\usepackage{tabularx} 

\usepackage{xcolor}

\definecolor{teal}{HTML}{0097AD}
\def\teal#1{{\color[HTML]{0097AD}{#1}}}

\def\red#1{{\color{red}{#1}}}

\definecolor{lorg}{HTML}{FFB887}
\definecolor{gfruit}{HTML}{ff5b59}

\def\nn{\nonumber}
\def\ph{\phantom}
\def\vph{\vphantom}
\newcommand{\bpm}{\begin{pmatrix}}
\newcommand{\epm}{\end{pmatrix}}
\newcommand{\bsm}{\begin{smallmatrix}}
\newcommand{\esm}{\end{smallmatrix}}
\newcommand{\bspm}{\left(\begin{smallmatrix}}
\newcommand{\espm}{\end{smallmatrix}\right)}
\newcommand{\deq}{\doteq}
\newcommand{\xmark}{\ding{55}}
\allowdisplaybreaks[2] 
\def\<{{\langle}}
\def\>{{\rangle}}

\def\bar{\overline}
\def\til{\widetilde}
\def\hat{\widehat}

\def\v{\vee}
\def\^{\wedge}
\def\del{{\partial}}

\def\Im{\mathrm{Im}}

\def\im{\mathrm{im}}
\def\GL{\mathrm{GL}} 
\def\SL{\mathrm{SL}}
\def\PSL{\mathrm{PSL}}
\def\O{\mathrm{O}}
\def\SO{\mathrm{SO}}
\def\Ss{\mathrm{Ss}}
\def\Sc{\mathrm{Sc}}
\def\Spin{\mathrm{Spin}}
\def\U{\mathrm{U}}
\def\SU{\mathrm{SU}}
\def\Sp{\mathrm{Sp}}

\def\so{\mathfrak{so}}
\def\u{\mathfrak{u}}
\def\su{\mathfrak{su}}
\def\sp{\mathfrak{sp}}

\def\bone{\mathbf{1}}
\def\diff{\mathrm{d}}
\def\Aut{\mathrm{Aut}}
\def\Lie{\mathrm{Lie}}
\def\Out{\mathrm{Out}}
\def\Inn{\mathrm{Inn}}
\def\rfl{\mathrm{refl.}}
\def\Ext{\mathrm{Ext}}
\def\Fix{\mathrm{Fix}}
\def\gcd{\mathrm{gcd}}
\def\Mat{\mathrm{Mat}}

\def\cA{{\mathcal A}} 
\def\ua{{\underline a}}
\def\cB{{\mathcal B}} 
\def\ub{{\underline b}}
\def\C{{\mathbb C}} 
\def\cC{{\mathcal C}} 
\def\scC{{\mathscr C}} 
\def\cCt{{\til\cC}} 
\def\uc{{\underline c}}
\def\cD{{\mathcal D}} 
\def\cDt{{\til\cD}} 
\def\fD{{\mathfrak D}}
\def\ud{{\underline d}}
\def\he{{\hat e}}

\def\fg{{\mathfrak g}}

\def\scH{{\mathscr H}}
\def\fh{{\mathfrak h}}
\def\cI{{\mathcal I}} 
\def\fI{{\mathfrak I}}
\def\scI{{\mathscr I}}
\def\bj{{\bf j}}

\def\hm{{\hat m}}
\def\cM{{\mathcal M}}
\def\bom{{\bf m}}

\def\cN{{\mathcal N}}
\def\bon{{\bf n}}

\def\bop{{\bf p}}
\def\Q{{\mathbb Q}} 
\def\boq{{\bf q}}
\def\cR{{\mathcal R}} 
\def\R{{\mathbb R}} 
\def\scS{{\mathscr S}}
\def\sS{\textsf{S}}

\def\sT{\textsf{T}}

\def\cU{{\mathcal U}}
\def\cY{{\mathcal Y}}
\def\Z{{\mathbb Z}}

\def\a{{\alpha}}

\def\ual{{\underline\a}}
\def\b{{\beta}}
\def\ube{{\underline\b}}
\def\g{{\gamma}}
\def\G{{\Gamma}}
\def\uga{{\underline\g}}
\def\d{{\delta}}
\def\D{{\Delta}}
\def\ude{{\underline\d}}
\def\e{{\epsilon}}

\def\th{{\theta}}
\def\iw{{\iota_\vp}}
\def\k{{\kappa}}
\def\l{{\lambda}}
\def\L{{\Lambda}}
\def\m{{\mu}}
\def\n{{\nu}}

\def\vp{{\varpi}}
\def\r{{\rho}}
\def\s{{\sigma}}
\def\S{{\Sigma}}
\def\t{{\tau}}
\def\tym{{\tau_{\textrm{YM}}}}
\def\bt{{\bm{\tau}}}

\def\f{{\phi}}

\def\ps{{\psi}}
\def\pst{{\til \psi}}

\def\Om{{\Omega}}

\preprint{IFT-UAM/CSIC-25-103}

\title{Special K\"ahler geometries of $\cN=4$ superYang-Mills}

\author[1, 2]{Philip C. Argyres,}
\author[3]{Antoine Bourget,}
\author[4]{Julius F. Grimminger,}
\author[5, 6]{Matteo Lotito,}
\author[5]{and Mitch Weaver}
\affiliation[1]{University of Cincinnati, Physics Department, PO Box 210011, Cincinnati OH 45221, USA}
\affiliation[2]{Instituto Balseiro, Centro At\'omico Bariloche 8400-S.C. de Bariloche, R\'io Negro, Argentina}
\affiliation[3]{Institut de physique théorique, Université Paris-Saclay,  CEA, CNRS, 91191, Gif-sur-Yvette, France}
\affiliation[4]{Mathematical Institute, University of Oxford, Andrew Wiles Building, Woodstock Road, Oxford, OX2 6GG, UK}
\affiliation[5]{Department of Physics, Korea Advanced Institute of Science and Technology 291 Daehak-ro, Yuseong-gu, Daejeon 34141, Republic of Korea}
\affiliation[6]{Instituto de F\'isica Te\'orica IFT-UAM/CSIC, C/ Nicol\'as Cabrera 13-15, Universidad Aut\'onoma de Madrid, Cantoblanco, Madrid 28049, Spain}

\emailAdd{philip.argyres@gmail.com}
\emailAdd{antoine.bourget@ipht.fr}
\emailAdd{grimminger@maths.ox.ax.uk}
\emailAdd{matteo.lotito@ift.csic.es}
\emailAdd{mtw7497@gmail.com}

\abstract{
The low energy effective theory on the moduli space of vacua of 4d superYang-Mills (sYM) theory defines a special K\"ahler geometry.
For simple sYM gauge algebras, $\fg$, we classify all compatible special K\"ahler structures by showing that they are in one-to-one correspondence with certain equivalence classes of integral symplectic representations of the Weyl group of $\fg$.  
We further demonstrate that, for principal Dirac pairing, these equivalence classes are in one-to-one correspondence with the S-duality orbits of the global structures of the corresponding $\fg$ sYM gauge theory, after a mistake in the field theory literature is corrected.
This provides a low-energy test of S-duality.
We also discuss twisted product geometries made from factors with special K\"ahler structures with non-principal Dirac pairings.
}

\begin{document} 

\maketitle

\section{Introduction and summary}
 
Information-rich and often exactly computable observables of a supersymmetric quantum field theory are the leading terms in the low-energy effective action of the massless degrees of freedom on $\cM$, its moduli space of vacua.
In the case of $4d$ theories with $\cN{=}2$ supersymmetry, this moduli space is a continuum of states, and the low-energy effective action is encoded in a singular complex geometry -- the moduli space geometry -- that incorporates the moduli space of vacua.
With more supersymmetry, the moduli space geometries become more constrained, and therefore more amenable to classification.

In this paper, we classify all possible moduli space geometries (with principal polarization) of theories with $\cN{=}4$ superconformal symmetry and a weak coupling limit described by a superYang-Mills (sYM) theory with simple gauge algebra, $\fg$.
The answer to this problem was seemingly given long ago \cite{Seiberg:1997ax}: it is specified completely by gauge algebra of the sYM theory.
This is due to the the orbifold description of the the moduli space of vacua which is solely determined by the Weyl group of $\fg$, 
\begin{align}\label{N4 orbi}
    \cM_{\fg} \deq \C^{3r} / (W(\fg) \otimes_\R \C^3), 
\end{align}
where $r \deq {\rm rk} (\fg)$ is the rank of $\fg$ and $W(\fg)$ is its Weyl group which forms a subgroup of the gauge group.
$W(\fg)$ is a real reflection group, so it is finite group acting on $\R^r$ via its defining real reflection representation.
Here, it is ``triply complexified'' by tensoring with $\C^3$ to get its reflection action on $\C^{3r}$.
$\cM_{\fg}$ is a K\"ahler manifold with K\"ahler metric given by the one inherited from a flat metric on $\C^{3r}$.
But the moduli space geometry has more structure than just the K\"ahler orbifold geometry describing $\cM_{\fg}$.
In particular, it has a \emph{special K\"ahler (SK) structure}.%
\footnote{In the $\cN{=}4$ class, it is a generalization of an $\cN{=}2$ SK structure, called a ``triple-SK structure'' in \cite{Argyres:2019yyb}, where it is described in detail.}
In appendix \ref{appA: N4 SKS}, we review the origin of the orbifold description of $\cM_{\fg}$ relative to a $\fg$ sYM theory, describe an $\cN{=}4$ special K\"ahler structure, and relate it to the low-energy effective action defined on $\cM_{\fg}$.

Our main result is the classification of SK structures that are compatible with the K\"ahler geometry of the $\cN{=}4$ sYM moduli space of vacua \eqref{N4 orbi}.%
\footnote{
By assuming the existence of a weak-coupling limit described by an $\cN{=}4$ sYM theory, the underlying K\"ahler geometry that describes the moduli space of vacua is determined to take the form \eqref{N4 orbi}. 
In particular, this assumption tells us that the Weyl group must act on $\C^{3r}$ through its reflection representation that defines it as a real reflection group.
Beyond fixing the K\"ahler geometry, we do not take into account any additional, dynamical data of an $\cN{=}4$ sYM theory that could be deduced from this assumption and attached to the SK geometry, such as its BPS charge lattice of physical, finite energy states.
This additional data can be combined with an $\cN{=}4$ SK structure and subsequently classified. 
We expect this would lead to some slight refinements of our classification results, but we leave these interesting considerations for future work.
}
We show (see appendix \ref{appA: SKS and orbit triple}) that an $\cN{=}4$ SK structure on $\cM_{\fg}$ with polarization $J$ (representing the Dirac pairing on the charge lattice of the low-energy theory) is defined by the equivalence class of a pair $(S, \bt)$ where $S: W_{\C} \to \GL(2r,\Z)$ is an integral representation of the complexified Weyl group that is symplectic with respect to $J$, and $\bt \in \Fix(S) \subset \scH_r$ is a symmetric $r\times r$ matrix of low-energy $\u(1)^r$ gauge couplings that is compatible with $S$.
We show how to construct all such triples $(J,S,\bt)$ and compute the resulting equivalence classes, which thereby classifies all possible $\cN{=}4$ sYM SK structures on $\cM_{\fg}$.
While the irreducible representations of Weyl groups into $\GL(r,\Z)$ are well-studied, our classification of SK structures requires the classification of Weyl group representations into $\GL(2r,\Z)$ that are reducible but not necessarily decomposable.%
\footnote{The irreducible $\Z$-reflection reps of the Weyl group that are $\Q$-equivalent to its real reflection representation have been classified \cite{feit1998integral}. 
Our required use of this result is a consequence of the sYM assumption; see appendix \ref{appA: N4 SKS}.}
The solution is summarized in table \ref{tab:lattices} and involves a module extension problem that is explained in \ref{sec:bindingsAndExtGroups}.

For fixed polarization $J$, $\bt$ is compatible with $S$ if it resides in the fixed point set of $S$ acting on the Siegel upper half-space of degree $r$, $\scH_r$.
This set, denoted $\Fix(S)$, forms a $1$-parameter family of compatible $\bt(\t)$ matrices in $\scH_r$ where the parameter $\t$ is valued in $\scH_1$.
Thus, the inequivalent triples $(J, S, \bt(\t))$ organize into the equivalence classes of fixed-point set triples $(J, S, \Fix(S))$ which are labeled by the equivalence class of the representation $S$.  
We call such classes \emph{SK orbits} and use them to present the classification of $\cN{=}4$ SK structures in table \ref{tab:summary}.
When $J$ is \emph{principal},%
\footnote{
Principal $J$ means the charge lattice is comprised of a maximal set of massive states that are mutually local.
Correspondingly, there always exists an associated symplectic basis of $J$ in which its matrix form is the ordinary symplectic form defining the symplectic matrix group $\Sp(2r,\Z)$.
}
and for a given Weyl group $W$, each row in the second column of table \ref{tab:summary} represents an inequivalent SK orbit defined by the equivalence class of some representative $\GL(2r,\Z)$, symplectic representation $S$ of $W$. 
The 3rd column of the table is a certain modular group, called the \emph{self-duality group of the SK orbit}, which are reviewed in appendix \ref{app Hecke}.
It is defined by (the equivalence class of) $S$, acts on $\t$ through fractional linear transformations, and encodes the geometry of the 1-dimensional parameter space whose points are the inequivalent $\cN{=}4$ SK structures $(J,S,\bt(\t))$ within the fixed-point set triple $(J, S,\Fix(S))$.
This geometry can be realized as the modular curve formed by quotienting $\scH_1$ by this self-duality group, and so is parameterized by $\t\in \scH_1$ relative to the identifications determined by the self-duality group.
For this reason, we refer to this geometry as the \emph{conformal manifold of the SK orbit}.

How does the classification of principal $\cN{=}4$ SK structures compare with the predicted classification of the corresponding $\cN{=}4$ sYM SCFT?
As $4d$ gauge theories, the local space-time dynamics are specified by the gauge algebra $\fg$ and the complex gauge coupling $\tym \in \scH_1$,%
\footnote{Not to be confused with the boldface matrix $\bt(\t)$ of low-energy couplings on the moduli space or its complex parameter $\t$!} 
but they also carry discrete ``global structure'' data that is sensitive to the global topology of space-time \cite{Aharony:2013hda}.
This data can be expressed in terms of a maximal spectrum of Wilson-'t Hooft line operators that are mutually local, i.e. they are genuine line operators \cite{Kapustin:2005py, Gaiotto:2010be}.
Such a spectrum is specified by a rank-$2r$ Weyl-invariant lattice with a principal Dirac pairing, and points in this lattice correspond to the $\fg \times \fg^{\mathrm L}$ representations of the Wilson-'t Hooft operators.%
\footnote{Here, $\fg^{\mathrm L}$ is the Langlands dual Lie algebra to $\fg$.
For simply laced $\fg = A_r, D_r, E_{6,7,8}$, $\fg^{\mathrm L} \cong \fg$, while for non-simply laced $\fg$ we have $B_r^{\mathrm L} \cong C_r$, $C_r^{\mathrm L} \cong B_r$, and the duals of $G_2$ and $F_4$ each have a rotated root system that leaves the algebra invariant \cite{Argyres:2006qr}.
}
Once this global structure data is specified, the gauge theory defines an absolute QFT, i.e. one that can be consistently placed on space-time manifolds with non-trivial topology.
However, the set of consistent line operator spectra is not unique since there are distinct global structures that each define an absolute variant of the sYM gauge theory --- these different choices are often called the \emph{global variants} of the gauge theory.
Furthermore, the conjectured S-duality of $\cN{=}4$ sYM acts on these global variants to combine them into various orbits -- \emph{S-duality orbits} -- with an associated \emph{self-duality group}.
An absolute $\cN{=}4$ theory is thus labeled by one of these orbits along with the marginal coupling $\tym \in \scH_1$.
But because S-duality also acts on $\tym$, the inequivalent absolute theories are labeled by a global variant orbit with the equivalence class of $\tym \in \scH_1$ relative to the action of the self-duality group of that orbit.
In other words, absolute theories are labeled by a global variant S-duality orbit and a point in a 1-dimensional geometry that is formed from the quotient of $\scH_1$ by the self-duality group of the orbit; this geometry is the \emph{conformal manifold of the S-duality orbit}.
In summary, the inequivalent absolute $\cN{=}4$ sYM SCFTs are in 1-to-1 correspondence with the families of global variant S-duality orbits that are parameterized by the \emph{conformal manifold of that orbit}.
These S-duality orbits and their self-duality groups have been worked out in detail for most simple $\fg$ in \cite{Aharony:2013hda, Bergman:2022otk}, and we summarize these results in table \ref{tab:RBTL}.

For fixed gauge algebra $\fg$, the relation between the data describing the global variants of the $\cN{=}4$ sYM theory; the global variants of the low-energy $\u(1)^r$ gauge theory on its Coulomb branch \cite{Argyres:2022kon,DelZotto:2022ras}; and the \emph{principal} $\cN{=}4$ SK geometries of the CB suggest there is a general relation between the global variants of an $\cN{=}2$ SCFT, the global variants of the IR theory on its CB, and the principal SK geometries of that CB.%
\footnote{The distinct sets of Wilson-'t Hooft line operators defining the global variants of the sYM theory can be put in 1-to-1 correspondence with the distinct sets of Wilson-'t Hooft line operators defining the global variants of the low-energy $\u(1)^r$ gauge theory on its Coulomb branch \cite{Argyres:2022kon}.
The latter are specified by \emph{line lattices}, which are refinements of the (non-principal) sYM BPS charge lattice to ones that are principal, i.e. maximal and mutually local. 
When the BPS charge lattice is non-principal, as is true for most simple $\fg$, the additional charges in the refined lattice represent genuine probe line operators.
A line lattice is identified with the $1$-homology lattice of the pav fiber, so it is part of the CB SK structure; see \cite{Argyres:2022kon} for more details.
}
Thus, we are led to compare the number of inequivalent principal SK orbits versus S-duality orbits of absolute $\fg$ $\cN{=}4$ sYM theories, as well as their self-duality versus S-duality groups, respectively.%
\footnote{This is the appropriate data to compare in each case because it is gauge-invariant and independent of the conformal manifold parameters $\t$ and $\tym$ of an SK orbit and S-duality orbit, respectively.}  

For a given Weyl group $W$, table \ref{tab:summary} shows the number of distinct SK orbits (describing the inequivalent moduli space geometries) and their self-duality groups.
These agree with the number of S-duality orbits and their self-duality groups except in those cases marked with \red{\xmark}'s in the table, c.f. table \ref{tab:RBTL}.
These disagreements come in two types that are correlated with the column the \red{\xmark} is in. 
The first type (second column \red{\xmark}) is a disagreement in the number of SK orbits, and the second type (third column \red{\xmark}) is a disagreement between the self-duality groups.
For all the disagreements that occur in the $\fg=D_{2k}$ cases, they are due to a mistake in the field theory literature of not properly taking into account the (outer) automorphism interchanging spinor and conjugate-spinor representations.
Doing so leads to additional equivalences between absolute $\cN{=}4$ theories that either reduces the number of distinct S-duality orbits or increases the size of a self-duality orbit. 
For the $A_1$ and $BC_r$ cases, although the number of orbits agree, one SK orbit always has a larger self-duality group.

\begin{table}[!ht]
\normalsize
\centering
\begin{tabular}{|c|c|c|}
\hline 
$W$ & 
distinct $\bj =\bone$ SK orbits & self-duality group of SK orbit \\ 
\hline\hline 
$A_1$ & $\{(\Z_1,0)\}$ & \ph{\xmark} $\PSL(2,\Z)$ \red{\xmark} \\ 
\hline 
$A_r$ & one SK structure orbit $\{(\Z_s, 0),\cdots\}$ & \multirow{2}{*}{$\G_0((r{+}1)/s^2)$} \\
$(r{\ge}2)$ & for each $s^2|(r{+}1)$ (see section \ref{suN sec}) & \\
\hline
\multirow{2}{*}{$BC_2$} 
& $\teal{\lcirclearrowright} \{(\Z_2,1), (\Z_2^\v,1)\}$ & \teal{$H_4$} 
\\
& $\{(\Z_2,0)\} \teal{\leftrightarrow} \{(\Z_2^\v,0)\}$  & \ph{\xmark} $\PSL(2,\Z)$ \red{\xmark}
\\  
\hline 
\multirow{3}{*}{\begin{tabular}{c} $BC_{2k}$ \\ $(k{\ge}2)$ \end{tabular}} & $\{ (\Z_1, 0), (\Z_2, 0) \}$ & $\G_0(2)$ 
\\ 
& $\{(\Z_2,1),(\Z_2^\v,1)\}$ & $\G_0(2)$ 
\\ 
& $\{(\Z_2^\v,0)\}$ & \ph{\xmark} $\PSL(2,\Z)$ \red{\xmark} 
\\ \hline 
\multirow{2}{*}{$BC_{2k+1}$} & $\{ (\Z_1, 0), (\Z_2, 0), (\Z_2^\v, 1) \}$ & $\G_0(4)$ 
\\ 
& $\{ (\Z_2^\v, 0) \}$ & \ph{\xmark} $\PSL(2,\Z)$ \red{\xmark} 
\\ \hline 
\multirow{3}{*}{$D_4$} & $\{ (\Z_1, 0), (\Z_2^2, 00) \}$ & $\G_0(2)$
\\  
& \red{\xmark} $\ \scriptstyle{\{(\Z_2^V,1),(\Z_2^2,01)\} \teal{\leftrightarrow} \{(\Z_2^S,1) ,(\Z_2^2,10)\} \teal{\leftrightarrow} \{(\Z_2^C,1),(\Z_2^2,11)\}}$ & $\G_0(2)$ 
\\ 
& \red{\xmark}  $\quad \{(\Z_2^V,0)\} \teal{\leftrightarrow} \{(\Z_2^S,0)\} \teal{\leftrightarrow} \{(\Z_2^C, 0)\}$ & $\PSL(2,\Z)$
\\ \hline 
\multirow{5}{*}{\begin{tabular}{c} $D_{4k}$ \\ $(k{\ge}2)$ \end{tabular}} & $\{ (\Z_1, 0) , (\Z_2^2, 00) \}$ & $\G_0(2)$ 
\\  
& $\{ (\Z_2^V, 1), (\Z_2^2, 01) \}$ & $\G_0(2)$ 
\\  
& \red{\xmark} $\quad \scriptstyle{\{ (\Z_2^S,1),(\Z_2^2,10)\} \teal{\leftrightarrow} \{ (\Z_2^C,1),(\Z_2^2,11)\}}$ & $\G_0(2)$
\\  
& $\{ (\Z_2^V,0) \}$ & $\PSL(2,\Z)$ 
\\  
& \red{\xmark} $\quad \{(\Z_2^S,0)\} \teal{\leftrightarrow} \{(\Z_2^C,0)\}$  & $\PSL(2,\Z)$ 
\\ \hline 
\multirow{5}{*}{$D_{4k+2}$} & $\{ (\Z_1, 0), (\Z_2^2, 00) \}$ & $\G_0(2)$ 
\\  
& $\{ (\Z_2^V, 1), (\Z_2^2, 01) \}$ & $\G_0(2)$ 
\\  
& $\teal{\lcirclearrowright} \{ (\Z_2^S, 0), (\Z_2^C, 0), (\Z_2^2, 10) \}$ & \ph{\xmark} \teal{$\G_0(2)$} \red{\xmark} 
\\  
& $\{ (\Z_2^2, 11) \}$ & $\D$ 
\\  
& $\{ (\Z_2^V, 0) \}$ & $\PSL(2,\Z)$ 
\\ \hline 
\multirow{2}{*}{$D_{2k+1}$} & $\{ (\Z_1, 0), (\Z_2, 1), (\Z_4, 0)\}$ & $\G_0(4)$ 
\\  
& $\{ (\Z_2, 0) \}$  & $\PSL(2,\Z)$ 
\\ \hline 
$E_6$ & $\{ (\Z_1, 0), (\Z_3, 0) \}$ & $\G_0(3)$ 
\\ \hline 
$E_7$ & $\{ (\Z_1, 0), (\Z_2, 0) \}$ & $\G_0(2)$ 
\\ \hline 
$E_8$ & $\{ (\Z_1, 0) \}$ & $\PSL(2,\Z)$ 
\\ \hline 
$F_4$ & $\teal{\lcirclearrowright} \{ (\Z_1, 0), (\Z_1^\v, 0) \}$ & \teal{$H_4$}  
\\ \hline 
$G_2$ & $\teal{\lcirclearrowright} \{ (\Z_1, 0), (\Z_1^\v, 0) \}$ & \teal{$H_6$}  
\\ \hline 
\end{tabular}
\caption{Summary of classification results for $\cN{=}4$ sYM SK structures.
In the second column, each row containing (a set of) curly brackets represent an inequivalent SK orbit for the Weyl group, $W$, in the first column.
As explained in \ref{sec Z equiv}, an SK orbit is determined by the $\Sp(2r,\Z)$ orbit of an integral, symplectic representation of $W$.
We label such representations by a pair, $(\Z_s^X,n)$, where first entry refers to subgroups of the center of a Lie algebra with Weyl group $W$, and the second entry refers to an element of their corresponding Ext group, which is  summarized in table \ref{tab:lattices}. 
The set of representations enclosed in curly brackets are all $\Z$-equivalent.
The \teal{teal} arrows between curly brackets denote further $\Z$-equivalences that incorporate Weyl group (outer) automorphisms; $\teal{\lcirclearrowright}\{\cdots \}$ indicates self-equivalences within an SK orbit while $\{ \cdots \} \teal{\leftrightarrow} \{ \cdots \}$ indicates equivalences between SK orbits that are not $\Z$-equivalent while using the identity automorphism of $W$.
The 3rd column gives the self-duality group of each SK orbit as a modular subgroup (table \ref{tab:mod grp}).
Since circular arrows $\teal{\lcirclearrowright}$ are self-equivalences of an SK orbit, they contribute to the self-duality group of that orbit, so we color the group in \teal{teal} to reflect this fact.
SK orbits or self-duality groups with \red{\xmark}'s indicate those cases where either the number of orbits or the self-duality group, respectively, does not agree with the field theory S-duality conjectures as stated in \cite{Kapustin:2005py, Aharony:2013hda, Bergman:2022otk}, c.f. table \ref{tab:RBTL}.}
\label{tab:summary}
\end{table}
\FloatBarrier

We interpret this latter disagreement as resulting from the moduli space geometry simply failing to distinguish between inequivalent absolute $\cN{=}4$ field theories because they lack sensitivity to do so as an IR observable.
This is not a contradiction: there is no \emph{a priori} reason that a low-energy observable such as the moduli space geometry should be different for all microscopically inequivalent field theories, i.e. the mapping from distinct SCFTs to a distinct (scale-invariant) moduli space geometry is allowed to be many-to-one.%
\footnote{For example, it is known that there are distinct $4d$ $\cN{=}2$ SCFTs with the same scale-invariant CB geometry. 
This already occurs at rank-$1$: the $\cN{=}4$ $\su(2)$ superYang-Mills theory (which has a hypermultiplet in the adjoint) and the $\su(2)$ gauge theory with four fundamental hypermultiplets.
Both theories have the same scale-invariant CB geometry (with an $I_0^*$ singularity).
They're distinguished by their mass deformations which is reflected in their flavor symmetries: the former has $\SU(2)$ while the latter has $\SO(8)$ flavor symmetry \cite{Argyres:2015ffa}.}
Presumably, if we include additional data from the field theory beyond the K\"ahler geometry of the moduli space specified by \eqref{N4 orbi}, such as the BPS charge lattice of massive states out on the moduli space or the mass deformations of the SK geometry, then this discrepancy between self-duality groups would be resolved because such data would provide a non-trivial compatibility constraint on the self-duality transformations of the SK orbit.%
\footnote{Note that it is not consistent for the opposite disagreement between self-duality groups to occur: the self-duality group of an SK orbit cannot be smaller than the self-duality group of the SCFT it is the moduli space geometry of. 
This would imply there are self-identifications within the marginal coupling space of the SCFT that are not respected by their corresponding moduli space geometries.
This is a contradiction because any identifications among SCFTs implies their low-energy observables, such as their moduli space geometries, must also be the same.
We find no such discrepancies in our classification.}
Conversely, when classifying SK geometries, there is no \emph{a priori} reason that all of them should be realized as the moduli space of a physical QFT.
This is because we might lack knowledge of a sufficient set of physical consistency conditions to impose on the geometries to render them physical.
Therefore, in our present classification, we can conclude that we are not lacking such knowledge if we find that each distinct SK geometry occurs as the moduli space of an absolute $\cN{=}4$ sYM theory.
Despite the enlarged SK orbit self-duality groups for the $A_1$ and $BC_r$ Weyl groups, evidence for this conclusion comes from the fact that we always find the same number of SK orbits as there are S-duality orbits.

With the understanding that the S-dualities of $\cN{=}4$ sYM theories are (formally) conjectures, our results can also be seen as a test of these equivalences because we compute the duality structure of one of their low-energy observable without direct reference to the duality structure of the field theory.
We provide a more detailed comparison between the duality structure of SK orbits and S-duality orbits of absolute sYM theories in section \ref{sec LE test}.
We find complete agreement with the field theory results apart from the discrepancies mentioned above, and we further elaborate on their explanations.
How the moduli space geometry of an SCFT is reflected in and derived from the operator content of the field theory is a vexed question.
Both $\cN{=}4$ sYM theories and their moduli space geometries are well-studied, so the classification problem we address in this paper provides a concrete setting where this relationship can be studied.
In table \ref{tab:dictionary}, we describe the relationship between some conformal field theory (CFT) observables and geometric quantities associated with the K\"ahler geometry of its moduli space of vacua \eqref{N4 orbi} that are motivated by our results; it also serves to introduce some terminology that we use in the rest of the paper.
\begin{table}[ht]
\centering
\begin{tabularx}{0.95\textwidth}{| >{\centering\arraybackslash}X | c | >{\centering\arraybackslash}X |} 
\hline
$\cN{=}4$ $\fg$ sYM CFT & relation & $\cN{=}4$ $\fg$ sYM moduli space of vacua \\
\hline\hline
Chiral ring of scalar BPS operators & 1-to-1 & Coordinate ring of moduli space K\"ahler geometry \\ 
\hline
Global structures of sYM gauge theory & many-to-1 & Ext classes of pairs $(R_A, R_A^{\v})$; $R_A \deq$ $\GL(r,\Z)$ irrep of $W(\fg)$ \\ 
\hline
S-duality orbit of absolute $\cN{=}4$ sYM theory $^\red{*}$ & 1-to-1 & principal SK orbit $\deq$ equivalence class of principal, $\GL(2r,\Z)$ rep of $W(\fg)$  \\
\hline
S-duality group of absolute $\cN{=}4$ sYM theory & subgroup of & Self-duality group of principal SK orbit \\ 
\hline
Conformal manifold of absolute $\cN{=}4$ sYM theory & covers & Conformal manifold of principal SK orbit \\ 
\hline
\end{tabularx}
\caption{Relationship between some observables in the sYM conformal field theory with geometric quantities associated with its moduli space of vacua.
The terminology in the right-most column is defined in section \ref{sec Weyl Z reps} of the text.
The asterisk $^\red{*}$ is to indicate that the S-duality orbits and SK orbits are in 1-to-1 correspondence only after a correction to the list of global structure S-duality orbits, described in section \ref{sec results}, is incorporated.
}
\label{tab:dictionary}
\end{table}

\noindent\paragraph{Outline of the paper.}

Appendix \ref{appA: SK Geo} describes generic $\cN{=}2$ SK geometries and subsequently derives the properties the SK geometry of the moduli space of an $\cN{=}4$ sYM theory must obey in appendix \ref{appA: N4 SKS}.
This content is meant to serve as background material for the paper, but much of the discussion is probably well-known to experts.
In particular, appendix \ref{appA: equivalenceofN=4SKstructures} establishes the relations equivalent $\cN{=}4$ SK structures satisfy in terms of their defining data, it contains numerous subsections meant to discuss some background material and derivations in a pedagogical manner.  

Section \ref{sec Weyl Z reps} describes our procedure for classifying all possible $\cN{=}4$ sYM SK structures in terms of certain integral representations of the Weyl group.
We keep the discussion general by not assuming that the Dirac pairing, $J$, is principal.
Some of the mathematics we need concerning representations of finite groups over the integers is summarized in appendix \ref{app Z rep}.

Section \ref{sec results} is devoted to explaining our classification results for principal $J$ through the computational details of the minimal examples that present the variety of distinguished results.
Table \ref{tab:lattices} summarizes the intermediate computation of Ext groups while table \ref{tab:summary} summarizes our final results in terms of those Ext groups.
A Mathematica notebook automating these calculations for every Weyl group is available from the authors upon request.
Among the infinite series of Weyl groups corresponding to simple Lie algebras, we prove our results uniformly in $r$ for the most intricate case in section \ref{suN sec}: the Weyl groups $W(A_r)$ of the $A_r = \su(r{+}1)$) series.
We do not actually prove the classification results for all $r$ of the $BC_r$ and $D_r$ series of Weyl groups.
We instead report the pattern found by brute force computation, though a proof along the lines of section \ref{suN sec} should not be too hard, just lengthy.
We detail the full classification computations for $\fg=\so(12)$ in section \ref{sec:so12} and $\fg=\so(5)=\sp(4)$ in section \ref{BC2example}; these examples were chosen to illustrate the peculiar features of the most involved computations that occur for the cases in which outer automorphisms of the Weyl group must be taken into account.

Section \ref{sec LE test} compares our SK orbit classification results to the field theory classification results of absolute sYM orbits that are predicted by S-duality, and discusses the relationship between global structures of the sYM gauge theory and SK geometries.

Section \ref{sec npp} raises the question of whether there are ``exotic'' principally polarized CB geometries corresponding to $\cN{=}4$ sYM with non-simple gauge algebras which are not just the product of the simple ones.
This hinges on finding reducible symplectic representations which are indecomposable, and involves the use of non-principally polarized symplectic representations as building blocks.
We show how $\fg = \u(N) = \u(1) \oplus \su(N)$ gives an example of this, albeit with a free factor.

\noindent \paragraph{Further directions.}

While considering the possible SK geometries that are consistent with unbroken $\cN{=}4$ supersymmetry but not necessarily a weak-coupling limit corresponding to an $\cN{=}4$ sYM theory, some obvious questions are:
\begin{itemize}
    \item For simple $\fg$, the relative field theories of all $\cN{=}4$ sYM theories have non-principal Dirac pairing on their BPS charge lattice besides $\fg = G_2$ and $F_4$ \cite{Argyres:2022kon}.%
    \footnote{By relative field theory we mean the version of an $\cN{=}4$ sYM theory that is defined by $\fg$ and its marginal coupling $\tym$, which determine its BPS charge lattice, but is not equipped with a choice of global structure that completes it to an absolute theory.}
    For each $\fg$, is there always a unique SK geometry with the corresponding non-principal polarization of the BPS charge lattice that could correspond to this minimal/``relative'' version of the field theory?
    More generally, do all non-principally polarized SK geometries compatible with \eqref{N4 orbi} have physical interpretations?  
    \item Besides the $\u(N)$ examples of section \ref{sec npp}, are there other example of ``exotic'' principally polarized geometries built on reducible (non-principal) symplectic representations of Weyl groups?
    \item How are generalized global symmetries, such as $1$- and $2$-form symmetries \cite{GarciaEtxebarria:2019caf} and non-invertible duality defects \cite{Kaidi:2022uux}, reflected in the SK geometry, and how do the geometries change upon the (discrete) gauging of these symmetries? (See \cite{Bourget:2018ond, Argyres:2018wxu, Arias-Tamargo:2019jyh} for a related discussion.)
    \item If we drop the requirement of a weak-coupling limit, so we are not forced to assume the K\"ahler geometry of the moduli space of vacua described by the orbifold \eqref{N4 orbi}, are there other non-Weyl group isotrivial SK orbifold geometries compatible with $\cN{=}4$ supersymmetry? 
    More generally, are there non-orbifold isotrivial geometries? (See \cite{Argyres:2019ngz} for a related discussion.)
\end{itemize}

The project of classifying and constructing isotrivial moduli space geometries can also be extended to the more difficult cases that possess unbroken $\cN{=}3$ or $\cN{=}2$ supersymmetry.
Many of the above questions also apply to these theories, or, in some cases, must be addressed.
As the results of the present paper demonstrate, such making progress on these constructions promises to shed light on defining the analog of global structures in sYM theories to non-lagrangian theories where there is no notion of gauge group.

\paragraph{Note added:}

After the completion of this work, the paper \cite{cecotti2025} appeared, and we received a preliminary copy of \cite{Moscrop:2025mtl}, both of which have overlap with this paper.
Where our results overlap, they agree.

\section{Integer representations of Weyl and complexified Weyl groups}
\label{sec Weyl Z reps}

\subsection{Ingredients of an $\cN{=}4$ SK structure}
\label{sec: SKS ing}

We collect and briefly describe some ingredients of SK structures of $\cN{=}4$ CB geometries that are needed to describe their classification in the rest of this section.%
\footnote{See appendix \ref{appA: notation} for a collection of our notational conventions.}

For a given Weyl group $W(\fg) = W_{\fg} = \mathrm{Weyl}(\fg) \equiv W$ for a (semi-)simple, compact Lie algebra $\fg$ (with Cartan subalgebra $\fh$), and for a given integral symplectic form $J$ on $\fh \oplus \fh \cong \R^{2r}$, we consider pairs $(S, \bt)$ where  
\begin{itemize}
    \item $S : W \to \Sp_{J^\v}(2r, \Z)$ is a faithful, integral, symplectic reflection representation that is $\R$-equivalent to two copies of the fundamental reflection representation of $W$ (as a finite real reflection, i.e. Coxeter, group),
    \item $\bt$ is an $r \times r$ matrix in the Siegel upper half-space such that $\bt \in \Fix(S)$, i.e., for all $w \in W$, $S(w) \circ \bt = \bt$, given in \eqref{M act}.  This presupposes that we work in a basis where $J$ takes the skew-block form \eqref{J def}.
\end{itemize}   
We then define the concept of $\cN{=}4$ sYM SK structures associated with $\fg$ and fixed choice of $J$ as follows: 
\begin{itemize}
\item An \emph{$\cN{=}4$ SK structure} for Lie algebra $\fg$ and integral symplectic form $J$ is an equivalence class of pairs $(S, \bt)$, defined as above, under the \emph{twisted integral equivalence relation}: $(S, \bt) \sim (S', \bt')$ for all \emph{reflection automorphisms} of the Weyl group, $\f,\psi \in \Aut_\rfl(W)$, and all integral intertwiners $M^{(\f,\psi)} \in \Sp_{J^\v}(2r,\Z)$ such that%
\footnote{In this expression, it is sufficient to take $\f$ or $\ps$ to be the identity automorphism since, if this expression is valid for all $w\in W$, it is also valid for all $w \in \ps^{-1}\circ W$.
Thus, upon evaluating this expression for the image of $W$ under $\ps^{-1}$, $\f \to \f' \deq \f\circ \ps^{-1}$ and $\ps \to \mathbf 1_W$.
The version of twisted equivalence using two elements of $\Aut_{\rfl}(W)$ is useful for understanding the algebraic structure of self-duality groups that include twisted self-intertwiners.}
\begin{align}\label{eq:Def twist SKS}
    S'(\f\circ w) M^{(\f,\psi)} = M^{(\f,\psi)} S(\psi\circ w) \quad \text{for all}\,\, w \in W(\fg), \quad \bt' = M^{(\f,\psi)} \circ \bt.
\end{align}

\item  An \emph{SK structure orbit} is an equivalence class of just the integral symplectic representation $S$, i.e., without specifying a compatible value of $\bt$.
It represents the \emph{moduli space} or \emph{conformal manifold} of inequivalent SK structures with representation $S$.
\end{itemize}
Note that upon taking $\f = \ps = \mathbf 1_W$ to be the trivial reflection automorphism in $W_{\fg}$, we recover the (perhaps more recognizable) \emph{untwisted integral equivalence relation} between SK structures:
\begin{align}\label{eq:DefSKstructure}
    (S, \bt) &\sim (MSM^{-1} , M \circ \bt) , &
    &\text{for all} &
    M &\in \Sp_{J^\v}(2r, \Z).
\end{align}

The definitions of these ingredients and their motivation are described in detail in appendix \ref{appA: SK Geo}.
We now synthesize a brief description of these ingredients.  
The SK structure of a rank-$r$ $\cN{=}4$ sYM moduli space, is specified by an $r$-dimensional complex orbifold
\begin{align}\label{cplx orbi}
    \cC = \C^r / W_\C ,
\end{align}
that carries a canonical K\"ahler structure but is also equipped with some extra structure that comprises a special K\"ahler structure.
Here $W_\C \subset \GL(r,\C)$ is the complexification of the Weyl group, a finite group acting linearly, holomorphically, and faithfully on $\C^r$.
Because $W_\C$ is a reflection group, it is generated by reflections, elements $r_I \in W_\C$ of order 2 which fix a codimension-1 hyperplane in $\C^r$.
The collection of these hyperplanes is the fixed point set of the orbifold action, and $\cC^*$ is the smooth, non-simply connected component of $\cC$ that is obtained by removing this fixed point set from $\cC$.

Part of the extra structure is the rank-$2r$ lattice, $\L$, of electric and magnetic charges of states and probes under the low energy $\U(1)^r$ gauge group.
This lattice carries an integral symplectic form, $J$, giving the Dirac pairing between charges.
The charge lattice may suffer a monodromy $S_{\Z}(\g) \in \Sp_{J^\v}(2r,\Z)$ upon being dragged around the fixed point hyperplanes of $W_\C$ along some closed path $\g$.%
\footnote{Here $J^\v \deq J^{-t}$ denotes the inverse transpose; the reason for its occurrence is explained in appendix \ref{appA: SK Geo review}.} 
These are basis changes of the lattice which preserve the symplectic pairing $J^{\v}$ according to the matrix relation $S_{\Z}^t(\g) J^{\v} S_{\Z}(\g) = J^{\v}$.
Such monodromies define an integral symplectic representation $S_{\Z}$ of $\pi_1(\cC^*)$ into $\Sp_{J^{\v}}(2r,\Z)$ that we call the {\it monodromy map}. 
The image $S \deq \Im S_{\Z} \subset \Sp_{J^{\v}}(2r,\Z)$ of the monodromy map forms a subgroup we call the \emph{monodromy group}.
As shown in appendix \ref{appA: Mono equiv Weyl}, $\pi_1(\cC^*)$ surjects as a group onto $W_{\C}$, which implies the monodromy group is isomorphic to the orbifold group $W_{\C}$ as abstract groups, so it is generated by the reflection elements, $r_I \in W_\C$.
Through this isomorphism, we obtain an integral symplectic reflection representation of $W_{\C} \cong W(\fg)$ that we simply denote by $S$ as used in the above definition of an $\cN{=}4$ SK structure (orbit).
As a matrix subgroup, $W_\C$ is the complexification of the real reflection group $W_\R \subset \GL(r,\R)$, so relative to a real basis of $\C^r$, $W_\C$ is the image of a reducible $2r$-real-dimensional representation with image $W_\R \oplus W_\R$.
Thus, for an $\cN{=}4$ sYM CB, we have
\begin{align}\label{orbi ref real}
    S = \Im S_{\Z} \cong_\R W_\R \oplus W_\R ,
\end{align}
or, in words, the monodromy group $S$ defined by the integral symplectic reflection representation $S_{\Z}$ of $W(\fg)$ must be equivalent over the reals to two copies of the real reflection representation image of $W(\fg)$.

A \emph{symplectic basis} of $\L$ is a choice of splitting into lagrangian sublattices (``magnetic'' and ``electric'') with respect to $J$, $\L = \L_m \oplus \L_e$, with respect to which $J$ is skew block off-diagonal,
\begin{align}\label{J def}
    J &= \bpm 0 & \bj \\ -\bj^t & 0 \epm,
\end{align}
with $\bj$ an integral $r\times r$ matrix.
Even with the matrix form of $J$ fixed, there is still a large group of elements $M \in  \Sp_{J^{\v}}(2r,\Z) \subset \GL(2r,\Z)$ generating basis changes of $\L$ that preserve $J^{\v}$ under the action $ J^{\v} \mapsto M^t J^{\v} M = J^{\v}$ which implies $J \mapsto M J M^t = J$.
We define this group to be the {\it EM duality group} of the low-energy theory on $\cC$.
Two integral symplectic representations, $S$ and $S'$, satisfying \eqref{orbi ref real} and with respect to Dirac pairings $J$ and $J'$ taking the same matrix form (so they have the same invariant factors) in a choice of their symplectic bases, are \emph{untwisted integral equivalent},
\begin{align}\label{int equ}
    S \sim_\Z S',
\end{align}
iff there is an $M \in \Sp_{J^\v}(2r,\Z)$ such that
\begin{align}\label{untwisted_equiv}
    S'(w) = M S(w) M^{-1} \qquad  \forall w\in W.
\end{align}
We call $M$ an \emph{integral intertwiner} of the representations $S$ and $S'$.
Therefore, two integrally equivalent symplectic $\Z$-reflection representations of the Weyl group describe the same SK structure orbit, since they differ only by a symplectic change of charge lattice basis which is an EM duality transformation of the low-energy effective action.

Let us comment on the slightly more general notion of twisted integral equivalence between symplectic representations that is referenced in \eqref{eq:Def twist SKS}. 
As we show in appendix \ref{appA: equivalenceofN=4SKstructures}, equivalence between SK structures allows for their $\Z$-reflection reps to be \emph{twisted $\Z$-equivalent}, where ``twisted" means the SK structures are $\Z$-equivalent up to ``twisting" by a reflection outer automorphism of $W(\fg)$.
(We show in appendix \ref{appA: equivalenceofN=4SKstructures} that this relation is the most general, i.e. minimally sufficient, condition that characterizes when two SK structures are isomorphic.)
Relative to the usual untwisted $\Z$-equivalence relation between group representations, twisted $\Z$-equivalence occurs when a representation is $\Z$-intertwined with another representation that has been composed with a group automorphism.%
\footnote{For our purposes, we are concerned with Weyl groups as finite real reflection groups, so we only want to consider their reflection automorphisms.} 
Concretely, for any \emph{reflection automorphisms} of the Weyl group, $\f,\psi \in \Aut_\rfl(W)$, then $S \cong_\Z S'$ if there is an $M^{(\f,\psi)} \in \Sp_{J^\v}(2r,\Z)$ such that
\begin{align}\label{twisted_equiv}
    S'(\f\circ w) M^{(\f,\psi)} = M^{(\f,\psi)} S(\psi\circ w)
\end{align}
for all $w\in W$.
In this case we say $M^{(\f,\psi)}$ is an integral \emph{(reflection) twisted intertwiner}.
Note that when $\ps = \f = \mathbf 1_W$ are the identity element in $W$, or when they are both inner automorphisms of $W$ (i.e. equal to conjugation by some element in $W$), then the twisted $\Z$-equivalence relation reduces to untwisted $\Z$-equivalence relation defined in \eqref{untwisted_equiv}.

Relative to a choice of symplectic basis, \eqref{J def}, the low-energy effective action also specifies at each point on the moduli space an $r \times r$ complex matrix of gauge couplings,
\begin{align}\label{tau def}
    \bt &\in \scH_r \deq \{\, \bt\in \GL(r,\C), \ \bt = \bt^t, \ \Im \bt >0 \,\} .
\end{align}
$\scH_r$ is the Siegel upper half-space of degree $r$.
The EM duality group acts on $\scH_r$ via a fractional linear action (see appendix \ref{appA: SK Geo review}) which we denote $\bt \mapsto M \circ \bt$ for $M \in \Sp_{J^\v}(2r,\Z)$.
Two $\bt$ related in this way are equivalent since they differ only by a change of basis of the charge lattice.
In the case of $\cN{=}4$ sYM moduli spaces, the SK structure is \emph{isotrivial}, which means that $\bt$ is constant over the moduli space.
Thus the action $\bt \mapsto S \circ \bt$ of the monodromy group must leave $\bt$ invariant.
Call the fixed point set of this $S$ action $\Fix(S)$.
Therefore, for $\cC$ isotrivial we must have
\begin{align}\label{fix S}
    \bt \in \Fix(S) \subset \scH_r .
\end{align}
$\cN{=}4$ sYM theories with simple gauge algebra have one exactly marginal complex coupling, $\t_{YM} \in \scH_1$, so we expect $\Fix(S)$ to depend on a complex parameter $\t \in \scH_1$ as well.
Indeed, we find this to be the case, (see section \ref{sec sympl reps}) so $\bt \equiv \bt(\t) \in \Fix(S)$.

We must also consider the twisted $\Z$-self-equivalence relation, 
\begin{align}\label{twisted_self_equiv}
    S(\f\circ w) M^{(\f,\psi)} = M^{(\f,\psi)} S(\psi\circ w), \qquad \forall \, w \in W(\fg),
\end{align}
whose solutions are determined by at least one non-trivial $\f, \ps \in \Aut_{\rfl}(W)$ and an EM duality intertwiner matrix, $M^{(\f,\ps)}\in \Sp_{J^\v}(2r,\Z)$.
These represent the \emph{twisted automorphisms} of the SK orbit defined by $S$ because they leave $S$ invariant as a representation and its fixed point set $\Fix(S)\subset \scH_r$ invariant as a set.
Denote the set of these EM duality self-intertwiners $M^{(\f,\ps)}$ by
\begin{align}\label{Sdual cover}
    \til\scS_S \deq \bigcup_{\f,\psi\in\Aut_\rfl(W)} \bigl\{ M \in \Sp_{J^\v}(2r,\Z) \ |\ (S{\circ}\f) \, M = M \, (S{\circ}\psi) \bigr\} .
\end{align}
Each pair $((\f,\ps), M^{(\f,\ps)})$ is called a \emph{twisted $\Z$-self-intertwiner} of $S$.
The structure of this set is described in more detail in section \ref{sec SK twist Z equiv}, where we show that it is a subgroup in $\Sp_{J^\v}(2r,\Z)$ that is parameterized by $\PSL(2,\R)$ matrices which form modular groups $\scS_S$ as discrete subgroups of $\PSL(2,\R)$.

Fixed point sets of twisted $\Z$-equivalent representations for some intertwiner $M \in \Sp_{J^\v}(2r,\Z)$ are conjugate under the intertwiner $M$.
The twisted self-intertwiners, $\til\scS_S$, induce identifications on $\Fix(S)$.
Likewise, since $\Fix(S)$ is parameterized by $\t\in \scH_1$, their preimages in $\scS_S$ induce identifications on $\scH_1$ that defines the modular curve $\scC(S) \deq \scH_1 / \scS_S$.
Physically, each point $\t$ in $\scC(S)$ represents a distinct SK structure defined by the triple $(J, S, \bt(\t))$ within the SK orbit of $S$. 
We therefore call the modular subgroup $\scS_S$ the \emph{SK structure orbit self-duality group} and $\scC(S)$ its corresponding \emph{conformal manifold}, in light of their expected relation to the analogous quantities of the sYM SCFT.
The conformal manifold of the set of SK structures with a given $\fg$ and $J$ is the disjoint union $\coprod_i\scC(S_i)$ of all integrally inequivalent SK orbits $S_i$. 
These two quantities are both invariants of an SK orbit and are described in section \ref{sec SK twist Z equiv}.
Their untwisted versions are discussed beforehand in section \ref{sec SK Z equiv}.

In summary, an $\cN{=}4$ moduli space SK structure is determined by an equivalence class of pairs $(S,\bt)$ under EM duality transformations as claimed in \eqref{eq:DefSKstructure}.
All other properties of the moduli space geometry (its special coordinates and its metric) are determined by the $(S,\bt)$ equivalence class, as explained in appendix \ref{appA: N4 SKS}.%
\footnote{In appendix \ref{appA: notation}, we list our notational conventional and the identifications (abuses) between various notations that are frequently used throughout the text.}

Thus, to classify all $\cN{=}4$ sYM moduli spaces, for each gauge Lie algebra $\fg$ of rank $r$ we need to compute all integrally twisted-inequivalent $\Sp_{J^\v}(2r,\Z)$ reflection representations $S$ of $W(\fg)$, and also compute all $\bt\in\Fix(S)$.
We  show how to carry this out in the rest of this section.
For most of this section we  keep our discussion general by not making any assumption on $J$.
But, along the way, we will see that, modulo the overall normalization of $J$ which cannot be determined in a scale-invariant isotrivial SK geometry,%
\footnote{
The normalization of $J$ can be determined by turning on mass deformations, as was shown, for example, in rank 1 SCFTs \cite{Argyres:2015ffa, Argyres:2015gha}.
Also, an induction argument in the rank \cite{Amariti:2024yjg} constrains the allowed $J$ to a finite list at each rank for theories with characteristic dimension \cite{Cecotti:2021ouq} not equal to 1 or 2.} there are only a finite number of allowed inequivalent polarizations.

\subsection{Steps for classifying integral symplectic representations of a Weyl group }\label{sec2:method}

In this section, we provide an explicit algorithm which, given a Weyl group $W(\fg)$ of a simple, compact Lie algebra $\fg$ and a symplectic pairing $J$, fully classifies the corresponding $\cN{=}4$ SK structures into equivalence classes we call \emph{SK orbits}. 
We summarize here the steps and the main results, referring to the subsections below for the details. 
\begin{itemize}
\item \textbf{Step 1: $W$-invariant lattices. } We construct a finite number of $W$-invariant lattices $\Gamma_A$ for $A \in \cI$ an index set. These are in one-to-one correspondence with $\Z$-equivalence classes of $\GL(r,\Z)$ representations of $W$ which are $\Q$-equivalent to the reflection representation. 
\item \textbf{Step 2: A sufficient class of symplectic $\Z$-reflection representations. } Given two such representations $A,B \in \cI$ we construct a family of $\Sp_{J^\v} (2r,\Z)$ representations $S_{\cA}$ with label $\cA\deq (A,B;D)$ of $W$, which depend on a \emph{binding} $D$, and show that every SK structure contains at least one $(S,\bt)$ where $S \sim_{\Z} S_{(A,B;D)}$. 

\item \tbf{Intermezzo: twisted vs untwisted integral equivalence for $S_{\cA}$ reps.}

Having established a sufficient reference set $\mathfrak R \deq \{ S_{\cA} \, | \, \cA = (A,B;D)\}$ of appropriate representations, we can begin to classify these representations relative to the two notions of equivalence: twisted and untwisted $\Z$-equivalence.
We first evaluate the untwisted $\Z$-equivalence relation \eqref{untwisted_equiv} because it is simplest to implement in practice.
This process constitutes \tbf{Step 3}, \tbf{Step 4}, and \tbf{Step 5}, described below, and it significantly reduces the number of untwisted $\Z$-inequivalent $S_{\cA}$ reps we have to consider for \tbf{Step 6}, which tests for twisted $\Z$-equivalence among these remaining representations.
The choice of $(A,B)$ for $S_{\cA}$ is always finite, since there are only a finite number of irreducible $\Z$-reflections reps $R_A$ that can be constructed.%
\footnote{This follows from the fact that there are only a finite number of $W(\fg)$-invariant rank-$r$ lattices, and as we show in section \ref{sec inv latts}, they correspond to the subgroups of the center of $\fg$, $Z(\fg)$.}
However, for fixed $(A,B)$ in $\cA$, the choice of binding $D$ is much more unconstrained: it is an $r \times r$ real symmetric matrix only constrained to satisfy \eqref{binding def}, so we focus on constraining it first.
This is done in \tbf{Step 3} in terms of an extension problem which classifies such bindings in the form of a (typically) cyclic group $\Z_n(\cA)$ defined in \eqref{eq:ext_group}.
This reduces the reference set $\mathfrak R$ down to a finite number of equivalence classes that depend on $(A,B)$ and the $|\Z_n(\cA)|$ choices of binding. 
In \tbf{Step 4}, we evaluate the untwisted equivalence relation among the different pairs $(A,B)$ and for each choice of binding.
This further reduces the reference set to a finite number of SK orbits, and in \tbf{Step 5} we evaluate the untwisted $\Z$-self-equivalence relation among what's left to define their untwisted self-duality groups and conformal manifolds.
In \tbf{Step 6}, we evaluate the twisted $\Z$-equivalence relation between the remaining representations and among themselves (corresponding to twisted self-equivalences).
The number of such equivalences to test for is constrained by the size of the reflection outer automorphism group of $W(\fg)$, given in table \ref{tab:outer}, which are finite groups whose orders are independent of $|W(\fg)|$ since they are subsets of the Coxeter diagram automorphisms of $W(\fg)$.
Solutions to these relations will either further reduce the number of inequivalent SK orbits, or it will increase the size of a self-duality group (equivalently, shrink the size of their conformal manifold).
We give more substantive guidance on these remaining \tbf{Steps} and their results below.

\item \textbf{Step 3: Untwisted integral equivalence class of bindings.}

Relative to the untwisted $\Z$-equivalence relation \eqref{untwisted_equiv}, we now set about determining when two bindings are integrally inequivalent when the $\GL(r,\Z)$ $\Z$-reflection reps $R_A, R_B^{\v}$ are fixed. 
The details of this \tbf{Step} are contained in the first portion of section \ref{sec SK Z equiv}.

Given $A,B \in \cI$, we find for which $D,D'$ are $S_{(A,B;D)}$ and $S_{(A,B;D')}$ $\Z$-equivalent, with no reflection outer automorphism twist. 
We show that the set of equivalence classes is parametrized by an extension group $\Ext^1_{\Z W} (R_A, R_B^\v)$ that we can explicitly compute. Hence we introduce a new finite index set 
\begin{align}
\fI \deq \bigsqcup\limits_{A,B\in\cI} \Ext^1_{\Z W} (R_A, R_B^\v) \, . 
\end{align}
Elements of $\fI$ are denoted by 
\begin{align}\label{notation}
\boxed{
    \tikzmarknode{a0}{\cA} \deq (\tikzmarknode{a1}{A}, \tikzmarknode{a2}{B}; \tikzmarknode{a3}{D}) \in \fI \, ,
}
\end{align} 
\begin{tikzpicture}[remember picture, overlay]
    \draw[blue, thick, <-, shorten <=1.5pt](a1) -- ++(0,-.5) -- ++(-1.4,0) -- ++(0,.5) --++(-.3,0) node[left]{{\tiny labels of Weyl reps $R_A$, $R^\v_B$}};
    \draw[blue, thick, <-, shorten <=1.5pt](a2) -- ++(0,-.5) -- ++(-1,0) ;
    \draw[blue, thick, <-, shorten <=1.5pt](a3) -- ++(0,-.5) -- ++(1.4,0) --++(0,.5) --++(.3,0) node[right]{{\tiny element of $\Ext^1_{\Z W}(R_A,R^\v_B)$}};
\end{tikzpicture} 
$\!\!\!$where we are now taking $D$ to stand for an equivalence class of bindings, which is an element of the corresponding $\Ext^1_{\Z W}$ group.
(We have boxed and annotated this definition because it is a notation we  employ extensively in the rest of the paper.)
For each $\cA$, its fixed point set in the rank-$r$ Siegel half space, $\Fix(S_\cA) \subset \scH_r$, is 1-dimensional (as expected).

\item \textbf{Step 4: Untwisted integral equivalences between inequivalent bindings.}

We now know when the $\Z$-reflection reps $S_{(A,B;D)}$ and $S_{(A,B;D')}$ are $\Z$-equivalent for fixed $A,B \in \cI$ by classifying the possible bindings $D, D'$.
We move on to evaluating the untwisted $\Z$-equivalence relation \eqref{untwisted_equiv} between $S_{(A,B;D)}$ and $S_{(A',B';D')}$ for choices of inequivalent bindings $D, D'$ while $(A,B) \neq (A',B')$ vary over the allowed pairs of irreducible rank-$r$ $\Z$-reflection reps $R_A$ and $R_B^{\v}$.
The details of this \tbf{Step} are contained in second portion of section \ref{sec SK Z equiv}.

We construct explicitly a family of maps labeled by a pair $(\cA_i, \cA_j) \in \fI \times \fI$ 
\begin{equation}
    \cM_{\cA_i \cA_j} : \SL(2,\R) \to \Sp(2r,\R)
\end{equation} 
such that for all $w \in W$, $\cA_i \in \fI$, and $\g_a \in \SL(2,\R)$, they satisfy intertwiner and morphism properties,
\begin{align}\label{M algebra}
   \cM_{\cA_1\cA_2}(\g) S_{\cA_2}(w) &= S_{\cA_1} (w) \cM_{\cA_1\cA_2}(\g) \, ,\\
   \cM_{\cA_1\cA_2}(\g_1) \cM_{\cA_2\cA_3}(\g_2) &= \cM_{\cA_1\cA_3}(\g_1\g_2) \, .  \nn  
\end{align} 
$\SL(2,\R)$ acts by fractional linear transformations on the natural coordinate, $\t\in \scH_1$, parameterizing $\Fix(S_\cA)$.
These maps and \eqref{M algebra} thus encapsulate the algebra of equivalences among SK structures $(S_\cA,\bt)$.

In particular, this allows us to parameterize the set of intertwiners implementing integral equivalences between $S_{\cA_1}$ and $S_{\cA_2}$ by
\begin{equation}\label{inner SL2R}
    \scS_{\cA_1 \cA_2} \deq \cM_{\cA_1 \cA_2}^{-1} (\Sp_{J^\v}(2r,\Z)) \subset \SL(2,\R) ,
\end{equation}
where $\cM_{\cA_1\cA_2}^{-1}$ denotes the preimage in $\SL(2,\R)$. 
It is fully characterized in terms of divisibility conditions on the $\SL(2,\R)$ matrix elements. 
We show that the set of intertwiners is exactly 
\begin{equation}\label{inner intertwiners}
\{ M \in \Sp_{J^\v}(2r, \Z) | M S_{\cA_1}  = S_{\cA_2} M\} = \cM_{\cA_1   \cA_2} (\scS_{\cA_1 \cA_2 }),
\end{equation} 
which is to say, $M \in \Sp_{J^{\v}}(2r,\Z)$ intertwines $S_{\cA_1}$ and $S_{\cA_2}$ iff there exists a $\g\in \SL(2,\R)$ such that $M = \cM_{\cA_1\cA_2}(\g)$.
Thus the maps $\cM_{\cA_1\cA_2}(\cdot)$ capture all untwisted intertwiners between $S_{\cA_1}$ and $S_{\cA_2}$.

So we define an equivalence relation on $\fI$ by 
\begin{equation}
    \cA_1 \sim \cA_2 \qquad \text{if and only if} \qquad  \scS_{\cA_1 \cA_2 } \neq \emptyset \, . 
\end{equation}
Then the set of SK structure orbits associated with $\fg$ and $J$ are in bijection with $\fI/\sim$.
So, in particular, for fixed $\fg$ and $J$,  
\begin{equation}
    \# \, \textrm{SK structure orbits} \quad \equiv \quad\begin{array}{c}
       \# \textrm{connected components of} \\
         \textrm{the conformal manifold of SK} \\
         \textrm{structures with given $\fg$ and $J$}
    \end{array}  \quad= \quad | \fI/\sim |   \, . 
\end{equation}

\item \textbf{Step 5: Self-duality groups from untwisted integral self-equivalence. }

For a given SK structure corresponding to $\cA \in \fI$, the set of untwisted integer self-intertwiners implementing self-equivalences,
\begin{equation}\label{eq:defSgroups}
    \scS_\cA \deq  \cM_{\cA\cA}^{-1} (\Sp_{J^\v}(2r,\Z)) \subset \SL(2,\R) ,
\end{equation}
is a \emph{group}.
In particular, it is a discrete (modular) subgroup of $\SL(2,\R)$ that parameterizes the automorphisms of an SK structure, i.e. the intertwiners that leave $S_{\cA}$ invariant and map $\Fix(S_{\cA})$ to itself.
Thus, it is by definition the untwisted \emph{self-duality group of the SK structure orbit} $S_{\cA}$, or, for short, the \emph{S-duality group of} $S_{\cA}$.
To each self-duality group corresponds a \emph{conformal manifold}, $\scH_1/\scS_{\cA}$, that is the modular curve associated to $\scS_{\cA}$.
This space parameterizes the \emph{moduli space of an SK orbit} in the sense that each point in it corresponds to an untwisted $\Z$-inequivalent SK structure with the same self-duality group.
We explain these structures in the last portion of section \ref{sec SK Z equiv}.

This notion of SK orbit self-duality group is related to, but in principle distinct from, the self-duality group of the associated $\cN=4$ sYM theory.
As explained in section \ref{sec LE test}, we find the two do not always agree, and such cases are marked by red \red{\xmark}'s in the summary table \ref{tab:summary}.

\item \tbf{Step 6: Reflection automorphism twisted integral equivalence.}

After we've classified the set of inequivalent symplectic $\Z$-reflection representations $S_{\cA}$ and their self-duality groups $\scS_{\cA}$ relative to untwisted $\Z$-equivalence, the last step is to evaluate the twisted $\Z$-equivalence relation between the untwisted $\Z$-inequivalent $S_{\cA}$, and among themselves.%
\footnote{A more thorough discussion of the origin of this relation -- both mathematically and in terms of SK structure equivalences -- is explained throughout appendix \ref{appA: equivalenceofN=4SKstructures}.}
We only have to do this for the reflection outer automorphisms, which are described in section \ref{sec SK twist Z equiv}.
In the process, we identify which Weyl groups have such automorphisms and state the result in table \ref{tab:outer}.

When evaluating this relation between the untwisted $\Z$-inequivalent $S_{\cA}$, this can identify their SK orbits if it is satisfied for some reflection outer automorphism.
If there exist solutions to the twisted $\Z$-self-equivalence relation \eqref{twisted_self_equiv}, 
the definition of the untwisted self-duality group $\scS_{\cA}$ in \eqref{eq:defSgroups} has to be modified to a twisted version \eqref{Sdual cover} (or equivalently, \eqref{scSA def}). (The definition of the conformal manifold is analogously modified.)
This modification, and the resulting algebraic structure formed by the corresponding EM duality intertwiners, is explained in section \ref{sec SK twist Z equiv}.
\end{itemize}

\noindent \tbf{Implementation of all Steps for principal polarization.} For $J$ principal, $\cA = (A,A; D)$, and we carried these \tbf{Steps} to classify all $S_{\cA}$ and their corresponding SK structure orbits and (twisted) self-duality groups relative to twisted $\Z$-equivalence.
These computations are discussed in section \ref{sec PrinPolResults}. 
In table \ref{tab:summary}, we present the results of these computations for all simple, compact $\fg$.
Section \ref{sec results} discusses the computational details of the \tbf{Steps} in terms of the minimal rank $\fg$ whose results possess a distinguishing feature. 
In particular, for an explicit example of the calculations, the reader is urged to look at the $\fg=\so(12)$ example worked out in section \ref{sec:so12}.  

\subsection{Weyl group invariant lattices}
\label{sec inv latts}

Our first task is to list $\Z$-equivalence classes of $\GL(r,\Z)$ representations of $W$ which are $\Q$-equivalent to the reflection representation of $W$. 
These are in one-to-one correspondence with lattice representations, i.e. representations formed by restricting the reflection representation of $W$ on $\R^r$ to $W$-invariant lattices in $\R^r$, as a change of basis for a lattice is given by a matrix in $\GL(r,\Z)$, and have been classified by Feit \cite{feit1998integral}.%
\footnote{Note that \cite{feit2003some} corrects an error in the classification of \cite{feit1998integral}.}
Hence, \textbf{Step 1} amounts to reviewing the classification of the $\Z$-equivalence classes of rank-$r$ $W$-invariant lattices from which we can construct the corresponding lattice representations.
We first introduce some notation. 

\begin{table}[ht]
\centering
\begin{tabular}{|c|c|c|c|}
\hline 
Weyl$(\fg)$ & $A\subset Z(\fg)$ & Lattice $\G_A$ & $\Ext^1_{\Z W}(R_A,R_A^\v)$
\\ \hline \hline
$A_1$ & $\Z_1$ & $\G_{\rm{root}}$ & $\Z_1=\{0\}$
\\ \hline 
\begin{tabular}{c} $A_r$\\ $(r\ge2)$\end{tabular} & \begin{tabular}{c} $\Z_d$\\ $d|(r{+}1)$\end{tabular} & $\G_d$ & $\Z_{\gcd(d,\, (r+1)/d)}$
\\ \hline 
$BC_2$ & $\Z_2 \teal{\simeq} \Z_2^\v$ & $\G_{\rm{co-wt}} {=} \G_{\rm{root}} \teal{\simeq} \G_{\rm{coroot}} {=} \G_{\rm{wt}}$ & $\Z_2=\{0,1\}$
\\ \hline  
\multirow{3}{*}{\begin{tabular}{c} $BC_{2k}$\\ ($k\geq2$) \end{tabular}} & $\Z_2$ & $\G_{\rm{root}}^C {=} \G_{\rm{coroot}}^B$ & $\Z_2=\{0,1\}$ \\  
& $\Z_2^\v$ & $\G_{\rm{root}}^B {=} \G_{\rm{coroot}}^C$ & $\Z_2=\{0,1\}$ \\ 
& $\Z_1$ & $\G_{\rm{wt}}^B {=} \G_{\rm{co-wt}}^C {=} \G_{\rm{wt}}^C {=} \G_{\rm{co-wt}}^B$ & $\Z_1=\{0\}$ 
\\ \hline 
\multirow{3}{*}{$BC_{2k+1}$} & $\Z_2$ & $\G_{\rm{root}}^C {=} \G_{\rm{coroot}}^B$ & $\Z_1=\{0\}$ \\  
& $\Z_2^\v$ & $\G_{\rm{root}}^B {=} \G_{\rm{coroot}}^C$ & $\Z_2=\{0,1\}$ \\ 
& $\Z_1$ & $\G_{\rm{wt}}^B {=} \G_{\rm{co-wt}}^C {=} \G_{\rm{wt}}^C {=} \G_{\rm{co-wt}}^B$ & $\Z_1=\{0\}$
\\ \hline
\multirow{3}{*}{$D_4$} & $\Z_2 {\times} \Z_2$ & $\G_{\rm{root}}$ & $\Z_2{\times}\Z_2 = \{00,01,10,11\}$
\\ 
& $\Z_{2\,V} \teal{\simeq} \Z_{2\,S} \teal{\simeq} \Z_{2\,C}$ & $\G_V \teal{\simeq} \G_S \teal{\simeq} \G_C$ & $\Z_2=\{0,1\}$
\\
& $\Z_1$ & $\G_{\rm{weight}}$ & $\Z_2=\{0,1\}$
\\ \hline 
\multirow{4}{*}{\begin{tabular}{c}$D_{4k}$ \\ ($k \geq 2$)\end{tabular}} & $\Z_2 {\times} \Z_2$ & $\G_{\rm{root}}$ & $\Z_2{\times}\Z_2 = \{00,01,10,11\}$
\\ 
& $\Z_{2\,V}$ & $\G_V$ & $\Z_2=\{0,1\}$
\\
& $\Z_{2\,S} \teal{\simeq} \Z_{2\,C}$ & $\G_S \teal{\simeq} \G_C$ & $\Z_2=\{0,1\}$
\\
& $\Z_1$ & $\G_{\rm{weight}}$ & $\Z_2=\{0,1\}$
\\ \hline 
\multirow{4}{*}{$D_{4k+2}$} & $\Z_2 {\times} \Z_2$ & $\G_{\rm{root}}$ & $\Z_2{\times}\Z_2 = \{00,01,10,11\}$
\\ 
& $\Z_{2\,V}$ & $\G_V$ & $\Z_2=\{0,1\}$
\\
& $\Z_{2\,S} \teal{\simeq} \Z_{2\,C}$ & $\G_S \teal{\simeq} \G_C$ & $\Z_1=\{0\}$
\\
& $\Z_1$ & $\G_{\rm{weight}}$ & $\Z_1=\{0\}$
\\ \hline 
\multirow{3}{*}{$D_{2k+1}$} & $\Z_4$ & $\G_{\rm{root}}$ & $\Z_1=\{0\}$
\\ 
& $\Z_2$ & $\G_V$ & $\Z_2=\{0,1\}$
\\
& $\Z_1$ & $\G_{\rm{weight}}$ & $\Z_1=\{0\}$
\\ \hline 
\multirow{2}{*}{$E_6$} & $\Z_3$ & $\G_{\rm{root}}$ & $\Z_1=\{0\}$
\\  
& $\Z_1$ & $\G_{\rm{weight}}$ & $\Z_1=\{0\}$
\\ \hline 
\multirow{2}{*}{$E_7$} & $\Z_2$ & $\G_{\rm{root}}$ & $\Z_1=\{0\}$
\\  
& $\Z_1$ & $\G_{\rm{weight}}$ & $\Z_1=\{0\}$
\\ \hline 
$E_8$ & $\Z_1$ & $\G_{\rm{root}}$ & $\Z_1=\{0\}$
\\ \hline 
$F_4$ & $\Z_1 \teal{\simeq} \Z_1^\v$ &$\G_{\rm{root}} \teal{\simeq} \G_{\rm{coroot}}$ & $\Z_1=\{0\}$
\\ \hline 
$G_2$ & $\Z_1 \teal{\simeq} \Z_1^\v$ & $\G_{\rm{root}}\teal{\simeq} \G_{\rm{coroot}}$ & $\Z_1=\{0\}$
\\ \hline 
\end{tabular}
\caption{For each Weyl group (1st column, using Killing-Cartan notation) are listed the subgroups of the center of the corresponding Lie algebra(s) (2nd column).
The 3rd column gives the names from Lie algebra theory of the corresponding Weyl-invariant lattices.
The 4th column computes the Ext group \eqref{eq:ext_group} of bindings between $R_A$ and $R_A^\v$, and labels each Ext group element for use in table \ref{tab:summary}.
Some subgroups $A$ appear twice in the 2nd column as $A$ and $A^\v$ when the dual lattice $\G_A^\v$ is not integrally equivalent to $\G_{A'}$ for any subgroup $A'$.
Also, equivalences between subgroups/lattices that are due to outer reflection automorphisms of the Weyl groups are shown using \teal{$\simeq$}.
The 2nd column should be seen as attributing a name/label to each distinct lattice given by its corresponding subgroup with additional notation to distinguish lattices with identical subgroups. 
}
\label{tab:lattices}
\end{table}

\begin{itemize}
\item Consider a simple real Lie algebra $\fg$ of rank $r$, $\fh$ a Cartan subalgebra, and $\fh^\ast$ its linear dual. 
We denote by $K : \fg \times \fg \to \R$ the Killing form, which is a non-degenerate symmetric bilinear form. 
$K$ will also denote the induced inner products on $\fh$ and $\fh^\ast$. 

\item We pick a basis $(\a_i)_{i=1,\dots ,r}$ of simple roots, and denote by $\G_{\textrm{root}} \subset \fh^\ast$ the lattice they generate. 
The long roots are normalized such that they have norm squared equal to 2, i.e., $K( \a_{\rm{long}} , \a_{\rm{long}}) = 2$.  
Let $(\varpi_i)_{i=1, \dots, r}$ be the corresponding basis of fundamental weights,%
\footnote{Recall that to each root $\a$ is associated the one dimensional eigenspace $\fg_\a \subset \fg$, and there is a unique element $H_\a \in [\fg_\a , \fg_{-\a}]$ such that $\a (H_\a) = 2$.
The fundamental weights are defined uniquely by the relation $\varpi_i (H_{\a_j}) = \d_{ij}$.} 
and $\G_{\rm{weight}} \subset \fh^\ast$ the lattice they generate.  
$Z \deq \G_{\rm{weight}} / \G_{\rm{root}}$ is the center of the simply connected Lie group with Lie algebra $\fg$. 
\item As an abstract group, the Weyl group $W$ is a real reflection group, i.e. Coxeter group, so can be realized as the subgroup of $\O(\fh^\ast) \cong \O(r, \R)$ generated by reflections about hyperplanes perpendicular to the simple roots. 
We call $w_i$ the (simple) reflection about the simple root $\a_i$.  
This defines an action of $W$ on $\fh^{*}$ that renders $\fh^\ast$ an irreducible representation of $W$ that we call the \emph{reflection representation} and denote by $R_{\R}$ with image $W_{\R} \deq \Im R_{\R}$. 
It is a real, faithful, orthogonal representation with respect to the positive definite scalar product $K$.
We will often implicitly identify the abstract group $W(\fg)$ with its real reflection image, $W_{\R}$.

\item For any subgroup $H \subseteq Z$, there is a lattice $\G_H \subset \fh^\ast$ such that $H = \G_{\rm{weight}} / \G_H$.
There is also a corresponding dual lattice $\G_H^\v \subset \fh^\ast$, where by dual we mean with respect to the scalar product $K$. 
In particular, 
\begin{align}\label{co-latt}
    \begin{array}{ccccl}
        \G_{\rm{weight}} \deq \G_{\{1\}} & \supseteq & \G_H & \supseteq & \G_Z \deq\G_{\rm{root}} \\
        \G_{\rm{coroot}} \deq \G_{\{1\}}^\v & \subseteq & \G_H^\v &  \subseteq & \G_Z^\v \deq \G_{\rm{coweight}}  
    \end{array}
\end{align}
For simply laced algebras the dual lattices do not provide new lattices; in particular, $\G_{\rm{weight}} \cong \G_{\rm{coweight}}$ and $\G_{\rm{root}} \cong \G_{\rm{coroot}}$.
For non-simply-laced algebras, however, the dual lattices can give non-isomorphic representations. 

\item The restrictions of the representation $R_{\R}$ to the lattices $\G_H$ and $\G_H^\v$ constructed above, for a subgroup $H \subseteq Z$, provide \emph{integral} reflection representations $R_{\Z}^H$ of $W$.
To ease notation, we will simply denote $R_H \deq R_{\Z}^H$ with dual representation $R_H^\v$. 
These representations are all equivalent over $\R$ but they may or may not be equivalent over $\Z$. 

\item If we pick a basis $\bm{\g}^H = (\g^H_i)_{i=1,\dots,r}$ of $\G_H$, we can associate to it the dual basis $\bm{\g}_H = (\g_H^i)_{i=1,\dots,r}$ defined by $K(\g^H_i , \g_H^j) = \d_i^j$. 
When expressed in these bases, the representations $R_H$ and $R^\v_H$ induce explicit group morphisms $W \to \GL(r,\Z)$ that we denote $R_{H, \bm{\g}^H}$ and $R_{H, \bm{\g}_H}^\v$, 
\begin{align}
   R_{H, \bm{\g}^H} : w &\mapsto \Mat_{\bm{\g}^H}(R(w)) ,&
   R_{H, \bm{\g}_H}^\v : w &\mapsto \Mat_{\bm{\g}_H}(R(w)) . 
\end{align}
Then we have $R_{H, \bm{\g}_H}^\v (w) = \left( R_{H, \bm{\g}^H} (w) \right)^{-t}$. 

\item To lighten the notations, we assume a choice of basis $\bm{\g}^H = (\g^H_i)_{i=1,\dots,r}$ of $\G_H$ has been made once and for all, and we suppress it from the notations. 
Furthermore, we index the set of inequivalent representations and invariant lattices by some label set $\cI$, so $R_A$ for $A\in \cI$ runs over the inequivalent $R_H$ and $R^\v_H$ for all $H\subset Z$, and similarly for lattices $\G_A$ along with their chosen bases, $\bm{\g}^A$.
We also denote the inverse transpose of a matrix by a $^\v$ superscript,
\begin{align}
    R_A^\v \deq R_A^{-t} .
\end{align}
Thus $R^\v$ denotes what is commonly called the dual or contragredient representation to $R$.

\item
Since the set $\{ R_A, A\in \cI\}$ has, by definition, all inequivalent representations, the contragredient set $\{ R^\v_A, A\in \cI\}$ does too, though generally permuted and relative to different bases.
In the case of simply-laced Lie algebras, we can (and do) choose representatives $A \leftrightarrow \G_H$, i.e., without using any dual lattices $\G^\v_H$.
But for non-simply-laced Lie algebras there are necessarily be inequivalent dual representations, $R^\v_H \ncong_\Z R_{H'}$ for any $H'$.
Consequently in these cases there are distinct labels $A \leftrightarrow R_H$ and $B \leftrightarrow R_H^\v$, and their associated representations are contragredient,
\begin{align}
    R_A^\v &\deq R_A^{-t} 
    = \left( R_{H, \bm{\g}^H} \right)^{-t} 
    = R_{H, \bm{\g}_H}^\v = R_B \, . 
\end{align}
The full list of lattices is given in table \ref{tab:lattices} \cite{feit1998integral, feit2003some}. 

\item We call $I_{AB}$ the matrix of the identity map acting on $\fh$, mapping from $\fh$ in basis $\bm{\g}^B$ to $\fh$ in basis $\bm{\g}^A$. Similarly, we call $K_{AB}$ the matrix of the Killing form in bases $\bm{\g}^A$ and $\bm{\g}^B$. 
(In particular, the subscript $AB$ labels these matrices and not their matrix components.)
This means we have the matrix equations,  
\begin{align}\label{eq:intertwiner}
  I_{AB} R_B(w) = R_A(w) I_{AB}  \, ,
  \qquad \text{for all } w \in W,
\end{align}
and 
\begin{align}\label{eq:intertwiner2}
  K_{AB} R_B(w) = R^\v_A(w) K_{AB}  \, ,
  \qquad \text{for all } w \in W .
\end{align}
Thus $I_{AB}$ and $K_{AB}$ are intertwiners (a.k.a., $W$-equivariant maps) between $R_B$ and $R_A$ or $R_A^\v$, respectively.
We  use the defining properties \eqref{eq:intertwiner} and \eqref{eq:intertwiner2} of these intertwiners extensively in all computations to follow. Note that  
\begin{align}\label{Inv Kt}
    I_{AA} &= \bone_r , &
    I_{AB}^{-1} &= I_{BA} , &
    K_{AB}^t &= K_{BA}  ,
\end{align}
and 
\begin{align}\label{IK alg}
    I_{AB} I_{BC} &= I_{AC} , &
    K_{AB} I_{BC} &= K_{AC} , &
    K^\v_{AB} K_{BC} &= I_{AC} ,
\end{align}
but, in general, $I^t_{AB} \neq I_{BA}$, $K^{-1}_{AB} \neq K_{BA}$ and $I_{AB} K_{BC} \neq K_{AC}$.

\end{itemize}

The identification of the integer representations, or, equivalently, invariant lattices are familiar from gauge theory. But in application to non-lagrangian theories a description in terms of gauge theory does not exist, so we are interested in their properties without regard to any chosen basis.
The intertwiner matrices, $I_{AB}$ and $K_{AB}$, defined above carry this information.
By defining them as the matrices of the identity map and Killing form on $\fh$ with respect to our conventional lattice bases $\bm{\g}^A$ and $\bm{\g}^B$, we have chosen a particular normalization for $I_{AB}$ and $K_{AB}$.
Basis changes of the $\G_A$ and $\G_B$ lattices change $I_{AB}$ and $K_{AB}$ by multiplication on the left and right by arbitrary $\GL(r,\Z)$ matrices.
Thus invariant information in the individual intertwiner matrices can be encoded by normalizing them to integer matrices by dividing them by the rational gcd of their entries, i.e., $K_{AB}/\gcd(K_{AB})$ and $I_{AB}/\gcd(I_{AB})$.
Then, by multiplication on the left and right by $\GL(r,\Z)$ matrices they can be put in unique invariant factor form (a.k.a., Smith normal form), for example, $K_{AB}/\gcd(K_{AB}) \sim {\rm diag} \left\{ 1, f_1, f_1 f_2, \ldots, \textstyle{\prod_{i=1}^{r-1}} f_i \right\}$, for positive integers $f_i$.
(The first entry is 1 because we have factored out the gcd.)
Note that this implies that $K^\v_{AB}/\gcd(K^\v_{AB}) \sim {\rm diag} \{ 1, f_{r-1}, f_{r-1} f_{r-2}, \ldots, \textstyle{\prod_{i=1}^{r-1}} f_i \}$, and that 
\begin{align}\label{kktil}
    n_{AB} \deq \gcd(K_{AB})^{-1} \cdot \gcd(K^\v_{BA})^{-1} 
    &= \textstyle{\prod_{i=1}^{r-1}} f_i ,
\end{align}
is an integer invariant.
We will see below that this invariant governs the S-duality groups of many of the SK structures associated to the pair of representation $(R_A, R^\v_B)$.

Similar definitions apply to the $I$ intertwiners.
Furthermore, since, by definition, $R_A \ncong_\Z R_B$ for $A \neq B$, it follows that $I_{AB}$ cannot be invertible over the integers for $A\neq B$, and thus their invariant factors cannot all be 1.
Conversely, if $A=B$, then $I_{AA} = \bone$, and its invariant factors are all obviously 1.

We emphasize, however, that the invariant factors of the $I$'s and $K$'s do not exhaust the invariant information about the representations that they encode.
For example, the $\G_A$ and $\G_B$ basis changes which put, say, $I_{AB}$ in invariant factor form need not put $K_{AB}$ in that form, nor allow $I_{BC}$ or $K_{BC}$ to be put in that form by a basis change of $\G_C$.
Thus there is much more invariant information in the whole set of intertwiners than in their individual invariant factors.

\subsection{Symplectic representations of Weyl groups and their fixed point sets} 
\label{sec sympl reps}

We now perform \textbf{Step 2} of the method outlined in section \ref{sec2:method}, that is, we construct a sufficient class of symplectic integral representations. 
Consider an integral symplectic form $J$ of the block skew-diagonal form \eqref{J def}, where the skew block, $\bj$, is a non-degenerate $r \times r$ matrix with integer coefficients. 
The group $\Sp_{J^\v}(2r, \Z)$ is the group of matrices $M \in \Mat(2r, \Z)$ such that $M^t J^\v M = J^\v$.%
\footnote{
As reviewed in appendix \ref{appA: SK Geo review}, it is $\Sp_{J^\v}$, and not $\Sp_J$, which governs how the special coordinates and $\bt$ transform under EM duality transformations.
The two matrix groups are related by the inverse transpose of their elements, i.e. $M \in \Sp_{J^{\v}}(2r,\Z)$ iff $M^{-t} \in \Sp_J(2r,\Z)$.
}
The usual, or \emph{principally polarized}, symplectic group is recovered by taking $\bj = \bone$, in which case $J=J^\v$.

Introduce for $A,B \in \mathcal{I}$, following \cite{curtis1966representation}, a \emph{binding} of $R_A$ with $R_B^\v$, which is a map $L_{(AB, D)}: W \to \Mat(r,\Z)$ of the form
\begin{align}\label{binding def}
    L_{(A,B ; D)} (w) &\deq 
    R_A(w) I_{AB} D - I_{AB} D R_B^\v(w) 
    \ \in\ \Mat(r,\Z)
\end{align}
for all $w\in W$ for some symmetric matrix
\begin{align}\label{D symm}
    D &= D^t \ \in \ \Mat(r,\R) .
\end{align}
Note that a non-zero binding is not a group morphism.
And to each binding we associate the symplectic representation
\begin{equation}\label{symp rep}
    \boxed{
   \begin{aligned}
      S_{(A,B;D)} :  W &\to \Sp_{J^\v}(2r, \Z)  \\ 
    w & \mapsto \bpm R_A(w) & L_{(A,B;D)} (w) \\
    0 & R_B^\v (w) \epm  
   \end{aligned}
   } \ , 
\end{equation}
with $J$ given by \eqref{J def} with
\begin{align}\label{bj = I}
    \bj = I_{AB} .
\end{align}

Interpreted as the Dirac pairing in the low energy theory on the CB, $\bj$ is integral (the Dirac quantization condition) and its invariant factors are physical observables.
On the other hand, $I_{AB}$ is generally not an integral matrix as we have defined it.
But its normalization does not affect the SK geometry since $J$ only enters linearly in the definition of the EM duality group.
We are thus free to normalize it to be integral by multiplying it by $n/\gcd(I_{AB})$ for any non-zero integer $n$, though we cannot determine the value of $n$ by our methods.
So we ignore the physical normalization of $\bj$ and just take it to be \eqref{bj = I} from now on.

The fact that \eqref{symp rep} is a representation follows upon noticing that 
\begin{align}\label{SD sim S0}
    S_{(A,B;D)} &= 
    \bpm 1\ \  & I_{AB} D \\ 0 & 1 \epm^{-1}
    \bpm R_A & 0\\ 0& R_B^\v \epm
    \bpm 1\ \  & I_{AB} D \\ 0 & 1 \epm .
\end{align}
That it is symplectic follows from direct calculation, where Schur's lemma implies that $S_{(A,B;D)}$ preserves $J^\v$ iff $D-D^t \propto K_{BB}^\v$.
It then follows from \eqref{Inv Kt} that the factor of proportionality vanishes and so $D$ must be symmetric.

A theorem, reviewed in appendix \ref{app Z rep}, implies that every SK structure has a representative of the form $S_{(A,B;D)}$ given by \eqref{symp rep}.
More precisely, it says that any representation $W \to \Sp_{J^\v}(2r, \R)$ that is $\R$-equivalent to the direct sum of two copies of the fundamental reflection representation is $\Z$-equivalent to some $S_{(A,B;D)}$.%
\footnote{For the $\Z$-reps $S_{(A,B;D)}$, this is captured by the statement $\R \otimes_{\Z} S_{\Z} \cong_{\Z} R_{\R} \oplus R_{\R}$.}
So it is sufficient to determine when two $S_{(A,B;D)}$ are integrally equivalent to find all the inequivalent symplectic representations of $W$. 

The condition that $\bt \in \Fix(S)$ is easily translated, using the action \eqref{M act} of $S$ on $\scH_r$, to the condition that for all $w \in W$, $R_A(w) (\bt  + I_{AB} D I_{BA}^\v) = (\bt  + I_{AB} D I_{BA}^\v) R^\v_A(w)$. 
So, by Schur's lemma and \eqref{eq:intertwiner2}, there exists a complex number $\t$ (not to be confused with the $r\times r$ complex matrix $\bt$) such that $\bt + I_{AB} D I_{BA}^\v = \t K_{AA}^\v$, implying $\Fix(S)$ is the one-complex-dimensional set,
\begin{align}\label{bt Fix}
\boxed{
    \bt = \t K_{AA}^\v - I_{AB} D I_{BA}^\v \, , \qquad
    \t \in \C .
}
\end{align} 
$\bt \in \scH_r$ iff $\bt$ is symmetric, which is automatic for all $\t$ from \eqref{bt Fix}, and if $\Im\bt >0$.
This latter condition follows if $\t$ is restricted to the upper half-plane, $\t \in \scH_1$, i.e., 
\begin{align}
    \Im \t > 0 ,
\end{align}
because $K_{AA}$ is positive definite since the Killing form on $\fh^*$ from which it is derived is.
We will eventually identify $\t$ with the exactly marginal gauge coupling of the $\cN{=}4$ sYM theory.

For a given symplectic $\Z$-reflection rep $S \equiv S_{(A,B;D)}$ of $W(\fg)$, polarization $J$, and Weyl group $\fg$ there is an SK structure defined by the triple $(J, S, \bt(\t) \in \Fix(S) \subset \scH_r)$ and parameterized by $\t \in \scH_1$.
We refer to the whole 1-dimensional family of SK structures associated to $S$ as an \emph{SK structure orbit} or \emph{SK orbit} for short.
The SK orbit of $S$ is given by the triple, $(J, S, \Fix(S))$, which is completely fixed by $S$, i.e. independent of a choice of $\t \leftrightarrow \bt(\t)$.
Note that for any $M \in \Sp_{J^\v}(2r,\Z)$, $(S_M, \bt_M)  \deq (M S M^{-1}, M\circ \bt)$ satisfy the untwisted $\Z$-equivalence relation \eqref{untwisted_equiv}, so are equivalent SK structures, and $S$ and $S_M$ define the same SK structure orbit.
Such a transformation is implemented by an EM duality transformation which is a basis change in the $1$-homology lattice $\L$ of the polarized abelian variety (pav) fiber $A_{\bt} \sim \C^r / \L_{\bt}$ after we have chosen a symplectic basis.
Mathematically, these transformations represent isomorphisms between the corresponding pavs $A_{\bt}$ and $A_{M \circ \bt}$ of the respective SK geometries, so they determine the equivalence classes of pavs with respect to the Siegel half-space $\scH_r$.
This can be seen as the mathematical motivation for identifying SK structures that are related by untwisted integral equivalence.
Physically, these equivalences are just the low-energy EM duality frame equivalences, which identify $\Fix(S)$ and $\Fix(MSM^{-1})$ in $\scH_r$.
But there can exist non-trivial EM duality \emph{automorphisms} of the representation $S$.
In particular, such automorphisms exist if there exists an $M\neq 1$ such that $MS(w)M^{-1}=S(w)$ for all $w \in W(\fg)$.
Because it is an automorphism of $S$, it implies an identification between $\bt(\t) \equiv M\circ\bt(\t)$ in $\Fix(S)$.
This induces an identification on $\t\in\scH_1$, which is an identification within the SK orbit of $S$ between the two SK structures that are defined by $\bt(\t)$ and $M \circ \bt(\t)$.
The set of such automorphisms form a subgroup $\til\scS_S$ in $\Sp_{J^\v}(2r,\Z)$ that have a preimage as a (modular) subgroup $\scS_S$ in $\PSL(2,\R)$.
The preimage elements in $\scS_S$ define the untwisted identifications on $\scH_1$ that we call the \emph{untwisted self-duality group of the SK structure orbit}.
The action of $\scS_S$ on $\scH_1$ is given by the usual M\"obius action, i.e. via fractional linear transformations. 
We then define the set of inequivalent $\t \in \scH_1$ to be the \emph{untwisted conformal manifold} of the SK structure orbit represented by $S_{(A,B;D)}$.
Geometrically, this is the modular curve of the self-duality group, so it is the quotient space $\scH_1 / \scS_S$ with covering space $\scH_1$ relative to the M\"obius action by $\scS_S$.  
Physically, each point $\t$ in this space represents a distinct SK structure defined by the triple $(J, S, \bt(\t))$ within the SK orbit of $S$.  

We describe these SK structure self-equivalences, the self-duality groups they form, and their conformal manifolds relative to the untwisted $\Z$-equivalence relation \eqref{untwisted_equiv} in detail in section \ref{sec SK Z equiv}.
The subsequently define and analyze these same quantities after including self-equivalence solutions to the more general reflection automorphism twisted $\Z$-equivalence relation \eqref{twisted_equiv} in section \ref{sec SK twist Z equiv}.

\subsection{SK structure equivalence: untwisted integral intertwiners }\label{sec SK Z equiv}

In this section we discuss \tbf{Step 3}, \tbf{Step 4} and \tbf{Step 5} of the classification method described in section \ref{sec2:method}.
\tbf{Steps 3} and \tbf{4} determine which symplectic $\Z$-reflection representations $S_{(A,B;D)}$ of $W(\fg)$ are \emph{untwisted $\Z$-equivalent}, which means they are intertwined by an EM duality transformation $M \in \Sp_{J^\v}(2r,\Z)$, and their corresponding SK structures are related by this EM duality transformation.
In particular, given two such representations $(S_{\cA},\bt)$ and $(S_{\cA'},\bt')$ for some $\cA = (A,B;D)$ and $\cA'=(A',B';D')$, $(S_{\cA},\bt)$ and $ (S_{\cA'},\bt')$ are untwisted $\Z$-equivalent if there exists an $M \in \Sp_{J^\v}(2r,\Z)$ such that
\begin{align}\label{untwist Z equiv}
M S_{\cA}(w) M^{-1} &= S_{\cA'}(w) & & \text{and} & M\circ\bt &= \bt' & &\forall w \in W .
\end{align}
We start with \tbf{Step 3} by considering the case where $A, B$ are fixed, and we classify the $\Z$-inequivalent bindings, $D$. 
Then, for \tbf{Step 4}, we consider the remaining set of $S_{(A,B;D)}$ for different $A,B$ but with a choice of inequivalent binding $D$, and determine when they are untwisted $\Z$-equivalent.
For a fixed polarization $J$ and $\fg$, this completes the classification of $\cN{=}4$ SK structures up to EM duality transformation, i.e. \emph{untwisted $\Z$-equivalence}, and it defines their corresponding SK structure orbits up to automorphisms of the orbit, which is addressed in \tbf{Step 5}. (We still have to take into account equivalences coming from reflection outer automorphisms of $W(\fg)$, which are not EM duality transformations.
This constitutes \tbf{Step 6} and is addressed in section \ref{sec SK twist Z equiv}.)
In \tbf{Step 5} we evaluate the untwisted $\Z$-equivalence relation for $\cA = \cA'$ which correspond to automorphisms of the corresponding SK orbit that $S_{\cA}$ is in.
Such automorphism will define the \emph{self-duality group} and \emph{conformal manifold} of that SK orbit.

\subsubsection*{Step 3: Untwisted $\Z$-equivalent bindings}
\label{sec:bindingsAndExtGroups}

For a given $A,B \in \mathcal{I}$, we now determine which bindings $D$ give integrally equivalent symplectic representations $S_{(A,B;D)}$.
A binding can be thought of as specifying an extension of the $R_A$ representation (module) by $R_B^\v$,
\begin{align}
    1 \to R_B^\v \to S_{(A,B;D)} \to R_A \to 1 .
\end{align}
Such extensions are classified by Ext groups.
In particular, introduce the notion of \emph{inner binding} as a binding of the form \eqref{binding def} but with $I_{AB} D \in \Mat(r, \Z)$.
In this case, by \eqref{SD sim S0}, $S_{(A,B;D)} \cong_\Z S_{(AB,0)}$.
Let $\cB$ be the group of bindings (under addition), and $\cB_0$ be the subgroup of inner bindings.
Then 
\begin{align}\label{eq:ext_group}
    \Ext^1_{\Z W} (R_A, R_B^\v) = \cB / \cB_0
\end{align}
classifies the inequivalent $S_{(A,B;D)}$ extensions.
The largest denominator appearing in a rational entry in $D$ is bounded by the order of $W$, thus giving an upper bound on the size of the $\Ext^1_{\Z W}(R_A,R_B^\v)$ groups; however, for Weyl groups, the denominators are typically much smaller.%
\footnote{For Weyl groups of rank $r$, the largest denominator grows only as $\sim \sqrt r$ even though the order of the groups grow as $\sim r!$.}
These results are reviewed in appendix \ref{app Z rep}.

Note that if $D \propto K_{BB}^\v$, then $L\equiv 0$ by \eqref{eq:intertwiner} and \eqref{eq:intertwiner2}.
So the groups $\cB$ and $\cB_0$ can be computed as the groups of rational and integer $I_{AB} D$, respectively, satisfying \eqref{binding def}, as long as the ideal in $\Mat(r,\Q)$ generated by $K_{BB}^\v$ is quotiented out in computing $\cB$.
Upon clearing denominators, \eqref{binding def} for all $w \in W$ becomes a set of linear diophantine equations for the numerators and common denominator of entries of $D$.

\paragraph{Practical computation.}

Such equations can be solved by any of various efficient algorithms for computing the Hermite normal form of the coefficient matrix.
The solutions depend on the relative divisibility properties of the matrix elements of the $R_A$ and $R_B^\v$ representations, and seems complicated in general.
In particular, the solutions for $D$, and thus the $\Ext^1_{\Z W}(R_A,R_B^\v)$ groups, do not seem to be expressed in terms of the $K$ and $I$ intertwiners in any simple way. 

Instead, we list the results in the fourth column of table \ref{tab:lattices} in the case where $B=A$, which are the symplectic representations which preserve a principal polarization.
We prove these results for the most intricate case, the $A_r$ (i.e., $\su(r{+}1)$) series of Weyl groups, in section \ref{suN sec}.
But we do not actually prove the results listed in table \ref{tab:lattices} for the $BC_r$ and $D_r$ series of Weyl groups; instead, we have reported the pattern found by brute force computation at low ranks.
We implemented these computations for any Weyl group in a Mathematica notebook which is available from the authors upon request.
A proof of the $BC_r$ and $D_r$ series results along the lines of that of the $A_r$ series results of section \ref{suN sec} should not be too hard, just lengthy.

In slightly more detail, in order to compute the extension group, we first consider an arbitrary symmetric $r \times r$ matrix $D = (D_{\mu \nu})$, with $D_{\m\n} = D_{\n\m} \in \Q$. 
In order to impose condition \eqref{binding def}, we compute the $r$ matrices $R_{A} (w_i) D - D R_{A}^\v (w_i)$, whose entries are $r^3$ linear forms of the $r(r{+}1)/2$ variables $D_{\m\n}$ with integer coefficients (as the matrices $R_{A} (w_i)$ have integer entries). 
These linear forms can be repackaged as a linear map of lattices $\cY : \Z^{r(r+1)/2} \to \Z^{r^3}$, along with a chosen basis (corresponding to the parametrization of $D$). 
Hence $\cY$ can be seen as a $r^3 \times r(r+1)/2$ matrix with integer entries. 
Using the Hermite decomposition algorithm, one can find a \emph{unimodular} map $\cU : \Z^{r^3} \to \Z^{r^3}$ (equivalently, an $r^3 \times r^3$ matrix with integer entries and determinant 1) such that $\cU\cY$ is \emph{upper triangular}. 
It is then straightforward to compute $(\cU\cY)^{-1} (\Z^{r^3})$, as the upper-triangularity reduces the problem to solving $r^3$ linear equations in \emph{one} variable. 
This allows to write down the generic solution to \eqref{binding def} as a linear combination  
\begin{equation}
    D = \b I_{BA} K_{AB}^\v + n_1 \fD_1 + \dots + n_s \fD_s  
\end{equation}
with $\b \in \Q$ and $n_1, \dots, n_s \in \Z$. 
Using this form, it is then straightforward to compute the group of bindings, that of inner bindings, and thus the quotient. 
See section \ref{sec:so12} for an explicit example of such a computation. 

\subsubsection*{Step 4: Untwisted $\Z$-equivalence between inequivalent bindings}\label{sec Z equiv}

Once we have found the inequivalent bindings $D$ for a choice of integral Weyl reflection representations $R_A$ and $R_B^{\v}$ which preserve a given $J^\v$, it still remains to determine which $S_\cA$ with $\cA =(A,B;D)$ are integrally equivalent among all the different choices of $\cA$.
At this point, the different choices of $\cA$ are given by the combinations of $(A,B)$ and their inequivalent bindings $D$.
This is \textbf{Step 4} of the classification strategy described in section \ref{sec2:method}.

Mathematically, this amounts to classifying the following equivalence classes.
We have seen that a pair $(S,\bt)$ determines an SK geometry, where $S: W \to \Sp_{J^\v}(2r,\Z)$ is a faithful representation of the Weyl group $W_{\C}(\fg)$ that is $\Q$-equivalent to two copies of the reflection representation of $W$, and $\bt \in \text{Fix}(S) \subset \scH_r$.
And, based on the EM duality of the IR effective action, two such descriptions define the same geometry, $(S,\bt) \simeq (S',\bt')$, if there exists an $M \in \Sp_{J^\v}(2r,\Z)$ such that
\begin{align}\label{equiv1}
M S(w) M^{-1} &= S'(w) & & \text{and} & M\circ\bt &= \bt' & &\forall w \in W .
\end{align}

The key to classifying these equivalence classes is the following technical lemma that allows us to characterize the $\Sp_{J^\v}(2r,\Z)$ matrices $M$ that satisfy the intertwining condition in \eqref{equiv1}:

\paragraph{Lemma 1. } 

\emph{Any $\GL(2r,\R)$ matrix that intertwines $S_{\cA_1}$ and $S_{\cA_2}$ \eqref{equiv1} is of the form 
\begin{equation}\label{eq:defMAA}
\boxed{   
 \cM_{\cA_1\cA_2} \left( \g \right) \deq 
 \bpm 1 & -I_{A_1B_1} D_1 \\ 0 & 1\epm
 \bpm \ua I_{A_1A_2} & \ub K_{A_1B_2}^\v \\ 
    \uc K_{B_1A_2} & \ud I_{B_1B_2}^\v \epm
 \bpm 1 & I_{A_2B_2} D_2\\ 0 & 1\epm  
} \, ,
\end{equation}
where
\begin{align}
    \g \deq \bspm \ua&\ub\\ \uc&\ud\espm & \in \GL(2,\R) \, ,
\end{align}
and}
\begin{align}
    \det \left[ \cM_{\cA_1\cA_2} 
    \left( \g    \right) \right] 
    = ( \det\g )^r \det I_{A_2A_1} \det I_{B_1B_2} \, . 
\end{align}
\smallskip
Here $\cA_i=(A_i,B_i;D_i)$ as usual, and we underline $\ua, \ub, \uc$, and $\ud$ only because we are running out of letters.
The expression \eqref{eq:defMAA} comes from a direct computation and several uses of Schur's lemma. 
The determinant comes from the formula for a block matrix determinant, and from the identity 
$K_{B_2A_1} I_{A_1A_2} K_{A_2B_1}^\v = I_{B_2B_1}^\v$, 
which follows from \eqref{Inv Kt} and \eqref{IK alg}.

The $\cM_{\cA_1\cA_2}$ so defined are thus maps
\begin{align}
    \cM_{\cA_1\cA_2} &: \GL(2,\R) \to \GL(2r,\R) ,
\end{align}
whose image intertwines $S_{\cA_1}$ and $S_{\cA_2}$,
\begin{align}\label{cM intertwiner}
    S_{\cA_1}(w) \, \cM_{\cA_1\cA_2}(\g) &=  \cM_{\cA_1\cA_2}(\g) \, S_{\cA_2}(w) ,  
    \qquad \forall w \in W ,\ \g \in \GL(2,\R) .
\end{align}
These maps also obey the important composition property:

\paragraph{Lemma 2. } \emph{For any $\g, \g' \in \GL (2,\R)$ and any $\cA, \cB, \cC \in \fI$, 
\begin{equation}\label{comp prop}
    \cM_{\cA\cB}(\g) \cM_{\cB\cC}(\g') = \cM_{\cA\cC} (\g\g') \, . 
\end{equation}
}
This follows from the identities \eqref{IK alg}. 

Such matrices give an equivalence between the two symplectic representations if $\cM_{\cA_1\cA_2}$ is integral and has determinant $\pm 1$.
By virtue of \eqref{cM intertwiner}, they satisfy $\cM_{\cA_1\cA_2}^t J^\v_1 \cM_{\cA_1\cA_2} \propto J^\v_2$, where $J_i$ are the symplectic forms preserved by the $S_{\cA_i}$ representations.
If $J_1$ and $J_2$ have different invariant factors then there does not exist an integral invertible $\cM_{\cA_1\cA_2}$ intertwining them.
Thus there can only be an equivalence between representations with equivalent symplectic forms.
Furthermore, if $J_1$ and $J_2$ are equivalent symplectic forms, then any $\cM_{\cA_1\cA_2}$ intertwining them has determinant $+1$.
So we introduce the sets of $2\times2$ matrices
\begin{equation}\label{eq:dualityGroupS}
   \scS_{\cA_1\cA_2} \deq   
   \left\{\, \g\in \GL(2,\R) \ |\ \cM_{\cA_1\cA_2}(\g) \in \SL(2r,\Z) \,\right\} \, , 
\end{equation}
that yield such integral intertwiners.
Note that in general the composition property \eqref{comp prop} does not require $\scS_{\cA_1\cA_2}$ to form a group.
The determinant condition can only have a solution if $( \det I_{A_2A_1} \det I_{B_1B_2} )^{1/r}$ is rational.
Since by \eqref{bj = I} the normalized integral pairings preserved by the two representations are $\bj_1 = I_{A_1B_1}/\gcd(I_{A_1B_1})$ and $\bj_2 = I_{A_2B_2}/\gcd(I_{A_2B_2})$, it follows that
\begin{align}
    \det I_{A_2A_1} \det I_{B_1B_2} 
    = \det I_{A_2B_2} \det I_{B_1A_1} 
    &= (\det \bj_2 / \det \bj_1) 
    \left[\gcd(I_{A_1B_1})/\gcd(I_{A_2B_2}) \right]^r , \nn
\end{align}
where we have used that $I_{B_2A_2} I_{A_2A_1} = I_{B_2B_1} I_{B_1A_1}$ by \eqref{IK alg}.
$\bj_1$ and $\bj_2$ are equivalent iff they have the same invariant factors, in which case their determinants are the same.
So, a refinement of \textbf{Lemma 1} is 

\paragraph{Lemma 3.  }
    
\emph{Let $S_{\cA_1}$ and $S_{\cA_2}$ be symplectic representations with respect to equivalent symplectic forms. Then they are $\Z$-equivalent representations iff there is an \underline{integral} $\cM_{\cA_1\cA_2}(\g)$ of the form \eqref{eq:defMAA} with $\g \in \GL(2,\R)$ and $\det\g = \gcd(I_{A_2B_2}) / \gcd(I_{A_1B_1})$.}\\
\smallskip

This gives a computable condition for the pair of symplectic representations $S_{\cA_1}$ and $S_{\cA_2}$ to be $\Z$-equivalent.  
For, since we are assuming that the $I$'s, $K$'s, and $D$'s are given (because we have solved the extension problem described in the previous subsection), integrality of $\cM_{\cA_1\cA_2}(\g)$ amounts to some divisibility criteria on $\ua$, $\ub$, $\uc$, and $\ud$ which can be solved algorithmically as for the Ext problem.

\subsubsection*{Step 5: SK orbit self-duality groups: untwisted integral self-equivalence}\label{sec SKO untwist auto grp}

We now turn to {\bf Step 5} of the method described in section \ref{sec2:method} where we characterize the automorphism group of untwisted integral self-equivalences of an SK structure orbit.
These are the non-trivial EM duality transformations that map and SK orbit back to itself which, at the level of their defining $S_{\cA}$ representation, corresponds to untwisted $\Z$-intertwiners that leave $S_{\cA}$ invariant.
This will allow us to define its untwisted \emph{self-duality group} and \emph{conformal manifold}.
The refinements of this group due to reflection outer automorphisms are discussed in section \ref{sec SK twist Z equiv}.

Consider the untwisted $\Z$-equivalence relation \eqref{untwist Z equiv} that is obtained from \eqref{twisted_equiv} after setting $\f = \ps = \mathbf 1_W$.
We should allow for the possibility that the SK structures defined by $\cA$ and $\cA'$ in this expression are equal, $\cA = \cA'$.
The resulting equation is called the \emph{untwisted integral self-equivalence relation} for $S_{\cA}$.
Because non-trivial solutions to the resulting equation fix $\cA$, they define the \emph{automorphisms} of the SK orbit $(J, S_{\cA}, \Fix(S_{\cA}) \subset \scH_r)$ that the SK structure for $S_{\cA}$ resides within once we choose a point $\bt(\t)\in \Fix(S_{\cA})$.
Such a point will depend on a complex parameter $\t \in \scH_1$ that we want to parametrize the \emph{moduli space} or \emph{conformal manifold} of the SK orbit, which means we want each value of $\t$ to correspond to a distinct SK orbit.
To do this we have to ensure there are no additional identifications we need to make on $\scH_1$.
As we will explain, such identifications are generated by the automorphisms of $S_{\cA}$ or, equivalently, by the solutions to untwisted integral self-equivalence relation. 

Solutions to the untwisted self-equivalence relation are solely determined by the corresponding intertwiner $M\in \Sp_{J^{\v}}(2r,\Z)$, and they will naturally form a subgroup $\til\scS_{\cA}$ of $\Sp_{J^{\v}}(2r,\Z)$ because they fix $S_{\cA}$. 
The induced M\"obius action on $\bt(\t) \in \Fix(S_{\cA})$ maps to another point $\bt'(\t') \in \Fix(S_{\cA})$ so leaves $\Fix(S_{\cA})$ invariant.
We therefore identify the points $\t \leftarrow \t'$ within the moduli space $\scH_1$ of the SK structure orbit.
Such points are related by fractional linear transformations, $\t' = \g \circ \t$ for corresponding group elements $\g \in \PSL(2,\R)$ that are the preimages of the self-interwiners $M_{\g} \in \Sp_{J^{\v}}(2r,\Z)$ under a group morphism $\cM_{\cA\cA}$,  $\cM_{\cA\cA}(\g) \deq M_{\g}$.
The preimages $\g = \cM_{\cA\cA}^{-1}(M)$ for $M \in \til\scS_{\cA}$ will form a subgroup $\scS_{\cA} \subset \PSL(2,\R)$ that we call the \emph{(untwisted) self-duality group} of the SK orbit.
Since points $\t \in \scH_1$ related by $\g \in \scS_{\cA}$ are identifications between the corresponding SK structures, we define the \emph{(untwisted) conformal manifold} of the SK orbit to be the modular curve of $\scS_{\cA}$, $\scH_1 / \scS_{\cA}$.%
\footnote{Since this section is dedicated to analyzing the untwisted $\Z$-equivalence relation, we will often drop the additional adjective ``untwisted" and just refer to the ``self-duality group" and ``conformal manifold" of the SK orbit.}
The self-duality group $\scS_{\cA}$ is an invariant of the SK orbit in the sense that any other SK orbit $\cA'$ that is untwisted $\Z$-equivalent to $\cA$ will have the same self-duality group and conformal manifold.

Physically, the self-duality group $\scS_{\cA}$ is useful in characterizing an SK orbit not only because it is gauge-invariant data, but because it is independent of the SK orbit conformal manifold $\scH_1/\scS_{\cA}$.
In particular, for two points $\t, \t' \in \scH_1/ \scS_{\cA}$ corresponding to two distinct SK structures $(J, S_{\cA}, \bt(\t))$ and $(J, S_{\cA}, \bt'(\t'))$ in the same SK orbit of $S_{\cA}$ for $\bt(\t), \bt'(\t') \in \Fix(S_{\cA})$, they will have the same self-duality group. 
Our claim is that this is the right geometric quantity provided by the SK geometry of the low-energy theory on the $\cN{=}4$ sYM SCFT Coulomb branch that captures the image of the SCFT self-duality group, which is commonly known as the \emph{S-duality group}.
We compare the two quantities explicitly in section \ref{sec LE test}.

To define the untwisted self-duality group, we can start from the untwisted equivalence relation \eqref{cM intertwiner} for $\cA_1 = \cA_2 \equiv \cA = (A,B;D)$. 
The untwisted self-intertwiners $\cM_{\cA\cA}$ take the form, 
\begin{equation}
    \cM_{\cA\cA} \left( \g \right) 
    = \bpm 1 & -I_{AB} D \\ 0 & 1\epm
    \bpm \ua  & \ub K_{AB}^\v \\ 
    \uc K_{BA} & \ud \epm
    \bpm 1 & I_{AB} D\\ 0& 1\epm \, ,
\end{equation}
where $\g \deq \bspm \ua&\ub\\ \uc&\ud\espm \in \SL(2,\R)$.
From \textbf{Lemma 2}, $\cM_{\cA\cA}$ is a group morphism; in fact, it is an isomorphism onto its image.
We introduce the group $\scS_\cA$ of $2\times2$ real determinant 1 matrices which parameterize the self-intertwiners of the SK structure orbit $\cA$.
It is therefore defined as
\begin{equation}\label{scSAA group}
    \scS_\cA \deq  \scS_{\cA\cA} = \cM_{\cA\cA}^{-1}(\Sp_{J^\v}(2r,\Z)) \subset \SL(2,\R) ,
\end{equation}
where $\cM_{\cA\cA}^{-1}$ denotes the preimage in $\SL(2,\R)$. 
It is some discrete subgroup defined by divisibility conditions. 
If there were no additional automorphism-twisted intertwiners, then $\scS_\cA$ would be the duality group of the SK orbit $\cA$; we will refine the definition of S-duality group below when we incorporate twisted intertwiners.

$\til\scS_\cA \deq \cM_{\cA\cA}(\scS_{\cA})$ is thus the group of intertwiners of the $\cA$ SK structure orbit.%
\footnote{
It is only a subgroup of the $\til\scS_\cA$ defined in \eqref{Sdual cover} since it does not (yet) contain the automorphism twisted self-intertwiners.
}
$\til\scS_\cA$ acts on $\bt \in \scH_r$ via the action $\bt \mapsto \cM_{\cA\cA}(\g)\circ \bt$ given in \eqref{M act}.
If this $\bt$ is in $\Fix(S)$ then a calculation shows that this maps the $\t\in\scH_1$ coordinate on $\Fix(S)$ via fractional linear transformations,
\begin{align}\label{g flt}
    \g: \t \mapsto \bspm \ua & \ub \\ \uc & \ud \espm \circ \t = \frac{\ua\t+\ub}{\uc\t+\ud} .
\end{align}
Thus $\scS_\cA$ acts as a group of identifications of the coupling $\t$ in the upper half-plane, giving a conformal manifold of the SK structure orbit $\cA$ as $\scH_1/\scS_\cA$, the modular curve of $\scS_\cA$.%
\footnote{Again, upon including automorphism-twisted identifications, as we do below, the S-duality group may be enlarged, and the conformal manifold correspondingly reduced.}
(More properly, $\scS_\cA$ should be defined as a subgroup of $\PSL(2,\R)$ because there is no difference between the action of $\g, -\g \in \scS_{\cA}$ on $\t$; for convenience, we stick with the realization of $\scS_{\cA}$ as $\SL(2,\R)$ matrices where this $\Z_2$ identification is understood.)

In the special case where the binding vanishes, $D=0$, we can be even more specific about what the group $\scS_{\cA}$ is.
In this case $\cM_{\cA\cA}(\g) \in \Sp_{J^\v}(2r,\Z)$ iff
\begin{align}\label{D0 S duality}
    \g &= \bpm \ua&\ub\\ \uc&\ud\epm 
    = \bpm \ual & \ube/\gcd(K_{AB}^\v) \\  \uga/\gcd(K_{AB}) & \ude \epm \in \SL(2,\Q) ,& 
    \ual, \ube, \uga, \ude &\in \Z .
\end{align}
By conjugating $\g \mapsto N \g N^{-1}$ by $N = \bspm \sqrt{\gcd(K^\v_{AB})} & 0\\ 0 & 1 / \sqrt{\gcd(K^\v_{AB})} \espm$, we see that the self-duality group of the $\cA = (AB,0)$ orbit is equivalent to the modular congruence group
\begin{align}\label{AB0 S duality}
    \scS_{(AB,0)} &\simeq \G_0(n_{AB})
    \deq \left\{ \bspm \ua & \ub \\ \uc & \ud \espm \in \SL(2,\Z) , \ \uc=0 \ ({\rm mod}\ n_{AB}) \right\} ,
\end{align}
where $n_{AB}$ is the integer invariant defined in \eqref{kktil}.

We have carried out the calculations for \tbf{Steps 3}, \tbf{4}, and \tbf{5} for all pairs of representations symplectic with respect to a principally polarized symplectic form $J = J^{\v}$.
As we discuss in section \ref{sec PrinPolResults}, these are the representations $S_\cA$ with $\cA = (A,A;D)$.
The results are recorded in the 2nd column of table \ref{tab:summary} which lists the (EM duality) equivalence class (SK orbit) of $S_{\cA}$ in terms of the naming system for the invariant lattices and Ext group elements that define the labels $A$ and $D$ of $\cA$; these are summarized in table \ref{tab:lattices}.
The identifications of the $\cA$ equivalence classes indicated by \teal{teal} arrows in that table do not follow from the above lemma since they do not correspond to equivalences generated purely by EM duality transformations.
Instead, they follow from the solutions to the \emph{twisted $\Z$-equivalence relation} given in \eqref{twisted_equiv}.
The last column of table \ref{tab:summary} provides the self-duality group of each SK orbit $\cA$, each of which takes into account the EM duality automorphisms computed in \tbf{Step 5}.
For those in black, the EM duality automorphisms are sufficient, but for those in \teal{teal}, there are additional automorphisms coming from \emph{twisted $\Z$-self-equivalences} of the SK orbit, which are twisted $\Z$-equivalences for $\cA = \cA'$. 
We will discuss these additional twisted equivalences in section \ref{sec SK twist Z equiv}.

\subsection{SK structure equivalence: automorphism twisted integral intertwiners}\label{sec SK twist Z equiv}

We now arrive at \tbf{Step 6} of the classification method given in section \ref{sec2:method} where we will take into account solutions to the (reflection outer automorphism) twisted $\Z$-equivalence relation among the symplectic $\Z$-reflection representations of $W(\fg)$.
At this point, we have classified a sufficient set of such representations and defined the self-duality groups and conformal manifolds of the SK orbits they represent according the untwisted equivalence relation \eqref{untwisted_equiv}.
We now want to complete this classification by considering the minimally sufficient, i.e. most general, condition that two such representations must satisfy for their corresponding SK structures to be equivalent.
This condition is the \emph{twisted integral equivalence relation} defined in \eqref{twisted_equiv}, which is the $\Z$-equivalence relation \eqref{untwisted_equiv} between a representation and another representation that has been composed (i.e. twisted) by an automorphism of $W(\fg)$.
In particular, for fixed symplectic pairing $J$, two SK structures $(S_\cA,\bt)$ and $(S_{\cA'},\bt')$ are $\f$-twisted equivalent, $(S_\cA,\bt) \cong_{\Z,\f} (S_{\cA'},\bt')$ if there exists an $M_{\f} \in \Sp_{J^\v}(2r,\Z)$ and an $\f \in \text{Out}_\rfl(W)$ such that
\begin{align}\label{twisted equiv}
	M_{\f} S_\cA(w) M_{\f}^{-1} &= S_{\cA'}(\f(w)) & & \text{and} & M_{\f}\circ\bt &= \bt' & &\forall w \in W,
\end{align}
where $\Out_{\rfl}(W)$ is the set of outer automorphism of $W$ that map reflections to reflections.
When $\f = \mathbf 1_W$ is the identity automorphism, this reduces to the usual notion of $\Z$-equivalence between representations/SK structures.
In our case, we are only concerned with reflection outer automorphisms because we want such automorphisms to map between reflection representations of $W(\fg)$. (See appendix \ref{app Z rep} for a mathematical formalization of this relation, and appendix \ref{appA: auto twist reps} for a general discussion.)

Let us briefly describe the origin of this equivalence relation in terms of the geometric equivalence between SK structures to explain why it is minimally sufficient.
(This question is more thoroughly addressed in appendix \ref{appA: equivalenceofN=4SKstructures}.)
This condition is sufficient because, as we show in appendix \ref{appA: SKS isometries}, any reflection automorphism $\f$ induces a map from an SK geometry to an SK isometric one with an identical orbifold (K\"ahler) geometry defining the CB.
This is simply because it defines an isomorphism $f_{\f}:\cC\to\cC$ of the orbifold K\"ahler structure that has a $W_{\C}$-equivariant lift to $\til f_{\f}: \cCt \to \cCt$.
However, the induced SK structure isometry maps the $\Z$-reflection representation $S_\cA$ to another one $S_{\cA'}$ given by the pullback of $S_{\cA}$ by $\f^{-1}$, $S_{\cA'} = (\f^{-1})^* S_\cA = S_{\cA} \circ \f^{-1}$ up to an integral intertwiner $M_{\f} \in \Sp_{J^\v}(2r,\Z)$.%
\footnote{Equivalently, if we view $S_{\cA}: \pi_1(\cC^*) \to \Sp_{J^\v}(2r,\Z)$ as the corresponding monodromy map, then $S_{\cA'}$ is given by the pullback of $S_{\cA}$ by $f_{\f}^{-1}$, $S_{\cA} = (f_{\f}^{-1})^* S_\cA = S_{\cA} \circ f_{\f}^{-1}$, up to the intertwiner $M_{\f} \in \Sp_{J^\v}(2r,\Z)$.}
As we detail in appendices \ref{appA: twist R_C action}-\ref{appA: SKS twist equiv}, working this out yields the twisted equivalence relation defined in \eqref{twisted_equiv} where $S_{\cA'}$ acts on the pav fiber $A_{\bt}$ over $\cC$ and $S_{\cA'}$ acts on the equivalent pav fiber $A_{\bt'}$ of the isometric CB where $\bt' = M_{\f} \circ \bt$.
Thus, the two SK geometries are the same.

To understand more concretely the geometric relation between two SK structures $S_{\cA}$ and $S_{\cA'}$ that are $\f$-twisted integral equivalent, it is useful to consider their monodromy groups, $\Im S_{\cA}$ and $\Im S_{\cA'}$.  
The twisted $\Z$-equivalence relation implies the monodromy groups of are isomorphic as subgroups of matrices in $\Sp_{J^\v}(2r,\Z)$, where the isomorphism is given by the intertwiner $M_{\f}$.
However, relative to the monodromies $\g \in \pi_1(\cC^*)$ linking discriminant components in $\cC^*$, (with preimages given by reflection hyperplanes in the cover, $\cCt$, that are labeled by the simple reflection in $W(\fg)$ that fixes them) they generate monodromy transformations about different discriminant components.
Furthermore, these different discriminant components have reflection hyperplane(s) preimages that are fixed by simple reflections of $W(\fg)$ which themselves are mapped to each other $\f \in \Out_{\rfl}(W)$.%
\footnote{Said in an equivalent, more geometric fashion, the fixed hyperplanes are mapped to each other under the induced isometry of the cover $\til f_{\f}$.}

In more physical terms, we can say that there is no low-energy way to physically distinguish the effective actions on two CBs whose fiber actions are related by twisted $\Z$-equivalence.
We summarize this by saying that every Weyl group automorphism induces a fiber-preserving symplectomorphism (or equivalently, an SK structure isometry) of the SK geometry that defines the total space of the CB.
Furthermore, we argue in appendix \ref{appA: equivalenceofN=4SKstructures} that, when combined with the SK isometries corresponding to EM duality transformations (giving rise to the usual untwisted $\Z$-equivalence relation), such automorphism SK isometries constitute \emph{all} non-trivial equivalences between $\cN{=}4$ SK structures, so by classifying their defining symplectic $\Z$-reflection reps according to twisted $\Z$-equivalence, we are implicitly classifying all such SK structure orbits. 

\subsubsection*{Interlude on outer automorphisms of Weyl groups}\label{}

We begin by summarizing a few facts about automorphisms of Weyl groups, mostly taken from 
\cite{bannai1969automorphisms, franzsen2001automorphisms}.
 
For each Weyl group $W$, one can define: 
\begin{itemize}
    \item $\Aut(W)$ is the group of all group automorphisms of $W$.
    \item $\Inn(W)$ is the group of inner automorphisms, i.e., those which act on $W$ by conjugation, $W \mapsto v W v^{-1}$ for some $v \in W$. 
    \item $\Out(W) = \frac{\Aut(W)}{\Inn(W)}$ is the group of outer automorphisms. 
    They are listed in table \ref{tab:outer}.  
    \item $\Aut_\rfl(W)$ is the group of all $\phi \in \Aut(W)$ preserving reflections, i.e., such that $\phi (w)$ is a reflection whenever $w$ is a reflection.  All inner automorphisms are reflection automorphisms, $\Inn(W) \subset \Aut_\rfl(W)$.
    \item $\Out_\rfl(W) = \frac{\Aut_\rfl(W)}{\Inn(W)}$ is the group of outer automorphisms preserving reflections, or \emph{reflection outer automorphisms} for short, also listed in table \ref{tab:outer}. 
\end{itemize}
In the next paragraph, we will see that in the computation of the modifications of SK structure orbits and their self-duality groups that come from the inclusion of automorphism intertwiners, the relevant group is $\Out_\rfl(W)$. 

Symmetries of the Coxeter diagram%
\footnote{Recall that Coxeter diagrams of Weyl groups are undirected versions of the Dynkin diagrams of the corresponding Lie algebras, i.e., without the arrow or marking differentiating long from short roots.} for $W$ correspond to reflection automorphisms of $W$, since these symmetries interchange nodes of the diagrams which corresponds to interchanging the corresponding simple generating reflections which define $W$ as a finite reflection group.
These diagram symmetries are shown in table \ref{tab:outer}.
Though they are reflection automorphisms, they are not necessarily outer automorphisms: only the diagram automorphisms of the $BC_2$, $D_{2r}$, $F_4$, and $G_2$ Weyl groups are outer.
These turn out to be all the reflection outer automorphisms of Weyl groups.

Finally, as a quotient of $\Aut_\rfl(W)$, $\Out_\rfl(W)$ may fail to be a subgroup of $\Aut_\rfl(W)$.
But in the Weyl group case, the quotient splits as $\Aut_\rfl(W) = \Inn(W) \rtimes \Out_\rfl(W)$, so we can (and do) realize $\Out_\rfl(W) \subset \Aut_\rfl(W)$.

\begin{table}[ht]
    \centering
\begin{tabular}{|c|c|c|c|} \hline 
$W$ & $\Out(W) = \frac{\Aut(W)}{\Inn(W)}$ & $\Out_\rfl(W) = \frac{\Aut_\rfl(W)}{\Inn(W)}$ & $\Aut(\textrm{Cox}(W))$  \\
\hline
$A_{r \neq 5}$ & $1$ & $1$ & $\Z_2$  \\
$A_5$ & $\Z_2$ & $1$ & $\Z_2$ \\
$BC_2$ & $\Z_2$ & $\Z_2$ & $\Z_2$  \\
$BC_{{\rm odd}>2}$ & $\Z_2$ & $1$ & 1  \\
$BC_{{\rm even}>2}$ & $\Z_2 \times \Z_2$ & $1$ & 1 \\
$D_{4}$ & $S_3 \times \Z_2$ & $S_3$ & $S_3$ \\
$D_{{\rm odd}>4}$ & $1$ & $1$ & $\Z_2$  \\
$D_{{\rm even}>4}$ & $\Z_2 \times \Z_2$ & $\Z_2$ & $\Z_2$  \\
$E_6$ & $1$ & $1$ & $\Z_2$  \\
$E_7$ & $1$ & $1$ & $1$ \\
$E_8$ & $\Z_2$ & $1$ & $1$  \\
$F_4$ & $D_8$ & $\Z_2$ & $\Z_2$ \\
$G_2$ & $\Z_2$ & $\Z_2$ & $\Z_2$ \\
\hline 
\end{tabular}
    \caption{Weyl groups $W$ (first column), their outer automorphisms $\Out(W)$ (second column), their outer automorphisms preserving reflections $\Out_{\rfl}(W)$ (third column), and automorphisms of the $W$ Coxeter diagram $\Aut(\textrm{Cox}(W))$ (fourth column). 
    $D_8$ is the dihedral group of order $8$.
    }
    \label{tab:outer}
\end{table}

It is useful to apply the automorphism twisted equivalence both to the integral irreducible Weyl group reflection representations, $R_A$, and to the integral reducible symplectic complexified Weyl group reflection representations, $S_\cA$.
Written out for the irreducible reflection representations this is
\begin{align}\label{RA equiv}
    R_A & \cong_{\Z,\f} R_B \quad \text{if} \quad 
    M R_A = (R_B\circ\f) M \quad \text{for} \quad M {\in} \GL(r,\Z) \quad \text{and} \quad \f {\in} \Aut_\rfl(W) .
\end{align}
If $\f \in \text{Inn}(W)$ is an inner automorphism, so $\f(w) = vwv^{-1}$ for some $v\in W$, then $\til M \deq R_B(v^{-1}) M$ gives an equivalence between $R_A$ and $R_B$ with $\f = \text{id}$.
So setting $\f=$ id in the equivalence relation \eqref{RA equiv} automatically covers the case of $\f\in\text{Inn}(W)$, and \eqref{RA equiv} can be modified by restricting to outer automorphisms without loss of generality,
\begin{align}\label{out cond}
\f \in \Out_\rfl(W) .
\end{align}
If $\f={\rm id}$, $\cong_{\Z,{\rm id}}$ is just the old $\cong_\Z$; call such an equivalence an \emph{inner equivalence}.

Since the set $\{R_A\}$ with $A\in\cI$ constructed in section \ref{sec inv latts} consists of all the $\Z$-inequivalent irreducible reflection representations of a given $W$, the only inner equivalences are $R_A \cong_\Z R_A$, by definition.
The outer equivalences are $R_A \cong_{\Z,\f} R_{\f(A)}$, which serves to define the action of the outer automorphism $\f$ on the set $\cI$ of reflection representations.

This action of $\Out_\rfl(W)$ on the representations is indicated in the second and third columns of table \ref{tab:lattices} by the blue arrows in the cases where it acts non-trivially in $\cI$.
These actions are easy to see from the description of the outer automorphism as a symmetry of the Coxeter diagram.
This induces an action on the weight space of the associated Lie algebra, and thus on the weight lattices we used to define the representations.
In the $D_{2\ell}$ cases the automorphism acts on the weight space as the reflection which interchanges the roots corresponding to the spinor and conjugate spinor, leaving the other basis vectors unchanged.
Its action on the representation lattices then follows: $\G_{\rm root}$ and $\G_{\rm weight}$ are invariant almost by definition, $\G_V$ is also invariant by inspection, and the spinor and conjugate spinor lattices are interchanged, $\G_S \teal{\leftrightarrow} \G_C$.
In the $BC_2$, $G_2$, and $F_4$ cases the automorphism is the reflection which exchanges long root directions with short root directions, so results in interchanging the root and co-root lattices.

\subsubsection*{SK structure equivalence: reflection twisted integral intertwiners}\label{sec SKS twist equiv}

We now arrive at the SK structure equivalence part of {\bf Step 6}.
We apply the expanded definition of twisted $\Z$-equivalence of SK structures through the twisted $\Z$-equivalence of their symplectic $\Z$-reflection rep and fixed points sets:  $(S_\cA,\bt) \cong_{\Z,\f} (S_{\cA'},\bt')$ if there exists an $\f \in \text{Out}_\rfl(W)$ and an $M_{\f} \in \Sp_{J^\v}(2r,\Z)$ such that
\begin{align}\label{SK twist equiv}
	M_{\f} S_\cA(w) M_{\f}^{-1} &= S_{\cA'}(\f(w)) & & \text{and} & M_{\f}\circ\bt &= \bt' & &\forall w \in W .
\end{align}
This form of the twisted $\Z$-equivalence relation is equivalent to that given in \eqref{twisted_equiv} upon setting $\ps \to \mathbf 1_W$, which is always valid to do.
The automorphism can be taken to be outer without loss of generality by the same argument as given above for the irreducible Weyl $\Z$-reflections reps, $R_{A}$.

When $\cA= \cA'$ in \eqref{SK twist equiv}, the solutions change the self-duality groups of SK structure orbits, but when $\cA \neq \cA'$, the solutions will change the number of untwisted inequivalent SK orbits.
We discuss the former scenario in the next portion of this section.

Note that as it stands, \eqref{SK twist equiv} is a proposed definition of SK structure equivalence that we claim is sufficient to capture all possible equivalences between $\cN{=}4$ SK structures based on their defining data in terms of a polarization $J$, its symplectic $\Z$-reflection rep $S_{\cA}$, and a point in the fixed point set $\Fix(S_{\cA})\subset \scH_r$. (See appendix \ref{appA: SKS and orbit triple}.)
In appendix \ref{appA: equivalenceofN=4SKstructures}, we prove it is minimally sufficient by showing, if the symplectic $\Z$-reflection reps $S_{\cA}$ and $\bt$ of two SK structures are related in this way, then they are necessarily equivalent.
That it is minimally sufficient follows from the fact that the most general SK isometry one can perform on an $\cN{=}4$ SK geometry --- which is a combination of an EM duality transformation in the pav fiber coupled with a particular coordinate transformation in the base of the CB corresponding to an element of $\Out_{\rfl}(W)$ --- will induce this transformation on $S_{\cA}$ and $\bt$, which are its defining data.

\subsubsection*{SK orbit self-duality groups: twisted integral self-intertwiners}\label{sec SKO twist self-duality}

We now turn to the part of {\bf Step 6} in the classification method described of section \ref{sec2:method} that characterizes the group of twisted $\Z$-self-equivalences of an SK structure (orbit) relative to the twisted $\Z$-equivalence relation defined in \eqref{twisted_equiv}.

Recall that the definition \eqref{scSAA group} of self-duality groups is not quite correct, since it misses the (preimages of the) \emph{reflection automorphism twisted $\Z$-self-equivalences} that satisfy \eqref{twisted_equiv} for $S = S'$ and are discussed in appendix \ref{appA: equivalenceofN=4SKstructures}.
These more general twisted self-equivalence solutions are determined by at least one non-trivial reflection outer automorphism $\f, \psi \in \Out_{\rfl}(W)$ with a corresponding twisted EM duality intertwiner matrix, $M^{(\f,\ps)} \in \Sp_{J^{\v}}(2r,\Z)$.
Including them into the set of all twisted or untwisted self-intertwiners, $ \til \scS_{\cA} \deq \{ M^{(\f,\ps)} \in \Sp_{J^\v}(2r,\Z) \, | \, \f, \ps \in \Out_{\rfl}(W) \, \}$, still produces a subgroup of $\Sp_{J^{\v}}(2r,\Z)$, but their composition properties are those of a groupoid (see \eqref{s-dual gpoid}).
Despite this structural change, each $M^{(\f,\ps)}$ has a unique preimage in $\g\in \PSL(2,\R)$ defined by a map $\cM_{\cA\circ \f, \cA\circ \ps}$, and the set of such preimages will form a subgroup $\scS_{\cA} \subset \PSL(2,\R)$ with the standard group law that follows from their composite action on $\scH_1$ via M\"obius transformations.
Furthermore, this action of $\scS_{\cA}$ on $\scH_1$ will continue to define the group of identifications in $\Fix(S_{\cA})$ corresponding to twisted self-equivalences between the SK structures $(J, S_{\cA}, \bt(\t) \in \Fix(S_{\cA}))$ that are in the SK structure orbit $(J, S_{\cA}, \Fix(S_{\cA}))$ for $S_{\cA}$.
Therefore, we continue to refer to $\scS_{\cA}$ so defined as the \emph{self-duality group} of the SK structure orbit defined by $S_{\cA}$, and the modular curve $\scH_1 / \scS_{\cA}$ is its conformal manifold whose points label the distinct SK structures in the orbit.

Given this more general definition of twisted $\Z$-equivalence, we formalize the notion of the self-duality group, $\scS_\cA$, of an SK structure orbit defined by the symplectic $\Z$-reflection rep $S_{\cA}$ to be the set of all equivalences $(S_\cA,\bt) \cong_{\Z,\f} (S_\cA,\bt')$ of $S_\cA$ to itself for some $\f \in \Out_{\rfl}(W)$ and $M_{\f}\in \Sp_{J^\v}(2r,\Z)$.
Each such self-equivalence represents an automorphism of the SK structure orbit.
For each pair $\f, \psi \in \Out_\rfl(W)$, define the sets 
\begin{align}\label{SK twist_self-intertwiner}
    \til\scS_\cA^{\ (\f,\psi)} \deq \left\{ \, M \in \Sp_{J^\v}(2r,\Z) \ \,|\ \, M \,S_\cA(\psi(w)) = S_\cA(\f(w))\, M, \ \ \forall\ w\in W \,\right\} ,
\end{align}
which is \eqref{SK twist_intertwiner} evaluated for $\cA = \cB$.
They are subsets of $\Sp_{J^\v}(2r,\Z)$, but are generally not subgroups when $\f\neq \psi$.
Instead, they satisfy the groupoid properties
\begin{align}\label{s-dual gpoid}
    \til\scS_\cA^{\ (\f,\psi)} \til\scS_\cA^{\ (\psi,\chi)} &= \til\scS_\cA^{\ (\f,\chi)}, &
    \left( \til\scS_\cA^{\ (\f,\psi)} \right)^{-1} &= \til\scS_\cA^{\ (\psi,\f)},
\end{align}
in a notation where we multiply or invert the elements of each set.
The set of all self-intertwiners of $S_{\cA}$ is the union over all pairs $\f,\psi \in \Out_\rfl(W)$ of $\til\scS_\cA^{\ (\f,\psi)}$, as in \eqref{Sdual cover}, and together the algebraic structure they form is a still a subgroup of $\Sp_{J^{\v}}(2r,\Z)$ according to the above composition properties.
Note that, from its definition, the sets $\til\scS_\cA^{\ (\f,\psi)}$ only depend on $\f$ and $\psi$ in the combination $\f\psi^{-1}$.
That is, $\til\scS_\cA^{\ (\f,\psi)} = \til\scS_\cA^{\ (\f',\psi')}$ if $\f\psi^{-1}=\f'(\psi')^{-1}$.
Interestingly, as opposed to the groupoid properties their images in $\Sp_{J^{\v}}(2r,\Z)$ satisfy, the union of the self-intertwiner preimages in $\SL(2,\R)$ under the $\cM_{\cA\cA}$ maps will form an $\SL(2,\R)$ subgroup with the usual group law under composition.
The reason is that all automorphism images of $\cA$ have the same fixed point set in $\scH_r$, i.e., $\Fix(S_{\cA\circ\f}) = \Fix(S_{\cA\circ\psi})$ for all $\f,\psi\in\Out_\rfl(W)$, as shown in appendix \ref{appA: equivalenceofN=4SKstructures}.
Thus the induced identifications act on the same $\t \in \scH_1$ coordinate irrespective of the automorphism labels, so the groupoid multiplication law collapses to a group law relative to this action.

Note that the union of all the $\til\scS_{\cA}^{\ (\f,\psi)}$ forms a subgroup of $\Sp_{J^\v}(2r,\Z)$.
However, this group need not be isomorphic to the self-duality group of the SK structure orbit (c.f. \eqref{eq:defSgroups}) because the $\cM_{\cA\cA}$ maps \eqref{eq:defMAA} of {\bf Lemma 1} are different for each $\til\scS_\cA^{\ (\f,\psi)}$ subset.

So, for the SK structure orbit defined by $S_{\cA}$, we denote the self-intertwiner preimages in $\PSL(2,\R)$ by
\begin{align}
    \scS^{(\f,\psi)}_\cA &\deq \cM_{\cA\circ\f,\cA\circ\psi}^{-1}(\til\scS^{\ (\f,\psi)}_\cA) ,
\end{align}
and define the (twisted) self-duality group of the SK structure orbit to be the union over $|\Out_\rfl(W)|$ many sets:
\begin{align}\label{scSA def}
    \scS_\cA \deq \bigcup_{\overset{\f,\psi\ \in}{\Out_\rfl(W)}} \scS^{(\f,\psi)}_\cA = \bigcup_{\overset{ \psi\ \in}{\Out_\rfl(W)}} \scS^{(1,\psi)}_\cA .
\end{align}

We illustrate this in the case $\Out_\rfl(W)=\Z_2 \deq \{0,1\}$.
There are just two symplectic representations related to $S_\cA$ by an outer automorphism: $S_\cA \circ0 = S_\cA$ and $S_\cA \circ1$.
Then their intertwiner groupoid and self-duality group can be depicted as
\begin{equation}\label{groupoid1}
\raisebox{-.5\height}{
\resizebox{12cm}{!}{%
\begin{tikzpicture}[scale=1,->,>=stealth', shorten >=1pt, shorten <=1pt, auto, node distance=1.0cm, thick]
  \node[circle, draw] (SA1) {$S_\cA \circ 1$};
  \node (empty1) [below of=SA1] {};
  \node[circle, draw] (SA0) [below of=empty1] {$S_\cA \circ 0$};
  
  \draw[->,blue] (SA1) to[bend right] node [left] {$\til\scS^{\ (1,0)}_\cA$} (SA0);
  \draw[<-,blue] (SA1) to[bend left] node [right] {$\til\scS^{\ (0,1)}_\cA$} (SA0);
  \draw[->,blue] (SA0) to[loop right] node [right] {$\til\scS^{\ (0,0)}_\cA$} (SA0);
  \draw[->,blue] (SA1) to[loop right] node [right] {$\til\scS^{\ (1,1)}_\cA$} (SA1);

  \node (empty2) [right=3 of empty1] {$\xrightarrow{\cM^{-1}_{\cA\circ i , \cA\circ j}}$};
  \node[circle, draw] (21) [right=1 of empty2] {$\t\in\scH_1$};
  \draw[->,red] (21) to[loop right] node [right] {$\scS_\cA = \bigcup_{i,j\in\Z_2} \scS_\cA^{(i,j)}$} (21);
\end{tikzpicture}}}
\end{equation}
where the blue arrows denote the $\til\scS^{\ (i,j)}_\cA$ intertwiner actions.
Note, furthermore, that by the groupoid properties and the equivalence of SK structures under the isometry induced by the reflection automorphism, it follows that $\til\scS^{\ (0,0)}_\cA = \til\scS^{\ (1,1)}_\cA$ and $\til\scS^{\ (1,0)}_\cA = \left(\til\scS^{\ (0,1)}_\cA\right)^{-1}$.
In particular, if $\til\scS^{\ (1,0)} = \emptyset$ is empty, then the self-duality group $\scS_\cA$ does not change from its value \eqref{scSAA group} found just using $\Z$-equivalence (i.e., without taking account of automorphism equivalences).
Conversely, if $\til\scS^{\ (1,0)}_\cA \neq \emptyset$, then the self-duality group is enlarged from its naive value, and, correspondingly, the conformal manifold of the orbit is smaller.\\

\noindent \tbf{Example analysis.} 
For exposition, we provide an example analysis of the twisted $\Z$-(self-)equivalence relation in the simple (but relevant) case where $\Out_{\rfl}(W) = \Z_2$; the other case for $\fg=D_4$ with $\Out_{\rfl}(W) = S_3$, is similar.

We have a set of $\Z$-inequivalent representations, $\{S_\cA\}$, and we are seeking to understand their equivalences, a.k.a., SK structure orbits. 
Define sets of automorphism twisted intertwiners between representations $\cA$ and $\cB$ by 
\begin{align}\label{SK twist_intertwiner}
    \til\scS^{\ (\f,\psi)}_{\cA\cB} &\deq \left\{\, 
    M \in \Sp_{J^\v}(2r,\Z) \ \,|\ \, M \,(S_\cA\circ\psi ) = (S_\cA\circ\f) \, M \, \right\},
\end{align}
which are the generalization of \eqref{SK twist_self-intertwiner}, $\til \scS_{\cA} = \til \scS_{\cA\cA}$. 
Assume $\til\scS^{\ (0,0)}_{\cA\cB} = \emptyset$ for $\cA\neq\cB$: $\cA$ and $\cB$ label distinct SK structure orbits before the imposition of automorphism equivalences.
If we have the situation that there is no non-trivial outer automorphism intertwiners of $S_\cA$ with itself,  $\til\scS^{\ (1,0)}_\cA = \emptyset$, but there are automorphism intertwiners of $S_\cA$ with another representation $S_\cB$, $\cB \neq \cA$, then the two orbits are identified, but their (common) self-duality groups are not changed from their non-automorphism values:
\begin{align}\label{sit 1}
\text{if} \quad \til\scS^{\ (1,0)}_\cA = \emptyset \quad \text{and} \quad \til\scS^{\ (1,0)}_{\cA\cB} \neq \emptyset, 
\quad \text{then} \quad \scS_\cA = \cM^{-1}_{\cA\circ0,\cA\circ0} \big(\til\scS^{\ (0,0)}_\cA\big) = \scS_\cB,
\end{align}
and $S_\cA$ and $S_\cB$ are in the same orbit.
In other words, in this case the set $\til\scS^{\ (1,0)}_{\cA\cB}$ of equivalences do not contribute to the self-duality group, but rather serve to identify what might have been thought to be inequivalent orbits.
On the other hand, 
\begin{align}\label{sit 2}
\text{if} \quad \til\scS^{\ (1,0)}_\cA \neq \emptyset, 
\quad \text{then} \quad \scS_{S_\cA} = 
\cM^{-1}_{\cA\circ0,\cA\circ0} \big(\til\scS^{\ (0,0)}_\cA\big) \cup
\cM^{-1}_{\cA\circ1,\cA\circ0} \big(\til\scS^{\ (1,0)}_\cA\big).
\end{align}
That is to say, in this case the outer automorphism intertwiners do not enlarge the number of representations in an orbit, but they do impose further identifications within the $S_\cA$ orbit, and so enlarge its self-duality group.

\subsection{Principally polarized symplectic representations}
\label{sec PrinPolResults}

The results presented in tables \ref{tab:summary} and \ref{tab:lattices} are for the special case of SK structures with principally polarized symplectic forms.
This is the case relevant for absolute $\cN{=}4$ sYM theories with simple gauge algebras; though see the discussion in section \ref{sec npp} for the possible role of non-principal symplectic forms in absolute $\cN{=}4$ theories with semi-simple gauge algebras.
In this section we will summarize the results of the classification \tbf{Steps} given in the previous subsections for when $J$ is principal. 

Principally polarized symplectic $\Z$-reflection representations of $W(\fg)$ are those for which the invariant factors of $\bj$ are all 1.
From \eqref{bj = I} it follows that this occurs if and only if $A=B$ in \eqref{symp rep}, in which case $\bj=\bone$ and $J = J^\v$ is the canonical symplectic form.
We  drop the $J^\v$ subscript from the symplectic group $\Sp(2r,\Z)$ in this case.
The results of the last subsection simplify somewhat upon setting $A=B$.
Thus principally polarized representations are ones of the form
\begin{align}\label{S principal}
    S_{(AA,D)} &= 
    \bpm 1\ \  & D \\ 0 & 1 \epm^{-1}
    \bpm R_A & 0\\ 0& R_A^\v \epm
    \bpm 1\ \  & D \\ 0 & 1 \epm 
    \in \Sp(2r,\Z), 
\end{align}
with
\begin{align}\label{D quotient}
    D &= D^t \ \in\ \bigl(\Mat(r,\Q)/\langle K_{AA}^\v\rangle \bigr)/\Mat(r,\Z) .
\end{align}
The last is the solution to the extension problem, described above, and classifies the inequivalent bindings of $R_A$ with $R^\v_A$.

{\bf Lemma 1} \eqref{eq:defMAA} simplifies to the statement that $M \in \Sp(2r,\Z)$ intertwines two principal symplectic integral representations $S_{(A_1A_1,D_1)}$ and $S_{(A_2A_2,D_2)}$ iff $M$ has the form
\begin{align}\label{M prin int}
    \cM_{\cA_1\cA_2}(\g) &\deq 
    \bpm 1 & - D_1 \\ 0 & 1\epm
    \bpm \ua I_{A_1A_2} & \ub K_{A_1A_2}^\v \\ 
    \uc K_{A_1A_2} & \ud I_{A_1A_2}^\v \epm
    \bpm 1 & D_2\\ 0& 1\epm , 
\end{align}
for $\g \deq \bspm \ua&\ub \\ \uc&\ud  \espm \in \SL(2,\Q)$.
The structure of this set of matrices was computed, giving the results shown in black in column two of table \ref{tab:summary}.

One then checks for the existence of integral outer automorphism twisted self-intertwiners which satisfy \eqref{twisted_self_equiv} and were described in \ref{sec SK twist Z equiv}.%
\footnote{These computations have been implemented in a Mathematica notebook available from the authors on request.}
The additional equivalences they yield are shown as \teal{teal} arrows within column two of table \ref{tab:summary}.
Such equivalences are either between SK orbits that are not related by the usual untwisted $\Z$-equivalence, or self-equivalences for a fixed SK orbit in which case it represents twisted self-intertwiner.
These results are not hard to predict, following the rules \eqref{sit 1} and \eqref{sit 2} developed above.
In particular, for the $D_{4r}$ cases only case \eqref{sit 1} occurs since the spinor and conjugate spinor lattices always appear in different SKS orbits, so the outer automorphism simply identifies the orbits as indicated by the \teal{teal} double arrows $\teal{\leftrightarrow}$ in table \ref{tab:summary}.
This has the effect in these cases of simply reducing the number of inequivalent duality orbits relative to the naive (no outer automorphism) number.
The resulting self-duality groups \eqref{scSA def} were also computed, and are shown in column three of table \ref{tab:summary}, which gives the self-duality groups as certain modular subgroups or as Hecke groups.
(These are defined in appendix \ref{app Hecke}.)
In the $D_{4r+2}$ cases and in the non-simply-laced cases, the outer automorphism intertwiners are of the \eqref{sit 2} variety, so impose identifications within an orbit, as indicated in table \ref{tab:summary} by the \teal{teal} curved arrows $\teal{\curvearrowright}$.
($BC_2$ actually has both types of identification.)
These imply a 2-fold identification on their conformal manifolds, thus enlarging their self-duality groups by a factor of 2 to the ones shown.
Examples of these computations in the $A_r$, $D_6$, and $BC_2$ cases are given in the next section.

\section{Results and examples: SK structures with simple $\fg$ and principal $J$}
\label{sec results}

We now illustrate the general arguments of the last section in the case of principally polarized $\cN{=}4$ sYM CB geometries corresponding to simple gauge algebras $\fg$.
In addition to the five exceptional Weyl groups corresponding to the five exceptional simple Lie algebras, there are three infinite series of corresponding Weyl groups: $A_r$ corresponding to $\fg=\su(r{+}1)$; $BC_r$ corresponding to $\fg=\so(2r+1)$ or $\sp(2r)$; and $D_r$ corresponding to $\fg=\so(2r)$.
In this section we construct the SK structures for the $\su(N)$ series, and compute their orbits and self-duality groups, for all $N$.
We highlight this case both because of its intricacy --- it depends on divisibility properties of $N$ --- and because it can be treated with pleasing uniformity.

Similar arguments can be applied to the two other infinite series as well, but they separate into seven distinct special sub-cases, as indicated in table \ref{tab:summary}, so a general proof of their properties is tiresome.
Instead, we have automated these calculations in a Mathematica notebook which is available from the authors upon request, from which the patterns shown in table \ref{tab:summary} can be induced from examples.
In sections \ref{sec:so12} and \ref{BC2example} we illustrate these computations in two of the most intricate cases: the $D_6$ or $\fg=\so(12)$ case, and the $BC_2$ or $\fg=\so(5)=\sp(4)$ case.
For the $E_{6,7,8}$, $F_4$, and $G_2$ exceptional groups we simply calculated them explicitly.

\subsection{$\su(N)$ SK structures}
\label{suN sec}

As indicated in table \ref{tab:lattices}, the integrally inequivalent $(N-1)$-dimensional irreducible representations of the Weyl group $S_N$ are labeled by the divisors of $N$, when $N>2$. The $N=2$ special case is discussed at the end of this subsection (\ref{su2 case}).  Thus we denote these representations $R_d$, with $d|N$.
We use the basis 
\begin{equation}
    (\a_2 , \dots , \a_{N-1} , -d \varpi_{N-1}) 
\end{equation} 
for the lattice $\G_d$, where $(\a_1 , \a_2 , \dots , \a_{N-1})$ is a basis of simple roots and $(\varpi_1 , \dots \ \varpi_{N-1})$ the corresponding basis of fundamental weights. 
In these $\G_d$ bases, we find
\begin{align}\label{sun ik mats}
    I_{dd'} &= \bpm 1 &  &  & & \\
        & \ \,1 & & & \\
        & & \ \ddots & & \\ 
        & & & \ \,1 & \\ 
        & & & & d/d' \epm , &
    K_{dd'} &= \bpm 2 & -1 &  & & \\
        -1 & 2 & \ddots & & \\
        & \ddots & \ddots & -1 & \\ 
        & & -1 & 2 & -d' \\ 
        & & & -d & \frac{N-1}{N} d d' \epm ,
\end{align}
and $\gcd(K_{dd'}) = \gcd(N,dd')/N$.
If $dd'\neq N$, then it is easy to see that 
$K_{dd'}/\gcd(K_{dd'})$ is not invertible over the integers, while if $dd'=N$, then all its invariant factors are 1, and it is invertible over the integers.
Thus 
\begin{align}
    R_d \cong_\Z R^\v_{d'} \quad \text{iff} \quad dd'=N.
\end{align}
This result also follows from the Lie algebra definitions of the $\G_d$ invariant lattices as those corresponding to subgroups $\Z_d \subset \Z_N$ of the center, together with the co-lattice equivalences \eqref{co-latt}.

Principally polarized symplectic representations are those $S_{(dd',D)}$ with $d=d'$, as in \eqref{S principal}.  
Note that for $d=d'$, $\gcd(K_{dd}) = d s_d/N$, where we have defined the integer
\begin{align}\label{sd def}
    s_d &\deq \gcd\left(\frac Nd, d \right) .
\end{align}
$s_d$ plays an important role in what follows. 
Note that $s_d^2 |N$; we  call it the ``square-divisor of $N$ associated to $d$''.

A somewhat laborious calculation using the explicit forms of the $R_d$ representation matrices shows that the condition for a symmetric $D \in \Mat(r,\R)$ to give an integral $S_{(dd,D)}$ representation is
\begin{align}\label{sun D mat}
    D &= \beta K_{dd}^\v + \frac{d}{N} \bpm {\bf 0}_{r-1} & \vec 0\\ \vec 0 & z \epm
    \quad \mod 1 , & \beta &\in \R , & z &\in \Z_{N/d} .
\end{align}
Here the matrix is all zeros except for the lower right entry, $z$, which is an integer, and the ``mod 1'' means that an arbitrary integer symmetric matrix can be added to $D$.
These integer matrices give the inner bindings which should be quotiented out in, as in \eqref{D quotient}.
This is what restricts $z$ to be in in $\Z_{N/d}$.
The ideal generated by $K_{dd}^\v$ over the rationals should also be quotiented out since it gives an identically vanishing binding.
Note that, using the explicit form \eqref{sun ik mats} for $K^\v_{dd}$, if $z=nd$ for some integer $n$, then, by taking $q=n$ in \eqref{sun D mat}, $D$ becomes integral, resulting in an inner binding.
As a result, the inequivalent solutions to the extension problem \eqref{D quotient} are parameterized by 
\begin{align}
    D_z &\deq \frac{d}{N} \bpm {\bf 0}_{r-1} & \vec 0\\ \vec 0 & z \epm,&
    &\text{with}&
    z &\in \Z_{s_d},
\end{align}
since the resulting $D_z$'s are equivalent both modulo $N/d$ and modulo $d$ in $z$.
To summarize, for a given $d$, the inequivalent $\Sp(2N-2,\Z)$ representations are 
\begin{align}
    S_{(d,z)} & \deq S_{(dd,D_z)} , &
    & d|N, &  z& \in \Z_{s_d} .
\end{align}
($S_{(d,z)}$ is a short-hand notation which we  use for the rest of this subsection.)

Next, we determine which $S_{(d,z)}$'s are equivalent for different $d$'s.
A general analysis searching for integral intertwiners $M(\g)$ of the form \eqref{M prin int} is somewhat complicated, as it depends on the specific form of the $I_{dd'}$ and $K_{dd'}$ intertwiners given in \eqref{sun ik mats}, and results in quadratic diophantine equations.
But a less direct approach works.
We present it in two steps:
\begin{description}
    \item[(i)] First, show that for any $S_{(d,z)}$ there is a divisor of $N$, $d'$, such that $S_{(d,z)} \cong_\Z S_{(d',0)}$.
    This implies that each equivalence class of SK structures contains a representative $S_{(d,0)}$ for some $d|N$.
    \item[(ii)] Second, determine the integer congruences between $S_{(d,0)}$ and $S_{(d',0)}$ for all $d$, $d'$ pairs that each divide $N$.
\end{description}

\paragraph{Step (i).}

A solution for the equivalence $S_{(d,z)} \cong_\Z S_{(d',0)}$ is $d'=\gcd(s_d,z) = s_{\gcd(d,z)}$.
To see this, take $A=d$, $B=d'= \gcd(s_d,z)$, $E=0$, and $D=D_z$ in \eqref{M prin int}.
Since, for $N>2$, all the entries except for the lower right entry of $I_{d'd}$, $I^\v_{d'd}$, $K_{d'd}$, $K^\v_{d'd}$, and $D_z$ are integers with gcd 1, if follows that $\g \in \SL(2,\Z)$ (i.e., $\ua$, $\ub$, $\uc$, and $\ud \in\Z$ and $\ua\ud-\ub\uc=1$).
Then demanding that the lower right entry of each block is integral imposes the further constraints that
\begin{align}
    \ua &= \frac{d}{\gcd(s_d,z)} \ual , &
    \ub &= \frac{z}{\gcd(s_d,z)} \ual - \frac{N/d}{\gcd(s_d,z)} \ube, & \ual,\ube &\in \Z .
\end{align}
Note that the coefficients in these expressions are all integer since $s_d |d$ and $s_d | (N/d)$.
In deriving this result we have used the fact that $\gcd(N, d \, \gcd(s_d,z)) = d \, \gcd(s_d,z)$, which also follows from $s_d | (N/d)$.
The determinant condition can then be written
\begin{align}
    1 &= \ua\ud-\ub\uc = ( \ual \  \ube ) X
    \bspm \uc \\ \ud \espm &
    &\text{with}&
    X &\deq 
    \frac1{\gcd(s_d,z)} \bpm -z & d \\ N/d & 0\epm .
\end{align}
Since $X$ is an integer matrix, it can be put in Smith normal form, $X = UYV$, with $U,V\in\GL(2,\Z)$ such that $Y$ is integer diagonal, $Y = {\rm diag} \{ f_1, f_1 f_2 \}$.
Since it is a $2\times 2$ matrix, its invariant factors are determined to be $f_1 = \gcd(X) = 1$ and $f_2 = f_1^2 f_2 = \det(X) = N/\gcd(s_d,z)^2$.
Taking $(\ual \ \ube) U = (1\ 0)$ and $V\bspm \uc \\ \ud \espm = \bspm 1\\0\espm$ then gives the desired solution.

\paragraph{Step (ii).}

Now we want to determine when $S_{(d,0)}\cong_{\Z} S_{(d',0)}$ for two distinct divisors of $N$, $d$ and $d'$.
The argument of step (i) applies equally well when $z=0$, where $d'=\gcd(s_d,0) = s_d$.  
Thus, for every $d$, $S_{(d,0)} \cong_\Z S_{(s_d,0)}$, so we need to ask about the equivalence between $S_{(s,0)}$ and $S_{(s',0)}$ when $s$ and $s'$ are each a square-divisor of $N$.
We prove, by way of contradiction, that no equivalence exists between such representations.
For suppose $s \neq s'$ are two distinct square-divisors of $N$.
Then any potential intertwiner $M(\g)$ between $S_{(s,0)}$ and $S_{(s',0)}$ must have the form
\begin{align}
    M(\g) &= \bpm \ua I_{s's} & \ub K^\v_{s's} \\
    \uc K_{s's} & \ud I_{s's} \epm, &
    \g &= \bpm \ua&\ub\\ \uc&\ud\epm \in \SL(2,\R).
\end{align}
Since (for rank $r>1$, i.e., $N>2$) all but the lower right entries in the $I$, $K$, intertwiners are integers with gcd 1, we must, in fact, have $\g \in \SL(2,\Z)$.
The lower right hand entries of each block give the $2\times 2$ matrix
\begin{align}
    \frac{1}{Nss'}
    \bpm \ua N (s')^2 & \ub (N-1) (ss')^2 \\
    \uc N^2(N-1) & \ud N s^2 \epm,
\end{align}
whose integrality requires
\begin{align}
    \g &= 
    \bpm \ual \frac{s}{\gcd(s,s')} & \ube \frac{N}{ss'} \\
    \uga  & \ude \frac{s'}{\gcd(s,s')} \epm , &
    \ual, \ube, \uga, \ude &\in \Z, 
\end{align}
where we have used that $\gcd(N,ss')=ss'$ since $s$ and $s'$ are square-divisors of $N$.
The determinant condition is
\begin{align}\label{sun det 5}
    1 = \det\g = \ual\ude \frac{ss'}{\gcd(s,s')^2} 
    - \ube\uga \frac{N}{ss'} .
\end{align}
But the right side is divisible by $ss'/\gcd(s,s')^2$ because $ss'/\gcd(s,s')$ is a square-divisor of $N$ if $s$ and $s'$ are, and $ss'/\gcd(s,s')^2 >1$ because $s\neq s'$.
Thus there is no integer solution to \eqref{sun det 5}, and so no intertwiner $M(\g)$ between $S_{(s,0)}$ and $S_{(s',0)}$ invertible over the integers.

\bigskip

We have thus shown that each $S_{(d,z)}$ SK structure is equivalent to an $S_{(s,0)}$ SK structure, where $s$ is a square-divisor of $N$, and that the $S_{(s,0)}$ SK structures are all inequivalent for distinct $s$.
The conclusion is that the $\cN{=}4$ $\su(N)$ sYM theory has distinct SK structure orbits that are in 1-to-1 correspondence with the square-divisors of $N$.
Furthermore, by \eqref{AB0 S duality}, the self-duality group of the $S_{(s,0)}$ SK structure (and every other SK structure in its orbit) is 
\begin{align}\label{sun S duality}
    \scS_{(s,0)} \simeq\G_0(N/s^2) \subset \SL(2,\Z) .
\end{align}
Here we have used that $n_{ss} \deq \gcd(K_{ss})^{-1} \gcd(K^\v_{ss})^{-1} = N/s^2$.

\subsubsection{The $\su(2)$ special case}
\label{su2 case}

The case of $\fg=\su(2)$ (or $A_1$) requires special discussion, as it is a kind of degenerate case.
In the $\su(2)$ case the dual Cartan algebra is $\fh^\ast = \R$ and the $\Z_2$ Weyl group acts by reflection through its origin.
It has just a single inequivalent integral representation, $R(1)=1$, and $R(w)=-1$, so there is no distinction between $R_{d=1}$ and $R_{d=2}$.
As a result there is just a single SK structure orbit with S-duality group which is $\PSL(2,\Z)$, and conformal manifold which is the $\SL(2,\Z)$ fundamental domain in the upper half-plane.

\subsection{$\so(12)$ example}
\label{sec:so12}

The SK structures of the $\so(2k)$ series is in some ways more intricate than that of the $\su(r+1)$ series.
We  explain this here briefly, with the calculational details shown only in the $\fg=\so(12)$ example.

First, the $\so(4k)$ and $\so(4k+2)$ series are qualitatively different, as is already apparent from table \ref{tab:lattices}, reflecting the fact that the center of $\so(4k+2)$ is $\Z_4$ and so has three subgroups and corresponding invariant lattices, while that of $\so(4k)$ is $\Z_2\times\Z_2$ so has five inequivalent subgroups.

Next, and less obvious, is the pattern of Ext groups classifying the extensions of these lattices by their conjugate lattices.
These are recorded in table \ref{tab:lattices}, and show a different pattern for the $\so(8k)$, $\so(8k+4)$, and $\so(4k+2)$ cases.
We do not know how to predict this pattern aside from an examination of the detailed divisibility properties of the intertwiners of these lattices.

These different symplectic lattices fall into integral equivalence classes, indicated in table \ref{tab:summary}, which we determine by brute force calculation.
The pattern of the resulting orbit S-duality groups is equally unobvious from this point of view.
As in the $\su(r+1)$ cases, they occur in the calculation as certain subgroups of $\SL(2,\R)$, and can be conjugated to finite-index subgroups of $\SL(2,\Z)$, shown in table \ref{tab:summary}.
However, the relative way they are embedded in $\SL(2,\R)$ has physical significance, since it controls the values that the matrix of low-energy couplings, $\bt_{ij}$, takes for these SK structures.
The freedom to embed the S duality groups in $\SL(2,\R)$ is needed to realize the multiple S-duality orbits with isomorphic S-duality groups.
For example, in the $\so(8k)$ case, there are four distinct orbits each with S-duality group $\G_0(2)$, but there are not four inequivalent ways of embedding a group isomorphic to $\G_0(2)$ in $\SL(2,\Z)$.
Finally, in the $\so(8k+4)$ case there is one S-duality orbit whose S-duality group is not (isomorphic to) a modular congruence subgroup.
It is, instead, the index-2 subgroup of $\SL(2,\Z)$ called $\D$, described explicitly in appendix \ref{app Hecke}.

To illustrate our construction, we work out in detail the principally polarized SK structures of type $D_r$ with $r \equiv 2$ mod 4. 
We present the computations for $r=6$ in order to show explicit computations and to lighten notations, but everything can be extended straightforwardly to any $r \equiv 2$ mod 4. 
This example is the richest in terms of number of SK structures and types of duality groups. 
We follow the steps outlined in Section \ref{sec Weyl Z reps}. 

\paragraph{Step 1: $W$-invariant lattices. }

As the algebra is simply laced, the set of lattices $\mathcal{I}$ for $D_{r}$ with $r$ even contains 5 lattices, corresponding to the 5 subgroups of $\Z_2 \times \Z_2$. We use these subgroups to label the lattices, so 
\begin{equation}\label{eq:Iso12}
    \mathcal{I} = \{ \Z_2^2, \Z_2^V , \Z_2^S, \Z_2^C, \Z_1 \} \, . 
\end{equation}
The lattices can be constructed as embeddings in $\R^r$ as
\begin{equation}
\begin{array}{c|c|c} 
\text{ Lattice } & \text{ Symbol } & \text{Construction} \\
\hline 
\G_{\text{root}} & \Z_2^2 & \Z^r|_{\S \in 2\Z} \\
\G_{V} & \Z_2^V & \Z^r \\ 
\G_{S} & \Z_2^S & \Z^r|_{\S \in 2\Z} \cup (\Z+\frac12)^r|_{\S \in 2\Z} \\ 
\G_{C} & \Z_2^C & \Z^r|_{\S \in 2\Z} \cup (\Z+\frac12)^r|_{\S \in 1+2\Z} \\ 
\G_{\text{weight}} & \Z_1 & \Z^r \cup (\Z+\frac12)^r 
\end{array} \ ,
\end{equation}
which is a standard construction from Lie algebra theory.
Thus this completes \textbf{Step 1} in the classification procedure. To be explicit in the computations later on, one can choose bases for the different lattices in the above table. In terms of elements of $\Z^6$, they are given as matrices in which basis vectors are columns: 
\begin{equation}
\scalebox{.7}{$
    \begin{array}{ccccc}
\Z_2^2 & \Z_2^V & \Z_2^S & \Z_2^C & \Z_1 \\
 \left(
\begin{array}{cccccc}
 1 & 0 & 0 & 0 & 0 & 0 \\
 -1 & 1 & 0 & 0 & 0 & 0 \\
 0 & -1 & 1 & 0 & 0 & 0 \\
 0 & 0 & -1 & 1 & 0 & 0 \\
 0 & 0 & 0 & -1 & 1 & 1 \\
 0 & 0 & 0 & 0 & -1 & 1 \\
\end{array}
\right) & \left(
\begin{array}{cccccc}
 1 & 0 & 0 & 0 & 0 & 0 \\
 -1 & 1 & 0 & 0 & 0 & 0 \\
 0 & -1 & 1 & 0 & 0 & 0 \\
 0 & 0 & -1 & 1 & 0 & 0 \\
 0 & 0 & 0 & -1 & 1 & 0 \\
 0 & 0 & 0 & 0 & -1 & 1 \\
\end{array}
\right) & \left(
\begin{array}{cccccc}
 1 & 0 & 0 & 0 & \frac12 & 0 \\
 -1 & 1 & 0 & 0 & -\frac12 & 0 \\
 0 & -1 & 1 & 0 & \frac12 & 0 \\
 0 & 0 & -1 & 1 & \frac12 & 0 \\
 0 & 0 & 0 & -1 & \frac12 & 1 \\
 0 & 0 & 0 & 0 & \frac12 & 1 \\
\end{array}
\right) & \left(
\begin{array}{cccccc}
 1 & 0 & 0 & 0 & 0 & \frac12 \\
 -1 & 1 & 0 & 0 & 0 & -\frac12 \\
 0 & -1 & 1 & 0 & 0 & \frac12 \\
 0 & 0 & -1 & 1 & 0 & -\frac12 \\
 0 & 0 & 0 & -1 & 1 & \frac12 \\
 0 & 0 & 0 & 0 & -1 & \frac12 \\
\end{array}
\right) & \left(
\begin{array}{cccccc}
 1 & 0 & 0 & 0 & \frac12 & \frac12 \\
 -1 & 1 & 0 & 0 & -\frac12 & -\frac12 \\
 0 & -1 & 1 & 0 & \frac12 & \frac12 \\
 0 & 0 & -1 & 1 & \frac12 & -\frac12 \\
 0 & 0 & 0 & -1 & \frac12 & \frac12 \\
 0 & 0 & 0 & 0 & \frac12 & \frac12 \\
\end{array}
\right) \\
\end{array}$}
\label{eq:repsso12}
\end{equation}
Using these bases, each simple reflection is represented as a matrix $R_{A} (w_i) \in \GL(6,\Z)$, where $A \in \cI$ specifies the lattice and $i=1, \dots, 6$ the simple reflection $w_i \in W$; the simple roots are the columns of the matrix labeled $\Z_2^2$ above.  

\paragraph{Steps 2,3: Untwisted equivalence class of bindings.x}

To find which representations $S_{(AA,D)}$ to consider, we have to compute, for each $A \in \cI$, the group $\Ext^1_{\Z W} (R_A, R_A^\v)$ of bindings modulo inner bindings \eqref{eq:ext_group}, in which $D$ is picked. 
For this, we follow the general procedure outlined in section \ref{sec:bindingsAndExtGroups}. 
We find that the most general solution to \eqref{binding def} for $A = \Z_1 \in \cI$ is 
\begin{equation}\label{eq:Dso12Weight}
    D = \beta   K_{\Z_1}^\v + n_1 \fD_1 + n_2 \fD_2  + \dots + n_s \fD_s  
\end{equation}
where $\b \in \R$ and $n_1, \dots, n_s \in \Z$, and
\begin{equation}\label{eq:killingso12}
K_{\Z_1}^\v = \bspm
\vph{\frac12} 2 & 1 & 2 & -1 & 2 & -4 \\
\vph{\frac12} 1 & 2 & 2 & 0 & 2 & -2 \\
\vph{\frac12} 2 & 2 & 4 & -1 & 4 & -6 \\
\vph{\frac12} -1 & 0 & -1 & 2 & -2 & 4 \\
\vph{\frac12} 2 & 2 & 4 & -2 & 6 & -8 \\
\vph{\frac12} -4 & -2 & -6 & 4 & -8 & 14 
 \espm \, , \quad
\fD_1 = \bspm
 0 & -\frac12 & 0 & -\frac12 & 0 & 0 \\
 -\frac12 & 0 & 0 & 0 & 0 & 0 \\
 0 & 0 & 0 & \frac12 & 0 & 0 \\
 -\frac12 & 0 & \frac12 & 0 & 0 & 0 \\
\vph{\frac12} 0 & 0 & 0 & 0 & 0 & 0 \\
\vph{\frac12} 0 & 0 & 0 & 0 & 0 & 0 \espm
\, , \quad  
\fD_2 = \bspm
 -\frac14 & \frac38 & -\frac14 & -\frac38 & -\frac14 & -\frac12 \\
 \frac38 & -\frac14 & -\frac14 & 0 & -\frac14 & -\frac34 \\
 -\frac14 & -\frac14 & -\frac12 & -\frac38 & -\frac12 & -\frac14 \\
 -\frac38 & 0 & -\frac38 & -\frac14 & -\frac34 & -\frac12 \\
 -\frac14 & -\frac14 & -\frac12 & -\frac34 & -\frac34 & 0 \\
 -\frac12 & -\frac34 & -\frac14 & -\frac12 & 0 & \frac14 \espm \, .
\end{equation}
$K^\v_{\Z_1}$ is the inverse transpose of the matrix of the Killing form in the basis specified in \eqref{eq:repsso12}, and all other $\fD_i$ with $i>2$ have integer coefficients, and can therefore be removed in the Ext group computation (it turns out $s=20$ here, so there are 18 matrices $\fD_i$ with integer entries).
One can check that for \emph{every} $n_1, n_2, \dots \in \Z$, there exists $\b \in \R$ such that $D$ in \eqref{eq:Dso12Weight} has integer entries.
Indeed, one checks that $n_1 = 1$ can be canceled by $\b = \frac12$ since the odd entries in \eqref{eq:killingso12} match exactly the half integers in $\fD_1$. 
Similarly, $n_2 = 1$ can be canceled by $\b = \frac58$. 
This is equivalent to saying that $\Ext^1_{\Z W} (R_A, R_A^\v)$ is trivial for $A=\Z_1$. 

The same computation for $A = \Z_2^2$ gives 
\begin{equation}\label{eq:KZ2Z2}
K_{\Z_2^2}^\v = \bspm
 1 & 1 & 1 & 1 & \frac12 & \frac12 \\
\vph{\frac12} 1 & 2 & 2 & 2 & 1 & 1 \\
 1 & 2 & 3 & 3 & \frac32 & \frac32 \\
\vph{\frac12} 1 & 2 & 3 & 4 & 2 & 2 \\
 \frac12 & 1 & \frac32 & 2 & \frac32 & 1 \\
 \frac12 & 1 & \frac32 & 2 & 1 & \frac32 \espm
\, , \quad
\fD_1 = \bspm
 -\frac12 & 0 & -\frac12 & 0 & -\frac12 & 0 \\
\vph{\frac12} 0 & 0 & 0 & 0 & 0 & 0 \\
 -\frac12 & 0 & -\frac12 & 0 & -\frac12 & 0 \\
\vph{\frac12} 0 & 0 & 0 & 0 & 0 & 0 \\
 -\frac12 & 0 & -\frac12 & 0 & \frac12 & 0 \\
\vph{\frac12} 0 & 0 & 0 & 0 & 0 & 0 \espm
\, , \quad  
\fD_2 = \bspm
 -\frac12 & -\frac12 & -\frac12 & -\frac12 & -\frac14 & -\frac14 \\
 -\frac12 & -1 & -1 & -1 & -\frac12 & -\frac12 \\
 -\frac12 & -1 & -\frac12 & -\frac12 & -\frac34 & -\frac34 \\
 -\frac12 & -1 & -\frac12 & -1 & -1 & -1 \\
 -\frac14 & -\frac12 & -\frac34 & -1 & -\frac14 & 0 \\
 -\frac14 & -\frac12 & -\frac34 & -1 & 0 & -\frac14 \espm
\, .
\end{equation}
In this case, for every $n_1, n_2, \dots \in \Z$, there exists $\b \in \R$ such that $D$ in \eqref{eq:Dso12Weight} equals one the the four following matrices modulo 1: 
\begin{equation}\label{eq:bindingsso12}
\begin{array}{cccc}
     00 & 11 & 01 & 10 \\
\bspm
\vph{\frac12} 0 & 0 & 0 & 0 & 0 & 0 \\
\vph{\frac12} 0 & 0 & 0 & 0 & 0 & 0 \\
\vph{\frac12} 0 & 0 & 0 & 0 & 0 & 0 \\
\vph{\frac12} 0 & 0 & 0 & 0 & 0 & 0 \\
\vph{\frac12} 0 & 0 & 0 & 0 & 0 & 0 \\
\vph{\frac12} 0 & 0 & 0 & 0 & 0 & 0 
\espm \, , \ &
\bspm
 0 & \frac12 & 0 & \frac12 & \frac14 & \frac34 \\
 \frac12 & 0 & 0 & 0 & \frac12 & \frac12 \\
 0 & 0 & 0 & \frac12 & \frac34 & \frac14 \\
 \frac12 & 0 & \frac12 & 0 & 0 & 0 \\
 \frac14 & \frac12 & \frac34 & 0 & \frac14 & 0 \\
 \frac34 & \frac12 & \frac14 & 0 & 0 & \frac34 
\espm \, , \ &
\bspm
 \frac12 & 0 & \frac12 & 0 & 0 & \frac12 \\
\vph{\frac12} 0 & 0 & 0 & 0 & 0 & 0 \\
 \frac12 & 0 & \frac12 & 0 & 0 & \frac12 \\
\vph{\frac12} 0 & 0 & 0 & 0 & 0 & 0 \\
\vph{\frac12} 0 & 0 & 0 & 0 & 0 & 0 \\
 \frac12 & 0 & \frac12 & 0 & 0 & \frac12 
\espm \, , \ &
\bspm
 \frac12 & \frac12 & \frac12 & \frac12 & \frac14 & \frac14 \\
 \frac12 & 0 & 0 & 0 & \frac12 & \frac12 \\
 \frac12 & 0 & \frac12 & \frac12 & \frac34 & \frac34 \\
 \frac12 & 0 & \frac12 & 0 & 0 & 0 \\
 \frac14 & \frac12 & \frac34 & 0 & \frac14 & 0 \\
 \frac14 & \frac12 & \frac34 & 0 & 0 & \frac14 
\espm \, .
\end{array}
\end{equation}
These generate the group%
\footnote{Careful not to confuse the label $A = \Z_2^2$ with the Ext group $\Z_2 \times \Z_2$. 
The elements of the latter are denoted $\{ 00 , 11 , 01 , 10\}$ and appear in tables \ref{tab:summary} and \ref{tab:lattices}.} 
\begin{equation}
    \Ext^1_{\Z W} (R_A, R_A^\v) = \Z_2 \times \Z_2 = \{ 00 , 11 , 01 , 10\} \qquad \text{for} \qquad  A = \Z_2^2   \, .   
\end{equation} 

We proceed similarly for each element of $\cI$ in \eqref{eq:Iso12}, and we find that $\Ext^1_{\Z W} (R_A, R_A^\v)$ is trivial for $A = \Z_2^S$ and $A = \Z_2^C$ while $\Ext^1_{\Z W} (R_A, R_A^\v) = \Z_2$ for $A = \Z_2^V$. 
These results are reported in table \ref{tab:lattices}. 
We then have an explicit construction of $4+2+1+1+1=9$ representations $S_{\cA}$, for $\cA\in \fI$, with 
\begin{eqnarray}
   \fI &=& (\Z_2 \times \Z_2) \sqcup \Z_2 \sqcup \Z_1 \sqcup \Z_1 \sqcup \Z_1 \\ &=& \{ (\Z_2^2 , 00) , (\Z_2^2 , 01) , (\Z_2^2 , 10) , (\Z_2^2 , 11) , (\Z_2^V , 0) , (\Z_2^V , 1) , (\Z_2^S , 0) , (\Z_2^C , 0)  , (\Z_1 , 0) \} \, .  \nonumber
\end{eqnarray}
It might help at this point to recall our notation \eqref{notation}: for example, the $\Z_2\times\Z_2$ in the first line refers to the Ext group, while in the second line in the $(\Z_2^2,ij)$ entries, $\Z_2^2$ is the subgroup, $A=\Z_2^2$, of the center of $\so(12)$ which we are using to label the representation, while $ij$ labels an element of the $\Z_2\times\Z_2$ Ext group.
For instance, $S_{(\Z_2^2, 10)}$ is given by \eqref{symp rep} with $R_A$ the representation of the Weyl group in the basis specified by the first matrix in \eqref{eq:repsso12}, $R_B^\v = R_A^\v$ its inverse transpose, and $D$ the last matrix in \eqref{eq:bindingsso12}.
Hence for instance we have 
\begin{equation}
    S_{(\Z_2^2,10)}(w_1) = \left( 
\scalebox{0.7}{
$\begin{array}{rrrrrr|rrrrrr} 
 -1 & \ph{-}1 & \ph{-1} & \ph{-1} & \ph{-1} & \ph{-1} & \ph{-1} & -1 & -1 & -1 & \ph{-1} & \ph{-1} \\
   & 1 &   &   &   &   & 1 &   &   &   &   &   \\
   &   & 1 &   &   &   & 1 &   &   &   &   &   \\
   &   &   & 1 &   &   & 1 &   &   &   &   &   \\
   &   &   &   & 1 &   &   &   &   &   &   &   \\
   &   &   &   &   & 1 &   &   &   &   &   &   \\ \hline 
   &   &   &   &   &   & -1 &   &   &   &   &   \\
   &   &   &   &   &   & 1 & 1 &   &   &   &   \\
   &   &   &   &   &   &   &   & 1 &   &   &   \\
   &   &   &   &   &   &   &   &   & 1 &   &   \\
   &   &   &   &   &   &   &   &   &   & 1 &   \\
   &   &   &   &   &   &   &   &   &   &   & 1 
\end{array}$ }
\right)\, . 
\end{equation}

\paragraph{Steps 4,5: Untwisted integral (self-)equivalences.}

We know from appendix \ref{app Z rep} that every SK structure has a representative $(S, \bt)$ where $S$ is one of the 9 representations $S_{\cA}$ constructed above. 
Hence, it is enough to compute which of these 9 representations are integrally equivalent. 
This data is contained in the sets $\scS_{\cA_1 \cA_2}$ in \eqref{eq:defSgroups}, which are exactly the intertwiners between $S_{\cA_1}$ and $S_{\cA_2}$. 

\underline{Example 1. } 
Let us begin with $\scS_\cA \deq \scS_{\cA\cA}$ for $\cA = (\Z_2^2 , 00)$. 
The results of section \ref{sec Weyl Z reps} instruct us to consider the map defined by \eqref{eq:defMAA}, which reads here 
\begin{equation}\label{eq:Mso12}
    \cM_{(\Z_2^2 , 00),(\Z_2^2 , 00)} \left( \bspm \ua&\ub \\ \uc&\ud\espm \right) = \bspm  \ua I_6 & \ub K_{\Z_2^2}^\v \\ \uc  K_{\Z_2^2}  &\ud I_6 \espm  
\end{equation}
where $I_6$ is the identity $6 \times 6$ matrix, $K_{\Z_2^2}^\v$ is given in \eqref{eq:KZ2Z2}, and $K_{\Z_2^2}$ is its inverse transpose, which has only integer coefficients (it is the $D_6$ Cartan matrix).  
The set $\scS_{(\Z_2^2, 00)}$ is then the set of $\SL (2,\Z)$ matrices $\bspm \ua&\ub \\ \uc&\ud\espm$ such that \eqref{eq:Mso12} has integer entries. 
Given \eqref{eq:KZ2Z2}, the only constraint is that $\ub$ should be even. 
Hence 
\begin{equation}
    \scS_{(\Z_2^2 , 00)} = \G^0 (2) \, . 
\end{equation}
This result is reported in table \ref{tab:summary}, where we write $\G_0(2)$ and not $\G^0(2)$, but this is irrelevant, as only the group up to conjugation is basis independent.
Since the index of $\G^0 (2)$ in $\SL(2\Z)$ is 3, we say that the symplectic representation $S_{(\Z_2^2, 00)}$ belongs to an orbit of size 3.
This latter information will be useful in the next section when we compare to the field theory S-duality predictions.

\underline{Example 2. } 
Consider now $\scS_{(\Z_2^2 ,11)}$. 
The computation is similar, except that we should now take into account non-trivial bindings in \eqref{eq:defMAA}, namely $D_1 = D_2$ given by the third matrix in \eqref{eq:bindingsso12}. 
A brute force evaluation of \eqref{eq:defMAA} for $\cA_1=\cA_2=(\Z_2^2, \Z_2^2; 11)$ gives for the self-intertwiner
\begin{equation}
\scalebox{.7}{$
    \left(
\begin{array}{cccccccccccc}
 \ua{+}\frac{\uc}{2} & {-}\uc & \uc & 0 & 0 &
   {-}\uc & \ub{-}\frac{5 \uc}{4} &
   \frac{\ua}{2}{+}\ub{-}\frac{\uc}{4}{-}\frac{\ud}{
   2} & \ub{-}\frac{\uc}{4} &
   \frac{\ua}{2}{+}\ub{+}\frac{3
   \uc}{4}{-}\frac{\ud}{2} &
   \frac{\ua}{4}{+}\frac{\ub}{2}{+}\frac{3
   \uc}{8}{-}\frac{\ud}{4} & \frac{3
   \ua}{4}{+}\frac{\ub}{2}{-}\frac{5 \uc}{8}{-}\frac{3
   \ud}{4} \\
 {-}\uc & \ua{+}\frac{\uc}{2} & 0 & \uc &
   {-}\uc & {-}\uc &
   \frac{\ua}{2}{+}\ub{-}\frac{\uc}{4}{-}\frac{\ud}{
   2} & 2 \ub{-}\frac{3 \uc}{2} & 2
   \ub{-}\frac{\uc}{2} & 2 \ub{-}\frac{\uc}{2} &
   \frac{\ua}{2}{+}\ub{-}\frac{\uc}{4}{-}\frac{\ud}{
   2} & \frac{\ua}{2}{+}\ub{-}\frac{5
   \uc}{4}{-}\frac{\ud}{2} \\
 0 & 0 & \ua{+}\frac{\uc}{2} & 0 & {-}\uc & 0 &
   \ub{-}\frac{\uc}{4} & 2 \ub{-}\frac{\uc}{2} & 3
   \ub{-}\frac{3 \uc}{4} & \frac{\ua}{2}{+}3
   \ub{+}\frac{\uc}{4}{-}\frac{\ud}{2} & \frac{3
   \ua}{4}{+}\frac{3 \ub}{2}{+}\frac{\uc}{8}{-}\frac{3
   \ud}{4} & \frac{\ua}{4}{+}\frac{3
   \ub}{2}{+}\frac{\uc}{8}{-}\frac{\ud}{4} \\
 {-}\uc & \uc & {-}\uc & \ua{+}\frac{\uc}{2}
   & 0 & 0 & \frac{\ua}{2}{+}\ub{+}\frac{3
   \uc}{4}{-}\frac{\ud}{2} & 2 \ub{-}\frac{\uc}{2}
   & \frac{\ua}{2}{+}3
   \ub{+}\frac{\uc}{4}{-}\frac{\ud}{2} & 4
   \ub{-}\uc & 2 \ub{-}\frac{\uc}{2} & 2
   \ub{-}\frac{\uc}{2} \\
 0 & 0 & {-}\uc & \uc & \ua{-}\frac{\uc}{2} & 0 &
   \frac{\ua}{4}{+}\frac{\ub}{2}{+}\frac{3
   \uc}{8}{-}\frac{\ud}{4} &
   \frac{\ua}{2}{+}\ub{-}\frac{\uc}{4}{-}\frac{\ud}{
   2} & \frac{3 \ua}{4}{+}\frac{3
   \ub}{2}{+}\frac{\uc}{8}{-}\frac{3 \ud}{4} & 2
   \ub{-}\frac{\uc}{2} & \frac{\ua}{4}{+}\frac{3
   \ub}{2}{-}\frac{7 \uc}{8}{-}\frac{\ud}{4} &
   \ub{-}\frac{\uc}{4} \\
 {-}\uc & 0 & 0 & \uc & 0 & \ua{-}\frac{3 \uc}{2}
   & \frac{3 \ua}{4}{+}\frac{\ub}{2}{-}\frac{5
   \uc}{8}{-}\frac{3 \ud}{4} &
   \frac{\ua}{2}{+}\ub{-}\frac{5
   \uc}{4}{-}\frac{\ud}{2} & \frac{\ua}{4}{+}\frac{3
   \ub}{2}{+}\frac{\uc}{8}{-}\frac{\ud}{4} & 2
   \ub{-}\frac{\uc}{2} & \ub{-}\frac{\uc}{4} &
   \frac{3 \ua}{4}{+}\frac{3 \ub}{2}{-}\frac{15
   \uc}{8}{-}\frac{3 \ud}{4} \\
 2 \uc & {-}\uc & 0 & 0 & 0 & 0 &
   \ud{-}\frac{\uc}{2} & \uc & 0 & \uc & 0 &
   \uc \\
 {-}\uc & 2 \uc & {-}\uc & 0 & 0 & 0 & \uc &
   \ud{-}\frac{\uc}{2} & 0 & {-}\uc & 0 & 0 \\
 0 & {-}\uc & 2 \uc & {-}\uc & 0 & 0 & {-}\uc & 0 &
   \ud{-}\frac{\uc}{2} & \uc & \uc & 0 \\
 0 & 0 & {-}\uc & 2 \uc & {-}\uc & {-}\uc & 0 &
   {-}\uc & 0 & \ud{-}\frac{\uc}{2} & {-}\uc &
   {-}\uc \\
 0 & 0 & 0 & {-}\uc & 2 \uc & 0 & 0 & \uc & \uc
   & 0 & \frac{\uc}{2}{+}\ud & 0 \\
 0 & 0 & 0 & {-}\uc & 0 & 2 \uc & \uc & \uc & 0
   & 0 & 0 & \frac{3 \uc}{2}{+}\ud
\end{array}
\right)$}
\end{equation}
and one can check that this has integer entries if and only if there exist integers $n_1 , n_2 , n_3 , n_4 \in \Z$ such that 
\begin{equation}
    \bspm \ua&\ub \\ \uc&\ud\espm  = \bspm   4 n_1{-}2 n_2{-}2 n_3{-}\frac{3 n_4}{2} & n_2{-}\frac{n_3}{2}{-}\frac{n_4}{4} \\
 2 n_3{-}n_4 & \frac{n_4}{2}  \espm \, . 
\end{equation}
In other words, 
\begin{equation}
    \scS_{(\Z_2^2 , 11)} = \left\{ \bspm 4n_1{-}2n_2{-}2n_3{-}\frac{3n_4}{2}& n_2{-}\frac{n_3}{2}{-}\frac{n_4}{4}\\ 2n_3{-}n_4& \frac{n_4}{2}\espm 
    \ \bigg| \  n_j \in \Z  \ \textrm{and} \, \det = 1 \right\} \, . 
\end{equation}
What is this group? First of all, we can conjugate it to a subgroup of $\SL(2,\Z)$ as follows: 
\begin{align}
   \D &\deq \bspm2&0\vph{\frac12}\\1&\frac12\espm   \scS_{(\Z_2^2,11)} \bspm2&0\vph{\frac12}\\1&\frac12\espm^{-1} = \left\{ \bspm a{+}d&2b{+}d\\2c{+}d&a\espm \ \Big| \  a,b,c,d \in\Z \ \text{and}\ \det = 1 \right\} \, . 
\end{align}
This is a subgroup of $\SL(2,\Z)$ which contains neither $\sS =  \bspm 0 & -1 \\ 1 & 0 \espm$ nor $\sT =  \bspm 1 & 1 \\ 0 & 1 \espm$, but does contain $\sS\sT$. Since $\SL(2,\Z)$ is generated by $\sS\sT$ and $\sS$, this means the group $\D$ has index 2 in $\SL(2,\Z)$. In particular, it is not a congruence subgroup. This implies that the symplectic representation $S_{(\Z_2^2 , 11)}$ belongs to an orbit of size $2$. 

\underline{Example 3. } 
Finally, let us quickly go over the computation of  $\scS_{\cA_1 \cA_2 }$ for $\cA_1 \neq \cA_2$. Proceeding in a similar way as above, we find for instance 
\begin{equation}
    \scS_{(\Z_2^2,00), (\Z_1,0)} = 
    \left\{ \bspm2n_1&n_2\\n_3&n_4\espm \ \Big| \ 
    n_j \in \Z  \ \text{and} \ \det = 1 \right\}  \, ,  \label{eq:Sz21}
\end{equation}
which is one of the three $\G_0 (2)$ cosets, while 
\begin{equation}
    \scS_{(\Z_2^2,00), (\Z_2^C,0)} = 
    \left\{ \bspm2n_1&2n_2\\n_3&n_4\espm \ \Big| \ 
    n_j \in \Z  \ \text{and} \ \det = 1 \right\} = \emptyset \, , 
\end{equation}
which is clearly empty. 
This means 
\begin{equation}
    S_{(\Z_2^2 , 00)} \cong_\Z  S_{(\Z_1,0)} \qquad S_{(\Z_2^2 , 00)} \ncong_\Z  S_{(\Z_2^{C},0)} 
\end{equation}
These computations are summarized in table \ref{tab:summary}. 

The computations of the previous paragraph show that $\cI$ is partitioned into 5 orbits, and for each orbit we can compute a duality group (up to conjugacy), whose index in $\SL(2,\Z)$ give the size of the orbits. 
We summarized all this in the 5 lines of the $D_{4k+2}$ entry in table \ref{tab:summary}. 
In order to fully characterize the SK structures, according to our definition \eqref{eq:DefSKstructure}, one needs to compute the fixed points of the representations, in which $\bt$ varies. 
This is given explicitly in \eqref{bt Fix}. 
For instance, we compute the fixed loci 
\begin{align}
\Fix(S_{(\Z_2^2,00)}) &= \left\{ \t \bspm
 1 & 1 & 1 & 1 & \frac12 & \frac12 \\
\vph{\frac12} 1 & 2 & 2 & 2 & 1 & 1 \\
 1 & 2 & 3 & 3 & \frac32 & \frac32 \\
 1 & 2 & 3 & 4 & 2 & 2 \\
 \frac12 & 1 & \frac32 & 2 & \frac32 & 1 \\
 \frac12 & 1 & \frac32 & 2 & 1 & \frac32 
\espm \right\}, &
\Fix(S_{(\Z_1,0)}) &= \left\{ \t \bspm
\vph{\frac12} 2 & 1 & 2 & -1 & 2 & -4 \\
\vph{\frac12} 1 & 2 & 2 & 0 & 2 & -2 \\
\vph{\frac12} 2 & 2 & 4 & -1 & 4 & -6 \\
\vph{\frac12} -1 & 0 & -1 & 2 & -2 & 4 \\
\vph{\frac12} 2 & 2 & 4 & -2 & 6 & -8 \\
\vph{\frac12} -4 & -2 & -6 & 4 & -8 & 14 \\
\espm \right\} \, ,
\end{align} 
for $\t \in \scH_1$, where we denote the rank-$r$ Siegel upper half space by $\scH_r$.
These results are, of course, consistent with the duality groups we found above. 
For instance, the transformation $\sS \deq \bspm0&-1\\1&0\espm$ belongs to the set \eqref{eq:Sz21}, which is sent via the map $\cM_{(\Z_2^2 , 00),(\Z_1,0)}$ to the symplectic matrix 
\begin{equation}
    \cM_{(\Z_2^2 , 00),(\Z_1,0)} (\sS) = \bspm  0 & - K_{\Z_1 , \Z_2^2}^\v \\    K_{\Z_1 , \Z_2^2} & 0 \espm  \, ,
\end{equation}
and  
\begin{equation}
    \bspm  0 & -K_{\Z_1,\Z_2^2}^\v \\ K_{\Z_1,\Z_2^2} & 0 \espm \circ \Fix(S_{(\Z_2^2,00)}) = \Fix(S_{(\Z_1,0)}) \, , 
\end{equation}
as can be checked explicitly. 

\paragraph{Step 6: Including reflection outer automorphisms of the Weyl group.}

From table \ref{tab:outer} we have $\Out_\rfl(D_6) = \Z_2 \doteq \{0,1\}$, in an additive notation where we denote the trivial (identity) automorphism by ``$0$'' and the non-trivial element by ``$1$''.
The non-trivial element corresponds to the $D_6$ Coxeter diagram symmetry which interchanges the ``spinor'' and ``conjugate spinor'' nodes, leaving the rest unchanged.
As such, it has the effect of interchanging the $\Z_2^S$ and $\Z_2^C$ representations as indicated in table \ref{tab:lattices}.

The effect on the equivalence classes of principally polarized symplectic representations (SK orbits) is not hard to guess.
For those provisional orbits found by looking for integral equivalences, and shown as sets of representations denoted $\{ \dots \}$ in the 2nd column of table \ref{tab:summary}, if an orbit does not include either the $\Z_2^S$ or $\Z_2^C$ representation, then there are no further equivalences under twisted intertwiners.
Most of the orbits for the $D_{4k+2}$ cases shown in table \ref{tab:summary} are of this type.

But if there is an orbit including a $\Z_2^S$ representation and no $\Z_2^C$ representation, then there is a twisted intertwiner identifying it with the corresponding orbit with $S$ and $C$ labels interchanged.
(Such an orbit must exist since interchanging $S$ and $C$ is the expression of the automorphism of the Weyl group.)
In this case those two orbits are equivalent, and so the number of orbits is reduced relative to the number found using only untwisted intertwiners. 
Note that in this case there are no additional self-intertwiners, so the S-duality group of the orbit stays the same relative to the group found using only untwisted intertwiners.
This is the situation described in \eqref{sit 1} in the general discussion of section \ref{sec SK twist Z equiv}.
From table \ref{tab:summary} we see that this situation does not occur for $D_{4k+2}$, but it does occur in the $D_{4k}$ cases.
(The $D_4$ special case will be remarked upon in section \ref{sec LE test}.)

Finally, if there is an orbit including both a $\Z_2^S$ representation and a $\Z_2^C$ representation, then there is a twisted intertwiner identification within the orbit. 
This does not change the number of orbits relative to the number found using only untwisted intertwiners, but does enlarge the S-duality group of the orbit relative to the group found using only untwisted intertwiners.
This is the situation described in \eqref{sit 2} in the general discussion of section \ref{sec SK twist Z equiv}.
From table \ref{tab:summary} we see that this situation does occur for the $\{(\Z_2^S,0), (\Z_2^C,0), (\Z_2^2,10)\}$ orbit of $D_{4k+2}$.

Since $\Z_2^S$ and $\Z_2^C$ are within the same orbit, they are integrally equivalent, and since they are related by the outer automorphism, there is a twisted integral equivalence between them. 
Composing these two equivalences, there is therefore a twisted integral self-equivalence of $\Z_2^S$, i.e., there is an intertwiner between $S_{\Z_2^S}\circ0$ and $S_{\Z_2^S}\circ1$.
This affects the S-duality group of this SK structure orbit, and we illustrate its computation here.

One can also check mechanically for the non-existence of twisted self-intertwiners for all other representations, bearing out the pattern argued above and shown in table \ref{tab:summary}, but we do not give any more details of these checks here.
See the $BC_2$ example in the next subsection where all twisted intertwiners are described explicitly.

Here we just focus on the twisted self intertwiners for the $(\Z_2^S,0)$ representation in the orbit.
Since it has a trivial Ext group, we drop the irrelevant ``$0$'' and denote it as just $\Z_2^S$.
We call the untwisted and twisted versions $S_{\Z_2^S}\circ0=S_{\Z_2^S}$ and $S_{\Z_2^S}\circ1$.
The 6 reflection generators of $\Z_2^S$ are easy to calculate from \eqref{symp rep} in our chosen basis \eqref{eq:repsso12}; we do not list them here since they are so large.
The generators of $S_{\Z_2^C}\circ1$ are the same as those of $S_{\Z_2^S}$ but with the two reflections corresponding to the spinor and conjugate spinor roots interchanged.
Using these representation matrices, or from \eqref{bt Fix}, the fixed locus of their action on $\scH_6$ is the 1-dimensional set
\begin{align}\label{Fix spinor rep}
\Fix(S_{\Z_2^S}) &= \left\{ 2\t K_{\Z_2^S}^\v \ \big|\  \t \in \scH_1\right\} ,& &\text{with}&
K_{\Z_2^S}^\v &= \frac12 \bspm
 3 & 0 & 1 & 2 & -4 & 3 \\
 0 & 4 & 2 & 0 & 4 & -2 \\
 1 & 2 & 3 & 2 & 0 & 1 \\
 2 & 0 & 2 & 4 & -4 & 4 \\
 -4 & 4 & 0 & -4 & 12 & -8 \\
 3 & -2 & 1 & 4 & -8 & 7 \\
\espm \, .
\end{align}

Evaluation of \eqref{eq:defMAA} gives the (untwisted) $\Z_2^S$ self-intertwiners
\begin{align}
    M^0_{\Z_2^S}\left(\bspm a&b\\c&d\espm\right) &= \bspm  a I_6 & b K_{\Z_2^S}^\v \\ c  K_{\Z_2^S}  & d I_6\espm \, .
\end{align}
This is an element of $\Sp(6,\Z)$ iff $\bspm a&b\\c&d\espm\in\SL(2,\Z)$ and $b$ and $c$ are even.
Furthermore, the M\"obius action \eqref{M act} of this intertwiner descends to an action on $\t\in\scH_1$ parameterizing $\Fix(S_\cA)$ in \eqref{Fix spinor rep} given by
\begin{align}\label{PSL2Z spinor}
   \cM^0_{\Z_2^S}\left(\bspm a&b\\c&d\espm\right) \circ \t &= \frac{a\t+b}{c\t+d}, 
\end{align}
which is the usual M\"obius action.
With the conditions following from the integrality of $M^0_{\Z_2^S}$, the transformations on $\scH_1$ form a modular subgroup
\begin{align}
    \scS^0_{\Z_2^S} &\doteq \left\{ \bspm a&b\\c&d\espm \in \SL(2,\Z) \ \text{and $b$ and $c$ even } \right\},
\end{align}
isomorphic to $\G_0(4) \subset \PSL(2,\Z)$.
In appendix \ref{app Hecke} we define $\G_0(4)$ as the subgroup with the lower left entry divisible by 4; it is isomorphic to the subgroup with lower left and upper right entries divisible by 2 by conjugation by the $\SL(2,\R)$ matrix $\bspm 2&0\\0&1/2\espm$.
Recall that $\G_0(4)$ has index 6 in $\SL(2,\Z)$.

We now perform the same calculation but for twisted intertwiners $M^1_{\Z_2^S}$, i.e., integer determinant 1 matrices which intertwine the untwisted representation $S_{\Z_2^S}= S_{\Z_2^S}\circ0$ with the outer automorphism twisted representation $S_{\Z_2^S}\circ1$.
We compute
\begin{align}
M^1_{\Z_2^S}\left(\bspm a&b\\c&d\espm\right) &= 
\bspm
 a & 0 & 0 & 0 & -\frac a2 & -a & b & 0 & 0 & 0 & -b & 0 \\
 0 & a & 0 & 0 & 0 & 0 & 0 & 2 b & b & 0 & 2 b & -b \\
 0 & 0 & a & 0 & -\frac a2 & -a & 0 & b & b & 0 & b & -b \\
 0 & 0 & 0 & a & -a & -2 a & 0 & 0 & 0 & 0 & 0 & -b \\
 0 & 0 & 0 & 0 & 2 a & 2 a & -b & 2 b & b & 0 & 4 b & -b \\
 0 & 0 & 0 & 0 & -\frac{3a}2 & -2 a & 0 & -b & -b & -b & -b & -b \\
 2 c & -c & 0 & 0 & c & 0 & d & 0 & 0 & 0 & 0 & 0 \\
 -c & 2 c & -c & 0 & -c & 0 & 0 & d & 0 & 0 & 0 & 0 \\
 0 & -c & 2 c & -c & 0 & 0 & 0 & 0 & d & 0 & 0 & 0 \\
 0 & 0 & -c & 2 c & 0 & -c & 0 & 0 & 0 & d & 0 & 0 \\
 c & -c & 0 & 0 & c & 0 & -\frac d2 & 0 & -\frac d2 & -d & 2 d & -\frac{3d}2 \\
 0 & 0 & 0 & -c & 0 & 0 & -d & 0 & -d & -2d & 2d & -2d
\espm ,
\end{align}
which is in $\Sp(6,\Z)$ iff $\bspm a&b\\c&d\espm\in\SL(2,\Z)$ and $a$ and $d$ are even.
It also descends to an action on $\t\in\scH_1$ given by the usual M\"obius action \eqref{PSL2Z spinor}.
This set
\begin{align}
    \scS^1_{\Z_2^S} &\doteq \left\{ \bspm a&b\\c&d\espm \in \SL(2,\Z) \ \text{and $a$ and $d$ even } \right\},
\end{align}
does \emph{not} form a group.

But the union of these two sets of identifications on the conformal manifold of the $\Z_2^S$ SK structure orbit,
\begin{align}\label{G02 version}
    \scS_{\Z_2^S} &\doteq \scS^0_{\Z_2^S} \cup \scS^1_{\Z_2^S} 
    = \left\{ \bspm a&b\\c&d\espm \in \SL(2,\Z) \ \text{and $a$ and $d$ even, or $b$ and $c$ even} \right\},
\end{align}
\emph{does} form a group, as is easy to check.
Since it is twice the size of $\scS^0_{\Z_2^S} \cong \G_0(4)$ which has index 6 in $\SL(2,\Z)$, it follows that $\scS_{\Z_2^S}$ must have index 3, and so it must be $\cong \G_0(2)$ since that is the only index 3 modular subgroup.
More explicitly, conjugating the group \eqref{G02 version} in $\SL(2,\Z)$ by $\bspm1&0\\1&1\espm$ takes it to $\G_0(2)$.
This shows that, upon including outer automorphism twisted intertwiners, the S-duality group of the $\{(\Z_2^S,0), (\Z_2^C,0), (\Z_2^2,10)\}$ orbit of $D_6$ is $\G_0(2)$, as claimed in table \ref{tab:summary}.

\subsection{$\so(5) = \sp(4)$ example} 
\label{BC2example}

\noindent \tbf{Steps 2-5.} In this subsection, as in the previous one, we carry out the explicit computations for $W=BC_2$. 
There are two lattices to consider, which we call the root lattice and the weight lattice, defined by the following matrices (as above, the basis vectors are the columns of the matrices): 
\begin{equation}
\begin{array}{c|c|c|c} 
\text{ Lattice } & \text{ Symbol } & \text{ Construction } & \text{Basis} \\ \hline 
\G_{\text{root}} & \Z_2 & \Z^2|_{\S \in 2\Z} & \bspm1&0\\-1&2\espm \\
\G_{\text{weight}} & \Z_2^\v & \Z^2 & \bspm1&0\\-1&1\espm
\end{array} \ .
\end{equation}
Hence $\cI = \{\Z_2, \Z_2^\v\}$. 
The Ext group computations follows exactly the same lines as in the example in section \ref{sec:so12}. 
We find $\Ext^1_{\Z W} (R_A, R_A^\v) = \Z_2$ for both $A = \Z_2, \Z_2^\v$, so 
\begin{equation}
\fI = \Z_2 \sqcup \Z_2 = \left\{ (\Z_2, 0), (\Z_2, 1), (\Z_2^\v, 0), (\Z_2^\v, 1) \right\} \, . 
\end{equation}
The explicit matrix form of the generating reflections in all these cases is
\begin{equation}
\begin{array}{c|c|c}
\cA \in \fI & S_\cA(w_1) & S_\cA(w_2) \\
\hline 
\rule{0pt}{20pt} \ (\Z_2, 0) \ \ & 
\ \bspm-1&2&0&0\\ 0&1&0&0\\ 0&0&-1&0\\ 0&0&2&1\espm \ \ & 
\ \bspm 1&0&0&0\\ 1&-1&0&0\\ 0&0&1&1 \\ 0&0&0&-1\espm \ \
\\[10pt] 
(\Z_2, 1) & 
\bspm-1&2&0&1\\ 0&1&-1&0\\ 0&0&-1&0\\ 0&0&2&1\espm &
\bspm 1&0&0&0\\ 1&-1&0&0\\ 0&0&1&1\\ 0&0&0&-1\espm 
\\[10pt] 
(\Z_2^\v, 0) & 
\bspm-1&1&0&0\\ 0&1&0&0\\ 0&0&-1&0\\ 0&0&1&1\espm &
\bspm 1&0&0&0\\ 2&-1&0&0\\ 0&0&1&2\\ 0&0&0&-1\espm 
\\[10pt]
(\Z_2^\v, 1) & 
\bspm-1&1&0&0\\ 0&1&0&0\\ 0&0&-1&0\\ 0&0&1&1\espm &
\bspm 1&0&0&-1\\ 2&-1&1&0\\ 0&0&1&2\\ 0&0&0&-1\espm  
\end{array}
\end{equation}
The bindings can be read from the top right $2 \times 2$ matrices. 

The sets of $\bt \in \scH_2$ fixed by these representations is easily computed.
For instance, we find
\begin{align}\label{FixZ20}
    \Fix(S_{(\Z_2,1)}) &=
    \bigg\{ \t \bspm 1&\frac12\\ \frac12&\frac12\espm - \bspm 0&0\vph{\frac12}\\0&\frac12\espm \, ,\,  \t\in\scH_1\bigg\} \ .
\end{align}

From here we can compute the duality groups. 
The intertwiners $\cM_{\cA\cB}$ defined in \eqref{M prin int} and \eqref{eq:defMAA} are given in the untwisted section of table \ref{BC2 ints}. 
The duality groups \eqref{eq:dualityGroupS} are therefore defined by requiring that these matrices have integer entries.
For instance, the M\"obius action \eqref{M act} of $\cM^0_{\cA\cA}$ descends to a M\"obius action on $\t\in\scH_1$ parameterizing $\Fix(S_\cA)$ given by
\begin{align}\label{PSL2Z0}
   \cM^0_{\cA\cA}\left( \bspm a&b\\c&d\espm \right) \circ \t &= \frac{a\t+b}{c\t+d}, 
\end{align}
where, from table \ref{BC2 ints}, $\bspm a&b\\c&d\espm \in \SL(2,\Z)$ and $b$ is even in the $\cA=(\Z_2,1)$ case and $c$ is even in the $\cA=(\Z_2^\v,1)$ case.
Hence the provisional duality groups are isomorphic to $\G_0(2)$ for $\cA = (\Z_2, 1)$, $(\Z_2^\v, 1)$, while the provisional duality group for $\cA = (\Z_2, 0)$, $(\Z_2^\v, 0)$ is $\PSL(2,\Z)$. 
(These modular groups are reviewed in appendix \ref{app Hecke}.)
There also exist integral intertwiners between $\cA=(\Z_2,1)$ and $\cB=(\Z_2^\v,1)$, shown in the last untwisted row of table \ref{BC2 ints}.
There are no solutions for integral untwisted intertwiners between any other pair of symplectic representations.

\begin{table}[ht]
\centering
\begin{tabular}{|c|c|c|}
\hline
\rule{0pt}{6mm} untwisted intertwiners $\cA\to\cB$ & $\cM^{\f=0}_{\cA\cB}$ & for $ad{-}bc{=}1$ and \\[2mm]  
\hline \hline
\rule{0pt}{20pt} \raisebox{-.5\height}{%
\begin{tikzpicture}
\node (1) at (0,2) {$(\Z_2, 0)$};
\draw[->,very thick] (1) to[loop, in=150, out=210, looseness=4] node[left=2pt] {$\PSL(2,\Z)$} (1);
\end{tikzpicture}} & 
$\bspm a&0&2b&b\\ 0&a&b&b\\ c&-c&d&0\\ -c&2c&0&d\espm$ & 
$a,b,c,d \in\Z$
\\[6mm] 
\raisebox{-.5\height}{%
\begin{tikzpicture}
\node (1) at (0,2) {$(\Z_2,1)$};
\draw[->,very thick] (1) to[loop, in=150, out=210, looseness=4] node[left=2pt] {$\G_0(2)$} (1);
\end{tikzpicture}} &
$\bspm a&0&b&\frac{b}{2}\\ c&a{-}2c&\frac b2&\frac a2{+}\frac b2{-}c{-}\frac d2\\ 2c&-2c&d&-c\\ -2c&4c&0&2c{+}d\espm$ & 
$a,c,d \in\Z$, $b\in2\Z$ 
\\[6mm] 
\raisebox{-.5\height}{%
\begin{tikzpicture}
\node (2) at (4,2) {$(\Z_2^\v,0)$};
\draw[->,very thick] (2) to[loop, in=30, out=330, looseness=4] node[right=2pt] {$\PSL(2,\Z)$}(2);
\end{tikzpicture}} & 
$\bspm a&0&b&b\\ 0&a&b&2b\\ 2c&-c&d&0\\ -c&c&0&d\espm$ & 
$a,b,c,d\in\Z$
\\[6mm]
\raisebox{-.5\height}{%
\begin{tikzpicture}
\node (2) at (4,2) {$(\Z_2^\v,1)$};
\draw[->,very thick] (2) to[loop, in=30, out=330, looseness=4] node[right=2pt] {$\G_0(2)$}(2);
\end{tikzpicture}} & 
$\bspm a{-}c&\frac{c}{2}&\frac{a}{2}{+}b{-}\frac{c}{2}{-}\frac{d}{2}&b\\ 0&a&b&2b\\ 2c&-c&c{+}d&0\\ -c&c&-\frac{c}{2}&d\espm$ &
$a,b,d\in\Z$, $c\in2\Z$ 
\\[6mm]
\raisebox{-.5\height}{%
\begin{tikzpicture}
\node (3) at (0,0) {$(\Z_2,1)$};
\node (4) at (4,0) {$(\Z_2^\v,1)$}; 
\draw[->,very thick] (3) -- (4);
\end{tikzpicture}} & 
$\bspm a{-}c&c&b{-}\frac{d}{2}&\frac{b}{2}{+}\frac{c}{2}\\ 0&2a&b&a{+}b\\ 2c&-2c&d&-c\\ -c&2c&0&c{+}\frac{d}{2}\espm$ &
$a,b,c\in\Z$, $d\in2\Z$ \\[6mm]
\hline \hline
\rule{0pt}{6mm} \teal{twisted} intertwiners $\cA\circ0 \teal{\to} \cB\circ1$ & $\cM^{\f=1}_{\cA\cB}$ & for $ad{-}bc{=}1$ and \\[2mm]
\hline \hline
\rule{0pt}{20pt} \raisebox{-.5\height}{\begin{tikzpicture}
\node (1) at (0,2) {$(\Z_2,0)$};
\node (2) at (4,2) {$(\Z_2^\v,0)$};
\draw[->, teal, very thick] (1) -- (2);
\end{tikzpicture}} & 
$\bspm0&a&b&2b\\ a&0&b&b\\ -c&c&0&d\\ 2c&-c&d&0\espm$ & $a,b,c,d \in\Z$ 
\\[6mm] 
\raisebox{-.5\height}{\begin{tikzpicture}
\node (3) at (0,0) {$(\Z_2,1)$}; 
\draw[->,teal, very thick] (3) to[loop, in=160, out=200, looseness=3] (3);
\end{tikzpicture}} & 
$\tfrac1{\sqrt2} \bspm0&2a&b&a{+}b\\ a{-}2c&2c&b{-}d&\frac b2{+}c\\ -2c&4c&0&2c{+}d\\ 4c&-4c&2d&-2c\espm$
& $a,b,d\in\sqrt2\Z$, $c\in \frac{\Z}{\sqrt2}$ 
\\[6mm]
\raisebox{-.5\height}{\begin{tikzpicture}
\node (4) at (4,0) {$(\Z_2^\v,1)$}; 
\draw[->,teal, very thick] (4) to[loop, in=20, out=340, looseness=3] (4);
\end{tikzpicture}} & 
$\tfrac1{\sqrt2} \bspm c&a{-}c&b{+}\frac c2&2b{-}d\\ 2a&0&a{+}2b&2b\\ -2c&2c&-c&2d\\ 2c&-c&c{+}d&0\espm$
& $a,c,d\in\sqrt2\Z$, $b\in \frac{\Z}{\sqrt2}$
\\[6mm]
\raisebox{-.5\height}{\begin{tikzpicture}
\node (3) at (0,0) {$(\Z_2, 1)$};
\node (4) at (4,0) {$(\Z_2^\v, 1)$}; 
\draw[->,teal, very thick] (3) -- (4);
\end{tikzpicture}} & 
$\tfrac1{\sqrt2} \bspm0&a&b&2b\\ a{-}2c&c&\frac a2{+}b{-}c{-}d&b\\ -2c&2c&-c&2d\\ 4c&-2c&2c{+}2d&0\espm$
& $a,b,c\in\sqrt2\Z$, $d\in \frac{\Z}{\sqrt2}$
\\[6mm]\hline
\end{tabular}
\caption{The $\Sp(4,\Z)$ matrices $\cM_{\cA\cB}^{\f\in\{0,1\}}$ for untwisted and twisted intertwiners among the four symplectic representations $\cA,\cB \in \{ (\Z_2,0), (\Z_2,1), (\Z_2^\v,0), (\Z_2^\v,1)\}$ of Weyl($BC_2$).
In column $1$, the $\lcirclearrowright$ ($\teal{\lcirclearrowright}$) arrows indicate self-intertwiners between symplectic representations $\cA = \cB$. 
}
\label{BC2 ints}
\end{table}

The untwisted rows of table \ref{BC2 ints} thus partition the set of representations $\cA$ into provisional equivalence classes, or orbits:
\begin{equation}\label{BC2 untwisted}
\raisebox{-.5\height}{\begin{tikzpicture}
\fill[gray!15] (-0.6,1.) rectangle (.6,1.5);
\fill[gray!15] (1.9,1) rectangle (3.1,1.5);
\fill[gray!15] (-0.6,-.25) rectangle (3.1,.25);
\node (1) at (0,1.25) {\small$(\Z_2,0)$};
\node (2) at (2.5,1.25) {\small$(\Z_2^\v,0)$};
\node (3) at (0,0) {\small$(\Z_2,1)$};
\node (4) at (2.5,0) {\small$(\Z_2^\v,1)$}; 
\draw[very thick] (3) to[loop,in=150,out=210,looseness=4] node[left=2pt] {\small$\G_0(2)$} (3);
\draw[very thick] (1) to[loop,in=150,out=210,looseness=4] node[left=2pt] {\small$\PSL(2,\Z)$} (1);
\draw[very thick] (2) to[loop,in=30,out=330,looseness=4] node[right=2pt] {\small$\PSL(2,\Z)$} (2);
\draw[very thick] (4) to[loop,in=30,out=330,looseness=4] node[right=2pt] {\small$\G_0(2)$} (4);
\draw[very thick] (3) -- (4);
\end{tikzpicture}}
\end{equation}
The gray rectangles each indicates a provisional SK orbit.
We write the provisional duality groups on the lines connecting a representation to itself. 
There are thus three orbits: two with provisional duality group $\PSL(2,\Z)$ and one with provisional duality group $\G_0(2)$.
For the line connecting distinct representations within the $\G_0(2)$ orbit, the set of intertwiners is not a group, but is explicitly described in table \ref{BC2 ints}.

\noindent \tbf{Step 6: Including reflection outer automorphisms of the Weyl group.} 

The orbits and duality groups in the last paragraph were only provisional because the discussion has so far omitted outer automorphisms. 
For $BC_2$ the reflection outer automorphism group is $\Z_2$ whose elements we  denote in an additive notation as $\f\in\{0,1\} = \Z_2$.
The $\f=0$ superscript in table \ref{BC2 ints} refers to the fact that those intertwiners involve no outer automorphism twist, meaning that they are twisted by the identity element $\f\deq0$ of the outer automorphism group.
If we look for intertwiners between untwisted representations $\cA\circ(\f=0)$ and their twisted versions, $\cA\circ(\f=1)$, we find the additional intertwiners listed in the twisted part of table \ref{BC2 ints}.

The actions of these intertwiners and the resulting self-duality groups can be represented schematically as
\begin{equation}
\raisebox{-.5\height}{%
\begin{tikzpicture}[xscale=0.9,yscale=0.9]
\fill[gray!15] (-0.8,-.25) rectangle (3.35,1.5);
\node (3) at (0,0) {$\scriptstyle (\Z_2,1)\circ0$};
\node (4) at (2.5,0) {$\scriptstyle (\Z_2^\v,1)\circ0$};
\node (3a) at (0,1.25) {$\scriptstyle (\Z_2,1)\circ1$};
\node (4a) at (2.5,1.25) {$\scriptstyle (\Z_2^\v,1)\circ1$}; 
\draw[very thick] (3) to[loop,in=150,out=210,looseness=4] node[left=-2pt] {$\scriptstyle \G_0(2)$} (3);
\draw[very thick] (4) to[loop,in=30,out=330,looseness=4] node[right=-2pt] {$\scriptstyle \G_0(2)$} (4);
\draw[very thick] (3a) to[loop,in=150,out=210,looseness=4] node[left=-2pt] {$\scriptstyle \G_0(2)$} (3a);
\draw[very thick] (4a) to[loop,in=30,out=330,looseness=4] node[right=-2pt] {$\scriptstyle \G_0(2)$} (4a);
\draw[very thick] (3) -- (4);
\draw[very thick] (3a) -- (4a);
\draw[teal, very thick] (4) to (4a);
\draw[teal, very thick] (3) to (3a);
\draw[teal, very thick, transform canvas={yshift=-.1cm}] (3) -- (4a);
\draw[teal, very thick, transform canvas={yshift=-.1cm}] (3a) -- (4);

\fill[gray!15] (-0.8,2.25) rectangle (3.35,4.);
\node (1) at (0,2.5) {$\scriptstyle (\Z_2,0)\circ0$};
\node (2) at (2.5,2.5) {$\scriptstyle (\Z_2^\v,0)\circ0$};
\node (1a) at (0,3.75) {$\scriptstyle (\Z_2,0)\circ1$};
\node (2a) at (2.5,3.75) {$\scriptstyle (\Z_2^\v,0)\circ1$};=
\draw[teal, very thick] (1) -- (2a);
\draw[teal, very thick] (1a) -- (2);
\draw[very thick] (1) to[loop,in=150,out=210,looseness=4] node[left=-2pt] {$\scriptstyle \PSL(2,\Z)$} (1);
\draw[very thick] (2) to[loop,in=30,out=330,looseness=4] node[right=-2pt] {$\scriptstyle \PSL(2,\Z)$} (2);
\draw[very thick] (1a) to[loop,in=150,out=210,looseness=4] node[left=-2pt] {$\scriptstyle \PSL(2,\Z)$} (1a);
\draw[very thick] (2a) to[loop,in=30,out=330,looseness=4] node[right=-2pt] {$\scriptstyle \PSL(2,\Z)$} (2a);

\fill[gray!15] (7.4,2.9) rectangle (11.1,3.4);
\fill[gray!15] (7.4,.4) rectangle (11.1,0.9);
\node (c1) at (8,3.15) {$\scriptstyle (\Z_2,0)$};
\node (c2) at (10.5,3.15) {$\scriptstyle (\Z_2^\v,0)$};
\node (c3) at (8,.65) {$\scriptstyle (\Z_2,1)$};
\node (c4) at (10.5,.65) {$\scriptstyle (\Z_2^\v,1)$}; 
\draw[very thick] (c3) to[loop,in=150,out=210,looseness=4] node[left=-2pt] {$\scriptstyle \color{teal} H_4$} (c3);
\draw[very thick] (c4) to[loop,in=30,out=330,looseness=4] node[right=-2pt] {$\scriptstyle \color{teal} H_4$} (c4);
\draw[teal, very thick] (c4) to[loop,in=20,out=340,looseness=3] (c4);
\draw[teal, very thick] (c3) to[loop,in=160,out=200,looseness=3] (c3);
\draw[very thick, transform canvas={yshift=.1cm}] (c3) -- (c4);
\draw[teal, very thick, transform canvas={yshift=-.1cm}] (c3) -- (c4);
\draw[teal, very thick] (c1) -- (c2);
\draw[very thick] (c1) to[loop,in=150,out=210,looseness=4] node[left=-2pt] {$\scriptstyle \PSL(2,\Z)$} (c1);
\draw[very thick] (c2) to[loop,in=30,out=330,looseness=4] node[right=-2pt] {$\scriptstyle \PSL(2,\Z)$} (c2);

\node at (4.75,3.15) {\Large\red{$\Rightarrow$}};
\node at (4.75,.65) {\Large\red{$\Rightarrow$}};
\end{tikzpicture}} \nn
\end{equation}
The gray rectangles are to aid the eye in seeing which representations are equivalent (related by untwisted or twisted intertwiners) and so belong to the same orbit.
The left columns show both the representations $\cA\circ0$ and their outer automorphism twisted counterparts, $\cA\circ1$.
But, since a representation $\cA=\cA\circ0$ and its automorphism twisted version, $\cA\circ1$, describe isomorphic SK structures $(\cA\circ0) \red{\cong} (\cA\circ 1)$, they must be identified.%
\footnote{This isomorphism is proven in appendix \ref{appA: equivalenceofN=4SKstructures}.}
This isomorphism induces additional equivalences among the provisional $3$ orbits shown in \eqref{BC2 untwisted}.
Diagrammatically, this is represented by identifying the two rows of gray rectangle to one row and is indicated by the \red{red} arrows.
The resulting orbits of representations and their (self-)equivalences are shown in the right column.
We see that of the provisional 3 orbits shown in \eqref{BC2 untwisted}, the two with provisional self-duality group $\PSL(2,\Z)$ are identified by a twisted intertwiner and become a single orbit with unchanged self-duality group $\PSL(2,\Z)$.
This is the situation \eqref{sit 1} described in the general  discussion of section \ref{sec SK twist Z equiv}.
The third provisional orbit in \eqref{BC2 untwisted}, with provisional self-duality group $\G_0(2)$, remains an orbit of the same two representations, but now has the enlarged self-duality group $H_4$ by virtue of additional twisted self-intertwiners.
This is the situation described earlier in \eqref{sit 2} of section \ref{sec SK twist Z equiv}.

It may be useful to describe explicitly how $H_4$ arises from the $\cM^\f_{\cA\cA}$ intertwiners shown in table \ref{BC2 ints} with $\cA=(\Z_2,1)$.
We have already remarked that the $\cM^{\f=0}_{\cA\cA}$ action on $\Fix(\cA)$ \eqref{FixZ20} descends to the action \eqref{PSL2Z0} on $\t\in\scH_1$ by $\PSL(2,\Z)$ matrices.
Noting that $\Fix(\cA\circ0) = \Fix(\cA\circ1)$ (almost by definition), a similar calculation shows that the $\cM^{\f=1}_{\cA\cA}$ action on $\Fix(\cA)$ also descends to the action \eqref{PSL2Z0} on $\t\in \scH_1$ but by matrices in $\SL(2,\R)$.
(We chose the parameterization of the intertwiners in table \ref{BC2 ints} just so that this would be the case.)
Thus, we have computed
\begin{align}\label{BC2 H_4 orbit}
\scS_\cA^0 \deq (\cM^{\f=0}_{\cA\cA})^{-1} (\Sp(4,\Z)) &= 
\left\{ \bspm a&b\\c&d\espm \in \SL(2,\R) 
\ \Big|\ a,c,d \in \Z, b\in 2\Z  \right\} 
\cong \G^0(2), \\
\scS_\cA^1 \deq (\cM^{\f=1}_{\cA\cA})^{-1}(\Sp(4,\Z)) &= 
\left\{ \bspm a'&b'\\c'&d'\espm \in\SL(2,\R) 
\ \Big|\ a',b',d'\in \sqrt2\Z, c'\in\tfrac{1}{\sqrt2}\Z \right\} .\nn
\end{align}
As indicated in \eqref{BC2 H_4 orbit}, $\scS_{\cA}^0$ forms the congruence subgroup $\G^0(2)\subset \SL(2,\Z)$, but $\scS^1_\cA$ by itself is not a group.
Nonetheless, their union $\scS^0_\cA \cup \scS^1_\cA$ is a group generated by $\bspm1&0\\1&1\espm$ and $\bspm0&\sqrt2\\-1/\sqrt2&0\espm$, which is isomorphic to the Hecke group $H_4$, reviewed in appendix \ref{app Hecke}.

\section{SK structure as a low-energy test of S-duality of N=4 sYM}
\label{sec LE test}

A set of sharp S-duality conjectures for absolute $\cN{=}4$ sYM theories were proposed and worked out in \cite{Kapustin:2005py, Aharony:2013hda, Bergman:2022otk}.
These specify which of the possible global forms of the gauge group --- decorated with some further discrete labels (the ``discrete theta angles'' of \cite{Aharony:2013hda}) --- occur as weak coupling limits of a connected conformal manifold.
These connected pieces of the conformal manifold are the called \emph{S-duality orbits of global structures} of the sYM theory.
Furthermore, the group of self identifications of the conformal manifold of each global structure orbit are also predicted to be certain Hecke subgroups.
A summary of these predictions is given in the first three columns of table \ref{tab:RBTL}.

\begin{table}[ht]
\centering
\begin{tabular}{|c|c|c|c|}
\hline 
$\fg$ & 
S-duality orbits of global structures & \begin{tabular}{c}S-duality\\[-2mm] group\end{tabular} & \begin{tabular}{c}Index\\[-2mm] in $H_{q(\ell)}$\end{tabular}\\ 
\hline\hline 
$A_1$ & $\{\SU(2),\SO(3)_\pm\}$ & \red{$\G_0(2)$} & \red3\\ 
\hline 
$A_r\ (r{\ge}2)$ & one S-duality orbit & \multirow{2}{*}{$\G_0(N/s^2)$} & \multirow{2}{*}{$\frac{N}{s^2}\cdot \underset{{\scriptscriptstyle p\,|\,\frac{N}{s^2}}}{\prod} \big(1{+}\frac1p\big)$} \\
$N\doteq r{+}1$ & for each $s^2|N$ & & \\
\hline
\multirow{2}{*}{$B_2=C_2$} 
& $\{\SO(5)_-\}$ & $H_4$ & 1\\
& $\{\Sp(4), \SO(5)_+\}$  & \red{$\G_0(2)$} & \red2\\  
\hline 
$B_{2k} / C_{2k}$ & $\{(\Sp(N)/\Z_2)_+, \Spin(N{+}1) \}$ & $\G_0(2)$ & 2\\ 
$(k{\ge}2)$ & $\{(\Sp(N)/\Z_2)_-, \SO(N{+}1)_-\}$ & $\G_0(2)$ & 2\\ 
$N\doteq4k$ & $\{\Sp(N),\SO(N{+}1)_+\}$ & \red{$\G_0(2)$} & \red2\\ 
\hline 
$B_{2k+1} / C_{2k+1}$ & $\{(\Sp(N)/\Z_2)_\pm, \Spin(N{+}1), \SO(N{+}1)_-\}$ & $\G_0(4)$ & 4\\ 
$N\doteq 4k{+}2$ & $\{\Sp(N),\SO(N{+}1)_+\}$ & \red{$\G_0(2)$} & \red2\\ 
\hline 
& $\{ \Spin(N),(\SO(N)/\Z_2)_{\bsm\pm+\\[-1.5pt]+\pm\esm} \}$ & $\G_0(2)$ & 3\\  
& $\{ \SO(N)_-,(\SO(N)/\Z_2)_{\bsm\pm-\\[-1.5pt]-\pm\esm} \}$ & $\G_0(2)$ & 3\\  
$D_{4k}$ & \red{$\{ \Ss(N)_-,(\SO(N)/\Z_2)_{\bsm\pm+\\[-1.5pt]-\pm\esm} \}$} & $\G_0(2)$ & 3\\  
$N \doteq 8k$ & \red{$\{ \Sc(N)_-,(\SO(N)/\Z_2)_{\bsm\pm-\\[-1.5pt]+\pm\esm} \}$} & $\G_0(2)$ & 3\\ 
& $\{\SO(N)_+\}$ & $\PSL(2,\Z)$ & 1\\  
& \red{$\{\Ss(N)_+\}$} & $\PSL(2,\Z)$ & 1\\  
& \red{$\{\Sc(N)_+\}$} & $\PSL(2,\Z)$ & 1\\ 
\hline 
& $\{ \Spin(N),(\SO(N)/\Z_2)_{\bsm\pm+\\[-1.5pt]+\pm\esm} \}$ & $\G_0(2)$ & 3\\  
$D_{4k+2}$ & $\{ \SO(N)_-,(\SO(N)/\Z_2)_{\bsm\pm-\\[-1.5pt]-\pm\esm} \}$ & $\G_0(2)$ & 3\\  
$N \doteq 8k{+}4$ & $\{ \Ss(N)_\pm, \Sc(N)_\pm, (\SO(N)/\Z_2)_{\bsm\pm+\\[-1.5pt]+\mp\esm} \}$ & \red{$\G_0(4)$} & \red6\\  
& $\{ (\SO(N)/\Z_2)_{\bsm\pm-\\[-1.5pt]-\mp\esm} \}$ & $\D$ & 2\\  
& $\{\SO(N)_+\}$ & $\PSL(2,\Z)$ & 1\\ 
\hline 
$D_{2k+1}$ & $\{\Spin(N),\SO(N)_-,(\Spin(N)/\Z_2)_{1,2,3,4}\}$ & $\G_0(4)$ & 6\\  
$N\doteq4k{+}2$ & $\{\SO(N)_+\}$  & $\PSL(2,\Z)$ & 1\\ 
\hline 
$E_6$ & $\{ E_6, (E_6)_{1,2,3} \}$ & $\G_0(3)$ & 4\\ 
\hline 
$E_7$ & $\{ E_7, (E_7)_\pm \}$ & $\G_0(2)$ & 3\\ 
\hline 
$E_8$ & $\{ E_8 \}$ & $\PSL(2,\Z)$ & 1\\ 
\hline 
$F_4$ & $\{ F_4 \}$ & $H_4$ & 1\\ 
\hline 
$G_2$ & $\{ G_2 \}$ & $H_6$ & 1\\ 
\hline 
\end{tabular}
\caption{The first 3 columns summarize the S-duality orbit and group predictions from field theory  \cite{Aharony:2013hda, Bergman:2022otk}.
The red entries highlight the places where these predictions disagree with our classification of SK structures given in table \ref{tab:summary}.
The $\pm$ signs in the subscripts for the $\SO(N)/\Z_2$ global structures are correlated; thus, for example, $(\SO(N)/\Z_2)_{\bsm\pm+\\[-1.5pt]+\mp\esm}$ denotes two global structures, not four.
The last column gives the index of the S-duality group as a subgroup of the Hecke group $H_q$ where $q = 3,4,6$ is determined \eqref{ell2q} by the lacing number, $\ell = 1,2,3$, respectively, of the gauge algebra $\fg$.
It counts the number of global structures in each orbit.
Note that $H_3 =\PSL(2,\Z)$.}
\label{tab:RBTL}
\end{table}

A comparison of this table with our results for SK structures given in table \ref{tab:summary} shows broad agreement between the S-duality groups of SK structure orbits and of global structure orbits, and a looser qualitative agreement between our listing of SK structure representations and global structures of gauge groups.
Indicated in red in both tables are the entries where either S-duality groups do not agree, or where the counting of the number of orbits do not agree.

In this section we discuss these discrepancies and argue that
\begin{enumerate}
    \item The disagreement of the S-duality groups for certain orbits in the $BC_r$ series (including $A_1$) is an ``IR accident'': the SK structure is a low energy observable which, in these cases, simply fails to distinguish between inequivalent field theories.
    Indeed, it is a striking and unexpected fact that in so many cases the SK structure successfully distinguishes $\cN{=}4$ sYM theories, whereas there are many other examples of $\cN{=}2$ theories where the scale invariant CB geometry fails to distinguish them.

    \item The counting of orbits disagreement in the $D_{4k}$ cases, as well as the disagreement of the S-duality group for an orbit of the $D_{4k+2}$ cases, are due to an error in the field theory literature: the SK structure computation gives the correct answer.
    
    \item Finally, the pattern observed that, in general, the number of field theory global structures in a given orbit is greater than the number of SK structure representations in the corresponding orbit, is to be expected, and does not indicate a physical contradiction.
    Instead, only a correspondence between equivalence classes (orbits) of SK structures and S-duality orbits of global structures is expected.
\end{enumerate}

Before addressing these three points, we pause to emphasize the logic of comparing low energy effective actions on the moduli space of vacua (SK structures) to S-duality conjectures in the field theory.
Field theory S-duality conjectures the exact equivalence between CFTs described by various weak coupling limits of their exactly marginal coupling.
These conjectures are supported, of course, by an overwhelming amount of evidence by now, but have not been proven in field theory.
A way of testing these conjectures is to compute all inequivalent moduli space geometries consistent with unbroken $\cN{=}4$ supersymmetry without making assumptions about S-duality properties or about the spectrum of massive BPS states out on the moduli space.
But mismatches between moduli space geometries and field theories can occur for two reasons:
\begin{itemize}
    \item moduli space geometries fail to distinguish inequivalent field theories, or
    \item moduli space geometries exist which do not arise from any (known) field theory.
\end{itemize}
We limit the second possibility by putting a few additional constraints on the geometries.
These are:
\begin{enumerate}
    \item[(i)] Do not allow complex singularities on a CB slice of the $\cN{=}4$ moduli space.
    \item[(ii)] Assume the existence of at least one free field theory point with a simple gauge group on the conformal manifold of the structure.
    \item[(iii)] Restrict to principally polarized SK structures.
\end{enumerate}
These then restrict to the Weyl group orbifolds studied in this paper, as explained in appendix \ref{appA: N4 SKS}.

Constraint (i) is imposed since it is known that all $\cN{=}4$ sYM theories have such CBs.
Though there also exist CBs of $\cN{=}4$ sYM theories with complex singularities \cite{Bourget:2018ond, Argyres:2018wxu}, they all arise from gauging certain discrete 0-form symmetries of theories without such singularities. 
This constraint eliminates these geometries as well as other possible ``exotic'' singular geometries compatible with $\cN{=}4$ supersymmetry, but possessing no known field theory origin.

Constraint (ii) is imposed since all $\cN{=}4$ sYM theories have an exactly marginal coupling with, by definition of having a Lagrangian description, a weak coupling limit.
From the discussion in appendix \ref{appA: SK Geo review} and \ref{appA: N4 SKS}, the K\"ahler structure of an SK geometry which satisfies the constraint (i) and which has a weak coupling point on its conformal manifold where it is described by a sYM lagrangian with simple gauge group $\fg$, will be that of a Weyl group orbifold.
By adopting this constraint we are eliminating geometries that could correspond to possible ``exotic'' $\cN{=}4$ SCFTs which have no weak coupling limits \cite{Papadodimas_2010}.

The third and final restriction, (iii), derives from wanting to compare to absolute $\cN{=}4$ theories.
Absolute $4d$ gauge theories, such as $\cN{=}4$ sYM theories, are defined by a choice of maximal mutually local probe lines \cite{Gaiotto:2010be, Aharony:2013hda}, meaning that the Dirac pairing between them is principal.%
\footnote{This global data is often referred to as the global structure or global variant of the gauge theory.}
This choice of probe lines in the sYM SCFT corresponds to a choice of probe lines that renders the low-energy $\u(1)^r$ gauge theory at a generic point on the CB to be absolute. 
Such abelian probe lines are specified by a charge lattice that is maximal and mutually local with respect to the Dirac pairing and contains the physical BPS charge lattice of the sYM theory \cite{DelZotto:2022ras, Argyres:2022kon}.
A lattice of probe lines with its principal Dirac pairing is referred to as a {\it line lattices} in \cite{Argyres:2022kon}, and it's identified with the homology lattice and symplectic pairing, respectively, of the principally polarized abelian variety fiber of a (necessarily) principal SK structure.
It is for this reason that we focus on SK structures with principal polarization.

In section \ref{sec npp} we discuss some possible consequences of lifting constraint (iii) as well as of allowing non-simple $\fg$ in constraint (ii).

\subsection{The $BC_r$ disagreement}

In the $A_1=BC_1$ case, though there is a single SK structure orbit and a single S-duality orbit of $\fg=\su(2)$ global structures, the S-duality groups of the two orbits do not match.
The SK orbit duality group is $\PSL(2,\Z)$ while the field theory S-duality group is $\G_0(2)$, an index 3 subgroup of $\SL(2,\Z)$.
Correspondingly, the field theory conformal manifold is a 3-fold cover of that of the SK orbit.
This is not a contradiction:  the CB geometry, after all, is just some observable of the field theory, and there is no reason it need uniquely characterize the field theory, so may fail to distinguish inequivalent field theories.

Conversely, the SK structure result cannot be interpreted as a mistake in the field theory S-duality conjectures.
For there are three weak coupling global structures, $\SU(2)$ and $\SO(3)_\pm$ in the notation of \cite{Aharony:2013hda}, and two of them, namely $\SO(3)_\pm$, are continuously connected at weak coupling in $\cN{=}4$ sYM.
Thus there can only be either: a single S-duality orbit containing all three weak coupling limits (which is the usual S-duality conjecture); or two S-duality orbits with one containing the $\SU(2)$ limit and the other containing the $\SO(3)_\pm$ limits.
The SK structure result cannot distinguish between these two since it identifies the SK geometries of all three weak coupling limits.

The S-duality group discrepancies for the $BC_{r>1}$ theories are similar.
In these cases there is an SK structure which fails to distinguish between two global structures, $\Sp(N)$ and $\SO(N+1)_+$, instead of three as in the $A_1$ case.%
\footnote{In the $BC_2$ case $\sp(4)=\so(5)$ so global structure $\Sp(4)=\Spin(5)$ which is still distinct from $\SO(5)_+$.}
The SK orbit duality group is once again a 3-fold cover of the field theory S-duality group.
The fact that this 3-fold cover fails to distinguish only 2 weak coupling limits indicates that it actually makes a 2-fold self-identification of one of the weak coupling limits.
In this case it is the $\SO(N+1)_+$ limit, where the SK structure cannot distinguish between gauge couplings $\t$ and $\t+1$ in the normalization where the theta angle periodicity is $\t\sim \t+2$.
This purely weak coupling discrepancy makes it clear that the SK orbit result is simply an ``IR accident'' and cannot be taken as an indication of a mistake in the field theory S-duality conjecture.

In more detail, S-duality transformations of $\cN{=}4$ sYM theories form subgroups of the level-$q$ Hecke groups $H_q$ for $q=3,4,6$ corresponding to the lacing number for the gauge Lie algebra $\ell=1,2,3$, respectively \cite{Dorey:1996hx}.
As a reminder, the lacing number is $\ell=1$ for simply-laced $\fg$'s, $\ell=2$ for $BC_r$ and $F_4$, and $\ell=3$ for $G_2$.
Also, $H_3 = \PSL(2,\Z)$, but $H_4$ and $H_6$ are inequivalent discrete subgroups of $\PSL(2,\R)$; they are defined in appendix \ref{app Hecke}.
The maximal possible field theory S-duality group, $H_q$, would be one that fixes the gauge algebra $\fg$.
But precisely for the $BC_{r>2}$ cases, GNO duality \cite{Goddard:1976qe} relates two different Lie algebras, $\sp(2N)$ and $\so(2N+1)$.
So in these cases the maximal S-duality group is only the index-2 subgroup of $H_4$ \cite{Argyres:2006qr}, which happens to be isomorphic to $\G_0(2)$, which is an index-3 subgroup of $H_3 =\PSL(2,\Z)$ as reviewed in appendix \ref{app Hecke}.
So the fact that the $H_4$ group is not a subgroup of the SK orbit duality group, $\PSL(2,\Z)$, is not a contradiction since the predicted $\G_0(2)$ S-duality group is a subgroup.

These failures of the SK structure result to distinguish between inequivalent field theories raises the question of what other observables, beyond the SK structure of the moduli space, could faithfully distinguish them.
Since this failure is detected even at weak coupling, such additional observables are easy to provide.
One, described in \cite{Gaiotto:2010be, Aharony:2013hda, Argyres:2022kon}, is the lattice of EM charges of (finite energy) BPS states in the field theory as a sublattice of the probe line charge lattice.
This clearly distinguishes the weak coupling limits since it is used as the definition of the gauge theory global structure.
There are hints, coming from studies of rank-1 $\cN{=}2$ geometries \cite{Caorsi:2018ahl, Closset:2021lhd, Closset:2023pmc}, that the BPS charge sublattice of the probe line lattice can be detected from some subtle arithmetic properties of the SK structures we have presented.
We will comment on these in the context of non-principally polarized SK structures in section \ref{sec npp}.

A more speculative proposal for a moduli space observable which could distinguish field theory global structures are the SK structures of $\cN{=}2$ supersymmetry-preserving mass deformed moduli spaces.  
The evidence that this might be a rich enough observable comes mostly from the study of rank-1 CB geometries \cite{Argyres:2015ffa, Argyres:2015gha} where the 3-fold degeneracy of the (principally polarized) scale invariant $A_1$ SK structure is lifted, as are many similar degeneracies in other rank-1 $\cN{=}2$ geometries.

Finally, there is a disagreement of our SK structure results with a claim in \cite{Argyres:2023eij}, who, in their study of rank-2 $\cN{=}4$ Seiberg-Witten curves, claim that the two SW curve orbits for $BC_2$ have duality groups which are both $\G_0(2)$.
This is an index-2 subgroup of the $H_4$ duality group of the $\{(\Z_2,1),(\Z_2^\v,1)\}$ SK orbit reported in table \ref{tab:summary}, and a 3-fold cover of the $\PSL(2,\Z)$ duality group of the $\{(\Z_2,0),(\Z_2^\v,0)\}$ SK orbit reported there.
(Comparing to the field theory predictions of table \ref{tab:RBTL}, they disagree with the $\{\SO(5)_-\}$ orbit, but agree with the $\{\Sp(4),\SO(5)_+\}$ orbit: the opposite of what we have computed here.)
Both claims of \cite{Argyres:2023eij} are mistaken; in the case of the curve for the first orbit it is due to missing the isometry of the CB resulting from the $BC_2$ reflection outer automorphism, and in the case of the curve for the second orbit it is due to missing a set of identifications hidden by the particular algebraic form of the Seiberg-Witten curve that they used.

\subsection{The $D_{2k}$ disagreement}

The automorphism-twisted equivalences of SK structures are crucial in obtaining the Hecke group S-duality groups that appear in the $BC_2$, $G_2$ and $F_4$ non-simply laced cases, and this was illustrated in detail in the $BC_2$ case in section \ref{BC2example}.
But in the $D_{2k}$ Weyl groups also have a non-trivial  reflection outer automorphism, and including their associated equivalences leads to a modification of the counting of global structures relative to that given in the field theory literature, and also to a modification of the S-duality group of one orbit of the $D_{4k+2}$ theories relative to the field theory prediction.
This modification of the S-duality group in the case of $D_6$ was also illustrated in detail in section \ref{sec:so12}.

While one could interpret this mismatch between our geometric classification and the field theory S-duality conjectures as another instance of the geometry failing to distinguish between inequivalent field theories, in this case it turns out the geometrical classification is correct, and a mistake --- a double counting of certain field theories --- was made in the field theory literature.
The reason is very simple.
They Weyl group reflection outer automorphism in the $D_{2k}$ cases, unlike in the non-simply-laced cases, lifts to a gauge algebra outer automorphism.
In particular, it is the automorphism which interchanges the spinor and conjugate spinor representations of $\so(4k)$.
This symmetry means that at the level of the sYM gauge theory lagrangian there is no way of distinguishing, even in principle, between an $\cN{=}4$ sYM theory in which we put probe lines charged in the spinor representation or in the conjugate spinor representation: they differ just by the arbitrary choice of a name.
In the $\so(8k)$ theories this just means that the pairs of orbits in table \ref{tab:RBTL} with $\Ss(N)_\pm$ and $\Sc(N)_\pm$ global structures (shown in red) are indistinguishable, so two of the S-duality orbits should be removed from the list.
In the $\so(8k+4)$ theories the $\Ss(N)_\pm$ and $\Sc(N)_\pm$ global structures appear in the same orbit, so their equivalence implies a 2-fold identification of its conformal manifold, and so a 2-fold enlargement of its S-duality group from $\G_0(4)$ (shown in red in table \ref{tab:RBTL}) to $\G_0(2)$.

One might object that once we have a theory with, say, semi-simple gauge algebra $\so(8k) \times \so(8k)$ then though the overall distinction between spinor and conjugate spinor representations is a matter of convention, the \emph{relative} distinction (i.e., whether they are the same or opposite) between the two spinor representations in each factor is observable.
This is incorrect because for $\cN{=}4$ sYM theories the two simple gauge factors of the theory are decoupled, and so there is no observable that can distinguish between a (spinor $\times$ spinor) and a (spinor $\times$ conjugate spinor) theory.
If, on the other hand, there exist semi-simple versions of $\cN{=}4$ theories whose SK geometry does not factorize because of some twisting of their abelian variety fibers, then in these theories the relative spinor versus conjugate spinor distinction may be observable.
The possibility of such twisted geometries is discussed further in section \ref{sec npp}.

$D_4$ is a special case where the reflection outer automorphism group is enlarged to $S_3$.
However, these automorphisms, permuting the $S$, $C$, and $V$ representations as well as some of the bindings of the $\Z_2^2$ or root lattice representations, always act among the untwisted orbits, so only identify them without enlarging their self-duality groups, as shown in table \ref{tab:summary}.
This symmetry makes the $D_4$ global structure orbits in table \ref{tab:RBTL} with $\SO(8)_\pm$, $\Ss(8)_\pm$, and $\Sc(8)_\pm$ indistinguishable.

\subsection{SK structures versus global structures of the $\fg$ sYM gauge theory}

A result of the discussion so far in this section is that S-duality orbits of sYM global structures are in 1-to-1 correspondence to the orbits of CB SK structures under EM duality.
One may wonder whether there is a canonical way of matching subsets of SK structures to the individual global structures that make up their S-duality orbits.
Comparing our SK structure representations $\cA=(R_A,D)$ shown in the second column of table \ref{tab:summary} to the global forms of the sYM gauge group shown in the second column of table \ref{tab:RBTL}, we can see that, heuristically, different choices of the representation $R_A$ can be identified with the global form of the gauge group, while the binding matrix Ext class $D$ plays the role of the ``discrete theta angles''.
But these identifications are not unique, and in many cases there are more global structures than SK representations.

On the one hand, this mismatch is not physically significant, and reflects the fact that gauge invariance is merely a redundancy in the description of the field theory.
On the field theory side different gauge group global structures are related by S-duality equivalences, while on the SK geometry side, different constructions of $(R_A,D)$ symplectic representations are related by physically unobservable basis changes (integral equivalences).
Another way of saying this is that a division of an S-duality orbit into global structures corresponds to a tiling of their conformal manifolds by copies of $H_q$ fundamental domains in $\scH_1$.
There are infinitely many arbitrary choices that can be made in such a tiling.
As a simple example, even in a weak coupling limit, if $w$ copies of $H_q$ fundamental domains have a common weak coupling cusp on the conformal manifold corresponding to an enlarged $\th \sim \th+2\pi w$ theta angle periodicity there, there is a continuous family of tilings by $w$ intervals in $\th$ of width $2\pi$ as $\th \in [\th_0 +2\pi n,\th_0+2\pi(n{+}1)]$ parameterized by $\th_0$.

On the other hand, at a weak coupling limit, the global form of the gauge group is a physical property of the gauge theory that can be observed on the CB as a relation between the charge lattice of probe lines and its charge sublattice of BPS states \cite{Gaiotto:2010be, Aharony:2013hda, Argyres:2022kon}.
This does not uniquely specify the associated discrete theta angles, but does pick out the global form of the gauge group.
At a weak coupling point on the conformal manifold one relates the global form of the gauge group to the integral representation of the Weyl group, since they are both related to choices of subgroups of the center of the simply connected gauge group, as described in section \ref{sec inv latts}.
This explains the qualitative matching between global structures and symplectic Weyl representations within the orbits listed in tables \ref{tab:summary} and \ref{tab:RBTL}.

Furthermore, the counting of the number of global structures in each orbit is also an invariant observable.
Since the conformal manifolds described above are the fundamental domains of finite-index subgroups of the Hecke groups $H_q$ in the $\t$ upper half-space, they are (well-studied) \emph{modular curves}.
These are Riemann surfaces with marked points of three types: $\Z_2$ orbifold points, $\Z_3$ orbifold points, and cusps.
Physically the orbifold points (a.k.a., elliptic points) correspond to couplings where the 0-form symmetry group is enhanced by the orbifold group, while the cusps correspond to weak coupling limits.
Each cusp also has a positive integer, $w$, the \emph{width}, associated to it.
Physically, the width at a cusp corresponds to the $\th$ angle periodicity $\th \sim \th + 2\pi w$ at that weak coupling point in the standard normalization.
The geometry of the conformal manifold (modular curve) is described by the number and types of the orbifold points, the number and widths of the cusps, and the genus of the curve.
These are computed for the modular curves of many modular congruence subgroups; see e.g., \cite{diamond2006first}.
The sum of all the widths is the index of the modular subgroup in $H_q$ which counts the number of global structures in that S-duality orbit.
This number is recorded for the field results of \cite{Aharony:2013hda} in the 4th column of table \ref{tab:RBTL}. 
(As discussed above, this has one incorrect entry for the $D_{2k+1}$ case, shown in red, which should be 3 instead of 6.)

Nevertheless, although there is no \emph{canonical} way of relating symplectic representations of the Weyl group to global structures, we can still make such a correspondence ``by hand''.
We  illustrate this for $\su(N)$; similar identifications can be made for the other simply-laced algebras, but it does not work in as simple a way for the non-simply-laced algebras, where there is no longer a simple correspondence between representations and central subgroups, as mentioned in section \ref{sec inv latts}.
This illustration will also serve to clarify the arbitrary choices that are made in making such a correspondence.

For $\su(N)$ the lattices are labeled by divisors $d|N$, with $d=N$ the root lattice and $d=1$ the weight lattice, corresponding to the central groups $H=\Z_d$; see table \ref{tab:lattices}.
Let's associate symplectic representations to gauge group global forms by the rule
\begin{align}\label{magnetic S}
    S_{(d,z)}  &\leftrightarrow (\SU(N)/\Z_{N/d})_z, &
    d &| N, &
    z &\in \Z_d,
\end{align}
where the $z$ subscript on the right is the ``discrete theta angle'' introduced in \cite{Aharony:2013hda}.
Note that we are labeling the representations by divisors $d$ and by binding index $z$ mod $d$, even though we saw in section \ref{suN sec} that most of these representations are, in fact, integrally equivalent.

Why did we make the ``wrong'' correspondence, $S_{(d,0)} \leftrightarrow \Z_{N/d}$, which inverts $d$?
For no good (i.e., gauge invariant) reason: we could have as easily made instead a $S_{(d,0)} \leftrightarrow \Z_d$ correspondence, or even more complicated ones.
Another way of thinking about this is that the correspondence \eqref{magnetic S} is the ``natural'' one for the GNO dual ``magnetic'' gauge algebra instead of the ``electric'' gauge algebra.
All this is just to emphasize that the following correspondence to global forms is not in any way canonical.

With the choice \eqref{magnetic S}, the $\sS$ and $\sT$ generators of the $\PSL(2,\Z)$ group act on the $(d,z)$ representation labels as \cite{Bergman:2022otk}
\begin{align}\label{PSL ST}
    \sT &: (d,z) \mapsto (d',z') = \left( d,z+\frac{N}{d} \right), &
    z' &\in \Z_{d'},\\
    \sS &: (d,z) \mapsto (d',z') = \left( \frac{N}{\gcd(d,z)} , \frac{N\,(d-z)}{d\,\gcd(d,z)} \right), &
    z' &\in \Z_{d'},\nn
\end{align}
where in the $(d',z')$'s on the right sides, $z'$  is defined modulo $d'$. 
Indeed, it is possible to show that these actions on $(d,z)$ obey $\sS^2=(\sS\sT)^3=1$.
Thus $\sS$ and $\sT$ generate $\PSL(2,\Z)$.

We claim these $\sS$ and $\sT$ maps correspond to integral equivalences of the $S_{(d,z)}$ symplectic representations.
It is complicated to write the explicit forms of the $M(\g)$ in \eqref{M prin int} intertwining $S_{(d,z)}$ with $S_{(d',z')}$ for the $\sT$ and $\sS$ actions defined in \eqref{PSL ST}.
However, it is easy to show that such equivalences must exist, using the result of section \ref{suN sec} that $S_{(d,z)} \cong S_{(\gcd(s_d,z), 0)}$.
Indeed it is a short calculation to show that $\gcd(s_{d'},z') = \gcd(s_d,z)$ for the $(d',z')$ on the right side of \eqref{PSL ST}, and thus that they are equivalent representations.

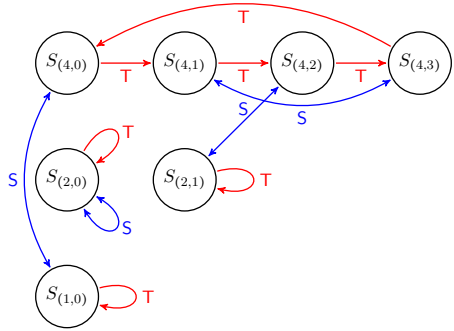
\begin{figure}
    \centering
    \resizebox{6cm}{!}{%
\begin{tikzpicture}[scale=0.5,->,>=stealth', shorten >=1pt, shorten <=1pt, 
  auto, node distance=2.5cm, thick]
  \node[circle, draw] (40) {$S_{(4,0)}$};
  \node[circle, draw] (41) [right of=40] {$S_{(4,1)}$};
  \node[circle, draw] (42) [right of=41] {$S_{(4,2)}$};
  \node[circle, draw] (43) [right of=42] {$S_{(4,3)}$};
  \node[circle, draw] (20) [below of=40] {$S_{(2,0)}$};
  \node[circle, draw] (21) [right of=20] {$S_{(2,1)}$};
  \node[circle, draw] (10) [below of=20] {$S_{(1,0)}$};

  \draw[->,red] (40) to node [below] {$\sT$} (41);
  \draw[->,red] (41) to node [below] {$\sT$} (42);
  \draw[->,red] (42) to node [below] {$\sT$} (43);
  \draw[->,red] (43) to[bend right] node [below] {$\sT$} (40);
  \draw[->,red] (20) to[above right, out=60, in=30, loop] node [right] {$\sT$} (20);
  \draw[->,red] (21) to[loop right] node [right] {$\sT$} (21);
  \draw[->,red] (10) to[loop right] node [right] {$\sT$} (10);

  \draw[<->,blue] (40) to[bend right] node [left] {$\sS$} (10);
  \draw[<->,blue] (20) to[below right, out=330, in=300, loop] node [right] {$\sS$} (20);
  \draw[<->,blue] (41) to[bend right] node [below] {$\sS$} (43);
  \draw[<->,blue] (42) to node [above] {$\sS$} (21);
\end{tikzpicture} 
}
\caption{The S-duality orbits of $\fg=\su(4)$ $\cN{=}4$ sYM global structures.
These are depicted by intertwiners $\sS$ and $\sT$ in \eqref{PSL ST} acting on the symplectic representations $S_{(d,z)}$ associated to the global forms by \eqref{magnetic S}.}
\label{fig su4}
\end{figure}
For a concrete example, for $\su(4)$ this set of representations are acted on by $\sS$ and $\sT$ as in figure \ref{fig su4}. 
This gives the two S-duality orbits and the action of $\PSL(2,\Z)$ on the global structures within each orbit, and reproduces the S-duality orbit diagram familiar from \cite{Aharony:2013hda}.

\section{Non-principally polarized SK geometries and twisted products}
\label{sec npp}

In addition to the principally polarized geometries we have discussed so far, there are also geometries whose Dirac pairing $J$ has invariant factors other than 1. 
These have the interpretation as corresponding to \emph{relative} QFTs \cite{Freed:2012bs}, i.e., field theories which are the boundary of a non-invertible topological field theory.
The low energy physics on the CB reflects the absolute versus relative distinction as follows. The Dirac pairing on the lattice, $\L_{\rm{charge}}$, of charges of finite energy states may be non-principal, and may be refined to a principally polarized ``line lattice'', $\L_{\rm{line}} \supset \L_{\rm{charge}}$, by including a maximal set of genuine probe line operators \cite{Gaiotto:2010be}.
So the CB geometry of the relative theory is that with the $\L_{\rm{charge}}$ pairing, $J$, so has an SK geometry in which the Weyl group is represented in $\Sp_{J^\v}(2r,\Z)$.
By contrast, the SK geometry of the CB of an absolute theory is one in which the Weyl group is represented in $\Sp(2r,\Z)$, as discussed in the previous three sections, where we saw that the different S-duality orbits of the global structures of \cite{Aharony:2013hda} correspond to the inequivalent $\Sp(2r,\Z)$ representations of the Weyl group.

The Dirac pairing on $\L_{\rm{charge}}$ is an observable property of a QFT with a Coulomb vacuum, so not all of the SK geometries we can construct using Weyl group orbifolds are physically realized as the geometries of absolute or relative QFTs.
Using semiclassical techniques, the Dirac pairing on $\L_{\rm{charge}}$ of $\cN{=}4$ sYM theories can be determined, as reviewed in \cite{Argyres:2022kon}.
For instance, the charge lattice Dirac pairing for the $\su(N)$ theory has invariant factors
\begin{align}\label{suN charge latt ifs}
    \text{invariant factors of }J_{\rm{charge}} = (1, \ldots, 1, N) ,
\end{align}
where there are $N{-}2$ ``1'' entries. 

On the other hand we can form symplectic representations $S_{(A,B;D)}$ \eqref{symp rep} with pairing given by \eqref{J def} with \eqref{bj = I}.

Here we chose to normalize the pairing so that the gcd of its invariant factors is 1, i.e., so that its smallest invariant factor is 1.
But since the SK structure does not depend on the overall normalization of the pairing, we can just as well work with $\bj = I^\v_{AB}$, with the understanding that we may have to multiply by an overall factor to clear denominators. 

For $\su(N)$, the possible inequivalent irreducible representations of the Weyl group are labeled by the divisors $d$ of $N$, i.e., $A, B \in \{ d \text{ such that } d|N \}$.
The invariant factors of the invariant Dirac pairing on $S_{(A,B;D)} = S_{(dd',D)}$ are the invariant factors of $I_{dd'}$ which are
\begin{align}\label{suN poss ifs}
    \text{invariant factors of } I_{dd'} = ( 1, \ldots, 1, d/d')
\end{align}
as computed from $I_{dd'}$ given in \eqref{sun ik mats}.
This shows there is a unique integral symplectic representation of the Weyl group with the physical charge lattice polarization, namely that with $(d,d')=(1,N)$.
Indeed, there is a unique $\Sp_{J^\v}(2r,\Z)$ representation of the Weyl group with these values of $(d,d')$ --- i.e., the Ext group is trivial, so we can set the binding matrix to $D=0$.
This corresponds to the expectation that there is a unique ``maximally relative'' $\su(N)$ $\cN{=}4$ sYM theory.
It therefore has a single connected conformal manifold with S-duality group $\PSL(2,\Z)$ (as opposed to a subgroup).

SK geometries with other values of $d'/d$ therefore do not correspond to the maximally relative field theory.
Clearly those with $d'/d=1$ are principally polarized, and correspond to the absolute theories.
If $1< d'/d <N$ is integral, then these geometries correspond to ``intermediate relative'' field theories, which are those for which the charge lattice has been refined to a line lattice, but without choosing a \emph{maximal} set of probe line charges.
By contrast, values of $d'/d \notin \Z$ cannot be found in this way.
Since the overall normalization of $J$ is not determined by the SK structure, we can clear denominators to make $J$ integral with invariant factors to $(1,d,\ldots,d,dd')$ (assuming $d$ and $d'$ coprime for simplicity).
In general, such Dirac pairings do not occur for any intermediate line lattices.
On the other hand, non-integral $d'/d$ could occur as the polarizations of lattices of non-mutually-local probe lines, i.e., of lines which are non-genuine line operators, which means lines which are boundaries of topological surface operators. 
These make sense as low energy CB $\u(1)^r$ gauge theories, but it is difficult to see how they can be defined as sensible kinds of local QFTs relative to a higher-dimensional symmetry TFT. 

Similar statements apply to the other simple Lie algebras.

Even if one is only interested in absolute field theories, non-principally polarized SK geometries are still relevant.
For there may be principally polarized geometries for \emph{non-simple} --- i.e., product --- gauge algebras which are built from non-principally polarized factor geometries.
(Indeed, examples of absolute product field theories which are formed from the product of relative field theory factors are known in the case of 6d (2,0) SCFTs \cite{Gukov:2020btk}.)

Consider two orbifold CB geometries
\begin{align}
    \text{CB}_i &\simeq_\C \C^{r_i} /(W_i)_\C, &
    i &= 1,2,
\end{align}
where $W_i$ are two Weyl groups.
To define these geometries we need to also specify integral symplectic representations
\begin{align}
    S_i : (W_i)_\C \to \Sp_{J^\v_i}(2r_i,\Z)
\end{align}
which leave invariant lattices $\L_i \simeq \Z^{2 r_i}$ with invariant symplectic pairings
\begin{align}
    J_i : \L_i \times \L_i \to \Z .
\end{align}
We can obviously form the direct product geometry, $\text{CB}_\times \deq \text{CB}_1 \times \text{CB}_2$, where $(W_1)_\C \times (W_2)_\C$ acts via the reducible representation $S_\times = (S_1,1) \oplus (1,S_2)$.
Then the induced invariant symplectic pairing is $J_\times = J_1 \oplus J_2$, whose invariant factors are, more or less, the union of those of $J_1$ and $J_2$.
In particular, if either of the $J_i$ are not principally polarized, then neither is $J_\times$.

We raise the question:  are there principally-polarized ``twisted'' product CBs,
\begin{align}
    \text{CB}_D &\deq \text{CB}_1 \times_D \text{CB}_2
\end{align}
which, when viewed as \emph{K\"ahler} geometries are product geometries, $\text{CB}_D \simeq_\C \text{CB}_\times$?
In other words, the ``twisting'' involves only the integral symplectic representation specifying the SK structure.
That means that the integral symplectic representation of the orbifold group,
\begin{align}
    S_D : (W_1)_\C \times (W_2)_\C \to \Sp(2(r_1{+}r_2),\Z),
\end{align}
should be principally polarized (as shown) and should be rationally equivalent to the product representation but not integrally equivalent,
\begin{align}
    S_D &\simeq_\Q S_\times ,&  &\text{but}& 
    S_D &\not\simeq_\Z S_\times.
\end{align}
That is, as should be familiar now and is explained in appendix \ref{app Z rep}, over the integers $S_D$ should be reducible but indecomposable,
\begin{align}\label{D twist rep}
    S_D(g_1,g_2) &= \bpm S_1 (g_1) & L(g_1,g_2) \\ 0 & S_2(g_2)\epm, &
    &\text{with} &
    L(g_1,g_2) &= S_1(g_1) D - D S_2(g_2)
\end{align}
with $D$ some fixed rational matrix independent of $(g_1,g_2) \in W_1 \times W_2$.

Now, a short calculation shows that a symplectic pairing invariant under \eqref{D twist rep} must have the form
\begin{align}\label{JD}
    J_D &= \bpm \k_1 J_1 + \k_2 D J_2 D^t & -\k_2 D J_2 \\ -\k_2 J_2 D^t & \k_2 J_2 \epm ,&
    \k_i &\in \Q .
\end{align}
Assuming there exists a non-vanishing $D$ such that $S_D$ is an integral representation, then there are some minimal value(s) of $\k_{1,2}$ such that $J_D$ is integral and has smallest invariant factor 1.
The question of the existence of twisted principally polarized representations $S_D$ is thus equivalent to the existence of a non-vanishing extension $D$ and rational numbers $\k_i$ such that $J_D$ in \eqref{JD} is principal, i.e., all its invariant factors are 1.

It follows from \eqref{JD} that
\begin{align}\label{detJD}
\det J_D = \k_1^{2r_1} \k_2^{2r_2} \det J_1 \det J_2 ,
\end{align}
independent of $D$.
We are interested in the case where $J_D$ is principally polarized, so $\det J_D=1$, while either or both $J_i$ are not, so $\det J_1 \det J_2 > 1$ and integral.  
This, \eqref{detJD}, rationality of the $\k_i$, and the integrality of $J_D$ put strong constraints on the possible invariant factors of $J_i$.
In general, this is an algebraically complicated question.
We leave the exploration of such ``twisted'' product $\cN{=}4$ sYM theories to a later work.
But we show now that non-trivial solutions can exist, by demonstrating a simple, familiar, though somewhat degenerate, example.

\subsection{$\u(N)$ $\cN{=}4$ sYM}

We now construct the twisted product \eqref{D twist rep} of the $\u(1)$ and $\su(N)$ $\cN{=}4$ sYM geometries.

Take $r_1=1$ and $r_2 = N-1$ with
\begin{align}\label{UN1}
J_1 &= \e \otimes {\rm diag}\{ N \} = N \e , &
J_2 &= \e \otimes {\rm diag}\{ 1, \ldots, 1, N \} \deq \e \otimes \bj ,
\end{align}
where $\e$ is the $2\times 2$ antisymmetric unit matrix.%
\footnote{Note that if you reverse the roles of $r_1$ and $r_2$, no solution exists.}  
Here we have chosen the non-principal pairing, $J_2$, of the $\su(N)$ theory to be that of the ``maximally relative'' $\su(N)$ sYM CB geometry described earlier --- i.e., that with symplectic representation $S_2 \deq S_{(d,d';D)} =  S_{(1,N;0)}$.
The normalization of the pairing $J_1$ is not constrained by physics since the $\u(1)$ theory is free.
We have chosen it to be proportional to $N$ so that a solution to \eqref{detJD} with square rational but non-integer $\k_1$ exists.  
We will see below that this allows for potentially non-trivial twistings.

Indeed, \eqref{detJD} implies 
$1 = \det J_D = \k_1^{2} \k_2^{2N-2} N^4$ so
\begin{align}\label{UN2}
\k_1 = N^{-2} \k_2^{1-N} .
\end{align}
Integrality of the lower right block of $J_D$ implies $\k_2$ is integral.  
Taking $\k_2 >1$ dramatically increases the denominator of $\k_1$ making arranging the integrality of the upper left block of $J_D$ increasingly difficult, especially if we are looking for a solution which is uniform in $N$.
This leads us to guess that 
\begin{align}\label{UN3}
\k_2 &= 1, & 
&\text{and so} &
\k_1 &= N^{-2} .
\end{align}
Now analyze the integrality of the upper right block of $J_D$.
Write the binding matrix in terms of $r_1 \times r_2 = 1 \times (N-1)$ row vectors as
\begin{align}\label{UN4}
D \deq \bpm v & w \\ x & y \epm ,
\end{align}
so that $DJ_2 = \bspm w\bj & -v\bj \\ y\bj & -x\bj \espm$.
Then, since the first $N-2$ diagonal entries of $\bj$ are 1, the first $N-2$ entries of the row vectors $v$, $w$, $x$, and $y$ are all integers.
And since integer entries of $D$ can be set to zero by a basis change, we can set the first $N-2$ entries of these rows to 0.
Since the last entry in $\bj$ is $N$, the last entry of the rows can have denominator $N$.  
So set the binding to the $2 \times (N-1)$ matrix 
\begin{align}\label{UN5}
D \deq \frac1N \bpm \vec0 & \a & \vec0 & \b \\ \vec0 & \g & \vec0 & \d \epm ,
\end{align}
where the $\vec 0$'s are $(N-2)$-component zero row vectors, and $\a, \b, \g, \d \in \Z_N$.

Now compute the $2\times 2$ upper left block in $J_D$ to find that it is $\e (1+\a\d-\b\g)/N$.
So for it to be integral we must have
\begin{align}\label{UN6}
\a\d-\b\g = -1 \ ({\rm mod}\ N) .
\end{align}
Putting \eqref{UN3} -- \eqref{UN6} together gives an integral and principal solution to \eqref{JD}.

We now have to check whether there are values of $\a, \b, \g, \d$ satisfying \eqref{UN6} which is a non-trivial binding for the appropriate Weyl$(\u(1))$ and Weyl$(\su(N))$ integral representations.
Weyl$(\u(1)) =1$ is trivial, so its only irreducible representation is $\r_1(1) = 1$, and its rank-2 symplectic representation is just $S_1 = \r_1\oplus\r_1$, so
\begin{align}\label{UN S1}
S_1(1) = \bpm 1 & 0 \\ 0 & 1 \epm.
\end{align}
On the other hand, the invariant Dirac pairing, $J_2$, of the ``maximally relative'' symplectic integral representation, $S_2$, of Weyl$(\su(N))$, is given in \eqref{UN1}, and
\begin{align}\label{UN S2}
    S_2 = S_{(1,N;0)} = \bpm R_1^\v & 0 \\ 0 & R_N \epm. 
\end{align}
Then from \eqref{D twist rep}, \eqref{UN S1}, and \eqref{UN S2}, the upper right $2\times 2(N-1)$ block of $S_D$ is
\begin{align}\label{UN8}
D(1-S_2) =  \frac1N 
\bpm \a (1-R_1^\v)_{N-1,j} & 0 \\
0 & \d (1-R_N)_{N-1,j} \epm ,
\end{align}
where only the last rows of the $1-R_1^\v$ and $1-R_N$ matrices appear because of the form \eqref{UN5} of the $D$ binding matrix.

Using the explicit form of the $R_d$ representation matrices computed in the bases used in section \ref{suN sec}, we find
\begin{align}\label{UN9}
(1-R_1^{-t}(w_k))_{N-1,j} &=   -\d_{j,N-2} \d_{k, N-1} \nn\\
(1-R_N(w_k))_{N-1,j} &=  +\d_{j,1}\d_{k, 1} .
\end{align}
These are not divisible by $N$, so we must have $\a=\d=0$ (mod $N$) for $S_D$ to be integral.
This implies from \eqref{UN6} that we must have $\b\g=1$ (mod $N$) if $S_D$ is to be principally polarized.
Since $\b$ and $\g$ do not appear in \eqref{UN8}, we can consistently choose them to be $\b=\g=1$.

Thus we have constructed a ``twisted'' $\u(N)$ SK structure suitable for describing an absolute $\u(N)$ sYM theory.
Moreover, it is likely that it is the unique solution which is uniform in $N$.
The existence of this solution is expected on physical grounds because the rank-$2(N{-}1)$ $\su(N)$ charge lattice is embedded in the rank-$2N$ charge lattice of the $\u(N)$ theory in such a way that the induced Dirac pairing on the $\su(N)$ sublattice is the non-principal one given by $J_2$ \cite{Argyres:2022kon}.

\acknowledgments

It is a pleasure to thank Antonio Amariti, Federico Bonetti, Sergio Cecotti, Cyril Closset,  Michele Del Zotto, Daniel Juteau, Mario Martone, Robert Moscrop, Sakura Sch\"afer-Nameki, Souradeep Thakur, Ben Webster, and Yu Zekai for helpful comments and discussions.
We also especially thank Robert Moscrop for sharing a preliminary version of his paper \cite{Moscrop:2025mtl}.
PCA is supported by DOE grant DE-SC0011784, and thanks Instituto Balseiro for their hospitality. AB was supported by the ERC Consolidator Grant 772408-Stringlandscape, and by the LabEx ENS-ICFP: ANR-10-LABX-0010/ANR-10-IDEX-0001-02 PSL*.
JFG is supported by the EPSRC Open Fellowship (Sch\"afer-Nameki) EP/X01276X/1 and the ``Simons Collaboration on Special Holonomy in Geometry, Analysis and Physics''.
ML is currently supported by a RyC grant RYC2023-043268-I of the Spanish State Research Agency (AEI) and also through the grants CEX2020-001007-S and PID2021-123017NB-I00, funded by MCIN/AEI/10.13039/501100011033 and by ERDF A way of making Europe.
ML also acknowledges the support of the National Research Foundation of Korea
(NRF) through grants 2021R1A2C2012350 and RS-2024-00405629.
MW is supported by the National Research Foundation of Korea
(NRF) through grants RS-2023-00208602 and RS-2024-00405629. 
This research was also supported in part by grant NSF PHY-2309135 to the Kavli Institute for Theoretical Physics (KITP), which AB thanks for their hospitality.

\appendix

\section{Coulomb branch SK structure of N=2 SCFTs and N=4 sYM}
\label{appA: SK Geo}

We recall the definition of an SK structure relative to the K\"ahler structure on the moduli space of vacua for a generic $4d$ $\cN{=}2$ SCFT.
The data of an SK structure defines the CB geometry of such SCFTs, so classifying such structures in equivalent to classifying CB geometries.
We then adapt our discussion to the SK structure of so-called \emph{isotrivial} SCFTs which brings about simplifications that enables them to be classified in an algorithmic fashion. 
Moduli space geometries that are orbifolds by crystallographic complex reflection groups (CCRGs) are examples that possess isotrivial SK structures. 
Of primary relevance to the paper are the SK structures of $\cN{=}4$ sYM CBs, which are examples of isotrivial CB geometries whose moduli space of vacua are complexified orbifolds by Weyl groups.
Weyl groups are real crystallographic reflection groups and their corresponding SK structures come in one-parameter families.
The weak-coupling limit of an $\cN{=}4$ sYM gauge theory with gauge algebra $\fg$ implies its moduli space of vacua is precisely an orbifold by the Weyl group of $\fg$, so
if we assume these one-parameter families have a weak-coupling limit where such a conformal gauge theory emerges, then they correspond to the CB geometries of such SCFTs.
In this way we are classifying the possible SK structures that an $\cN{=}4$ sYM theory could possess. %
\footnote{
This is why we refer to such orbifold geometries as $\cN{=}4$ ``sYM" geometries.
In general, $\cN{=}4$ supersymmetry only implies an isotrivial CB geometry, but not necessarily and one that is an orbifold by a Weyl group (or semi-direct product thereof upon discrete gauging).
Therefore, one might imagine there could exist ``non-sYM" CB geometries with $\cN{=}4$ supersymmetry that are not (Weyl group) orbifolds. 
We leave this interesting possibility to future work while noting that proving or disproving the existence of such moduli spaces will likely require leveraging the full constraints of $\cN{=}4$ SUSY on isotrivial $\cN{=}2$ SK structures which will define the $\cN{=}4$ enhancement of the triple SK structures that govern $\cN{=}3$ SCFTs 
\cite{Argyres:2019yyb}.
}
For a given $\cN{=}4$ sYM theory with simple gauge algebra $\fg$, we claim that such SK structures are determined by the triple $(J,S,\bt)$ where
\begin{itemize}
    \item $J$ is the Dirac pairing on the EM charge lattice; 
    \item $S$ is an integral representation of the orbifold group which preserves the Dirac pairing, and 
    \item  $\bt(\t)$ is a positive definite matrix that is the low-energy EM coupling matrix and is fixed by the action of this representation of the orbifold group for all values of the complex parameter $\t\in \scH_1$.
\end{itemize}
In particular, in this appendix we show that the SK structures of $\cN{=}4$ sYM theories are in 1-to-1 correspondence with the orbits of triples $(J,S,\bt)$ formed from equivalences that are defined by SK structure isomorphism.
In the body of the paper, we classify all distinct orbits of triples for simple $\fg$ with principal $J$.
Principal $J$ allows us to identify these SK structures as those corresponding to absolute sYM theories, i.e. those theories where a choice of global structure has been made in terms of a maximal, principal (mutually local) set of Wilson-'t Hooft line operators.

\subsubsection*{Notation.}\label{appA: notation}
Throughout the paper, the following notation is used to denote distinct mathematical quantities, e.g. certain groups, their group representations and the images of those representations.

\begin{itemize}
    \item Weyl group of a simple, compact Lie algebra $\fg$ with rank$(\fg) \deq r$: $W(\fg) \equiv W_{\fg} \equiv W$;
    \item Real reflection representation of $W_{\fg}$ that renders if a finite, real reflection (Coxeter) group, $R_{\R}: W_{\fg} \to \GL(r,\R)$, with image of $R_{\R}$: $W_{\R} \deq \Im R_{\R} \cong W_{\fg}$;
    \item Complex reflection representation $R_{\C} \deq \C \otimes_{\R} \R_{\R}$ of $W_{\fg}$ that renders it a rank $r$ complex reflection group (CRG), $R_{\C}: W_{\fg} \to \GL(r,\C)$, with image $W_{\C} \deq \Im R_{\C} \cong W_{\fg}$;
    \item Irreducible, rank-$r$ integral reflection representation of $W_{\fg}$ that is labeled by a subgroup $H \subseteq Z(\fg)$ of the center of $\fg$: $R_H : W_{\fg} \to \GL(r,\Z)$;
    \item $J^{\v}$-symplectic, rank-$2r$ integral reflection representation of $W_{\fg}$ that is typically reducible but indecomposable, $R_{\Z}: W_{\fg} \to \Sp_{J^{\v}}(2r,\Z)$, with image $W_{\Z} \deq \Im W_{\fg}$;
    \item $\cN{=}2$ rank-$r$ Coulomb branch of $\cN{=}4$ SK geometry, $C_{\fg} \equiv \cC \deq \cCt/W_{\fg}$ with covering space $\cCt \cong \C^r$ and smooth component $\cC^* \deq \cCt^* / W_{\fg}$;
    \item Fundamental group of the smooth Coulomb branch, $\pi_1(\cC^*)$, and its cover $\pi_1(\cCt^*)$;
    \item \emph{Monodromy map} of $\cC_{\fg}$, $S_{\Z}: \pi_1(\cC^*) \to \Sp_{J^{\v}}(2r,\Z)$, with image $S \deq \Im S_{\Z}$ called the \emph{monodromy group} and whose elements are called \emph{monodromy transformations}, $S(\g \in \pi_1(\cC^*))$;
    \item \emph{Complex monodromy map} of $\pi_1(\cC^*)$, $S_{\C} : \pi_1(\cC^*) \overset{S_{\Z}}{\rightarrow} S \to \GL(r,\C)$, that defines the monodromy transformation of the fiber coordinates ($1$-forms) and special coordinates through the monodromy group; (These become complex representations, i.e. group homomorphisms, of $\pi_1(\cC^*)$ for isotrivial SK geometries.)
    \item We denote the \emph{dual} or \emph{contragradient} of any linear representation $R$ by $R^{\v} \deq R^{-t}$ where $R^t$ is the transpose of the corresponding matrix element. 
\end{itemize}

\noindent\textit{A Note to the Reader:} For convenience and due to certain mathematical results, we sometimes abuse the notations stated above when writing by using it to label two (in principle) mathematically distinct quantities.
For example, we sometimes conflate/identify the notation used for an abstract group, its $\mathbb F$-representations, and the images of those representations as matrix subgroups.
We don't believe these abuses are too egregious, but nonetheless, this may cause confusion in some statements since they will appear to be to be mathematically imprecise.
We apologize to the reader in advance if it does so, but we believe these notational abuses can be untangled from context and don't detract from or nullify the truth value of any statements we make.  
Below, we try to indicate what notational identifications are made relative to the notational conventions stated above.

\begin{itemize}
    \item (Writing equivalence) We identify the Weyl group as an abstract group with its image under the various reflection representations, since they're all faithful: $W_{\fg} \equiv W_{\R}, W_{\C}, W_{\Z}$;
    \item (Writing equivalence) We conflate various $\mathbb F$-representations with their image. 
    In writing this means we replace $R_{\mathbb F} \to W_{\mathbb F}$ and $S_{\Z} \to S$;
    \item For all $\cN{=}2$ SK geometries, the monodromy map $S_{\Z}$ is canonically defined. 
    For the Weyl group orbifold SK geometries we consider, they are isotrivial, so $S_{\Z}$ is determined by $R_{\Z}$, the symplectic integral reflection representations of $W_{\fg}$.%
    \footnote{More precisely, as we prove in this appendix, the monodromy group $S$ is isomorphic to $W_{\Z}$, $R_{\Z}$ and $S_{\Z}$ reps are in $1$-to-$1$ correspondence with each other for $\cN{=}4$ SK structures.}
    Thus, classifying one implicitly classifies the other, and so in the main part of the paper we replace $R_{\Z} \rightarrow S_{\Z} \rightarrow S$.
\end{itemize}

\subsection{Review of SK geometry}\label{appA: SK Geo review}

The Coulomb branch (CB), $\cC$, is the subset of the moduli space of vacua of 4d $\cN{=}2$ supersymmetric theories which have $\u(1)^r$ gauge fields coupled to massive charged fields.\footnote{Here we mean by $\cC$ the \emph{smooth} locus of the CB, i.e., excluding for the moment subspaces where some charged states become massless on the CB.
We incorporate these massless particle subspaces --- which correspond to non-analyticities of the K\"ahler geometry --- after eqn.\ \eqref{SK metric} below.}
The vevs of the complex scalars in the $r$ vector multiplets are coordinates on $\cC$, and their kinetic terms endow $\cC$ with a K\"ahler structure in the usual way.
The relation of these scalar fields to the gauge fields via supersymmetry endows $\cC$ with an SK structure.
In particular, the gauge fields couple to massive electrically and magnetically charged states, encoded by the symmetric complex $r\times r$ matrix $\bt$ of coefficients of the gauge field kinetic terms.
This is in a basis of gauge fields with respect to which the charges of the massive states span a rank-$2r$ lattice, $\L = \L_e \oplus \L_m$, of integer electric and magnetic charges. 
This quantization of the charges is due to the Dirac quantization condition, which states that $\L$ carries a physically observable Dirac pairing, $\bj: \L_e \times \L_m \to \Z$, which is non-degenerate.

These physical properties of the low energy effective action on the CB can be encoded in an SK geometry of $\cC$, which is described in a coordinate-invariant way as the $r$-complex-dimensional base of an \emph{algebraic integrable system} $(\cA, \pi, J, \Om)$ \cite{Donagi:1995cf, Freed:1997dp, Donagi:1997sr, Neitzke:2014cja}.
We review how, with an appropriate choice of bases, the low energy effective action on the CB is derived from the integrable system data.

The complex phase space, $\cA$, of the integrable system is a connected $2r$-dimensional symplectic manifold with holomorphic symplectic form $\Om$ and a proper holomorphic lagrangian fibration $\pi: \cA \to \cC$.
Its fibers, $A_u = \pi^{-1}(u)$ for $u\in \cC$, are abelian varieties, and the fibration being lagrangian means that $\Om|_{A_u}=0$.
The rank-$2r$ 1-homology lattice of the fiber, $\L = H_1(A_u)$, is the EM charge lattice of the low-energy theory on the CB.
Since $\L$ is discrete, it is locally constant, so forms a linear system over $\cC$ which captures the monodromies $S_{\Z}(\g)$ it experiences after traversing a linking $1$-cycle $\g \in \pi_1(\cC)$ in $\cC$.
Specifically, if we fix a basis $(\hat\l_a, a=1,\ldots, 2r)$ of the charge lattice at a point $u\in \cC$, and then drag it around a closed path $\g \in \pi_1(\cC)$, it is allowed to come back to itself up to the linear action of the monodromy matrix $S_{\Z}(\g)$, $S_\Z(\g) \cdot \hat\l_a \deq S_{\Z}(\g)_a^{\ b} \hat\l_b$, that must produce a basis change of $\L$, so $S_{\Z}(\g) \in \GL(2r,\Z)$.
The set of these monodromies define a \emph{monodromy map} $S_{\Z}: \pi_1(\cC) \to \GL(2r,\Z)$ that produces an integral representation of $\pi_1(\cC)$.
We put the $\Z$ subscript on the monodromy $S_{\Z}$ to emphasize that it defines an integral representation of $\pi_1(\cC)$, and to contrast it with other representations we will introduce shortly.

Furthermore, the fibers come with a choice of a positive polarization, $J \in H^{1,1}(A_u) \cap H^2(A_u,\Z)$, which can be viewed as a nondegenerate integral skew-symmetric pairing on the charge lattice, $J: \L \times \L \to \Z$.
As it is discrete, it is also locally constant on $\cC$ like the charge lattice $\L$.
But because $J$ is identified with the Dirac pairing on the charge lattice in the low-energy theory on the CB, whose value is a physical observable, it must extend to a constant, and therefore global, section over $\cC$.
So, unlike the charge lattice, this implies it does not experience monodromies, which constrains $S_{\Z}(\g)$ to be valued in the subgroup $\Sp_{J^{\v}}(2r,\Z)$.

More concretely, define the symplectic form associated to $J$ by the $2r\times 2r$ matrix with entries $J_{ab} \deq J(\hat\l_a, \hat\l_b)$ relative to a charge lattice basis $(\hat\l_a, a=1,\ldots, 2r)$, which we'll often denote just by $J$ when the choice of basis is understood.
Since the polarization $J$ is globally defined, this matrix form of $J$ must be preserved by the monodromy $S_{\Z}(\g)$ associated to a linking $1$-cycle $\g$: $J \mapsto J' = S_{\Z}(\g)\cdot J \cdot S_{\Z}(\g)^{t} = J$.
This restricts the image of the monodromy map to $\Sp_{J^\v}(2r,\Z)$, the set of automorphisms of the symplectic form $J^{\v}$,%
\footnote{Note that this definition of $\Sp_{J^{\v}}(2r,\Z)$ is convention and defines for us what we call the {\it symplectic group} $\Sp_X(2r,\Z)$ of a generic symplectic form $X$.\label{symp group}}
\begin{align}\label{SK monod}
    S_{\Z} &: \pi_1 (\cC) \to \Sp_{J^\v}(2r,\Z), &
    \Sp_{J^\v}(2r,\Z) &\deq \{ M \in \GL(2r,\Z) \ | \ M^tJ^\v M = J^\v \} .
\end{align}
The image of the monodromy map, $S \deq \Im S_{\Z} \equiv S_{\Z}(\pi_1(\cC)) \subset \Sp_{J^\v}(2r,\Z)$, is the \emph{monodromy group} of $\cC$, and its elements are called \emph{monodromy transformations}. 
In writing throughout the text, we will sometimes conflate the monodromy group $S \leftrightarrow S_{\Z}$ with the monodromy map when the context is understood.
The monodromy group $S$ is in the symplectic group of the dual symplectic form $J^{\v}$ to $J$, where $J^\v \deq J^{-t}$.
This is an artifact of our definition of the symplectic group of a symplectic form, see footnote \ref{symp group}.
In particular, if $M$ is a basis change of the charge lattice $\L$ that preserves the matrix form of $J$, then $M\in \Sp_{J^{\v}}$ as opposed to $\Sp_J$ relative to our definition of the symplectic group $J$.
Note that since the special coordinates are valued in the (complexification of the) dual lattice of the homology (charge) lattice $\L$, under a monodromy $S_{\Z}(\g)$ they transform linearly under $S_{\Z}(\g)\in \Sp_{J^\v}$ as opposed to an element of $\Sp_J$.
Note that if $J$ is principal, there is no distinction between $J$ and $J^\v$, so $\Sp_{J^{\v}}(2r,\Z) = \Sp_J(2r,\Z) = \Sp(2r,\Z)$.
But for non-principal polarizations $\Sp_{J^\v}(2r,\Z)$ and $\Sp_{J}(2r,\Z)$, though isomorphic as abstract groups, are not integrally equivalent as subgroups of $\GL(2r,\Z)$ unless a condition on their invariant factors (that is specified below) is satisfied.%
\footnote{The two matrix groups are related by the inverse transpose of their elements, i.e. $M \in \Sp_{J^{\v}}(2r,\Z)$ iff $M^{-t} \in \Sp_J(2r,\Z)$, which establishes the isomorphism between them as abstract groups.}

Because symplectic groups relative to non-principal symplectic forms may be less familiar, we pause here to review them and the way they appear in an SK geometry.
Write an element of the charge lattice as $\L \ni \l = \ell^a\hat\l_a$ for some integers $\ell^a$ --- the ``charge vector''.
Then the Dirac pairing is 
\begin{align}\label{J pair}
    J(\ell_1,\ell_2) &\deq (\ell_1)^a J_{ab} (\ell_2)^b
\end{align}
thought of as a skew form on the charge vectors.
This is the traditional definition of the \emph{matrix} of the Dirac pairing.
The value of the Dirac pairing between two charges is a physical observable, so the normalization of $J$ is physical.
However, the individual matrix elements $J_{ab}$ depend on the arbitrary choice of lattice basis.

We can choose a \emph{symplectic basis} $(\hm_i, \he^i)$, $i=1, \ldots, r$, of $\L$ such that the sublattices $\L_m \deq \langle \hm_i \rangle$ and $\L_e \deq \langle \he^i \rangle$ are lagrangian with respect to $J$, i.e., $J(\hm_i, \hm_j) = J(\he^i, \he^j) = 0$.
$J$ in this basis is the non-degenerate integral $2r \times 2r$ matrix \eqref{J def}, i.e., $J = \bspm 0 & \bj \\ -\bj^t & 0 \espm$, with $(\bj)^{\ i}_{j} \deq J(\hm_j, \he^i)$.
This symplectic basis can be specialized to \emph{invariant factor form} (a.k.a., Smith normal form),
\begin{align}\label{bj snf}
    \bj &= \mathrm{diag} ( d_1, \ldots, d_r ),&
    d_i &\in \Z_{>0}, &
    &d_i | d_{i+1}.
\end{align}
The $d_i$ are the \emph{invariant factors} of $J$, and uniquely characterize it.
In particular, there exist change of bases $\mathbf{u}, \mathbf{v} \in \GL(r,\Z)$ respectively of the $\L_e$ and $\L_m$ sublattices, such that $\mathbf{ujv} = \mathbf{d}$ with $\mathbf{d}$ of the form \eqref{bj snf}.
Then $\bspm 0& \mathbf{u}^t\\ -\mathbf{v} & 0\espm^t J \bspm 0& \mathbf{u}^t\\ - \mathbf{v} & 0\espm = \bspm 0& \mathbf{d}\\ -\mathbf{d} &0 \espm$.

We can define the EM duality monodromies to be either linear transformations acting on the charge vector components as $\ell^a \mapsto {N^a}_b \ell^b$, or as linear transformations acting on the components of the special coordinate vector as $\s_a \mapsto {M_a}^b \s_b$.
Since our main focus is on the special coordinates, we define EM duality transformations to be the linear transformations acting as basis changes on the special coordinate components which also define basis transformations acting on the basis elements of the charge lattice.
The special coordinates are a vector of holomorphic functions on the CB which appear in the central charge of the low energy $\cN{=}2$ super Poincar\'e algebra on the CB as $Z_\l(u) = \s_a\ell^a$; we will give a more detailed definition of the special coordinates that connects them to the algebraic integrable system, shortly.
But since the central charge is a physical observable, this implies that the special coordinates are coordinates on the dual vector space to the (complexification of) the charge lattice.
Thus, components of vectors in the special coordinate vector space inherit a symplectic pairing using $J^{\v}$,
\begin{align}\label{Jv pair}
    J^\v(\s_1,\s_2) &\deq (\s_1)_a (J^\v)^{ab} (\s_2)_b , &
    J^\v &\deq J^{-t} .
\end{align}
Since the monodromies $S_{\Z}(\g)$ preserve the dual symplectic pairing and act as basis changes on the special coordinates components, special coordinate monodromies are in $\Sp_{J^\v}(2r,\Z)$ while the charge monodromies are in $\Sp_J(2r,\Z)$.
Note, also, that \eqref{SK monod} incorporates the choice of defining $\Sp_{J^\v}(2r,\Z)$ to be the set of $M\in \GL(2r,\Z)$ such that $M^t J^\v M=J^\v$ rather than $M J^\v M^t = J^\v$.  
The second choice is, in our convention, equivalent to the definition of the group $\Sp_J(2r,\Z)$, as is easily seen by taking the inverse transpose of its defining relation and using the fact that the inverse of any element of a group is also in the group.

Since the definitions of $\Sp_{J^\v}(2r,\Z)$ \eqref{SK monod}, or the similar definition of $\Sp_J(2r,\Z) \deq \{ N \in \GL(2r,\Z) \ | \ N^tJ N = J \}$, are linear in $J^\v$ and $J$, the normalizations of $J$ or $J^\v$ are irrelevant.
So one can always normalize a rational $J$ by dividing by the rational gcd of its elements, so that $J/\gcd(J_{ab})$ is an integral symplectic form.
In this case its leading invariant factor in \eqref{bj snf} is $d_1 =1$. 
The invariant factors, $d^\v_i$, of a normalized $J^\v$ are related to those of $J$, \eqref{bj snf}, by $d_i^\v = d_r/d_{r-i+1}$.
So, in general, $\Sp_J$ and $\Sp_{J^\v}$ are not isomorphic over the integers for non-principal $J$.%
\footnote{$J$ and $J^\v$ are isomorphic over the integers if they have the same invariant factors, which happens when $d_{i+1} d_{r-i} = d_i d_{r-i+1}$ for all $1\le i \le r-1$.}
Relatedly, for non-principal $J^\v$, in general $M^t \notin \Sp_{J^\v}$ when $M \in \Sp_J$.

We now return to describing the connection of the algebraic integrable system data to the low energy effective action on the CB.
In a symplectic basis there is always a choice of basis $( \diff z_j )$ of $H^{1,0}(A_u)$ such that its period matrix takes the form
\begin{align}\label{SK periods}
    \int_{\hm_i}\diff z_j &= (\bt_u)_{ij}, &
    \int_{\he^i} \diff z_j &= (\bj)^{\ i}_{j},
\end{align}
where $\bt_u$ is symmetric and has positive definite imaginary part, so is in the degree-$r$ Siegel half space, $\bt_u \in \scH_r$. 
This is the content of the Riemann conditions following from $A_u$ being an abelian variety.
Physically, $\bt_u$ is the holomorphically varying matrix of low energy $\u(1)^r$ gauge couplings.
The condition that the $A_u$ fibers are lagrangian implies that 
\begin{align}\label{SK special}
    (\bt_u)_{ij} &= \frac{\del a^D_i}{\del a^j},&
    &\text{with} &
    \diff_u a^D_i &\deq \int_{\hm_i}\Om, &
    (\bj^t)^{i}_{\ j} \diff_u a^j &\deq \int_{\he^i}\Om, &
\end{align}
where $\diff_u$ is the exterior derivative on $\cC$ and the integrals are fiber-wise integrals of the symplectic form $\Om$, which are well-defined because the fibers are lagrangian.
These fiber periods of $\Om$ can be written locally as total derivatives because $\Om$ is closed.
In particular, $\{ \diff_u a^j \}$ forms a basis of $(1,0)$ forms at each point on $\cC$, and in this basis
\begin{align}\label{Om form}
    \Om = \diff_u a^j \^ dz_j .
\end{align}

For SCFTs, the CB has a complex scale symmetry which fixes only the superconformal vacuum and under which the special coordinates scale homogeneously with weight 1.
Thus the superconformal vacuum is the origin of the special coordinates, 
\begin{align}\label{SC vac}
    a^i = a^D_i &= 0 ,  \qquad
    i = 1, \ldots, r, \qquad
    \text{at the superconformal vacuum.}
\end{align}
They can be determined everywhere on $\cC$ by integrating along paths from this origin their differentials given by the fiber periods of $\Om$.
These so-defined (dual) special coordinates on the CB, $a^j$ ($a^D_i$), are the vevs of the complex scalar superpartners of the low energy $\u(1)^r$ gauge fields. 
The $2r$-component vector of dual special coordinates and special coordinates, $(a^D_i,a^j)$, is what we referred to as the vector of special coordinates $\s_a$ in \eqref{Jv pair}.
The metric on the CB --- describing the kinetic terms of the scalars --- is given by
\begin{align}\label{SK metric}
    g &= (\det\bj) \, (\Im\bt_u)_{ij} \, \diff_u a^i \bar{\diff_u a}^j .
\end{align}
In this way we have translated the data of an algebraic integrable system into that of an SK geometry in special coordinates.

Although the above definition treated $\cA$ and $\cC$ as smooth complex manifolds, in fact, the CB, as a metric space, has complex codimension-1 finite-distance non-analyticities along a subvariety $\cD \subset \cC$.%
\footnote{All singularities occur in the closure of the codimension-1 singularities for physical reasons described around \eqref{Z cond} below; see \cite{Argyres:2020wmq} for a detailed description of the stratification of physical SK geometries by singular submanifolds.}
The smooth part of $\cC$, which we now denote $\cC^* \deq \cC \setminus \cD$, is what is described by the algebraic integrable system.
In particular, the monodromy map \eqref{SK monod} is a map from the fundamental group of the non-simply connected $\cC^*$.

The image of $\g\in\pi_1(\cC^*)$ under the monodromy map $S_{\Z}$ is an element of $\Sp_{J^\v}(2r,\Z)$.
Write it in terms of $r\times r$ block matrices (which we denote by bold letters) as
\begin{align}\label{monod map}
    S_\Z(\g) 
    \deq \bpm \bom & \bon \\ \bop & \boq \epm 
    \in \Sp_{J^\v}(2r,\Z) .
\end{align}
The definition of $\Sp_{J^\v}$ in \eqref{SK monod} implies
\begin{align}
    (\bj^{-1} \bom)^t \bop &= \bop^t (\bj^{-1} \bom) ,&
    (\bj^{-1} \bon)^t \boq &= \boq^t (\bj^{-1} \bon) ,&
    \boq^t (\bj^{-1} \bom) - (\bj^{-1} \bon)^t \bop &= \bj^{-1} . 
\end{align}
$S_\Z(\g)$ acts linearly on the basis of the homology lattice of the fiber (by definition), as well as on the special coordinates (by \eqref{SK special}, using the invariance of $\Om$) as
\begin{align}\label{SZ rep}
    S(\g) &: & 
    \bpm \hm \\ \he \epm &\mapsto
    S_\Z(\g) \bpm \hm \\ \he \epm,&
    \bpm \diff_u a^D \\ \bj \, \diff_u a \epm &\mapsto
    S_\Z(\g) \bpm \diff_u a^D \\ \bj^t \, \diff_u a\epm,
\end{align}
in a notation where $\hm$, $\he$, $a^D$, and $a$ are treated as $r$-component column vectors.

Define also the following $\GL(r,\C)$ matrices
\begin{align}\label{SC rep}
    S_\C(\g) &\deq 
    \bj^{-t} \bigl( \bop(\g) \bt + \boq(\g) \bj^t  \bigr) ,&
    S^\v_\C(\g) &\deq 
    \bj \bigl( \bop(\g) \bt + \boq(\g) \bj^t  \bigr)^{-t} . 
\end{align}
Then the monodromies act on the $dz$ basis of $(1,0)$-forms on $A_u$, on the special coordinates on $\cC$, and on $\bt$ as 
\begin{align}\label{M act}
    && dz &\mapsto S^\v_\C(\g) dz, \nn\\
    &S(\g):& \diff_u a &\mapsto S_\C(\g) \diff_u a, \\
    && \bt & \mapsto S(\g) \circ \bt \deq 
    (\bom \bt+\bon \bj^t)(\bop\bt+\boq \bj^t)^{-1} \bj^t .\nn
\end{align}
The last gives a group action (homomorphism) on $\scH_r$ in the sense that the composition of this action will reproduce the monodromy group multiplication rules.
On the other hand, the first two are only group actions if combined with the third action on $\bt$ when they are composed.
These actions follow from the definitions \eqref{SK periods} and \eqref{SK special}, and from the $S_\Z$ actions \eqref{SZ rep}.
All these actions of the monodromy group --- integral, complex, and M\"obius --- will play an important role shortly.

There is an additional condition on the special coordinates that stems from the physical origin of the divisor $\cD$ as the locus where the $\u(1)^r$ low energy effective action on $\cC$ breaks down due to charged states becoming massless.
The central charge, $Z_\l(u)$, of the $\cN{=}2$ Poincar\'e supersymmetry algebra measures the BPS lower mass bound on states of charge $\l \in \L$ at $u \in \cC$. 
Symmetry considerations in a SCFT (where there are no dimensionful parameters and a broken complex scale symmetry) imply that $Z_\l(u) = m^t a^D(u)+ e^t a(u)$ in a notation where $m$ and $e$ are the $r$-component column vectors of magnetic and electric charges relative to the basis $(\hm_i, \he^i)$ of $\L$.
Thus $\l = m^i \hm_i + e^j \he_j$, and $m^i, e_j \in \Z$.%
\footnote{More invariantly, the $2r$-component complex vector field $(a^D,a)$ is naturally a global section of the vector bundle whose fiber is the complexification of the dual charge lattice, $\C \otimes_\Z \L^*$, and the central charge is its dual pairing with $\L$.}
Then the physical requirement is that there must be at least one non-zero charged state becoming massless along each component of $\cD$,
\begin{align}\label{Z cond}	
    \forall u &\in \cD , &
    \exists \l &\in \L , \quad \l \neq 0 &
    &\text{such that} &
    Z_\l(u) &= 0 .	
\end{align}
This is an additional requirement that needs to be imposed, in principle, on the algebraic integrable system data $(\cA, \pi, J, \Om)$ for an SK geometry for it to be physical.
We will see below that this condition is automatically satisfied by the SK structures of $\cN{=}4$ sYM theories.

Two monodromy maps, $S$ and $S'$, which differ just by a choice of charge lattice basis correspond to equivalent SK structures.
Since a change of lattice basis is given by an element $Z \in \GL(2r,\Z)$, we introduce the notion of \emph{integral equivalence}, denoted $\cong_\Z$, of representations by
\begin{align}\label{Zequiv}
    S & \cong_\Z S' & &\text{iff} &
    \bigl\{ \exists \ 
    Z \in \GL(2r,\Z) \ &\big| \ 
    Z S(\g) = S'(\g) Z \quad \forall \ \g\in\pi_1(\cC^*) \big\}. 
\end{align}
Note that if $S$ preserves the symplectic form $J$, then $S'$ preserves the integrally equivalent symplectic form $J'= Z J Z^t$.
Analogously, if $S\cong_{\Z} S'$ using $Z\in \GL(2r,\Z)$ and $\bt \in \Fix(S)$, then $Z \circ \bt \in \Fix(S')$. 
Furthermore, this implies if $Z = S(\gamma)$ for some $\g$ so $S$ and $S'$ are equivalent by an inner automorphism of $\Im S_{\Z}$, then $\Fix(S) = \Fix(S')$.

$\bt$ is only defined up to $\Sp_{J^\v}(2r,\Z)$ transformations that correspond to the changes of symplectic basis preserving a given block-skew symplectic form \eqref{J def}.
Points of the Siegel upper half-space are identified by the $J^\v$-symplectic action \eqref{M act} because they correspond to equivalent SK structures.
Therefore, the points in the quotient space ${\scH}_r/\Sp_{J^\v}(2r,\Z)$ parameterize the inequivalent values for the low-energy couplings $\bt$.
This space is an $r(r+1)/2$-dimensional connected complex space with orbifold and cusp-like singularities.
Then $[\bt(u)]$, thought of as the $\Sp_{J^{\v}}(2r,\Z)$ equivalence class of a holomorphic function on $\cC^*$, defines the holomorphic map 
\begin{align}\label{tau map}
    [\bt]: \cC^* \to {\scH}_r/\Sp_{J^\v}(2r,\Z) .
\end{align}

\subsection{N=4 sYM SK structures}\label{appA: N4 SKS}

We now adapt the above discussion to the case of $\cN{=}4$ sYM SK structures relative to a gauge algebra $\fg$ which we take to be a (semi-)simple, compact Lie algebra of a compact, connected Lie group $G$. 
The starting point is the moduli space of vacua, $\cM_{\fg}$, defined in \eqref{N4 orbi} that possesses a K\"ahler structure upon which the SK structure is built.
After stating the properties of $\cN{=}4$ SK structures, we will show that they are determined by the triple of data $(J, S, \bt)$ where $S$ is defined to be a $J$-symplectic integral representation of the Weyl group of $\fg$. 
This results from proving the EM monodromy group $S \equiv S(\pi_1(\cC^*))$ is isomorphic to $W(\fg)$ as abstract groups while its integral representation in EM duality group $\Sp_J(2r,\Z)$ determines and is determined by a faithful $J$-symplectic integral representation of $W(\fg)$.

\subsubsection{K\"ahler structure of $\cN{=}4$ Coulomb branch.} \label{appA: CB Kahler} 
The K\"ahler structure is a geometric component of the moduli space of vacua, $\cM_{\fg}$, that is the $\cN{=}4$ Coulomb branch. 
The description of $\cM_{\fg}$ comes from the Lagrangian description of the $\fg$ sYM which provides a weak-coupling limit.
In this sense, it is a purely relative quantity in the field theory sense that it is solely defined in terms of the local operator field content of the theory. 

The $\cN{=}4$ vector multiplet has a top component consisting of six $\fg$-valued real scalars transforming in the $\mathbf 6$ irrep of the $\SU(4)_R$ R-symmetry which can be combined into three complex scalars $a^I$, $I,J\in 1,2,3$, that transform in the fundamental $\mathbf 3$ irrep of the $U(3)_R$ subgroup. 
For an $\cN{=}2$ gauge theory, upon integrating out the auxiliary fields the D-term equations generate a scalar potential for these vector multiplet scalars whose vacuum expectation values (vevs) minimize the potential.
This implies the vevs of the $a^I$ take values in a choice of complexified Cartan subalgebra $\fh \subset \fg$, so they can be parameterized as $a^I = a_i^I H^i$ where $\{ H^i\}_{i=1, \cdots, r}$ form a (real) basis for the rank-$r$ Cartan subalgebra $\fh \subset \fg$ and the $r$ coefficients $a_i^I \in \C$ parameterize the vev.
Such vevs trigger the Higgs mechanism which transforms the gauge group to $U(1)^r \times W(\fg)$ where $U(1)^r$ is a maximal torus subgroup generated by a maximal abelian subalgebra containing the choice of $\fh$ and $W(\fg)$ is the Weyl group of $G$. (For $G$ compact and connected, the Weyl group of $G$ and $\fg = \Lie(G)$ are identical.)
The space of $\cN{=}4$ Coulomb vacua must be gauge-invariant.
The $U(1)^r$ gauge transformations commute with $\fh$, so they leave the vevs invariant, and we can take $\fh_{\C} \otimes \C^3 \cong \C^{3r}$ to be the naive space of Coulomb vacua.
But this conclusion is too quick because the vevs are not left invariant under the action of the Weyl group which implements basis transformations of the Cartan subalgebra. 
Such choices are not physical, so we must quotient the naive space of vacua by the action of the Weyl group -- this results in the orbifold of eq. \eqref{N4 orbi}:
$$\cM_{\fg} \deq \C^{3r}/(W(\fg)\otimes_{\R} \C^3).$$
The Weyl group $W(\fg)$ as an abstract group is a Coxeter group, but it is naturally realized as real reflection group through its relation to the gauge algebra: it is the finite subgroup of $O(r,\R)$ acting on the real linear dual space $\fh^{*}$ to the Cartan subalgebra (i.e., the space of linear maps from $\fh$ to $\R$) that is generated by the reflections of the root system of $\fg$.%
\footnote{The root system of $\fg$ lives in $\fh^*$ and is formed by the roots of $\fg$, i.e. the non-zero weights of the adjoint representation.
Their $\R$-span generate $\fh^*$ and their $\Z$-span generate the root lattice, $\G_{\textrm{root}}$, which is a rank-$r$ lattice in $\fh^*$.}

This action on $\fh^* \cong \R^r$ via reflections with respect to the root system provides an orthogonal, real, faithful, and irreducible representation of the abstract group $W(\fg)$ we call the (real) \emph{reflection representation} $R_{\R}$,
\begin{align}\label{real refl rep}
R_{\R}: W(\fg) \to \O(\fh^*) \cong \O(r, \R) \subset \GL(r,\R)
\end{align}
where orthogonality is defined with respect to the positive definite scalar product on $\fh^{*}$.
We denote the image of this representation by $W_{\R} \deq \Im R_{\R} \cong W(\fg)$.
In this representation, $W(\fg)$ is generated by reflections, so every element can be expressed as a product of such reflections.%
\footnote{For Weyl groups, all elements $w$ that get mapped to reflections $R$ have order 2, $w^2 =1$.
This is not true for generic finite complex reflection groups which can have reflection elements orders greater than $2$.
Also, note that products of reflections are not always reflections, so not every element in $W_{\R}$ is a reflection in $\O(r,\R)$.}
A generating set of reflections is formed by the so-called simple reflections $\{ w_i \}_{i=1, \cdots, r}$ which are reflection with respect to a choice of simple roots $\{ \a_i \}_{i=1, \cdots, r}$ that (by definition) form a basis for $\fh^*$ and the root system.
The reflection representation acts as an isometry of the root system sitting inside $\G_{\textrm{root}}$, so this $W(\fg)$ action also preserves $\G_{\textrm{root}}$ itself, which defines $W(\fg)$ to be a real crystallographic reflection group.
In general, we are concerned with $W(\fg)$ through its action on $\C^r$, $\R^r$ and rank-$2r$ lattices $\L \cong \Z^{2r}$ via reflections, i.e. as a finite reflection group.
Therefore, we will implicitly identify its abstract elements $w \in W(\fg)$ with their image $R(w) \in \O(r,\R)$ under the reflection representation $R$ that forms the matrix subgroup $W_{\R}$.%
\footnote{We are not always precise in distinguishing between the abstract Weyl group $W(\fg)$ from its various reflection representations $R_{\mathbb F}$ and their images $\Im R_{\mathbb F} \cong W(\fg)$.
When we are being sloppy, it should be clear from context which of these we mean.}
This action generating $R$ extends linearly over $\C$ to define a linear, holomorphic, and faithful representation $R_{\C} \deq \C \otimes R: W(\fg) \to \fh^*_{\C} \cong \C^r$, which defines $W(\fg)$ as a complex crystallographic reflection group. 
We denote its image by $W_{\C} \deq \Im R_{\C} \cong \C \otimes W_{\R} \cong W(\fg)$.
It is further extended to an action on $\C^{3r}$ given by $\mathbf{1}_{3} \otimes R_{\C}$ where $\mathbf{1}_{3}$ is the $3 \times 3$ identity matrix.

Given this definition of Weyl group through its action on the root space, it is most natural to describe $\cM_{\fg}$ with respect to $\fh^*$ as opposed to $\fh$.
We do this by identifying $\fh$ with $\fh^*$.
This isomorphism is not canonical, so we fix it using the induced Killing metric $K(\cdot, \cdot): \fg \times \fg \to \R$ on $\fh^*$ (with a choice of normalization defining the length of short and long roots).
This defines a positive definite scalar product from which we identify $\C^{3r} \cong \fh^*_{\C} \otimes \C^3$. 

From the perspective of $\cN{=}4$ supersymmetry, $\cM_{\fg}$ is an $\cN{=}4$ Coulomb branch since it's massless degrees of freedom are given by a single $\cN{=}4$ vector multiplet.
Thus, it represents the total moduli space of the theory, i.e. there is no $\cN{=}4$ Higgs branch, so all vacua are $\cN{=}4$ preserving Coulomb vacua.
It gets no quantum corrections.
Upon picking a choice of $\cN{=}2$ subalgebra of the $\cN{=}4$ superconformal algebra, it decomposes into $\cN{=}2$ components consisting of a Higgs, mixed, and Coulomb branch. 
In this paper we are only interested in the ``Coulomb branch stratum'' of this decomposition which is the orbifold, 
\begin{align}\label{cplx orbi2}
    \cC_\fg = \C^r / W_\C.
\end{align}
Throughout the rest of this appendix, we let $\til \cC \equiv \til \cC_{\fg} \cong \C^r$ be the covering space of the CB.

\subsubsection{Weyl group action on the Coulomb branch.}\label{appA: W action on CB}
We can be more explicit about the representation matrices by first choosing a set of simple roots $\{\a_i\}_{i = 1,\cdots, r}$ whose $\Z$-span generate $\G_{\textrm{root}}$.
We use these simple roots to label the simple reflections $w_i \in W(\fg)$ which are reflections with respect to the hyperplane that is orthogonal to the simple root $\a_i$.
They generate the Weyl group, $W(\fg) = \< w_i \>_{i =1, \cdots, r}$, and their action on a generic root vector $\l \in \fh^*$ is given by
\begin{align}\label{simple reflections}
    w_i(\l) \deq \l - 2 \frac{(\l, \a_i)}{(\a_i,\a_i)} \a_i . 
\end{align} 
Note that once we label the set of simple reflections relative to a set of simple roots $\{ \a_i \}$, the corresponding reflection hyperplanes are fixed in $\C^r$.
This is because the set of simple roots $\{\a_i\}$ are fixed as abstract vectors, and they determine the action of simple reflections according to \eqref{simple reflections} in a basis-independent way.%
\footnote{When constructing SK geometries and the various representations of the Weyl group that are involved, we assume that we have fixed a choice of simple roots to label the simple reflections of $W(\fg)$ by.
This is because such a choice cannot be modified using isometry transformations within an SK geometry, so it only makes sense to compare geometries where this data is fixed.
See the section SK structure equivalences below for more elaboration on this point.}
However, the reflection matrices and the coordinates of the reflection hyperplanes will of course be basis dependent. 
In particular, upon a choice of simple root basis $\{\b_i\}_{i=1,\cdots, r}$ for $\C^r$, the simple reflection matrices $R(w_i)$ satisfy $w_i(\b_j) = R(w_i)_{jk} \b_k$.%
\footnote{Relative to a choice of $\fh^*$-basis $\{\b_i\}_{i=1,\cdots, r}$ we expand $\a_i = c_i^{\ j} \b_j$ so the simple reflection matrices take the form
\begin{align}\label{real reflections}
R(w_i)_{jk} = \delta_{jk} - 2 \frac{(\b_j,\b_{\ell})}{(\a_i,\a_i)} c_i^{\ \ell} c_i^{\ k} 
\end{align}
where the sum over $\ell$ is implied.}
This basis of the root space does not have to be the same as the simple roots $\{ \a_i \}$ labeling the simple reflections of $W(\fg)$.
Under a change of basis $\b_i \mapsto \b_i' \deq (U \cdot \b_i) \equiv U_{ij} \b_j$ for $U \in \GL(r,\C)$, the (complexified) reflection matrices will transform accordingly under conjugation, $R(w_i) \mapsto R'(w_i) \deq U \cdot R(w_i) \equiv U R(w_i) U^{-1}$.
All such basis transformations are intertwiners of the real reflection rep, so the new representation matrices $R'(w_i)$ are equivalent to the $R(w_i)$, and the Weyl group action on $\C^r$ is preserved. 
This implies the configuration of fixed hyperplanes corresponding to the simple roots $\{\a_i\}$ is also preserved, so the resulting orbifolds $\cM_{\fg}$ and $\cC_{\fg}$ are invariant under all such basis transformations of $\C^{3r}$ and $\C^r$, respectively.

\subsubsection{Special K\"ahler structure of the Coulomb branch.}\label{appA: CB SKS}
The total geometry containing the $\cN{=}4$ moduli space $\cM_\fg$ defined in \eqref{N4 orbi} possesses an $\cN{=}4$ special K\"ahler structure that is the $\cN{=}4$ version of the triple-SK structure possessed by $\cN\geq 3$ SCFT CBs and described in \cite{Argyres:2019yyb}.%
\footnote{
From the results of our paper in combination with those of \cite{Argyres:2022kon}, it follows that a principal SK structure is not unique for most simple gauge algebras $\fg$. 
In particular, it depends on whether the physical BPS charge lattice of the underlying sYM SCFT is principally polarized or not.} 
This induces an $\cN{=}2$ SK structure on $\cC_{\fg}$ which are the $\cN{=}2$ SK structures we classify in this paper.
Conversely, the triple-SK structure of $\cM_\fg$ is uniquely reconstructed from the $\cN{=}2$ SK structure of $\cC_\fg$ using the $\SU(3) \subset \SO(6)_R$ R-symmetry action; see \cite{Argyres:2019yyb} for further details on this point. 
As a complex space, $\cC_{\fg}$ is just $\C^r$, and is thus regular with a freely generated chiral ring, although one should remember that it still possesses metric singularities (non-analyticities) at the fixed-point loci of the Weyl group action.\\

$\cN{=}4$ sYM theories with gauge algebra $\fg$ have an exactly marginal complex gauge coupling, $\t \in \C$ with $\Im \t > 0$, that is valued in the Siegel-Upper half space.
The SK structure depends on $\t$, which should not be confused with $\bt \equiv (\bt)_{ij}$ which is the $r\times r$ matrix of low energy $U(1)^r$ couplings.
We will often refer to $\t$ as the UV-coupling and $\bt$ as the electro-magnetic (EM) or IR-couplings to distinguish them.
Dependency on the UV-coupling $\t$ in the SK geometry is reflected in dependence of the EM couplings on the UV-coupling, $\bt = \bt(\t)$.
The EM couplings are invariant under the action of $W(\fg)$, and they don't depend on the value of vevs parameterizing the Coulomb moduli space, i.e. they are constant on $\cM_{\fg}$ and $\cC_{\fg}$.
SCFT CB geometries for which the IR-couplings are constant are called \emph{isotrivial} and they have \emph{isotrivial} SK geometries \cite{Cecotti:2021ouq}. 
This property implies the map \eqref{tau map} is constant which renders them more simple than generic $\cN{=}2$ CB geometries.
All $\cN \ge 3$ SCFT moduli spaces are isotrivial by virtue of an $\cN{=}2$ selection rule \cite{Argyres:1996eh} which says that hypermultiplet effective actions do not depend on vector multiplet scalar vevs;  by rotating the choice of $\cN{=}2$ subalgebra in $\cN\ge 3$ theories, this implies their isotriviality.
There are many $\cN{=}2$ SCFT moduli spaces which are isotrivial as well \cite{Cecotti:2021ouq, Argyres:2024hdn}.

Constancy of $\bt$ implies it is fixed by the monodromy group action \eqref{M act}, so 
\begin{align}\label{fix Sa}
    \bt \in \Fix(S) \subset \scH_r.
\end{align}
$\Fix(S)$ is determined by the set of quadratic matrix equations
\begin{align}\label{fix tau}
    \bt &\in \Fix(S) & &\text{iff}&
    \bom\bt+\bon \bj^t &= \bt \bj^{-t} (\bop\bt+\boq\bj^t)&
    &\text{for all}&
    \bspm \bom & \bon \\ \bop & \boq \espm &\in S,
\end{align}
and is connected \cite{Caorsi:2018zsq}.
Note that it is enough to solve \eqref{fix tau} for a generating set of $S$, which is always finitely generated for the Weyl group orbifolds we consider. 
This is because, as we show in detail below, $S \equiv \Im S_{\Z} \subset \Sp_J(2r,\Z)$ is isomorphic to $W(\fg)$ as abstract groups, so the generating set corresponds to the simple reflections generating $W(\fg)$ whose action on the cover $\til \cC$ is given by $W_{\C}$.
This action fixes hyperplanes in $\til \cC$, so in terms of $\pi_1(\cC^*)$ elements, the generating set equals the monodromy transformations that correspond to the images of the monodromies linking these hyperplanes under the monodromy map.
Through the isomorphism with $W_{\C}$, such elements implicitly define a symplectic $\Z$-reflection rep of $W(\fg)$, $R_{\Z}$.
Conversely, given $R_{\Z}$, we can consistently define the monodromy map $S_{Z}$, so $R_{\Z}$ implements a well-defined action of the orbifold group $W(\fg)$ on the pav fiber $1$-homology lattice.
In practice, this allows us to evaluate this fixed point using $R_{\Z}$.
If the dimension of the fixed point set is greater than zero, then the geometry is not isolated, and the associated SCFT has a conformal manifold (with singularities) of exactly marginal deformations.
We therefore expect fixed point sets of dimension 1 for $\cN{=}4$ sYM theories.

Furthermore, isotriviality implies the SK metric \eqref{SK metric} is flat on $\cC^*$ and the special coordinates are flat coordinates.
Also in $\cN{=}4$ sYM theories, the orbifold group acts linearly on the Coulomb branch (special) coordinates as a finite reflection group.
This follows from the existence of a weak coupling limit where the action of the Weyl group gauge identifications on CB vevs are calculable from the classical Higgs mechanism.
The monodromy map actions $S_\C$ and $S^\v_\C$ are now representations (group homomorphisms) of the monodromy group $S \subset \Sp_{J^\v}(2r,\Z)$ and its dual $S^\v \subset \Sp_J(2r,\Z)$, respectively, by virtue of \eqref{fix tau}.
They are faithful, dual, $r$-complex-dimensional representations of the monodromy group, e.g. $\Im S_{\C} \cong S$ as abstract groups.
Invariance of the symplectic form $\Om$ \eqref{Om form} requires that the orbifold group acts in the dual (a.k.a., contragredient) representation $W_\C^\v$ on the abelian variety fiber.
Also, because the CB geometry is isotrivial, the abelian variety fiber is constant, so $A_u \deq A_\bt$ for some fixed $\bt \in \Fix(S)$.
Thus the orbifold description of the CB extends to an orbifold description of the algebraic integrable system,
\begin{align}\label{alg orb}
    \cA = (A_\bt \times \C^r) / (W_\C^\v \oplus W_\C) .
\end{align}

\subsubsection{Relation between the CB monodromy group and the Weyl group.}\label{appA: Mono equiv Weyl} 

We now argue that the EM monodromy group $S$, given by the image of $\pi_1(\cC^*)$ in $\Sp_{J^\v}(2r,\Z)$ under the monodromy map $S_{\Z}$, is isomorphic to the Weyl group $W(\fg)$ as abstract groups and to its complex reflection representation image, $W_{\C}$.
This enables us to construct the monodromy map from a faithful symplectic $\Z$-reflection rep of $W(\fg)$, $R_{\Z}$.
In particular, we will show that there exists a group surjection $\ps: \pi_1(\cC^*) \to W_{\C}$ that implies the monodromy map $S_{\Z}$, as an integral, unfaithful representation of $\pi_1(\cC^*)$, defines and is defined by a faithful $J$-symplectic $\Z$-reflection rep of $W(\fg)$, $R_{\Z}: W(\fg) \to \Sp_{J^\v}(2r,\Z)$ through the relation $S_{\Z} \deq R_{\Z} \circ \ps$.
Thus, it is isomorphic to the Weyl group complex reflection subgroup $W_\C$ as abstract groups, which implies the action $R_{\C}$ coincides with the monodromy group action $S_\C$ on the special coordinates as matrix group representations in $\GL(r,\C)$.
We can then conclude that an $\cN{=}4$ SK structure is determined by the $\Z$-reflection rep, $R_{\Z}$, of the Weyl group up to choosing the Dirac pairing $J$ and specifying $\bt \in \Fix(W_{\Z})$.

The argument has two ingredients.
The first is a general result on the connection between the orbifold group $W_\C$ (viewed as the matrix subgroup formed by the action of $W(\fg)$ on $\cCt$ through $R_{\C}$) and $\pi_1(\cC^*)$; though elementary, we give a detailed argument for this below, since we do not know where to find it in the literature.
The second is the observation that the monodromy map is trivial in the connected covering space $\til \cC^*$ of the orbifold \eqref{alg orb}, i.e. the closed paths in $\pi_1(\cC^*)$ whose preimages are in $\pi_1(\til \cC^*)$ are all in the kernel of the monodromy map, so $S = \Im S_{\Z} \cong \pi_1(\cC^*) / \pi_1(\cCt^*)$.
This follows from the fact the total space geometry of the CB geometry cover is a direct product of the CB covering space, $\til \cC \cong \C^r$ with a fixed abelian variety.
We start with some setup for the first ingredient.

The singular locus, $\cD\subset \cC$, is given by the fixed points of non-identity elements of $W_\C$.
Weyl groups are reflection groups, so are generated by reflections, which are elements, $r_i \in W_\C$, which fix a codimension-1 hyperplane $\til\cD_i \subset \C^r$.
The reflection hyperplanes $\til\cD_i$ are the preimages under the orbifold quotient map of the singular points of $\cC$ with respect to the polarized abelian variety fiber.
Denote their union by $\til\cD \deq \cup_i \til\cD_i$.
The singular locus of the CB is thus $\cD = \til\cD/W_\C$.%
\footnote{Note that the reflection hyperplanes in $\til \cC$ don't necessarily get mapped to hyperplanes in the CB, $\cC$, under the quotient map.
In general, the images of reflections hyperplanes will be more complicated, e.g. non-linear, codimension-1 hypersurfaces in $\cC$.}
Define $\cCt^* \deq \C^r \setminus \til\cD$, an $r$-dimensional complex vector space minus some finite number of distinct $(r-1)$-dimensional subspaces.
Then the smooth points of the CB are $\cC^* = \cCt^* / W_\C$.
Since we have removed the fixed points, $W_\C$ acts freely on $\cCt^*$, and so $\til \cC^*$ is a connected cover of $\cC^*$ with covering map $\vp: \cCt^* \to \cC^*$.

For a point $p \in \cC^*$, pick one preimage $\til p \in \vp^{-1}(p)$.
Then $\vp^{-1}(p)$ is the discrete set $\{ W_\C \cdot \til p \} \subset \cCt^*$.
By forming the fundamental groups of $\cCt^*$ and $\cC^*$ with base points $\til p$ and $p$, respectively, an elementary argument shows that there is an exact sequence of groups
\begin{align}\label{orb 3}
    1 \to \pi_1(\cCt^*) 
    \xrightarrow{\ \iw\ }
    \pi_1(\cC^*) \xrightarrow{\ \psi \ } \pi_1(\cC^*)/\pi_1(\cCt^*) 
    \overset{\pst}{\cong} W(\fg)\to 1 
\end{align}
where $\iw: \pi_1(\til \cC^*) \to \pi_1(\cC^*)$ is the induced map of fundamental groups coming from the covering map. 
Note that, since the sequence is exact, $\iw$ is injective so $\Im \iw \cong \pi_1(\til\cC^*)$ and $\psi: \pi_1(\cC^*) \to \pi_1(\cC^*)/\pi_1(\til\cC^*) $ is a surjection with kernel, $\ker \psi = \pi_1(\cCt^*)$. 
Therefore, $\Im \ps \cong \pi_1(\cC^*) / \pi_1(\cCt^*)$ which, after identifying the quotient with $W(\fg)$ as abstract groups through $\pst$, implies $\Im \ps \cong W(\fg)$ by group isomorphism theorem.
To show this we need that: (1) the map $\iw$ induced by the covering map $\vp$ is an injective group morphism; and (2) the resulting quotient is isomorphic to the orbifold group $W_\C$.
\begin{enumerate}
    \item 
Since $\cCt^*$ is a cover of $\cC^*$, $\vp$ maps homotopic paths to homotopic paths, and maps concatenations of paths to concatenations, so $\iw$ is a group morphism.
The identity in $\pi_1(\cC^*)$ is homotopic to the trivial path based at $p$, and so lifts by $\vp^{-1}$ to the trivial path based at $\til p$.  
Thus $\iw^{-1}([\mathbf 1_{\g}]) = [\mathbf 1_{\til \g}]$ in $\pi_1(\cCt^*)$, and $\iw$ is injective.
    \item
Any $\g\in\pi_1(\cC^*)$ lifts to a unique homotopy class of paths $\til\g$ in $\cCt^*$ with endpoints $\til\g(0)=\til p$ and $\til\g(1) = R_{\C}(g_\g) \cdot \til p$ for some $R_{\C}(g_\g) \in W_\C$.%
\footnote{This characterization of elements of $\pi_1(\cC^*)$ in terms of open paths in $\cCt^*$ with endpoints related by $W(\fg)$ follows from the fact that $\vp$ is a (normal) covering map (which satisfies the homotopy lifting property) and the deck group of $\cCt^*$ is the orbifold group, so the action of $W(\fg)$ on the covering space $\cCt^*$ (which is not simply-connected despite $\pi_1(\cCt^*)$ generating trivial monodromies) is a so-called ``covering space action", as defined in \cite{hatcher2002algebraic} (see section 1.3).}
Any representative of coset element $[\g] \in \pi_1(\cC^*)/\pi_1(\cCt^*)$ can be written $\g \n$ for some $\n \in \im(\iw)$.
Its pre-image in $\pi_1(\cCt^*)$ is $\iw^{-1}( \g \n) = \iw^{-1}(\g) \, \iw^{-1}(\n) = \til\g \til\n$ which is a path starting at $\til p$ and ending at $R_{\C}(g_\g) \cdot \til p$.
Furthermore, any path $\til\d$ in $\cCt^*$ starting at $\til p$ and ending at $R_{\C}(g_\g) \cdot \til p$ can be written as $\til\d = \til\g \til\n$ with $\til\n \deq \til\g^{-1} \til\d \in \pi_1(\cCt^*)$ since it starts and ends at $\til p$.
Thus the cosets $[\g] \in \pi_1(\cC^*)/\pi_1(\cCt^*)$ are in 1-to-1 correspondence with elements of $R_{\C}(g_\g) \in W_\C$.%
\footnote{Note that $\pst$ defines a \emph{group} isomorphism because, relative to a base point in $\cC^*$ with choice of preimage in $\cCt^*$, for two open paths $\til \g_1$ and $\til \g_2$ with endpoints related by two distinct elements $R_{\C}(w_i) \in W_{\C}$, they can be concatenated end-to-end to get an open path $ \til \g_3 \deq \til \g_1 * (R_{\C}(w_1)\cdot \til \g_2) $ (with respect to the same base point preimage) with endpoints related by the product $R_{\C}(w_2)R_{\C}(w_1)$.
The concatenated open path $\til \g_3$ will correspond to the closed path concatenation $\g_3 \deq \g_2 * \g_1$ of the two closed paths $\g_i$ in $\pi_1(\cC^*)$ that $\til \g_i$ are the preimages of.
Therefore, the mapping of $\g \in \pi_1(\cC^*)/\pi_1(\cCt^*)$ to its corresponding open path $\til \g$ in $\cCt^*$ with endpoints related by an element of $W_{\C}$ is a group morphism.}
\end{enumerate}

This completes the argument for the first ingredient. 
The second ingredient follows because of the direct product structure of the orbifold covering space in \eqref{alg orb}, any closed path $\til\n \subset \C^r$ in the $\cC$ orbifold covering space is accompanied by a trivial monodromy on $\L$ (the 1-homology of $A_\bt$), and therefore also for the image of this path in the $\cC$ orbifold \eqref{cplx orbi}.
The image of $\pi_1(\cCt^*)$ in $\pi_1(\cC^*)$ is governed by the injection $\iw$, so
\begin{align}\label{im mufG}
    \Im(S_{\Z} \circ \iw) = \mathbf 1,
\end{align}
the trivial subgroup of $\Sp_{J^\v}(2r,\Z)$.%
\footnote{This implies $\ker(S_{\Z} \circ \iw) \cong \pi_1(\cCt^*)$ by first group isomorphism theorems, so we have $\ker \ps \cong \ker(S_{\Z} \circ \iw)$.)}

With these two ingredients in hand, it follows that the monodromy map $S_{\Z}$ determines and is determined by an integral symplectic reflection representation of the orbifold group,
\begin{align}\label{orb 6}
    R_\Z: W(\fg) \to \Sp_{J^\v}(2r,\Z).
\end{align}
$R_\Z$ determines $S_{\Z}$ using the surjection $\psi$, $S_{\Z} \deq R_\Z \circ\psi$.
On the other hand, $S_{\Z}$ determines $R_\Z$ using the (right) inverse of $\psi$, $R_\Z(g) \deq S_{\Z}\circ\psi^{-1}(g)$, where $\g=\psi^{-1}(g)$ is any representative of the quotient equivalence class.%
\footnote{Note that this definition of $R_{\Z}$ is consistent with it being a $\Z$-reflection representation of $W(\fg)$, which means it should be faithful.
This is because every non-trivial element of $W_{\C}$ corresponds to a linking monodromy in $\pi_1(\cC^*)$, which means it is non-trivial under $S_{\Z}$.
So the only element of $W_{\C}$ in $\ker(S_{\Z} \circ \ps^{-1})$ is the identity element $R_{\C}(\mathbf 1_W)$ corresponding to the trivial loop in $\pi_1(\cC^*)$.}
This makes sense since if $\g$ and $\g'$ are in the same class, so $\psi(\g)=\psi(\g')=g$, then $S(\g') = S(\g'\g^{-1}\g)
= S(\g'\g^{-1})S(\g) = S(\g)$ where we used that $\g'\g^{-1} \in \ker\psi = \im\iw$ together with \eqref{im mufG}.%
\footnote{Using the fact that $\psi$ is a group homomorphism, we have $\psi(\g' \g^{-1}) = \psi(\g')\psi(\g^{-1}) = g (\psi(\g))^{-1} = g g^{-1} = 1 \in W_{\C}$.} 
Thus, having established the existence of $\psi$, we learn that if we are given $S_{\Z}$ or $R_{\Z}$, we can always define the other using $\psi$ such that the following diagram commutes,
\begin{center}
\begin{tikzcd}
\pi_1(\cC^*) \arrow[dr,
"S_{\Z}"] \arrow[d, "\psi"] \\
W(\fg) \arrow[r,
"R_{\Z}"]
& \Sp_{J^\v}(2r,\Z)
\end{tikzcd}
\end{center}

Moreover, we learn not only that $S \equiv \Im S_{\Z} \cong W_{\C} \cong W(\fg)$ as abstract groups, but also, since the monodromy group $S$ has the linear action $S_\C$ on the same $\C^r$ space of special coordinates, (so $S \cong_{\C} \Im S_{\C}$ are equivalent as complex reps over $\C$) that $\Im S_\C \cong_\C W_\C$ as matrix subgroups in $\GL(r,\C)$, i.e.  $S_{\C}$ and $R_{\C}$ yield isomorphic representations on $\C^r$.%
\footnote{Recall that linear complex representations are equivalent/isomorphic over the complexes, $\r \cong_\C \r'$, if there exists a $C\in\GL(r,\C)$ such that for all $g\in G$, $\r(g) = C \, \r'(g) \, C^{-1}$.
The mapping $C$ implementing the isomorphism is called an intertwiner, and the $\C$-equivalence relation says it is a $G$-equivariant map.}
Although the $S_\C$ representation \eqref{SC rep} depends on the value of $\bt$, it is easy to see that the condition $\Im S_\C \cong_\C W_\C$ is satisfied for any $\bt\in\Fix(S)$ as long as
\begin{align}\label{orbi ref Ra}
    S_{\Z} \cong_\R R_\R \oplus R_\R \quad  \Longrightarrow \quad S \cong_{\R} W_{\R} \oplus W_{\R}.
\end{align}
But this result follows by Matschke's theorem,%
\footnote{Matschke's theorem states that representations of a finite group are completely decomposable over $\R$.
But this result also follows from the fact $S = W_{\Z}$ according to $S_{\Z} = R_{\Z} \circ \ps$.}
so the result follows after combining \eqref{orbi ref Ra} with the the $\R$ equivalences, $\Im S_\C \cong_\R \R \otimes_\Z S$ and $R_\C \cong_\R R_\R \oplus R_\R$.
Thus, the integral symplectic representation $S_{\Z}$ of the monodromy group $S = \Im S_{\Z}$ is equivalent over the reals to two copies of the real reflection representation $R_{\R}$ of the orbifold group $W(\fg)$.

Conversely, since the linear coordinates of the $\C^r$ covering space upon which the Weyl group acts are (the pullback of) the special coordinates on the CB, given a choice of $\bt \in \Fix(S)$ for $S = \Im (R_{\Z} \circ \ps)$, the monodromy transformations in $S$ then determine the special coordinates, since their form can be fixed from their required monodromy transformations properties under the action of $S_{\C}$ as determined from the triple $(J, S_{\Z}, \bt)$.

\subsubsection{Defining data of an $\cN{=}4$ SK structure and their SK structure orbit.}\label{appA: SKS and orbit triple}

For a gauge algebra $\fg$, we have shown that the triple of data consisting of a choice of symplectic pairing $J$, the $W(\fg)$ integral symplectic reflection representation $R_{\Z}$, and a fixed point $\bt(\t) \in \Fix(R_{\Z})$ for $\t \in \scH_1$, determines the whole SK geometry of an $\cN{=}4$ CB.
So, we define an $\cN{=}4$ SK structure for an $\cN{=}2$ CB geometry $\cC_{\fg} = \C^r / W(\fg)$ with simple compact Lie algebra $\fg$ to be the triple of data,
\begin{align}\label{N4 SKS triple}
    \cN{=}4 \text{ SK structure for Weyl group } W(\fg) \deq  (J, \, R_{\Z},\,  \bt(\t) \in \Fix(R_{\Z}))
\end{align}
where $J$ is the polarization of the pav fiber $A_{\bt}$. 
In the paper, we classify the isomorphism classes of this data for $J$ principally polarized according to SK structure equivalence that is determined by the twisted integral equivalence relations the $R_{\Z}$ representations satisfy; this relation is established in the next section.
The SK structure given by $(J, R_{\Z}, \bt(\t) \in \Fix(R_{\Z}))$ is parametrized by $\t \in \scH_1$ up to identifications $\t \sim \scS_{R_{\Z}} \circ \t$ that are generated by the automorphism group $\scS_{R_{\Z}}$ of the SK structure.
We call $\scS_{\R_{\Z}}$ the \emph{self-duality group} of the SK orbit, and we show that they are parameterized by modular subgroups of $\PSL(2,\R)$.
The identifications on $\t \in \scH_1$ define the associated modular curves $\scC(R_{\Z}) \deq \scH_1 / \scS_{R_{\Z}}$ which represent the \emph{moduli space of the SK structure triple} in the sense that, for each point $\t \in \scC(R_{\Z})$, the corresponding SK structure are distinct.
Therefore, we call $\scC(R_{\Z})$ the \emph{conformal manifold} of the SK structure defined by the triple $(J, R_{\Z}, \bt(\t) \in \Fix(R_{\Z}))$.
It is useful to talk about SK structures in a way that doesn't depend on a point in their conformal manifold, so one can establish what quantities are invariant on $\scC(R_{\Z})$.
To this end, we define the \emph{SK structure orbit} or \emph{SK orbit} to be the set of all triples $(J, R_{\Z}, \bt(\t) \in \Fix(R_{\Z}))$ that are parameterized by $\scC(R_{\Z})$, denoted $(J, R_{\Z}, \Fix(R_{\Z}))$,
\begin{align}\label{N4 SKS orbit}
    \cN{=}4 \text{ SK structure orbit for Weyl group } W(\fg) \deq  (J, \, R_{\Z},\, \Fix(R_{\Z})).
\end{align}
Each point in an SK orbit is necessarily an inequivalent SK structure with symplectic integral reflection representation $R_{\Z}$ of $W(\fg)$.
Therefore, it is completely determined by the pair $(J, R_{\Z})$ and the self-duality group is necessarily an invariant of it.

\subsection{Equivalences of N=4 sYM SK structures}
\label{appA: equivalenceofN=4SKstructures}

In this section we explain the origin of the twisted integral equivalence relation \eqref{twisted_equiv} by showing that it is the minimally sufficient condition that implies two $\cN{=}4$ SK structures are equivalent/isomorphic.
We call such equivalences ``SK structure isometries".
In in appendix \ref{appA: SKS isometries}, we first describe what it means for two SK structures to be isomorphic in terms of the physical data describing the SK geometry, and we characterize a class of isomorphisms that can be canonically defined for $\cN{=}4$ SK structures in terms of the automorphisms of the orbifold group, $W(\fg)$.
From this we can argue that the set of all $\cN{=}4$ SK isometries we need to consider are given by these Weyl group automorphism isometries combined with an EM duality transformation.
In the appendices that follow, we take our time to concretely describe the affect of such automorphisms on the triple $(J,S,\bt)$ defining an $\cN{=}4$ SK structure/orbit.
In the process, we cover some basic mathematical background material on composing group representations with automorphisms of that group.
Because we are interested in the Weyl group as a finite reflection group, we are only concerned with the automorphisms that preserve reflections.
We conclude with a summary of the material in this appendix \ref{appA: equivalenceofN=4SKstructures} and a proof of the following statement (which establishes the completeness of our classification approach): two SK structures are isomorphic iff their integral symplectic reflection representations are related by this twisted integral equivalence relation for some reflection outer automorphism and EM duality transformation.

\subsubsection{SK structure isometries.}\label{appA: SKS isometries}
Two integral symplectic representations, $S$ and $S'$, correspond to equivalent SK structure orbits if they are integrally equivalent, $S \cong_\Z S'$.
In particular, a map implementing untwisted integral equivalence is just a change of basis of the charge lattice defining the pav fibers, and therefore has no effect on any low energy physical observables on the CB.
This is the equivalence that was taken into account in \textbf{Steps 3-5} of our classification algorithm and is described in section \ref{sec SKS twist equiv}.
It corresponds to the trivial automorphisms $\f = \ps = \mathbf 1_W$ in the twisted equivalence relation \eqref{twisted_equiv}.

But there can be other sources of equivalence between SK structures.
In particular, two CB geometries may be equivalent under a map which has an action on the points of the CB in addition to a change of basis of the homology lattice of the fibers.
To be considered equivalent, all physical observables in the low energy effective action on the CB, or, equivalently, all the coordinate- and basis-independent ingredients of the CB SK structures should be isomorphic.
In coordinate-free language, such an equivalence of SK structures is a polarization- and fiber-preserving holosymplectomorphism of the algebraic integrable system.

Relative to a choice of charge lattice basis (and a choice of base point on the CB), this means that an equivalence of CB geometries is a bijection of the CB to itself, $f: \cC \to \cC$, which maps the charge lattice, symplectic pairing, monodromy map, and special coordinates as
\begin{align}\label{isoeq1}
    f :&\ \L_u \to \L'_{u} \deq Z \L_{fu}, & u &\in \cC\nn\\
    f :&\ J \to J' \deq Z^{-1} J Z^{-t}, \\
    f :&\ S(\g) \to S'(\g) \deq Z S(f\g) Z^{-1}, & \g &\in \pi_1(\cC^*) ,\nn\\
    f :&\ \s(u) \to \s'(u) \deq Z^\v \s(fu) , & u &\in \cC, \nn
\end{align}
for some $Z \in \GL(2r,\Z)$.
The subscript on $\L_u$ labels the point in $\cC$ over which the charge lattice is a fiber.
Since the charge lattice experiences nontrivial monodromies, \eqref{isoeq1} should be understood to apply on simply connected subsets of $\cC^*$ containing the base point; similarly for the special coordinate maps.
%
But $Z$ implements an ``ordinary'' $\Z$-equivalence \eqref{Zequiv} of SK structures, so, up to an ordinary $\Z$-equivalence, we are free to take $Z=\bone$.
This then clearly leaves all physical observables --- like the EM duality conjugacy classes of the EM monodromies, the matrix $\bt$ of low energy EM couplings, the CB metric, and the $\cN{=}2$ central charge function --- invariant.
We refer, somewhat sloppily, to such equivalences which act on the points of the CB as \emph{SK structure isometries}.

In the case of $\cN{=}4$ sYM CBs, a class of SK structure isometries are easy to describe explicitly: those coming from reflection automorphisms of the Weyl group.
We have seen that their CBs are orbifolds by Weyl groups and their monodromy maps are essentially integral symplectic representations of the Weyl group $W$ (lifted by the quotient map in \eqref{orb 3}).
Any automorphism, $\f:W\to W$, of $W$ as a reflection group (that is, any group automorphism which maps reflections to reflections) induces a map $f:\cC \to \cC$ which is an SK structure isometry.
In particular, $\f$ determines a $W_\C$-equivariant map $\til f \in \GL(r,\C)$ of the vector space covering the CB orbifold, $\cC=\C^r/W_\C$, which descends to the isometry $f:\cC\to\cC$ upon quotienting by $W_\C$. 

This follows because, as we review in section \ref{sec Z equiv}, all reflection automorphisms of $W$ are either inner automorphisms or are Coxeter diagram automorphisms, so any reflection outer automorphism is necessarily a diagram automorphism, but a diagram automorphism isn't necessarily outer; see table \ref{tab:outer}.
Inner automorphisms, $\f(W) = vWv^{-1}$ for some $v\in W$, act as $\til f = R_{\C}(v)$ on $\C^r$, where $R_{\C}: W_{\fg} \to \GL(r,\C)$ is a complexified reflection representation of $W_{\fg}$.
(This representation is unique as a $\C$-equivalence class in the sense that such representations are all equivalent over $\C$; equivalently, different embeddings in $\GL(r,\C)$ differ by linear coordinate changes.)
By taking $R_{\C}$ to be the complexification of one of the integral representations $R_A$, it is clear that these inner automorphism SK structure isometries are equivalent, up to a change of coordinates, to ``ordinary'' $\Z$-equivalences \eqref{Zequiv}.

Coxeter diagram automorphisms are defined to be certain permutations of a basis of simple roots of $W$.%
\footnote{For reflection groups, the roots are the directions in $\R^r$ perpendicular to the reflection hyperplanes; they are not vectors (with specific norms) as for Lie algebra roots.}
Such a permutation of $r$ linearly independent lines in $\R^r$ can always be implemented by an orthogonal transformation, $\til f(\f) \in \O(r)$, which is interpreted as an element of $\GL(r,\C)$ when acting on $\R^r\otimes\C=\C^r$.
This definition of $\til f(\f)$ is not unique: there are $|W|$ choices corresponding to composing any given choice of $\til f(\f)$ with all combinations of simple reflections.
This corresponds to composing the Coxeter diagram automorphism with the inner automorphisms.
Note, however, that Coxeter diagram automorphisms need not be inner automorphisms.
Even though both give (complexified) orthogonal actions on $\C^r$, the inner ones are characterized by these transformations being in $W_\C$, while it is possible that $\til f(\f) \notin W_\C$ for the Coxeter diagram automorphisms.
It turns out that only $\fg = BC_2, D_{2r}, F_4$ and $G_2$ have reflection outer automorphisms, and in each case all diagram automorphisms are of this automorphism type, c.f. table \ref{tab:outer}.

As we explained in appendix \ref{appA: SKS and orbit triple}, the defining data of an SKS is the triple $(J,S,\bt \in \Fix(S))$.
Furthermore, by definition all SK isometries will lead to isomorphisms of this data, so the task is to understand which ones are non-trivial.
Such SK isometries are the ones that generate equivalences between SK structures.
Since $J$ is something we choose to be fixed (because we don't consider isogenies of the fiber, which are the transformations that could change $J$), and since $\bt$ is determined up to a point in the conformal manifold that is determined $R_{\Z}$, the only piece of data that can be toggled through such transformations is $R_{\Z}$ itself.
Thus, a non-trivial SK isometry is one that transforms $R_{\Z}$ non-trivially.
And as we've just argued, the most general such transformation is a combination of a reflection automorphism with an EM duality transformation. 
Since all inner automorphisms are reflection preserving and correspond to EM duality transformations themselves, we can restrict ourselves the reflection outer automorphisms which don't correspond to EM duality transformations.
So the most general, non-trivial SK isometry is a combination of a reflection outer automorphism with an EM duality transformation.
As we will use the rest of this appendix to show explicitly, working out the consequences of this transformation leads to reflection automorphism twisted $\Z$-equivalence relation.
In conclusion, all non-trivial $\cN{=}4$ SKS isometries --- that is, all isometries which transform $R_{\Z}$ and therefore transform the defining triple of data ---  are classified by the reflection outer automorphisms of $W(\fg)$ and a choice of EM duality transformation.

\subsubsection{Automorphism twisting the Weyl group complex reflection action.}\label{appA: twist R_C action} 
The Weyl group is realized as a complex reflection group by defining an action on $\C^r$ in terms of $\GL(r,\C)$ matrices that form a matrix subgroup that is isomorphic to $W(\fg)$ as an abstract group and is generated by reflections. 
This realizes the complex reflection representation $R_{\C}$ of $W(\fg)$ which we construct from the real reflection representation $R_{\R}$ according to $R_{\C} \deq \C\otimes R_{\R}$.
As we showed in the last section, we can automorphism twist $R_{\C}$ using basis changes in $\cCt\cong \C^r$, which means we can also automorphism twist the complex reflection action of $W(\fg)$ on $\C^r$ that is defined by $R_{\C}$.
We will explain this automorphism twist of the action in pedantic detail for future reference.

It is a fact that the complex reflection representation $R_{\C}$ is unique up to $\C$-equivalence.
This means we can perform basis changes in the CB cover $\til \cC$ to obtain the composition of $R_{\C}$ with a reflection outer automorphism $\phi$.
This is simply because the composition of $R_{\C}$ with any such $\phi$ will yield another complex reflection rep of $W(\fg)$, so it must be conjugate to $R_{\C}$ for some intertwiner, i.e. basis change matrix in $\GL(r,\C)$.
This has the effect of twisting the action of $W(\fg)$ on $\C^r$ by such an outer automorphism.

The action of $W(\fg)$ is defined through the complexified real simple reflection matrices $ R_{\C}(w_i) \in W_{\C}$.
If we define the action of a simple reflection $w_i \in W(\fg)$ on a simple root basis element $\b_j$ to be $w_i \circ \b_j \deq R_{\C}(w_i)_{jk} \b_k$, then $W(\fg)$ will act on coordinate components $b^j$ of a generic point $x = b^j \b_j \in \C^r$ in the cover with the inverse transpose matrix $R^{\v}(w_i) \deq R_{\C}(w_i)^{-T}$, $w_i \circ b^j \deq R_{\C}^{\v}(w_i)^{jk}b^k$. 

Now let $\phi \in \Out_{\rfl}(W)$ be a reflection preserving outer automorphism with corresponding basis change $B_{\phi} \in \GL(r,\C)$.
Upon implementing the basis change by $B_{\phi}$, the simple reflection matrices are mapped to themselves as a set but permuted among themselves according to $\phi$,
\begin{align}\label{outer rfl rep}
    R_{\C}^{\v}(w_i) \overset{\f}{\mapsto}  B_{\f}^{\v} R_{\C}^{\v}(w_i)B_{\f}^T = R_{\C}^{\v}(\phi(w_i)) = R_{\C}^{\v}(w_j),
\end{align}
where $w_j = \phi(w_i)$ and $i,j = 1, \cdots, r$ are not necessarily equal.
This has the affect of twisting the action of $W(\fg)$ by $\phi$ on a generic point of the cover, $x = b^j_{\phi} \b^{\phi}_j \in \C^r$, written in the new basis of simple roots $\{\b_j^{\phi} \}$.
In particular, acting with $w_i$ implements the matrix transformation by $w_j$,
\begin{align}\label{twist action}
w_i \circ b^k_{\phi} = (R_{\C}^{\v}(\phi(w_i))b_{\phi})^k = (R_{\C}^{\v}(w_j)b_{\phi})^k.
\end{align}
As a consequence, if two points $p, q \in \C^r$ were initially related by the action of $w_j$, $q^k = w_j \circ p^k = (R_{\C}^{\v}(w_j)p)^k$, those same points are now related by the action of $w_i$ after twisting the $W(\fg)$ action by $\phi$,
\begin{align}\label{twist points}
w_i \circ p^k = (R_{\C}^{\v}(\phi(w_i))p)^k = (R_{\C}^{\v}(w_j)p^k) = q^k.
\end{align}
Furthermore, this also implies that the hyperplane fixed by the action $w_j$ is now fixed by the action of $w_i$.
If $v \in \C^r$ was initially in the hyperplane fixed by the reflection of $w_j$, then $w_j \circ v^k = (R_{\C}^{\v}(w_j) v)^k = v^k$, then after twisting the action of $W(\fg)$ it is now fixed by the action of $w_i$, 
\begin{align}\label{twist hyperplane}
    w_i \circ v^k = (R_{\C}^{\v}(\phi(w_i))v)^k = (R_{\C}^{\v}(w_j)v)^k = v^k.
\end{align}
To summarize, we have shown that implementing the basis change $B_{\f}\in \GL(r,\C)$ in $\cCt$ corresponding to $\f \in \Out_{\rfl}(W)$ has the effect of twisting the action of $W(\f)$ on $\C^r$ (defined by $w \circ p = R_{\C}(w) \cdot p$) through the $\f$-twist of the $\C$-reflection rep $R_{\C}$,
\begin{align}\label{C ref rep twist}
    R_{\C} \overset{\f}{\mapsto} B_{\f}R_{\C}B_{\f}^{-1} = R_{\C}\circ \f.
\end{align}
In terms of the orbifold geometry of the CB, this has the geometric effect of changing the reflection hyperplane that a simple reflection fixes.
In particular, if $R_{\C}(w_k)$ initially fixes the hyperplane $\cDt_k$ and $w_k = \f^{-1}(w_i)$, then after $\f$-twisting the reflection $w_k \deq \f^{-1}(w_i)$ now fixes the hyperplane $\cDt_i$. 
Importantly, because the complex reflection actions of $W(\fg)$ given by $R_{\C}$ or $R_{\C}\circ \f$ are equivalent over $\C$ (by construction), their quotients will be the exact same topologically, so the CB $\cC_{\fg} \deq \cCt / W_{\fg}$ remains invariant under the $\f$-twisting. 

\subsubsection{Monodromy transformations and the Weyl group.}\label{appA: Mono trans Weyl grp}

Monodromy transformations are $\Sp_{J^{\v}}(2r,\Z)$ elements forming the monodromy group $S = \Im S_{\Z}$
that are associated to the linking monodromies in $\cC^*$ that comprise $\pi_1(\cC^*)$.
The short exact sequence \eqref{orb 3} in appendix \ref{appA: N4 SKS} implies this data is captured by a symplectic $\Z$-reflection representation $R_{\Z}$ of the orbifold group $W(\fg)$ which, as a $\GL(2r,\Z)$ representation, naturally defines an action of $W(\fg)$ on the symplectic lattice of the pav fiber, $\L = H^1(A_{\bt}, \Z) \cong_{\Z} \Z^{2r}$.%
\footnote{In particular, $S_{\Z}$ and $R_{\Z}$ are related by the group surjection $\ps: \pi_1(\cC^*) \to W(\fg)$, $S_{\Z} \deq R_{\Z}\circ \psi$, which implies $W_{\Z} \cong_{\Z} S$ and $S \cong W_{\fg}$ as abstract groups.
As explained in \ref{appA: Mono equiv Weyl}, this implies $\R \otimes_{\Z} S \cong_{\R} W_{\C} \cong_{\R} W_{\R} \oplus W_{\R}$ as linear representations over $\R$.}
In this section we recall the way in which $R_{\Z}$ determines the monodromy map $S_{\Z}$ by describing the way in which particular linking monodromies in $\cC^*$ and their monodromy transformations are associated to particular (e.g., reflection) Weyl group elements.
In doing so, we will implicitly define the group surjection $\ps: \pi_1(\cC^*) \to W_{\C}$ in \eqref{orb 3} from which we can then define monodromy transformations through $S_{\Z} \deq R_{\Z} \circ \ps$.
Then, from the argument that defines $\ps$, it is straightforward to understand how twisting the $R_{\C}$ action on $\cCt$ by a reflection outer automorphism will transform the monodromy transformations that are associated to certain linking monodromies in the CB.
We address this twisting of the monodromy transformations in the next section.

In what follows, we will distinguish between the monodromy map $S: \pi_1(\cC^*) \to \Sp_J(2r,\Z)$, its image $S_{\Z} \deq \Im S$ and the corresponding symplectic $\Z$-reflection rep of $W(\fg)$, $R_{\Z}: W(\fg) \to \Sp_J(2r,\Z)$ with image $W_{\Z}$.
We assume we are given $R_{\Z}$ and define $S \deq R_{\Z} \circ \psi$.

The Weyl group orbifold geometry of the CB implies that all closed paths in $\g \in \pi_1(\cC^*)$ must come from a unique homotopy class of paths $[\til \g]$ in $\cCt^*$ such that their endpoints $\til\g(0)=\til p$ and $\til\g(1) = w_{\g} \circ \til p = R_{\C}(w_{\g})\cdot \til p$ are related by the action of some $w_{\g} \in W(\fg)$ on $\til \cC$ using the complex reflection representation $R_{\C}$.
In particular, if the end points are equal $\til p = w_{\g} \circ \til p$, then $\til \g \in \pi_1(\til \cC^*)$ is closed and $w_{\g} = \mathbf 1_W$ must be the identity element of $W(\fg)$.%
\footnote{This follows because $\til \g \subset \cCt^*$ and $W(\fg)$ acts transitively on $\cCt^*$, i.e. without non-trivial fixed points, once we have removed the reflection hyerplanes.}
It is then natural to identify $w_{\g}$ with $\g$, $w_{\g} \leftrightarrow \g$, and use this identification to define the group surjection $\psi$, $\g \mapsto \psi(\g) \deq w_{\g}$.
Given the $\Z$-reflection representation $R_{\Z}$ of $W(\fg)$, we obtain the induced monodromy transformation $S(\g) \in \Sp_{J^\v}(2r,\Z)$ associated to the linking monodromy $\g$ through $R_{\Z}$ by setting, $S(\g) \deq R_{\Z}(\psi(\g)) = R_{\Z}(w_{\g})$.

In summary, from the short exact sequence \eqref{orb  3}, we learn that the monodromy transformation $S(\g)$ associated to the linking monodromy $\g \in \pi_1(\cC^*)$ is given by $R_{\Z}(w)$ where $w \in W(\fg)$ is the Weyl group element that relates the endpoints of the homotopy class of paths $[\til \g]$ in $\til \cC^*$ whose image in the CB under the cover map $\vp$ is $\g$.
Moreover, if $\g$ links the discriminant component $\cD$ whose preimage is not a reflection hyperplane $\cDt_i$ in the cover, then the corresponding Weyl group element $w$ relating the endpoints of the open path preimage of $\g$ will not be a reflection element, and so its fixed point set in $\cCt$ will not be a single reflection hyperplane but will be some union of them.
Lastly, in this construction we have assumed that we know what homotopy class of open paths $[\til \g_w]$ is the preimage of a given linking monodromy $\g_w \in \pi_1(\cC^*)$, and we subsequently used this to consistently define the surjection $\psi(\g_w) \deq w$ that defines the monodromy transformation $S(\g_w) \deq R_{\Z}(w) = R_{\Z}(\psi(\g_w))$ in terms of $R_{\Z}$.
We could just as well have assumed we know what the surjection $\psi$ is to define the group isomorphism between $S_{\Z}$ and $W_{\Z}$ using $S \deq R_{\Z} \circ \psi$.
This gives us an association between linking monodromies $\g_w \in \pi_1(\cC^*)$ and Weyl group elements $w \in W(\fg)$ from which we can consistently define the homotopy class of paths $[\til\g_w]$ that are the preimages of the linking monodromies $\g_w$.

A crucial ingredient in the relation between linking monodromies and the Weyl group comes from the fact that the preimages of closed paths in $\pi_1(\cC^*)$ in terms of a homotopy class of paths in $\til \cC^*$ with end points related by Weyl group elements.
If we consider this fact in terms of the generating reflections of $W_{\R}$, then the preimages of the discriminant locus being linked are reflection hyperplanes, and we can explicitly describe the homotopy classes of paths that would serve as the preimage of such monodromies.
Take $\til p \in \vp^{-1}(p)$ to be in the preimage of a base point $p \in \cC^*$ with respect to which we can define the fundamental groups of $\til \cC^*$ and $\cC^*$, respectively.
Consider a reflection hyperplane $\til \cD_{i} \subset \til \cC$ fixed by the simple reflection $R_{\C}(w_i)$ whose image in the CB under the covering map is the discriminant component $\cD_{i}\subset \cC$, and let $\til \cD_i^{\perp}\cong \C$ be the $1$-dimensional orthogonal subspace to $\til \cD_i$.
We define $\til \g_i(\til p,\til q): [0,1) \to \til \cD_i^{\perp} \subset \til \cC$ to be the open path not containing the origin of $\til \cD_i^{\perp}$ and with endpoints $\til \g_i(0) = \til p$ and $\til \g(1) \deq \til q$ which are related by the action of the simple reflection $w_i$,
\begin{align}\label{}
    w_i\circ \til p = R_{\C}^{\v}(w_i)\til p = \til q.
\end{align}
Because $\til p$ and $\til q$ are related by $R_{\C}(w_i)$, upon passing to the CB orbifold quotient using the covering map $\vp: \til \cC^* \to \cC^*$, $\til \g_i$ becomes the closed path $\g_i \deq \vp \circ \til \g_i \in \pi_1(\cC^*)$ that resides in the image of $\til \cD_i^{\perp}$ under $\vp$ and links the discriminant component $\cD_i$ corresponding to the image of $\til \cD_i$, $\vp(\til \cD_i) = \cD_i$. 
The homotopy class of the closed paths $[\g_i] \in \pi_1(\cC^*)$ coming from the homotopy class of open paths $[\til \g_i]$ is precisely the set of linking monodromies that get assigned the monodromy transformation $S(\g_i) \deq R_{\Z}(\psi(\g_i)) = R_{\Z}(w_i)$ coming from the symplectic $\Z$-reflection rep of $W(\fg)$.

There are two potential ambiguities that could occur in assigning reflection elements to linking monodromies to generate the corresponding monodromy transformation through the $\Z$-reflection representation of the orbifold group.
The first potential ambiguity could come from CRGs with reflection elements $w \in G$ that have order $n >2$, $w^n = \mathbf 1_G$.%
\footnote{This is not an issue for Weyl groups because all reflections have order two.}
The second potential ambiguity comes about when there exist distinct reflections $w_1 \neq w_2$ whose corresponding reflection hyperplanes $\cDt_i$ both get mapped to the same discriminant component under the covering map $\vp$.
This second scenario comes about when there exists distinct reflections that are in the same conjugacy class of $G$.%
\footnote{This case is relevant for certain Weyl groups, e.g. $W(A_n) = S_{n+1}$, because their reflection elements do not always corresponding to distinct conjugacy classes.}
In both cases, the potential ambiguity arises from the fact that we have distinct reflections that are associated to monodromies linking the same discriminant component, so the question is how to understand the differences between the linking monodromies that correspond to these distinct reflections.
Both cases can be resolved by understanding how the linking monodromies, their preimages, and the associated monodromy transformations are all distinguished relative to the distinct reflections.

In the first case, let us assume $w$ is a simple reflection of maximal order, i.e. it can't be written as the power of another reflection.
Then, when this occurs, $w$ and $w^i$ for $i = 1, \cdots, n-1$ all fix the same reflection hyperplane $\cDt_w$ which gets mapped to the same discriminant component $\cD_w$ in $\cC$, so whatever linking monodromy they each get mapped to will link the same discriminant component.
Thus, one might ask what distinguishes between these linking monodromies and their associated monodromy transformations that are assigned to these different powers of $w$?
The answer is that, for each $w^i \neq \mathbf 1_G$, there still corresponds a homotopy class of open paths $[\g_w^i]$ whose endpoints are related by the action of $w^i$ and which serve as the preimages of linking monodromies $\g_w^i$ of the same discriminant component $\cD_w$.
In particular, these classes of open paths are distinct since they have different endpoints that are related by powers of $w$.
Therefore, if $\g_w \equiv \g_w^1$ is the linking monodromy corresponding to open paths related by $w$ with monodromy transformation $S(\g_w) \deq R_{\Z}(w)$, the monodromy transformations associated to the higher power linking monodromies $\g_w^i$ are given by the corresponding power of $R_{\Z}(w)$, $S(\g_w^i) \deq R_{\Z}(w^i) = R_{\Z}(w)^i$.
This implies the linking monodromies $\g_w^i$ correspond to $i$th order windings of the $\g_w$ monodromy, which are distinct as closed paths in $\pi_1(\cC^*)$.

Now let's transition to the second case where we have distinct reflections $w_i$ with distinct reflection hyperplanes $\til \cD_i$ that all get mapped to the same discriminant component $\cD$ upon passing to the orbifold quotient by a CRG $G$.
As mentioned in the footnote, this occurs when distinct reflections exist in the same conjugacy class of the CRG, $G$, so their hyperplanes get identified upon passing to the orbifold quotient and they are subsequently mapped to same discriminant component $\cD$.
In this case, for each such reflection $w_i \in G$, there still exists a distinct homotopy class of open paths $[\til \g_i]$ with endpoints related by $R_{\C}(w_i)$ which is the preimage of a monodromy $\g_i\in \pi_1(\cC^*)$ linking the discriminant component $\cD \subset \cC$.
Moreover, the associated monodromy transformations $R_{\Z}(w_i)$ will also be distinct because $R_{\Z}$ is faithful.
So, the question is, in what way are the linking monodromies of $\cD$ distinct?
The answer comes from the fact that a single discriminant component in the CB is allowed to possess multiple linking monodromies that are topologically inequivalent, i.e. not homotopic. 
In particular, the number of inequivalent linking monodromies it will possess is equal to (or at least, correlated to) the number of distinct reflection hyperplanes that are in its preimage, and this number is equal to the number of distinct reflections in the corresponding conjugacy class of the reflection elements fixing those hyperplanes.
So, in this scenario, for each distinct reflection of the CRG there is a distinct homotopy class of open paths with corresponding distinct linking monodromy and monodromy transformation. 

In conclusion, let us state what we have learned from these two scenarios in understanding the relation between linking monodromies, their monodromy transformations, and reflection elements of the CRG: distinct reflections always correspond to distinct homotopy classes of open paths in $\cCt^*$ (with endpoints related by the reflection) that serve as the preimages for distinct linking monodromies in $\pi_1(\cC^*)$ with distinct monodromy transformations determined by those reflections.
The key point is that, if the monodromies of distinct reflections link the same discriminant component, they will be distinguished in terms of their relative winding or by being topologically distinct loops in $\pi_1(\cC^*)$.

\subsubsection{Twisting the monodromy transformations of linking monodromies.}\label{appA: mono trans twist}
We now describe how twisting the Weyl group action $R_{\C}$ on the cover $\til \cC$ by a reflection outer automorphism $\phi$ transforms the monodromy transformations $S_{\Z}(\g) \in \Sp_J(2r,\Z)$ that are associated to the linking monodromies $\g \in \pi_1(\cC^*)$ in the smooth CB, $\cC^*$.
The discriminant components in $\cC^*$ are defined to be those that come from the image(s) of a reflection hyperplane in $\til \cC^*$ where such hyperplanes are fixed by a reflection element $w\in W(\fg)$.
The monodromies that link these discriminant components generate $\pi_1(\cC^*)$.
We will assume we have some fixed symplectic $\Z$-reflection rep $R_{\Z}$ such that $S \deq R_{\Z}\circ \psi$ with respect to the group surjection $\psi$.

In defining the Weyl group element that is associated to a given linking monodromy and its monodromy transformation, the action of the Weyl group on $\cCt$ via the complex reflection representation $R_{\C}$ played a prominent role. 
In particular, it determined which Weyl group element related the end points of the open path lift of a given closed path linking monodromy, and it was this Weyl group element that was used to define the associated monodromy transformation.
Thus, if we can change the definition of this action such that the Weyl group element relating the end points changes but without changing the configuration of fixed hyperplanes in $\cCt$, i.e. so the original linking monodromy and lift exist after changing the action, then we will change what Weyl group element is associated to that linking monodromy and its monodromy transformation. 
Changing the action through the basis change that twists the Weyl group action by a reflection outer automorphism is precisely the transformation that will permute Weyl group reflection elements but preserve the configuration of reflection hyperplanes.

Consider a monodromy $\g_i \in \pi_1(\cC^*)$ linking a simple discriminant component $\cD_i$ whose preimage is a hyperplane $\cDt_i$ that is fixed by a simple reflection $w_i \in W(\fg)$ under the action by $R_{\C}$,
\begin{align}
    w_i \circ v = R^{\v}_{\C}(w_i) v = v \quad \forall \,\, v \in \cDt_i.
\end{align}
Let $[\til \g_i]$ be the preimage homotopy class of paths with endpoints $\til\g_i(0) = \til p$ and $\til \g_i(1) = w_i \circ \til p = R^{\v}_{\C}(w_i)p$ being related by the action of $w_i$.
Then, as before, this defines the group surjection $\psi: \pi_1(\cC^*) \to W(\fg)$ for all simple discriminant linking monodromies according to $\psi(\g_i) \deq w_i$.
Now lets perform the basis change $B_{\phi}\in \GL(r,\C)$ corresponding to the composition of $R_{\C}$ with a reflection outer automorphism $\phi \in \Out_{\rfl}(W)$, so $R^{\v}_{\C}(w) \mapsto B_{\f}^{\v} R^{\v}_{\C}(w) B_{\f}^{T} = R^{\v}_{\C}(\phi(w))$ for all $w\in W(\fg)$, and the complex reflection action is now
\begin{align}
    w_i \circ p \deq R^{\v}_{\C}(\phi(w_i))p = R^{\v}_{\C}(w_j) p
\end{align}
where $\f(w_i) = w_j$ is another simple reflection in $W(\fg)$.
Note that, in general, basis transformations in the cover $\cCt$ have no affect on the $\Z$-rep $R_{\Z}$ we have chosen to define the monodromy transformations with respect to.
However, as we will show, only when the basis transformation corresponds to a reflection outer automorphism $\phi \in \Out_{\rfl}(W)$ can we interpret it as transforming $R_{\Z} \mapsto R^{\phi}_{\Z}\deq R_{\Z} \circ \phi^{-1}$.
Also, basis changing by $B_{\phi}$ is equivalent to permuting the simple reflections generating the image of $R_{\C}$ according to $\phi$ because it changes the labels of simple reflections, which are made with respect to a choice of simple roots $\{\a_i \}$. (Here we have $\a_i \mapsto \phi(\a_i) = \a_j$.) 
So, $R_{\C}$ and $R_{\C}\circ \f$ are isomorphic not just as matrix subgroups, but equal as sets.
Therefore, the reflection hyperplanes $\cDt_i \in \cCt$ are also permuted among each other in an analogous fashion, so the exact same set of fixed hyperplanes remain before and after the basis change, and $\cCt^*$ is unchanged topologically. 
Furthermore, the orbifold geometry $\cCt^* /(R_{\C}\circ \f)$ defining the new smooth CB is isomorphic to the orbifold geometry defining $\cCt^*/R_{\C}$, so the smooth CB and its topology, i.e. linking monodromies, are left invariant.%
\footnote{Of course, these statement are somewhat trivial because all basis transformations in the cover $\cCt$ are diffeomorphisms, so they will not change the topology of $\cCt^*$ or the orbifold geometry defining $\cC^*$.
What is special about the basis transformations corresponding to \emph{reflection outer automorphisms} is they simply permute the set of matrices defining image of $R_{\C}$ and their corresponding reflection hyperplanes in a way that corresponds to permuting the corresponding set of simple roots labeling them.
This essentially follows from the definition of reflection outer automorphisms as Coexter diagram automorphisms after one applies this transformation to the corresponding Dynkin diagram.
However, when acting on the latter, it may not be an automorphism because it doesn't necessarily leave the length of the simple roots invariant, i.e. a Coxeter diagram automorphism could map between long roots and shorts roots. 
}
We can therefore consider the exact same set of linking monodromies forming $\pi_1(\cC^*)$ and their preimage homotopy classes in $\cCt^*$ upon twisting the action by $\phi$.

Let us do this for $\g_i$ and its preimage homotopy class of open paths $[\til \g_i]$ and ask what Weyl group element relates the endpoints $\til\g_i(0) = \til p$ and $\til \g_i(1) = R^{\v}_{\C}(w_i)p$ of this class once we perform the twist by $\phi$?
To answer this, note that $R^{\v}_{\C}(w_i) = R^{\v}_{\C}(\phi(\phi^{-1}(w_i)) = \phi^{-1}(w_i) = R^{\v}(\phi(w_k))$, which implies
\begin{align}
    R^{\v}_{\C}(w_i)p = R^{\v}_{\C}(\f\circ \f^{-1}(w_i)) p = w_k \circ p.
\end{align}
Therefore, the endpoints of $\til \g_i$ are now related by the simple reflection element $w_k \deq \f^{-1}(w_i)$ instead of $w_i$, so we associate $\g_i \leftrightarrow w_k$ and define the monodromy transformation $S_{\phi}(\g_i)$ of $\g_i$ to now be given by $R_{\Z}(w_k)$ instead of $R_{\Z}(w_i)$.
Going through this process for all the linking monodromies of discriminant components $\cD_i$ with reflection hyperplane preimages and their associated open path uplifts, we will implicitly define a new group surjection given by the pullback of $\psi$ by $\f^{-1}$, $\psi_{\phi} \deq \ps^*\f^{-1} = \f^{-1} \circ \ps : \pi_1(\cC^*) \to W(\fg)$ which satisfies $\psi_{\phi}(\g_i) \deq \phi^{-1}(w_i)  \equiv w_k = (\phi^{-1} \circ \psi)(\g_i)$ where we used the fact that the old surjection satisfied $w_i = \psi(\g_i)$.
Moreover, $\psi_{\f}$ defined in this way will produce the new $\phi$-twisted monodromy map $S_{\Z}^{\phi} \deq R_{\Z} \circ \psi_{\f}$,
\begin{align}
   S_{\Z} \overset{\phi}{\mapsto} S_{\Z}^{\phi} \deq R_{\Z} \circ \psi_{\phi} = R_{\Z} \circ \phi^{-1} \circ \psi, \quad S_{\Z}^{\f}(\g_i) = (R_{\Z} \circ \phi^{-1} \circ \psi)(\g_i) = R_{\Z}(w_k).
\end{align}

From this result, we see that there are the two equivalent ways of viewing the twisting of the monodromy map $S_{\Z}$ by the $\cCt$ basis change for $\phi\in \Out_{\rfl}(W)$: we can view it as twisting $\psi$ by the pullback of $\phi^{-1}$, or we can view it as twisting symplectic $\Z$-reflection rep $R_{\Z}$.
If we take the view that $\psi \mapsto \f^{-1} \circ \ps$, then we can understand our ability to twist the monodromy map by a reflection outer automorphism as a consequence of the fact that the short exact sequence \eqref{orb 3} -- which is defined by the surjection $\psi$ -- is not unique, but is only determined up to composition by an automorphism in $\Out_{\rfl}(W)$.
Alternatively, if we view it as twisting $R_{\Z}$, then the ability to twist $S_{\Z}$ is a consequence of $R_{\Z}$ only being determined up to composition by an automorphism in $\Out_{\rfl}(W)$.
Given our goal of classifying SK structures through $R_{\Z}$ representations, it is most natural for us to take the latter view, which we will do.  
We now we need to understand how SK structures differing by such twists are related, i.e. when are they equivalent?
We turn to answering this question in a subsequent section after we first discuss the general notion of twisting a group representation by an automorphism.

\subsubsection{Interlude: Automorphism twists of group representations.}\label{appA: auto twist reps} 

Let us characterize the affect of a reflection preserving outer automorphism on the various reflection representations of the Weyl group that play a role in defining an $\cN{=}4$ SK structure.
This consists of the real, $R_{\R}$, complex $R_{\C}$, and integer $R_{\Z}$ reflection representations, which are related by 
\begin{align}\label{reflection reps}
    R_{\C} \cong_{\R} R_{\R} \oplus R_{\R} \cong_{\Z} \R \otimes_{\Z} R_{\Z}.
\end{align}
The complex reflection representation $R_{\C}$ acts on $\til \cC$, and the symplectic integral reflection representation $R_{\Z}$ acts on the pav fiber $1$-homology lattice $\L$.
We start by reviewing how automorphisms of a finite group $G$ act on $G$-representations more generally.

Given two representations $\r_i: G \to \GL(n,\cR)$ of a finite group $G$ over a ring $\cR$, we say they are $\cR$-equivalent if there exists an $\cR$-linear map $B \in \GL(n,\cR)$ such that 
\begin{align}\label{R equiv}
B\r_1(g)B^{-1} = \r_2(g)
\end{align} 
for all $g \in G$.
The more general notion of \emph{twisted equivalence} comes about naturally when asking about the action of an automorphism group of $G$ on its space of representations of fixed dimension over some ring, $\cR$. 
In general, given an $\cR$-representation $\r: G \to \GL(n,\cR)$ of a finite group $G$ over a ring $\cR$, composing it with an automorphism $\s \in \Aut(G)$ will yield another $\cR$-representation of dimension $n$, so automorphisms of $G$ permute the $n$-dimensional $\cR$-representations of $G$.
Thus, it is natural to ask when two such representations $\r_1, \r_2$ are related by an automorphism up to $\cR$-equivalence, i.e., when does there exist $\phi\in \Aut(G)$ and $B \in \GL(n,\cR)$ such that 
\begin{align}\label{G-rep twisted equiv}
B\r_1(g)B^{-1} = \r_2(\s(g))
\end{align}
for all $g \in G$?%
\footnote{Composing a $G$-representation $\r: G \to \GL(n,\cR)$ by an automorphism of $G$ will always yield another representation in $\GL(n,\cR)$, so there will necessarily exist some other representation $\r'$ for which the twisted equivalence relation is true.}
When the automorphism $\s = \mathbf{1}_G$ is the identity, this reduces to the usual notion of $\cR$-equivalence between representations, and we continue to define $\r_1$ and $\r_2$ to be $\cR$-equivalent in this scenario.
Twisted $\cR$-equivalence also reduces to $\cR$-equivalence when $\s \in \Inn(G) \subseteq \Aut(G)$ is an inner automorphism of $G$.
This is because an inner automorphism corresponds to conjugation by some element $s \in G$, $\s(g) = s g s^{-1}$, so the twisted equivalence relation becomes
\begin{align}\label{inner twisted equiv}
B\r_1(g)B^{-1} = \r_2(\s(g)) = \r_2(s g s^{-1}) = \r_2(s)\r_2(g)\r_2(s)^{-1} \,\, \Rightarrow \,\, B_s \r_1(g) B_s^{-1} = \r_2(g)
\end{align}
where we've defined $B_s \deq \r_2(s)^{-1} B \in \GL(n,R)$ and used the fact that the $\r_i$ are group homomorphisms.
Thus, it is only interesting to ask about twisted $R$-equivalence between representations when the twist is implemented by an outer automorphism of $G$, $\s \in \Out(G) \deq \Aut(G)/\Inn(G)$.%
\footnote{
We will henceforth implicitly assume this is the case when considering twisted $\cR$-equivalence.
}
When this is true, we say $\r_1$ and $\r_2$ are \emph{twisted $R$-equivalent} representations.
Conversely, If there exist two $R$-representations $\r_1, \r_2$ that are not twisted equivalent for the \emph{identity} automorphism, then we say they are $R$-inequivalent.
If they are also not twisted equivalent for \emph{any} automorphism, (which is stronger than being $R$-inequivalent) we say they are twisted $R$-inequivalent.

Note that it is possible for $\r_1 = \r_2$ for some non-trivial $\s\in \Out(G)$. 
When this occurs, the set of matrices $\{\r_1(g)\}_{g \in G}$ are mapped back into themselves, so the outer automorphism $\s$ is implementing a permutation on this set that is induced by some $\cR$-basis transformation.
If $\r_1 = \r_2$ is also injective, so $G \cong \r_1(G) \subset \GL(n,\cR)$ as abstract groups, the twisted $\cR$-equivalence relation is saying that $\s$ lifts to an outer automorphism of the matrix subgroup formed by $\r_1(G)$ in $\GL(n,\cR)$.
Also, note that it is not necessarily true that for all representations of a given type, they will be twisted $\cR$-equivalent by some outer automorphism of $G$.
In general, this will depend on the type of group $G$ is and the type of representations one is considering.

If we restrict ourselves to faithful (injective) representations, then a sufficient condition for twisted equivalence is that the images $\r_1(G)$ and $\r_2(G)$ -- each of which is a matrix subgroup of $\GL(n,\cR)$ that is isomorphic to $G$ as an abstract group -- are conjugate to one another in $\GL(n,\cR)$: $B \r_1(G) B^{-1}$ for some $B \in \GL(n,\cR)$.
In other words, if we consider the sets of matrices $\{ \r_i(g)\}_{g \in G} $ for each $\r_i$, then the two sets are conjugated into one another by $B$.
This is sufficient because $B$ conjugates $\r_1(g)$ to some unique element $\r_2(h)$ for all $g \in G$, $B\r_1(g)B^{-1} = \r_2(h)$, and conjugation preserves the group relations defining each $\r_i(G)$ as a matrix subgroup that is isomorphic to $G$ by injectivity.
Thus, the induced mapping $\s: G \to G$ given by $g \mapsto \s(g) \deq h$ is an automorphism of $G$, so we obtain the twisted equivalence relation, $B \r_1(g) B^{-1} = \r_2(\s(g))$ for all $g\in G$.
Note that in general, there is no guarantee the automorphisms will preserve all properties of the representations involved, i.e. if they can be extended to automorphisms of the matrix subgroups that the $\Im\r_i$ form.

In our current setting, we are concerned with a finite reflection group $G$ over $\cR$ for $\cR = \Z, \R$ or $\C$.
This means we are identifying the elements of $G$ with their image in a faithful $\cR$-reflection representation $\r$ such that each element is generated by a set of so-called \emph{reflection generators} which are the elements of $G$ that map to reflections in $\GL(n,\cR)$ under $\r$.
We want this identification of reflections generators under $\r$ to be preserved under the composition of $\r$ with an automorphism of $G$, so we only consider the affect of the subgroup of reflection preserving automorphisms $\Aut_{\rfl}(G) \subset \Aut(G)$ which map reflections to reflections under composition. 
Thus, given a reflection representation $\r$ -- which is an injective representation such that the image of an element in $G$ is generated by reflections in $\GL(n,\cR)$ -- composing it with some $\f \in \Aut_{\rfl}(G)$ will yield another reflection representation $\r'$ up to a similarity transformation $B \in \cR$.

It is possible that there only exists a single $\cR$-reflection representation $\r$ that is unique up to $\cR$-equivalence.
When this is true, it means composing $\r$ with any reflection automorphism will be $\cR$-equivalent to $\r$, so there exists $B_{\phi} \in \GL(n,\cR)$ such that
\begin{align}
    B_{\phi} \r(g) B_{\phi}^{-1} = \r(\phi(g)) \quad \forall \, g \in G.
\end{align}
In words, this says that any reflection preserving \emph{outer} automorphism $\phi \in \Out_{\rfl}(G)$ has a corresponding $\cR$-linear transformation $B_{\phi} \in \GL(n,\cR)$ such that composing $\r$ with $\phi$ is equal to performing the basis change by $B_{\phi}$.
These outer automorphism basis changes fix the matrices of the reflection rep as a set, but they permute them among themselves in such a way that corresponds to an automorphism of $G$, the abstract subgroup they generate in $\GL(n,\cR)$.

At this point we have established that composing an $\cR$-reflection rep with a reflection preserving automorphism $\phi$ leads to a twisted $\cR$-equivalence with another $\cR$-reflection representation.
This statement holds equally well for the real reflection representation of a Weyl group, $W_{\R}: W(\fg) \to \GL(r,\R)$, the corresponding complex reflection rep, $W_{\C} \deq \C \otimes_{\R} W_{\R} \subset \GL(r,\C)$, and for the symplectic integer reflections reps $W_{\Z} \equiv S: W(\fg) \to \Sp_J(2r,\Z)$.
However, differences emerge when we ask about how many $R$-inequivalent reflection reps exist since the answer depends on the ring $\cR$. 
For finite complex reflection groups (CRGs), they can have multiple reflection representations over a subring $\cR\subseteq \C$ that are $R$-inequivalent depending on $\cR$.
In particular, this can be true when $\cR = \C$ (see \cite{marin2009automorphismscomplexreflectiongroups}), but it turns out that all $\C$-reflection representations of a finite, irreducible CRG are twisted equivalent for some reflection automorphism (see \cite{thiel2014reflreps}, corollary 2.56).%
\footnote{Recall that for finite groups, $\mathbb K$-equivalence among their linear representations over a field $\mathbb K$ is completely determined by their characters: two $\mathbb K$-linear representations are $\mathbb K$-equivalent if and only if their characters agree for all group elements.}
For Weyl groups with $\cR = \R$ or $\C$, the situation is most simple: there exists a unique reflection representation up to $\cR$-equivalence.
These are precisely the reflection representations $R_{\R}$ and $R_{\C} \deq \C \otimes R_{\R}$.
However, as we have shown in the classification results of this paper, the situation can be even more drastic when $R = \Z$ for Weyl groups: there can exist multiple \emph{twisted inequivalent} symplectic $\Z$-reflection reps.

Below, we will we consider the action of $W(\fg)$ reflection outer automorphisms on $\cN{=}4$ SK structures in terms of their corresponding SK isometries and derive the corresponding transformation of their $R_{\Z}$ representation. 
What emerges is the statement that the twisted $\Z$-equivalence relation is the minimally sufficient condition two SK structure $R_{\Z}$ reps must satisfy in order to be deemed equivalent.
In the next section established this statement in detail.

\subsubsection{Twisted $\Z$-equivalence implies SK structure equivalence.}\label{appA: SKS twist equiv}

We now put together many results from the previous sections to describe their consequences for the equivalence between $\cN{=}4$ SK structures that are related by a reflection outer automorphism of the Weyl group.
Throughout, we continue to be pedantic about notation: we will explicitly distinguish between the Weyl group as an abstract group, its symplectic $\mathbb F$-reflection representations and their images, and the integer representation of $\pi_1(\cC^*)$ given by the monodromy map $S_{\Z}$, and its image $S = \Im S_{\Z}$.
The integer representation $R_{\Z}$ of the Weyl group is related to $S_{\Z}$ through the group surjection $\psi$, $S_{\Z} \deq R_{\Z} \circ \psi$, which is defined in the short exact sequence \eqref{orb 3}.

For a given Weyl group $W(\fg)$, we will show,
\bo
\1 \emph{Any $\cN{=}4$ SK structure $(J, R_{\Z}, \bt)$ is isomorphic to the $\phi$-twisted SK structure $(J,  R_{\Z}\circ \f^{-1}, \bt)$ where $\phi \in \Out_{\rfl}(W)$ is a reflection preserving outer automorphism of $W(\fg)$.}
\eo
We prove this by showing the two SK structures are related through SK isometries, in particular, by twisting the complex action of the Weyl group on the CB cover $\cCt$ using the $\GL(r,\C)$ coordinate transformation corresponding to $\f$.

Consider an $\cN{=}4$ SK structure for Weyl group $W(\fg)$ labeled by the triple $(J, R_{\Z}, \bt)$ for some fixed $J$ and $\bt \in \Fix(R_{\Z})$. 
Through the group surjection $\ps: \pi_1(\cC^*) \to W(\fg)$ (which is defined by the short exact sequence in \eqref{orb 3}), there is a corresponding monodromy map $S_{\Z} \deq R_{\Z} \circ \ps: \pi_1(\cC^*) \to \Sp_{J^{\v}}(2r,\Z)$ that defines the monodromy transformations $S_{\Z}(\g)$ for linking monodromies $\g \in \pi_1(\cC^*)$ in the CB of this SK geometry.
The Weyl group acts on the CB covering space $\cCt$ according to $w \circ p \deq R_{\C}(w) \cdot p$ for $p \in \cCt$.%
\footnote{In using the $R_{\C}(w) \cdot p$ notation, we are not specifying whether the action is on coordinate components of $p$ using $R_{\C}^{\v}$ or on the choice of basis elements using $R_{\C}$.
Specifying this does not affect the argument.} Let $\f \in \Out_{\rfl}(W)$ with corresponding basis change $B_{\f}\in \GL(r,\C)$ that we implement on the covering space $\cCt$ of the CB. 
Since $\f$ is reflection preserving, it maps reflections of $W(\fg)$ to reflections, $w_i \mapsto \f(w_i) \deq w_j$ for some $i,j = 1, \cdots, r$ labeling the simple reflections.
Since it is outer, this mapping cannot be reproduced by conjugation in $W(\fg)$.
As was shown in appendix \ref{appA: SKS isometries}, the coordinate transformation $B_{\f}$ for $\f$ is a valid SK structure isometry that descends to a K\"ahler isometry of the CB orbifold $\cC$ and does affect the fiber $A_{\bt}$, so it will necessarily transform $(J,R_{\Z}, \bt)$ to an equivalent SK structure $(J, R_{\Z}, \bt)_{\f}$ which is in the same SK structure orbit, so will have the same self-duality group and conformal manifold.
The results of the previous appendices allow us to work out what this $\f$-twisted SK structure is.

In appendix \ref{appA: twist R_C action}, we showed that performing the $\f$-basis change $B_{\f}$ twists the complex reflection rep $R_{\C}(w) \mapsto B_{\f} R_{\C}(w) B_{\f}^{-1} = R_{\C}(\phi(w))$ for all $w\in W(\fg)$.
This says that $R_{\C}$ and $R_{\C}\circ \f$ are $\C$-equivalent as representations, equal as sets of matrices, are isomorphic as matrix subgroups of $\GL(r,\C)$ (i.e. $W_{\C} \deq \Im R_{\C} \cong \Im(R_{\C} \circ \f)$), and are isomorphic as abstract groups to $W(\fg)$.
So the mapping $R_{\C}(w_i) \mapsto R_{\C}(\f(w_i)) =R_{\C}(w_j)$ is a reflection group isomorphism.
As a consequence, the quotient geometry of the smooth CB orbifold $\cC^* = \cCt^*/ W(\fg)$ is invariant, but the action of $W(\fg)$ on $\cCt$ using $R_{\C}$ is now $\f$-twisted to $w_i \circ p = R_{\C}(\f(w_i))\cdot p$.
In appendix \ref{appA: mono trans twist}, we showed that the construction of the group surjection $\ps: \pi_1(\cC^*) \to W(\fg)$ utilizes the complex reflection action of $W(\fg)$.
After basis changing by $B_{\f}$, this leads to it being transformed by the pullback of the inverse twist,   
\begin{align}\label{twist psi}
\ps \overset{\phi}{\mapsto} \ps_{\f} \deq \ps^*\f^{-1} = \f^{-1} \circ \ps.
\end{align}
Consequently, if $w_i = \psi(\g_i)$ for the linking monodromy $\g_i \in \pi_1(\cC^*)$ of the discriminant component $\cD_i$ (corresponding to image of the $w_i$ reflection hyperplane in $\cCt$) and $\f^{-1}(w_i) = w_k$ for a simple reflection $w_k \in W(\fg)$, then $\psi_{\phi}(\g_i) = (\phi^{-1} \circ \psi)(\g_i) = \f^{-1}(w_i) = w_k$.
From the definition, $S_{\Z} \deq R_{\Z} \circ \ps$, the monodromy map is also $\f$-twisted,
\begin{align}
   S_{\Z} \overset{\phi}{\mapsto} S_{\Z}^{\phi} \deq R_{\Z} \circ \psi_{\phi} = R_{\Z} \circ \phi^{-1} \circ \psi = R_{\Z}^{\f} \circ \ps.
\end{align}
It satisfies $S_{\Z}^{\f}(\g_i) = (R_{\Z}^{\f} \circ \psi)(\g_i) = R_{\Z}(w_k)$ where we've defined the $\f$-twist of $R_{\Z}$ by $\f^{-1}$, $R_{\Z}^{\f} \deq R_{\Z}\circ \f^{-1}$.
This shows that implementing the SK isometry $B_{\f} \in \GL(r,\C)$ corresponding to $\f \in \Out_{\rfl}(W)$ results in $\f$-twisting the symplectic $\Z$-reflection representation of $W(\fg)$ by $\f^{-1}$,
\begin{align}\label{W Zrep twist}
    R_{\Z} \overset{\phi}{\mapsto} R_{\Z}^{\f} \deq R_{\Z}\circ \f^{-1}.
\end{align}
As we discussed in appendix \ref{appA: auto twist reps} (and in section \ref{sec SK twist Z equiv}), twisting a $\Z$-reflection representation $R_{\Z}$ of the Weyl group defines a new $\Z$-reflection representation $R_{\Z}^{\f}$.
As was the case between $R_{\C}$ and $R_{\C}\circ \f$, $R_{\Z}$ and $R_{\Z}^{\f}$ are equal as sets of matrices (i.e., $W_{\Z} \deq \Im R_{\Z} = W_{\Z}^{\f} \deq \Im R_{\Z}^{\f}$), are isomorphic as matrix subgroups of $\Sp_{J^{\v}}(2r,\Z)$ (i.e., $W_{\Z} \deq \Im R_{\Z} \cong \Im R_{\Z}^{\f}$), are both isomorphic as representations over $\R$ to $R_{\R} \oplus R_{\R}$, and are isomorphic to $W(\fg)$ as abstract groups.
This implies that the fixed point sets of their M\"obius action on $\scH_r$ are the same, since a point $\bt \in \Fix(R_{\Z})$ is fixed by all elements of $W_{\Z}$, which is equal to $W_{\Z}^{\f}$ as sets of matrices.
So, $\Fix(R_{\Z}) \overset{\f}{\mapsto}  \Fix(R_{\Z}^{\f}) = \Fix(R_{\Z})$ as a set, and moreover the $\f$-twisting fixes $\Fix(R_{\Z})$ pointwise, so $\bt \overset{\f}{\mapsto} \bt$ and the pav fiber of the SK geometry, $A_{\bt}$, is invariant. 
In terms of the $\cN{=}4$ SK structure triple $(J, R_{\Z}, \bt)$, after $\f$-twisting it transforms to
\begin{align}\label{SKS twist I}
    (J, R_{\Z}, \bt) \overset{\phi}{\mapsto} (J, R_{\Z}, \bt)_{\f} \deq (J, R_{\Z}^{\f} = R_{\Z}\circ \f^{-1}, \bt).
\end{align}
This is our desired result.

Let us extend this scenario further to make it most relevant for the SK structure classification goals of this paper in terms of the $R_{\Z}$ representations. 
We will add the following ingredients: assume that we have a finite set $\scI_{\Z}(\fg)\deq \{R_{\Z}, R_{\Z}', \cdots \}$ of the symplectic integral $\Z$-reflection representations of $W(\fg)$ such that they are all (untwisted) $\Z$-inequivalent.

As we discussed in appendix \ref{appA: auto twist reps}, twisting an element of $\scI_{\Z}$ by $\f \in \Out_{\rfl}(W_{\fg})$ (via composition) will map $\scI_{\Z}$ to itself up to conjugation by an integral intertwiner $M_{\f} \in \Sp_{J^{\v}}(2r,\Z)$. 
Relative to our current scenario, let us assume $R_{\Z}$ is twisted $\Z$-equivalent to $R'_{\Z}$ by $\f^{-1}$ for some $\Z$-intertwiner $M_{\f}\in \Sp_{J^\v}(2r,\Z)$: 
\begin{align}\label{Rz twist Rz'}
    R_{\Z}'(w) = M_{\f}R_{\Z}(\f^{-1}(w))M_{\f}^{-1} \quad \forall \, w \in W(\fg).
\end{align}
(This is equivalent to saying $R_{\Z}^{\f}$ is $\Z$-equivalent to $R'_{\Z}$ by $M_{\f}$ so $R_{\Z}^{\f} \mapsto M_{\f} R_{\Z}^{\f} M_{\f}^{-1} = R'_{\Z}$.)
Going back to the $\f$-twisted SK structure $(J, R_{\Z}^{\f}, \bt)$, let us now perform the basis change by $M_{\f}$ in the pav fiber homology lattice $\L_{\bt} \overset{M_{\f}}{\mapsto} M_{\f} \L_{M_{\f}\circ \bt}$ of the SK geometry to intertwine $R_{\Z}^{\f}$ with $R_{\Z}'$.
Unlike before, the lattice basis change will transform $\bt \overset{M_{\f}}{\mapsto} \bt^{\f} \deq M_{\f} \circ \bt \equiv \bt' \in \Fix(R_{\Z}')$ (and likewise for the fixed point sets, $\Fix(R_{\Z}) \overset{M_{\f}}{\mapsto} \Fix(M_{\f}R_{\Z}M_{\f}^{-1}) = \Fix(R_{\Z}')$) so the SK structure triple becomes,
\begin{align}\label{SKS twist II}
    (J, R_{\Z}, \bt) \overset{M_{\f} \circ \phi}{\longmapsto} (J, R_{\Z}' = M_{\f}R_{\Z}^{\f} M_{\f}^{-1}, \bt^{\f}).
\end{align}
Since this is an EM duality transformation, it is an SK structure isometry, and the resulting SK structure will be in the same SK orbit as the original, so we have an equivalence of SK structures
\begin{align}\label{SKS twist III}
    (J, R_{\Z}, \bt) \overset{M_{\f} \circ \f}{\cong} (J, R_{\Z}' = M_{\f}R_{\Z}^{\f} M_{\f}^{-1}, \bt^{\f}).
\end{align}

This result holds for arbitrary $\f \in \Out_{\rfl}(W_{\fg})$.
So if our initial symplectic $\Z$-reflection rep $R_{\Z}$ is twisted $\Z$-equivalent by $\ps \in \Out_{\rfl}(W_{\fg})$ to another $\til R_{\Z}$ for a $\Z$-interwiner $M_{\ps}\in \Sp_{J^{\v}}(2r,\Z)$, 
\begin{align}\label{SK twist IV}
    \til R_{\Z} = M_{\ps}(R_{\Z}\circ\ps)M_{\ps}^{-1},
\end{align}
then the previous result indicates that we can transform the SK structure $(J, R_{\Z}, \bt)$ to the SK structure with the twisted $\Z$-equivalent rep $\til R_{\Z}$, $(J, \til R_{\Z}, \til \bt)$ for $\til \bt \deq M_{\ps}\circ \bt \in \Fix(\til R_{\Z})$, by performing SK isometries.
In particular, we do this by first performing the inverse basis change $B_{\ps}^{-1} \in \GL(r,\C)$ for the $\ps$-twist in $\cCt$, which transforms $R_{\Z} \overset{\ps^{-1}}{\mapsto} R_{\Z}\circ \ps$, and then we perform the lattice basis change by $M_{\ps}$, $\L_{\bt} \overset{M_{\ps}}{\mapsto} M_{\ps} \L_{M_{\ps}\circ \bt}$, which maps $R_{\Z} \overset{M_{\ps} \circ \ps^{-1}}{\mapsto} \til R_{\Z}$ and $\bt \overset{M_{\ps}}{\mapsto} \til \bt$.
This implies the SK structure defined by $R_{\Z}$ is equivalent to all the SK structures that are defined by the symplectic $\Z$-reflection reps $R_{\Z}$ is twisted $\Z$-equivalent to.
In other words, we have established the following corollary of the previous result using the twisted equivalence relation:
\bo
\1 \emph{Two $W(\fg)$ principally polarized SK structures determined by the symplectic $\Z$-reps, $R_{\Z}$ and $R_{\Z}'$, of $W(\fg)$ are isomorphic if they satisfy the \emph{twisted equivalence relation}: for any \emph{reflection automorphism} of $W(\fg)$, $\f \in \Aut_\rfl(W)$, there exists an $M_{\f} \in \Sp_{J^\v}(2r,\Z)$ such that
\begin{align}\label{twist Zequiv}
    R_{\Z}'(\f\circ w) M_{\f} = M_{\f} R_{\Z}(w)
\end{align}
for all $w\in W$.}
\eo
Thus, we have finally come full circle in understanding the origin of the twisted equivalence relation, and why it is the minimally sufficient condition that the symplectic $\Z$-reflection representations defining $\cN{=}4$ SK structures have to satisfy in order for them to be isomorphic and in the same SK orbit.

\subsubsection{Summary.}\label{appA: SK summary}

We end by telling the story of this appendix \ref{appA: equivalenceofN=4SKstructures} through a summary of its results and their consequences for the central goal of this paper: to classify the $\cN{=}4$ SK structures for a Weyl group $W(\fg)$ through the symplectic $\Z$-reflection representations of $W(\fg)$.\\

\noindent \tbf{Origins of twisted $\Z$-equivalence.}\label{appA: twist equiv origins} In appendix \ref{appA: Mono equiv Weyl} we showed that an $\cN{=}4$ SK structure is determined by the triple $(J, R_{\Z}, \bt)$ consisting of a polarization $J$ (lattice symplectic pairing on pav fiber $A_{\bt}$), symplectic $\Z$-reflection rep $R_{\Z}$ of the Weyl group $W(\fg)$, and an element $\bt \in \Fix(R_{\Z})\subset \scH_r$ in the Siegel-upper half-space representing the EM couplings between (probe) BPS states and the low-energy $U(1)^r$ gauge theory on the CB.
This result followed from the important fact that the monodromy map $S_{\Z}: \pi_1(\cC^*) \to \Sp_{J^\v}(2r,\Z)$ of the SK structure always takes the form $S_{\Z} \deq R_{\Z} \circ \ps$ for some $R_{\Z}$ of $W(\fg)$ where $\psi: \pi_1(\cC^*) \to W_{\C}(\fg)$ is the group surjection (defined in terms of the short exact sequences \eqref{orb 3}) that relates the fundamental group $\pi_1(\cC^*)$ of the smooth CB to the orbifold group $W(\fg)$ as abstract groups.
Thus, it is determined by $R_{\Z}$ and $\ps$, which means the monodromy transformations of all SK structure data is determined and their values can be fixed accordingly.
This conclusion was stated formally in appendix \ref{appA: SKS and orbit triple}.

We showed in appendix \ref{appA: SKS isometries} that all reflection automorphisms of the CB orbifold group, $W(\fg)$, extend to isometries/equivalences of its SK structure.
As a consequence, two SK structures related by such automorphisms are equivalent.
We also explained that, once combined with the EM duality transformations of $\Sp_{J^\v}(2r,\Z)$, such reflection automorphism induced isometries constitute all possible non-trivial $\cN{=}4$ SK structure isometries.
Hence, to classify $\cN{=}4$ SK structures we have to understand how these reflection automorphisms act on their defining data in the form of the triple $(J, R_{\Z}, \bt)$, in particular, for fixed $J$ and $W(\fg)$, this means understanding how such SK isometries affect $R_{\Z}$.
We subsequently established that these reflection automorphism SK isometries transform $R_{\Z}$ to another symplectic $\Z$-reflection rep of $W(\fg)$, $R'_{\Z}$, that is \emph{twisted $\Z$-equivalent} to $R_{\Z}$, and moreover that this condition is the minimally sufficient condition that must be satisfied for their corresponding SK structure to be equivalent.

EM duality transformations are by definition the elements of the EM duality group $M \in \Sp_{J^\v}(2r,\Z)$ that constitute the symplectic basis changes of the $1$-homology lattice $\L_{\bt} \deq H^1(A_{\bt}, \Z)$ of the pav fiber, $A_{\bt} \sim \C^r / \L_{\bt}$.%
\footnote{Symplectic means they preserve the polarization $J^{\v}$ after going to a symplectic basis of $\L_{\bt}$ with respect to $J$.
We make this choice because the EM duality group and the low-energy theory on the CB are canonically defined with respect to such a basis. However, it still makes mathematical sense to talk about the SK geometry when we are not in such a basis, in which case the elements of $\GL(2r,\Z)$ generate the analogous SK isometries that are due to lattice basis changes of the pav fiber.}
As a result, they simply transform the representation $R_{\Z}$ to its image $R'_{\Z}$ under the $\Z$-intertwiner $M$, $R_{\Z} M = M R'_{\Z}$, which implies $R_{\Z}$ and $R'_{\Z}$ satisfy the usual notion of untwisted $\Z$-equivalence. 
Inner automorphisms of $W(\fg)$ by some $v \in W(\fg)$ are reflection preserving, and they correspond to this usual notion of $\Z$-equivalence between $\Z$-reflections reps $R_{\Z}$ and $R'_{\Z}$ with $\Z$-intertwiner $M_v \deq R_{\Z}(v) \in \Sp_{J^\v}(2r,\Z)$ such that $R_{\Z} M_v = M_v R'_{\Z}$.
In terms of SK isometries, $M_v$ is a symplectic basis changes in the pav fiber lattice, $\L_{\bt}$, which necessarily transforms $R_{\Z}$ to $R'_{\Z}$.
So in the end, inner automorphisms of $W(\fg)$ form a subgroup of EM duality group and do not representation an SK isometry that is distinct from an EM duality transformation.
This leads to the $\Z$-equivalence criterion between SK structures: two SK structures whose defining $R_{\Z}$ reps are related by untwisted $\Z$-equivalence are isomorphic.
Thus, a necessary step in classifying $\cN{=}4$ SK structures is to classify the untwisted $\Z$-equivalences between their defining symplectic $\Z$-reflection reps, $R_{\Z}$.

We also stated in appendix \ref{appA: SKS isometries} that reflection outer automorphisms $\f \in \Out_{\rfl}(W)$ of $W(\fg)$ do not correspond to EM duality transformations of the fiber lattice.
By contrast, they are implemented by a basis change $B_{\f} \in \GL(r,\C)$ in the CB cover $\cCt$ that results in the complex reflection representation $R_{\C}$ of $W(\fg)$ being twisted by $\f$, $R_{\C} \overset{\f}{\mapsto} B_{\f}R_{\C}B_{\f}^{-1} = R_{\C} \circ \f$.
This twists the action of $W(\fg)$ on $\cCt$ defined using $R_{\C}$ such that, geometrically, it changed the reflection hyperplane that a simple reflection fixed before the $\f$ basis change.
In particular, this meant the hyperplane fixed by a simple reflection $w_i \in W(\fg)$ before $\f$-twisting was now fixed by the simple reflection $w_k \deq \f^{-1} \in W(\fg)$ after $\f$-twisting.
We then wanted to understand how this $\f$-twisting shift of reflection hyperplanes propagated through the SK geometry to affect the monodromy map $S_{\Z}$ and the monodromy transformations it defines through the symplectic $\Z$-reflection rep $R_{\Z}$.
We worked this out in appendices \ref{appA: Mono trans Weyl grp} and \ref{appA: mono trans twist} and found that $R_{\Z}$ was also $\f$-twisted.%
\footnote{The induced affect on $R_{\Z}$ and $S_{\Z}$ from a reflection outer auto SK isometry is already encoded in the results of appendix \ref{appA: Mono equiv Weyl}.
In the appendices \ref{appA: Mono trans Weyl grp}, \ref{appA: twist R_C action}, and \ref{appA: mono trans twist}, we more thoroughly discussed this encoding.}
In particular, by $\f$-twist we mean $R_{\Z}$ gets transformed to the pullback of $\f^{-1}$ by $R_{\Z}$, $R_{\Z} \overset{\f}{\mapsto} R_{\Z}^{\f} \deq R_{\Z} \circ \f^{-1}$. 
The induced transformation of the monodromy map is also a $\f$-twisting, $S_{\Z} \deq R_{\Z} \circ \ps \overset{\f}{\mapsto} R_{\Z}^{\f} \circ \ps$, which doesn't correspond to a basis change in $\L_{\bt}$ that would yield a $\Z$-equivalent rep of $\pi_1(\cC^*)$.
Overall, we see that the affect of a reflection outer automorphism SK isometry by $\f\in \Out_{\rfl}(W)$ is to $\f$-twist various representations of $W(\fg)$ and $\pi_1(\cC^*)$ via composition by $\f$ or $\f^{-1}$.
Because of this, in appendix \ref{appA: auto twist reps}, we provided a general discussion of twisting group representations by group automorphisms to better understand the origin and affect of this canonical action on representations.

The SK isometries corresponding to $\Out_{\rfl}(W)$ are distinct from EM duality transformations, so the most general SK isometry we can perform is one that combines and element of the two. 
If we consider first performing a $\f$-twist by $\f \in \Out_{\rfl}(W)$ followed by an EM duality transformation for some $\Z$-intertwiner $M\in \Sp_{J^\v}(2r,\Z)$, then $R_{\Z}$ and $S_{\Z}$ transform to
\begin{align}\label{Gen SK isometry}
    R_{\Z} \overset{M \circ \f}{\mapsto} M(R_{\Z}\circ \f^{-1})M^{-1} \Rightarrow S_{\Z} \overset{M \circ \f}{\mapsto} M(R_{\Z}\circ \f^{-1}\circ \ps)M^{-1}.
\end{align}
The transformation of $R_{\Z}$ under this combination of SK isometries is the most general transformation of $R_{\Z}$ we can generate using any SK isometries. 
Since $R_{\Z}$ is the defining data of a $J$-symplectic $\cN{=}4$ SK structure $(J, R_{\Z}, \bt\in \Fix(R_{\Z})$, if any two such $\cN{=}4$ SK structures are equivalent, then their symplectic $\Z$-reflection reps are necessarily related using some combination of a twist by $\f \in \Out_{\rfl}(W)$ with an EM duality transformation $M\in \Sp_{J^\v}(3r,\Z)$.
In other words, they must be \emph{twisted $\Z$-equivalent.}
As a consequence, the classification of $J$-symplectic $\cN{=}4$ SK structures is complete once we classify their $R_{\Z}$ representations up to twisted $\Z$-equivalence, which is precisely what we do in this paper for principal $J$. \\

\noindent\tbf{Classifying $\cN{=}4$ SK geometries.}\label{appA: SK classification}
If $\f \in \Aut_\rfl(W)$ is any reflection automorphism of $W$, then there is an associated SK structure isometry, relating two physically indistinguishable CB geometries, but which relates two \emph{a priori} distinct SK structures,
\begin{align}
    (J,S,\bt) \cong_\f (J,S\circ\f,\bt) .
\end{align}
If $(J,S\circ\f,\bt) \cong_\Z (J',S',\bt')$, by combining it with the reflection automorphism isometries we get a potentially larger set of equivalences
\begin{align}\label{atZe}
    (J,S,\bt) \cong_{\Z,\f} (J',S',\bt') .
\end{align}
We call this whole set of possible equivalences of $\cN{=}4$ sYM CB geometries \emph{automor\-phism-twisted $\Z$-equivalences}, and they are defined more generally and formally in appendix \ref{app Z rep}.
If $\f\in \Inn(W)$ is an inner automorphism, then the SK structure isometry $\cC \cong_\f \cC'$ can be traded by a change of coordinates for a $\Z$-equivalence between two associated symplectic representations, so $\Inn(W)$ do not give additional equivalences.
But if $\f \in \Out_\rfl(W) \deq \Aut_\rfl(W) / \Inn(W)$, then the associated SK structure isometry gives a new equivalence of geometries whose associated symplectic representations may or may not be integrally equivalent.
(An explicit description of all these equivalences in the $W=BC_2$ case is given in section \ref{BC2example}.)

Thus, we have shown in this appendix that:
\begin{quote}
    \emph{The possible SK structures of an $\cN{=}4$ sYM CB $\cC_{\fg}$, for simple gauge algebra $\fg$ and with Dirac pairing $J$, are in 1-to-1 correspondence with automor\-phism-twisted $\Z$-equivalence classes of pairs $(S,\bt)$ with $S$ an integral $J^\v$-symplectic representation of the orbifold group satisfying \eqref{orbi ref Ra}, and $\bt\in\Fix(S)$.}
\end{quote}

Finally, we mentioned earlier that the central charge condition \eqref{Z cond} is an additional condition that an algebraic integrable system must satisfy for it to describe a physical SK structure.
This condition requires that the monodromy $S(\g)$ for loops linking $\cD$ fixes integral linear combinations of special coordinates that must vanish there.
But this is automatic for $\cN{=}4$ sYM orbifold geometries, because the integral symplectic representation $W_\Z$ acts on the $2r$ component vector $(a^D\ a)$ of (dual) special coordinates, and $\cD$ is the fixed-point locus of this action.
Thus $\cD$ is the union of the eigenvalue $1$ eigenspaces of $W_\Z$, and since $W_\Z$ is integral, these eigenspaces are spanned by integral combinations of the (dual) special coordinates.


\section{Integer and rational representations}
\label{app Z rep}

\subsection{Basic definitions}

We first introduce the usual notions of finite group representations over rings, and equivalence. In what follows, let $W$ be a \emph{finite} group and $R$ be a commutative ring (think of $\Z$ or $\Q$).

\paragraph{Definition ($R$-representation). }

An \emph{$r$-dimensional $R$-representation} of $W$ is a group morphism $\r: W \to \GL(r,R)$. 

\paragraph{Definition (Reducible and decomposable representations). }

An $R$-representation $\r$ is \emph{reducible}, respectively \emph{decomposable}, if there exists a matrix $B \in \GL(r,R)$ such that for all $w \in W$, $B \r(w) B^{-1}$ is block triangular, respectively block diagonal, with at least two non-trivial blocks. 

\paragraph{Definition ($R'$-equivalence). }

Let $R' \subseteq R$ be a subring of $R$. 
Two $R$-representations $\r,\r'$ of $W$ are called \emph{$R'$-equivalent}, denoted $\r \sim_{R'} \r'$, if there is a matrix $B \in \GL(r,R')$ such that for all $w \in W$, $B \r'(w) = \r(w) B$. 

\medskip 

\noindent We also twist the above definition of equivalence using automorphisms of the group $W$. 

\paragraph{Definition (Automorphism-twisted $R'$-equivalence). }

Let $\Phi\subseteq\Aut(W)$ be a subgroup of the group of automorphisms of $W$. 
Two $R$-representations $\r,\r'$ of $W$ are called \emph{$(R',\Phi)$-equivalent}, denoted $\r \sim_{(R',\Phi)} \r'$, if there is a matrix $B \in \GL(r,R')$, and an automorphism $\f \in \Phi$, such that for all $w\in W$, $B \r'(w) = \r(\f(w)) B$. 

\subsection{Indecomposable reducible representations over $\Z$ and $\Q$}

We repeatedly use the two following results in the bulk of the paper. 

\paragraph{Theorem \cite[Th. 73.5]{curtis1966representation}. } 
Every $r$-dimensional $\Q$-representation of $W$ is $\Q$-equivalent to an $r$-dimensional $\Z$-representation of $W$. 

\paragraph{Theorem \cite[Th. 73.12]{curtis1966representation}. } 

Let $\rho$ be a reducible finite-dimensional $\Q$-representation of $W$ of the form 
\begin{align}
    \r &= \bspm\r_1 &*\\0&\r_2\espm
\end{align}
with $\r_{1,2}$ some $\Q$-representations of $W$. 
Then there exist $\Z$-representations $\til\r_{1,2}$ of $W$ such that 
\begin{align}
  \til\r_i  &\sim_\Q \r_i & &\textrm{and}&
  \r &\sim_\Z \bspm\r_1&*\\0&\r_2\espm.
\end{align}

\medskip

These follow from Burnside's argument, which we reproduce here since it is key to the arguments of this paper.

\paragraph{Lemma 1 (Burnside \cite{Burnside:1908}). } 

\emph{Let $G$ be a finite subgroup of $\GL(n,\Q)$. Then there exists a matrix of change of basis $A \in \GL(n,\Q)$ such that $A^{-1} \cdot G \cdot A \subset \GL(n,\Z)$. }

\paragraph{Proof. }

Let $g_k \in \GL(n,\Q) $ be generators of $G$,  indexed by $k \in K$, where $K$ is a finite set. Let $d$ be the greatest common divisor of all the entries of all the matrices $g_k$. The matrices $d g_k$ have only integer entries. Define the lattice 
\begin{equation}
I = \left\{ (x_1 , \dots , x_n) \in \R^n 
\ \Big| \  \forall k \in K \, , \quad 
g_k \cdot \bspm x_1 \\ \vdots \\ x_n \espm 
\in \Z^n \right\} \, . 
\end{equation}

Define, for $i=1, \dots , n$, 
\begin{equation}
  a_{i,i} = \mathrm{min} \left\{ x_i \in \R_{>0} \ \big|\ \exists (x_1 , \dots , x_{i-1}) \in \R^{i-1}  \, , \, (x_1 , \dots , x_i , 0 , \dots , 0) \in I \right\} \, .   
\end{equation} 
From the construction of $a_{i,i}$, there exists $(a_{i,1} , \dots , a_{i,i-1}) \in \R^{i-1}$ such that 
\begin{equation}\label{eq:aiInI}
\mathbf{a}_i := (a_{i,1} , \dots , a_{i,i} , 0 , \dots , 0) \in I \, . 
\end{equation}
Define then the matrix 
\begin{equation}
A = \bpm
a_{1,1} & a_{2,1} & \cdots & a_{n,1} \\
0 & a_{2,2} & \cdots & a_{n,2} \\
\vdots & \ddots  & \ddots & \vdots \\
0 & \cdots  &   0 & a_{n,n}\\
\epm \, . 
\end{equation}

The matrix $A$ has the following crucial property: 
\begin{equation}\label{eq:equivalence}
X \in \Z^n \quad \Leftrightarrow \quad A \cdot X \in I \, . 
\end{equation}
The direct implication immediately follows from the lattice structure of $I$. Consider then 
\begin{equation}\label{eq:xX}
x = A \cdot X \in I \, . 
\end{equation}
We want to show that $X \in \Z^n$, and we use induction on the components $X_i$ of $X$, starting from the last entry $X_n$. The $n$-th entry of equation \eqref{eq:xX} gives $x_n = a_{n,n} X_n$, and by definition of $a_{n,n}$, we have 
\begin{equation}
\frac{x_n}{a_{n,n}} = X_n \in \Z \, . 
\end{equation}
Assume now that $X_n , \dots , X_{r+1} \in \Z$. Then
\begin{equation}
x - X_{r+1} \mathbf{a}_{r+1} - \dots - X_n \mathbf{a}_n \in I
\end{equation}
since the $\mathbf{a}_i \in I$ using \eqref{eq:aiInI}. Moreover only the first $r$ entries of that vector are non vanishing, so by definition of $a_{r,r}$ we have 
\begin{equation}
\frac{x_r - X_{r+1} a_{r+1,r} - \dots - X_n a_{n,r}}{a_{r,r}} \in \Z \, . 
\end{equation}
This concludes the proof of \eqref{eq:equivalence}, and as a direct consequence,  $A^{-1} \cdot g_k \cdot A $ has integer coefficients, which proves the lemma. 

\paragraph{Lemma 2. } 

\emph{Using the notations of the previous lemma, if every $g \in G$ has a block diagonal form with blocks of sizes $n_1 , \dots , n_r$ such that $n_1 + \dots + n_r = n$, then so does the matrix $A$.  }

\paragraph{Proof. }

Assume the block decomposition is given by $n = n_1 + n_2$. The set $I$ then splits into $I = I_1 \oplus I_2$. We can then apply the proof of the previous lemma independently on $I_1$ and $I_2$, constructing matrices $A_1$ and $A_2$. The final matrix $A$ has $A_1$ and $A_2$ as its two blocks. 

\subsection{Symplectic representations}

Let $J$ be a non-degenerate antisymmetric $2r \times 2r$ matrix with integer coefficients. 
The skew normal form theorem asserts that there exists an integer matrix $U$ with determinant 1 such that 
\begin{equation}
    J = U^t \bpm 0&\d\\-\d&0\epm U 
\end{equation}
with $\d = \mathrm{diag} (\d_1, \dots, \d_r)$ and the $\d_i$ are positive integers such that $\d_i | \d_{i+1}$. 
These integers are uniquely determined by the matrix $J$. 
We say that $J$ is principally polarized if $\d_1 = \dots = \d_r = 1$. 
The symplectic group $\Sp_J(2r,\Z)$ is the group of $2r \times 2r$ matrices $M$ such that $M^t J M = J$, and we call $J$ its \emph{symplectic form}.

\paragraph{Theorem. } 

Let $W$ be a Weyl group of rank $r$. 
Let $\cR : W \to \Sp_J(2r, \Z)$ be a group morphism such that $\cR \simeq_\Q R_1 \oplus R_2$ with $R_i : W \to \GL(r,\Z)$ (for $i=1,2$) two representations which are $\Q$-equivalent to the standard reflection representation of $W$. 
Then there exists a matrix $P \in \GL(2r,\Z)$ such that 
\begin{equation}
\label{eq:SPrepReduced}
    P \cR(w) P^{-1} = \bpm S(w)&L(w)\\ 0&\d S^{-t}(w)\d^{-1}\epm
\end{equation}
for some representation $S: W \to \GL(r,\Z)$ and some matrix $L(w) \in \GL(r,\Z)$, and \eqref{eq:SPrepReduced} is a symplectic representation for the symplectic form $P^{-t} J P^{-1} = \bspm 0 & \d \\ -\d & 0 \espm$. 

\paragraph{Proof. }

\cite{curtis1966representation} defines a \emph{binding function} for the pair $(R_1, R_2)$ as a function $L: W \to \mathrm{Mat}(r,\Z)$ such that the mapping 
\begin{align}\label{binding rep}
\cR'&: W \to \GL(2r,\Z), &
\cR'(w) & = \bpm R_1(w) & L(w) \\ 0 & R_2(w) \epm ,
\end{align}
is a group morphism. 
This is equivalent to saying that for each $w, w' \in W$, we have $L(ww') = R_1(w) L(w') + L(w) R_2(w')$. 
The binding function $L$ is \emph{inner} if there exists a matrix $D \in \GL(r,\Z)$ such that for all $w \in W$, $L(w) = R_1(w) D - D R_2(w)$. 
Theorem 73.22 in \cite{curtis1966representation} asserts that $|W| L$ is an inner binding for any binding $L$.

This guarantees that there exists a matrix $D \in \GL(r,\Z)$ such that, if we define 
\begin{equation}
    L(w) :=\frac{1}{|W|} \left[R_1(w) D-D R_2(w)\right] \, , 
\end{equation}
then $L(w) \in \GL(r,\Z)$ for all $w \in W$, and $\cR$ is $\Z$-equivalent to the group representation \eqref{binding rep}.
Let $P\in \GL(2r,\Z)$ such that for all $w \in W$, $\cR(w) = P^{-1} \cR'(w) P$. 
Define 
\begin{equation}
J' = P^{-t} J P^{-1} = \bpm A & B\\-B^t & C\epm
\end{equation}
with $A^t = -A$ and $C^t = -C$. 
Then $\cR'(w) \in \Sp_{J'}(2r, \Z)$ for all $w \in W$.  
This implies 
\begin{eqnarray}\label{ABC eqns}
A &=& R_1^t(w) A R_1(w) , \nn\\
B &=& R_1^t(w) A L(w) + R_1^t(w) B R_2(w) , \\
C &=& L^t(w) A L(w) - R_2^t(w) B^t L(w) + L^t(w) B R_2(w) + R_2^t(w) C R_2(w) . \nn
\end{eqnarray}
We now use the fact that the $R_{i}$ are reflection representations of the real Weyl group $W$, so the matrices $R_{i}(w)$ are orthogonal. 
Intertwiners of orthogonal representations are symmetric so $A=0$, and the last equation in \eqref{ABC eqns} can be rewritten as 
\begin{equation}
    \left[ C + \frac{B^t D - D^t B}{|W|} \right] R_2^{-1}(w) = R_2^t(w) \left[ C + \frac{B^t D - D^t B}{|W|} \right] .
\end{equation}
This means the matrix in square brackets is symmetric, so $C = \frac{D^t B - B^t D}{|W|}$, and  we have found that 
\begin{equation}
    J' = \bpm 0 &B\\-B^t & \frac{D^t B -B^t D}{|W|} \epm \, . 
\end{equation}

Now let us put $B$ in Smith normal form, i.e., we write $B =  U^{t} \d V$ with $U,V$ integer matrices with determinant 1, and $\d$ integer and diagonal with each diagonal entry dividing the next one. 
Then define 
\begin{equation}
   J'' = \bpm U & 0\\0& V\epm^{-t} 
   J' \bpm U & 0\\0& V\epm^{-1} 
   = \bpm0 &\d\\-\d& \frac{(D'')^t \d -\d D''}{|W|} \epm
\end{equation}
\begin{equation}
   \cR''(w) = \bpm U&0\\0&V\epm \cR'(w) \bpm U&0 \\ 0&V \epm^{-1} 
   = \bpm R''_1(w) & L''(w) \\0 & R''_2(w)\epm
\end{equation}
with $R_1'' = U R_1 U^{-1}$, $R_2'' = V R_2 V^{-1}$, $D'' = U D V^{-1}$ and $L'':= \frac{1}{|W|} \left[ R''_1 D'' -D'' R''_2 \right] = U L V^{-1}$. 
In particular, the middle equation in \eqref{ABC eqns} becomes 
\begin{equation}
\label{eq:deltaR1R2}
    \d = (R_1'')^t(w) \, \d \, R''_2(w) \, . 
\end{equation}

We perform one more transformation. 
Write the lower right entry of $J''$ as 
\begin{equation}
    \frac{(D'')^t \d- \d D''}{|W|} = \D - \D^t \, , 
\end{equation}
with $\D$ some matrix with integer coefficients. 
Then we transform $J''$ and $\cR''(w)$ to
\begin{align}
    J''' &= \bpm 1&\d^{-1}\D^t\\0&1\epm^{-t} J''
    \bpm1&\d^{-1}\D^t\\0&1\epm^{-1} 
    = \bpm0&\d\\-\d&0\epm ,\\
    \cR'''(w) &= \bpm 1&\d^{-1}\D^t\\0&1\epm 
    \cR''(w) \bpm 1&\d^{-1}\D^t\\0&1\epm^{-1} 
    = \bpm R'''_1 (w) & L'''(w)\\0 & R'''_2(w) \epm .\nn
\end{align}
with $R'''_1= R''_1$, $R'''_2= R''_2$ and $D'''= D''-|W| \d^{-1} \D^t$. 
This finishes the proof, using \eqref{eq:deltaR1R2}. 

\paragraph{Corollary. } 

Assume $J$ is principally polarized. 
Any representation $\cR : W \to \Sp_{J^\v}(2r,\Z)$ of the rank $r$ Weyl group $W$ which is $\Q$-equivalent to two copies of the standard reflection representation is $\Z$-equivalent to a representation of the form 
\begin{equation}
\label{eq:SPrepReduced2}
    \bpm S (w) & L(w) \\ 0 & S^{-t} (w) \epm \, , \qquad 
    L(w) = \frac{1}{|W|} (S (w) D - D S^{-t} (w) ) \in \GL(r,\Z)
\end{equation}
for some representation $S : W \to \GL(r,\Z)$ and some matrix $D \in \GL(r,\Z)$. 

\section{Hecke groups and subgroups}
\label{app Hecke}

The level-$q$ Hecke groups for integer $q>2$ are
\begin{align}\label{ell2q}
    H_q &\deq \left\langle \sS_\ell, \sT \, \vert\, \sS_\ell^2 = (\sS_\ell \sT)^q = 1 \right\rangle, & &\text{where} & \ell &\deq 2[\cos(2\pi/q)+1] ,
\end{align}  
which act on $\t\in\scH_1$ as $\PSL(2,\R)$ M\"obius transformations%
\footnote{This presentation is equivalent to the ordinary definition of the Hecke groups \cite{Hecke:36}, with generators $\sS,\sT_\l$ acting as $\sS: \t'\to-1/\t'$, $\sT_\l: \t' \to \t' + \l$, where $\l \deq \sqrt\ell$, provided we rescale $\t' = \l \t$.} 
\begin{align}\label{H2q}
    \sS_\ell : \t &\mapsto -1/(\ell\t), &
    \sT : \t &\mapsto \t+1 .
\end{align}
A double cover in $\SL(2,\R)$ is
\begin{align}
    \sS_\ell &= \bspm 0&-1/\sqrt\ell\\ \sqrt\ell&0\espm, &
    \sT &= \bspm1&1\\0&1\espm \, ,
\end{align}
though in this case $\sS_\ell^2 = (\sS_\ell \sT)^q = -1$.
We often use the presentation as a $\SL(2,\R)$ subgroup rather than as a $\PSL(2,\R)$ one in what follows.

When $q=3,4,6$ then $\ell$ are the integers $\ell=1,2,3$, respectively.
Note that $H_3 = \PSL(2,\Z)$, while $H_4$ and $H_6$ are other inequivalent discrete subgroups of $\PSL(2,\R)$.
These groups and some of their finite index subgroups appear as S duality groups.
In particular, subgroups of $H_q$ appear as the S duality groups of theories whose gauge Lie algebra has lacing number 
$\ell(q)$ \cite{Dorey:1996hx, Argyres:2006qr, Aharony:2013hda}.

The proper subgroups which appear are the subgroups of $\SL(2,\Z)$
\begin{align}
    \D &\deq \left\{ \bspm a&b\\c&d\espm \in \SL(2,\Z) \, \Big|\, a+b+d = a+c+d = 0 \ (\rm{mod}\ 2)\right\} \, ,\nn\\
    \G_0(n) &\deq \left\{ \bspm a&b\\c&d\espm \in \SL(2,\Z) \, \Big|\, c = 0 \ (\rm{mod}\ n)\right\} \, .
\end{align}
Also, considered as a projective subgroup, $\G_0(\ell) \subset H_q$ is the subgroup generated by $\langle \sS_\ell \sT \sS_\ell, \sT \rangle$.
Their indices as subgroups of $H_q$ are computed for general $n$ in, for example, \cite{diamond2005first}.
The indices for most of the groups appearing in table \ref{tab:summary} are
\begin{equation}\label{tab:mod grp}
\begin{array}{r|cccc}
\text{group} & \ \D\ & \ \G_0(2)\ & \ \G_0(3)\ & \ \G_0(4)\ \\ \hline
\text{index in }\PSL(2,\Z) \simeq H_3 & 2 & 3 & 4 & 6 \\
\text{index in }H_4 & - & 2 & - & 4 \\
\text{index in }H_6 & - & - & 2 & - \\
\end{array} \ .
\end{equation}

\bibliographystyle{JHEP}
\bibliography{N4SKlines.bib}
\end{document}